\documentclass[11pt,english]{book}
\usepackage[T1]{fontenc}
\usepackage[latin1]{inputenc}
\usepackage{a4wide}
\usepackage{fancyhdr}
\usepackage{graphicx}
\usepackage{setspace}
\usepackage{amssymb}
\usepackage[authoryear]{natbib}

\makeatletter

\providecommand{\LyX}{L\kern-.1667em\lower.25em\hbox{Y}\kern-.125emX\@}
\newcommand{\lyxline}[1]{
  {#1 \vspace{1ex} \hrule width \columnwidth \vspace{1ex}}
}

 \usepackage{verbatim}
 \newenvironment{lyxlist}[1]
   {\begin{list}{}
     {\settowidth{\labelwidth}{#1}
      \setlength{\leftmargin}{\labelwidth}
      \addtolength{\leftmargin}{\labelsep}
      }}
   {\end{list}}

\usepackage{multicol}
\usepackage{sectsty}

\allsectionsfont{\sffamily}

\usepackage{babel}
\makeatother
\begin{document}
\thispagestyle{empty}

\begin{center}\textsf{\textbf{\LARGE Formation and Destruction of
Autocatalytic Sets}}\end{center}{\LARGE \par}

\begin{center}\textsf{\textbf{\LARGE in an Evolving Network Model}}\end{center}{\LARGE \par}

\vspace*{\fill}
\begin{center}\includegraphics[  width=7cm,
  keepaspectratio]{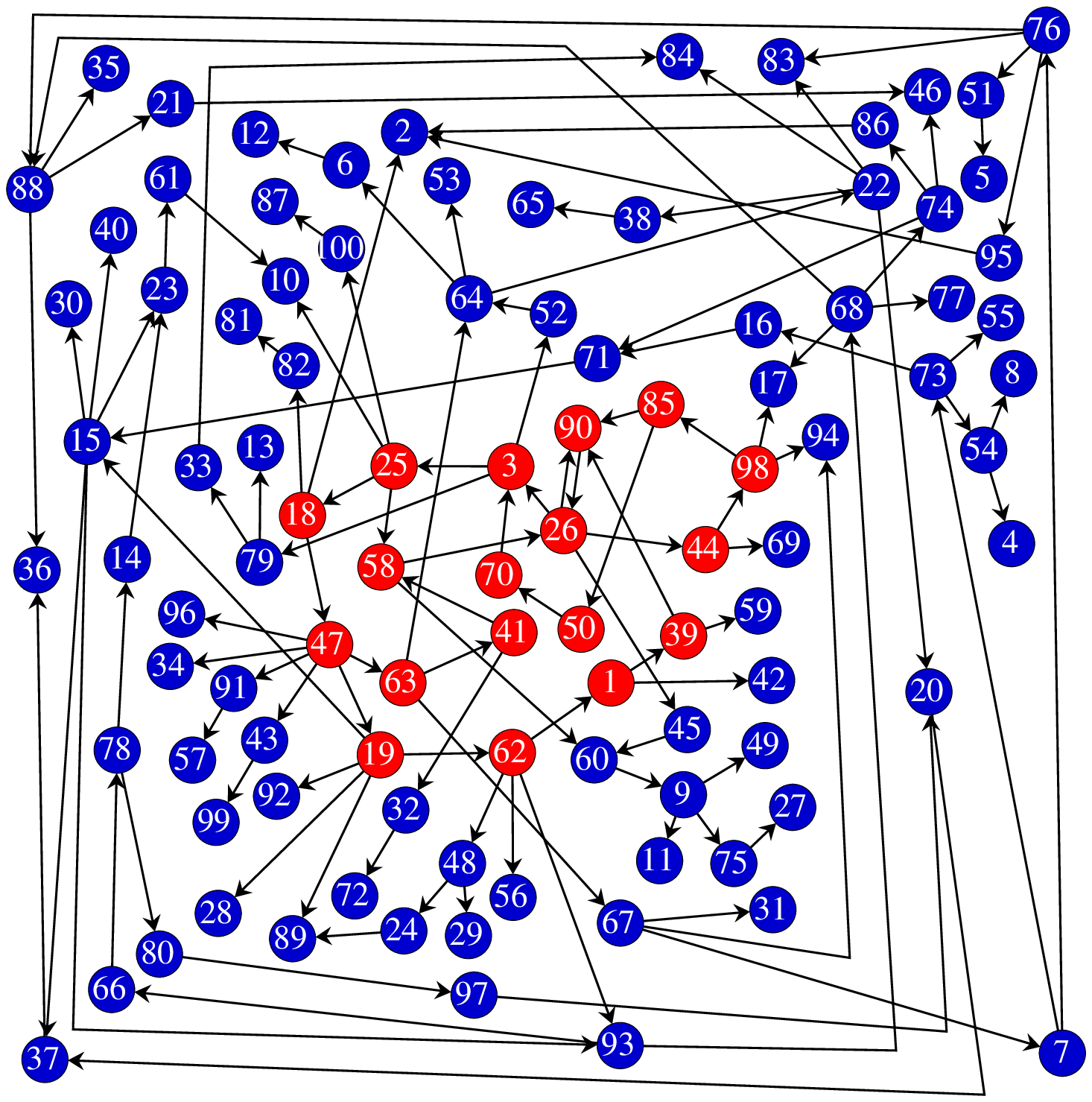}\end{center}

\bigskip{}
\begin{center}\textsf{A thesis submitted for the degree of}\\
\textsf{Doctor of Philosophy in the Faculty of Science}\end{center}

\vspace*{\fill}
\begin{center}\textsf{\textbf{\Large Sandeep Krishna}}\end{center}{\Large \par}

\begin{singlespace}
\vspace*{\fill}
\begin{center}\includegraphics[  scale=0.75]{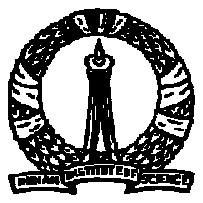}\textsf{\scriptsize }\\
\textsf{\scriptsize Centre for Theoretical Studies}\\
\textsf{\scriptsize Indian Institute of Science}\\
\textsf{\scriptsize Bangalore - 560012, India}\end{center}{\scriptsize \par}
\end{singlespace}

\begin{center}\textsf{\scriptsize August 2003}\end{center}{\scriptsize \par}
\newpage

~

\newpage\vspace*{\fill}
\begin{center}\textsf{\textbf{\huge Declaration}}\end{center}{\huge \par}
\bigskip{}

This thesis describes work done by me during my tenure as a PhD student
at the Centre for Theoretical Studies, Indian Institute of Science,
Bangalore. This thesis has not formed the basis for the award of any
degree, diploma, membership, associateship or similar title of any
university or institution.\\

\begin{flushleft}Sandeep Krishna\\
August 2003\end{flushleft}

\begin{flushleft}Centre for Theoretical Studies\\
Indian Institute of Science\\
Bangalore - 560012\\
India\end{flushleft}
\vfill{}\newpage

\tableofcontents{}

\chapter{\label{cha:Introduction-and-Summary}Introduction}

In this thesis I will analyze a model of an evolving set of catalytically
interacting molecules. The dynamical rules of the model attempt to
capture key features of the evolution of chemical organizations on
the prebiotic Earth. Such a set of molecules is an ideal example of
a `network' -- a system of several interconnected components. The
concept of a network is often used as a metaphor in describing a variety
of chemical, biological and social systems, as overviewed in section
\ref{sec:Networks-in-chemical,}. Graphs provide a natural way of
representing networks in a mathematical model. I describe how a graph
can be used to represent a network and the difficulties involved in
creating such a representation in sections \ref{sec:Graph-representation-of}
and \ref{sec:Difficulties-of-creating}. Sections \ref{sec:Structure-of-networks},
\ref{sec:Dynamical-systems-on} and \ref{sec:Dynamics-of-networks}
discuss several studies of network systems, broadly classified according
to whether they focus on the structure, function or evolution of networks.
This classification is not meant to be precise -- several studies
could be placed in more than one category. In analyzing the model
I will be mainly interested in the evolution of the graph representing
the chemical network. Some issues of interest are discussed in section
\ref{sec:Dynamics-of-networks}, including the spontaneous growth
of non-random graph structures, the effect of different graph structures
on the selective pressures that drive the evolution and sudden mass
extinctions of species. Sections \ref{sec:Framework-of-a} and \ref{sec:Extensions-of-the}
describe a framework for modeling an evolving network, within which
some of these issues can be addressed. The specific model I will analyze
will be an instance of this framework. Sections \ref{sec:The-origin-of}
and \ref{sec:Catastrophes-and-recoveries} mention some of the interesting
phenomena observed in the model and discuss its possible use in addressing
puzzles connected with the evolution of chemical networks on the prebiotic
Earth. Finally, I provide a `map' of subsequent chapters in section
\ref{sec:A-map-of}.

\section{\label{sec:Networks-in-chemical,}Networks in chemical, biological
and social systems}

The term `network' is often used to describe a variety of chemical,
biological and social systems: The metabolism of a cell is a network
of substrates and enzymes interacting via chemical reactions, ecosystems
are networks of biological organisms with predator-prey, competitive
or symbiotic interactions, the brain is a network of interconnected
neurons, and the Internet is a network of interconnected computers. 

Reviews of recent work on `networks' \citep{Watts,Strogatz2,Bose,AB,DM,DMbook,Handbook}
testify to the diversity of systems for which the term is used: 

\begin{figure}
\begin{center}\includegraphics[  width=16cm,
  keepaspectratio]{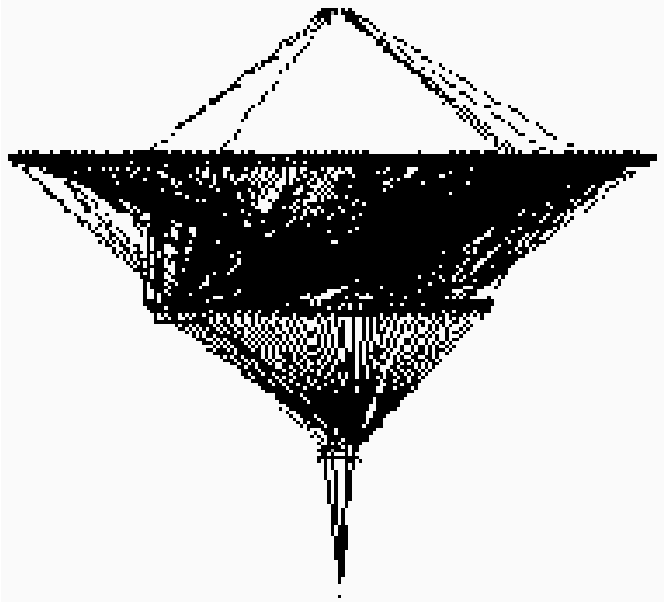}\end{center}

\caption{Ythan estuary food web. Each of the 135 nodes (circles) represents
a species and there are 601 directed links (arrows, pointing from
prey to predator). This data is freely available from the website
of the COSIN project (\emph{http://www.cosin.org}). \emph{}The graph
has been drawn using LEDA, the Library of Efficient Data types and
Algorithms, which is now a proprietary software distributed by Algorithmic
Solutions Software GmbH (\emph{http://www.algorithmic-solutions.com/enleda.htm}).
The vertical positioning of each node is decided by the length of
the shortest path leading to it from the basal species (the bottom
node).\label{cap:foodweb}}
\end{figure}

\begin{figure}
\begin{center}\includegraphics[  width=15cm,
  keepaspectratio]{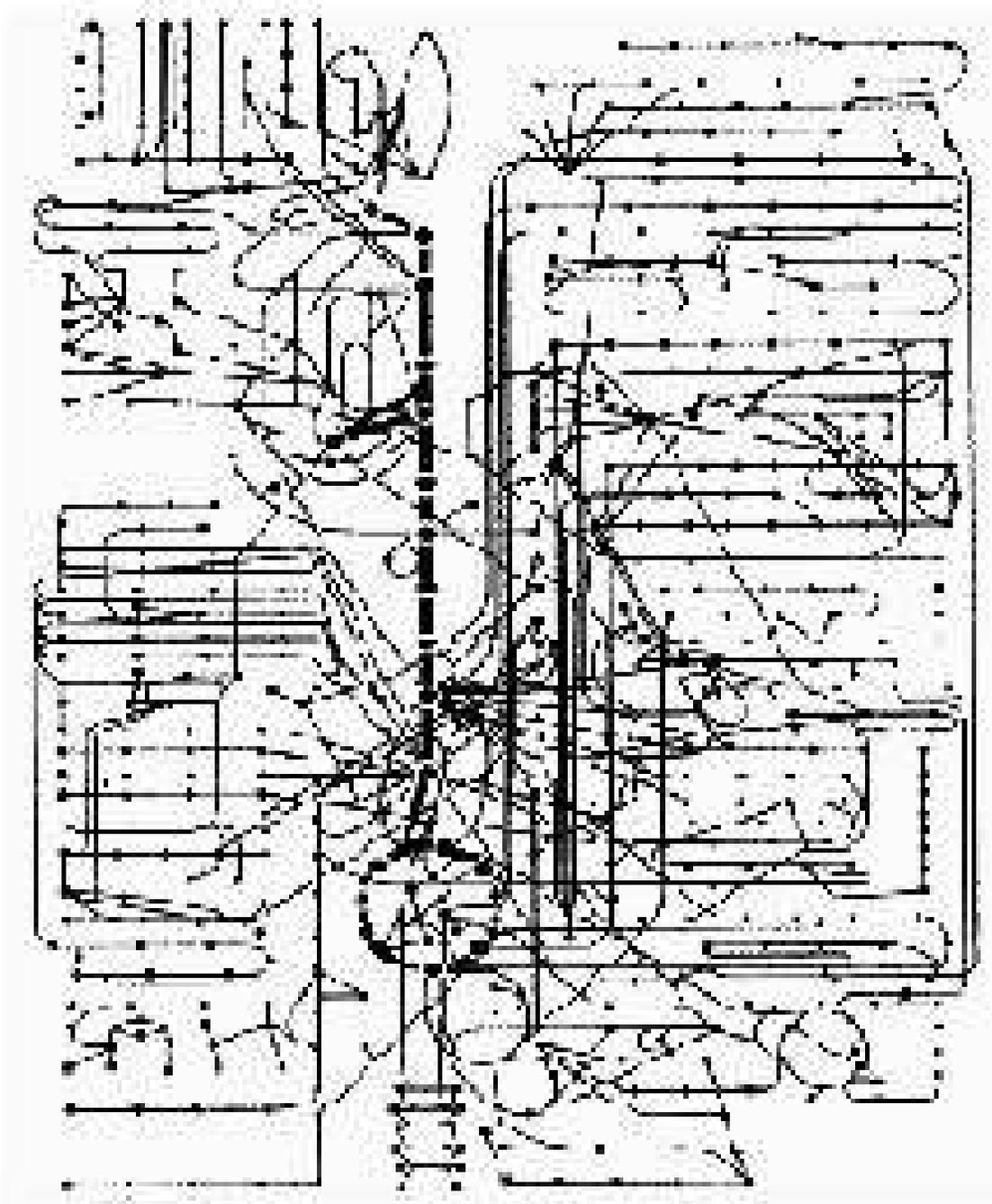}\end{center}

\caption{A graph of around 500 chemical reactions that take place in cells;
image by Dennis Bray, Cambridge Univ. Each node represents a chemical
species. Two species are linked if there is a reaction in which one
species is a reactant and the other a product.\label{cap:metabolism}}
\end{figure}

\begin{figure}
\begin{center}\includegraphics[  width=15cm,
  keepaspectratio]{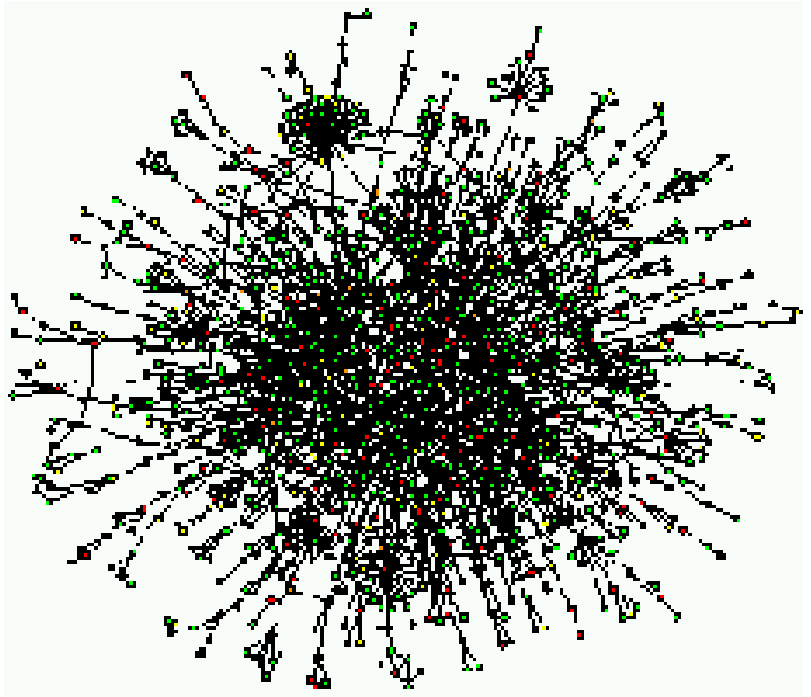}\end{center}

\caption{The protein interaction network for the yeast \emph{Saccharomyces
cerevisiae}. The nodes represent proteins. Pairs of proteins that
are known to interact are linked. There are 1870 nodes and 2240 links.
A node is coloured red if removing the corresponding protein is lethal
for the cell, green if non-lethal, orange if removing the protein
results in slower growth, and yellow if the effect of removing the
protein is not known. Image available at \emph{http://www.nd.edu/$\sim $networks/gallery.htm},
courtesy Hawoong Jeong, Korea Advanced Institute of Science and Technology.
\label{cap:protein}}
\end{figure}

\begin{figure}
\begin{center}\includegraphics[  width=15cm,
  height=18cm]{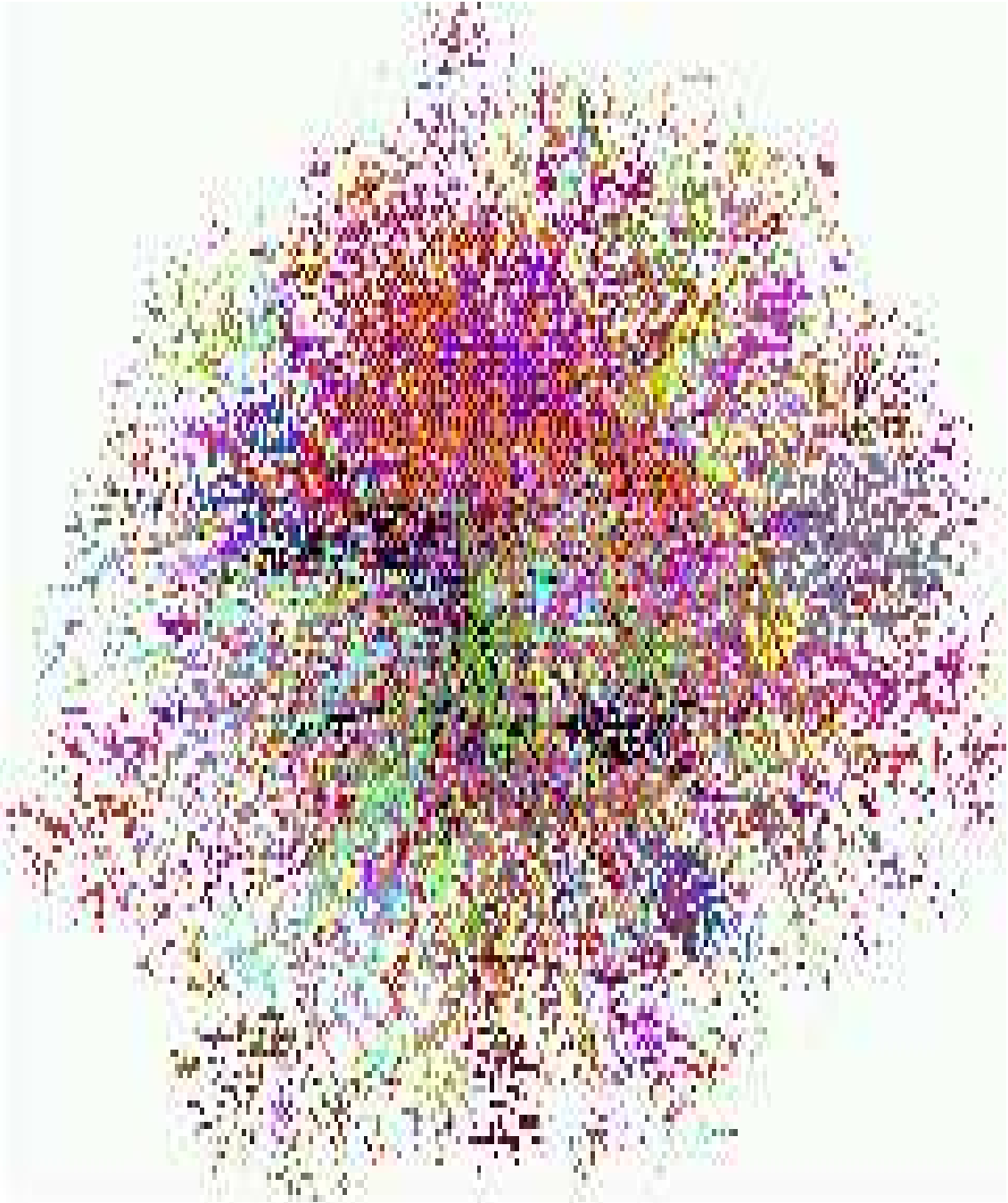}\end{center}

\caption{A graph of the Internet at the router level, as of September 1998.
Image by Hal Burt and Bill Cheswick, courtesy Lumeta Corporation.\label{cap:routerInternet}}
\end{figure}

\begin{itemize}
\item \textbf{Ecological networks}: \citet{WM}, \cite{MoS} and \citet{CGA}
have studied a number of food webs found in freshwater, marine-freshwater
interface and terrestrial environments. They have constructed graph
representations of these food webs. A graph consists of a set of `nodes',
pairs of which may be connected by `links'. It is usually drawn as
a set of circles (the nodes) with arrows (the links) connecting pairs
of nodes. In the graph representations created by Montoya and Sol\'{e},
the nodes represent species, genera and sometimes higher taxa, and
links represent predator-prey interactions. Figure \ref{cap:foodweb}
shows a graph for the Ythan estuary food web. It has 135 nodes and
601 links; each link points from a species to one of its predators. 
\item \textbf{Transcription regulatory networks} \citep{MSIKCA,SDS,SMMA,FJVBO,MSA}:
Transcription of a gene is the process of constructing an mRNA from
the DNA sequence of the gene. This process is regulated by proteins,
called transcription factors, that enhance or inhibit transcription
by binding to sites on the DNA, usually upstream of the gene, which
in turn helps or prevents RNA polymerase and the rest of the basal
transcription machinery from binding and initiating transcription.
The protein product of one gene may act as a transcription factor
for another gene -- this network of interacting genes is called a
transcription regulatory network. 
\item \textbf{Biochemical signaling networks}: A number of biochemical signaling
pathways exist in cells which receive and transmit information, and
perform computational functions that are necessary for regulating
various cellular activities. These pathways are highly interconnected,
with different pathways sharing several common components, and thus
forming a signaling network. One example is the mitogen-activated
protein kinase network, which is involved in regulating the cell cycle
\citep{BRI}.
\item \textbf{Neural networks}: The complete neural network of the nematode
\emph{Caenorhabditis elegans} has been mapped and can be represented
as a graph with 302 nodes, each corresponding to one neuron, and more
than 5000 links corresponding to synaptic or gap junction connections
between neurons \citep{WSTB,WS,KL}.
\item \textbf{Polymer chain networks}: The conformation space of a lattice
polymer chain can be represented as a graph, with the nodes representing
possible conformations of the polymer chain and the links representing
the possibility of transforming one conformation to the other through
local movements of the chain \citep{SAB}. \citet{SeC} consider a
different type of network -- they represent a linear polymer as a
regular one-dimensional graph with nearest neighbour links and some
long range links.
\item \textbf{Metabolic networks} \citep{FW,JTAOB}: The metabolism of an
organism, such as the bacterium \emph{Escherichia coli,} is typically
specified by a set of chemical reactions involving different substrates
that are catalyzed by various enzymes. Figure \ref{cap:metabolism}
shows a graph of around 500 metabolic reactions. Each node represents
one chemical species; a link connects two species if there is a reaction
in which one species is a reactant and the other a product. This graph,
unlike that in Figure \ref{cap:foodweb}, is an undirected one --
the links have no associated direction and are therefore drawn as
lines, not arrows.
\item \textbf{Protein interaction networks}: The protein-protein interactions
in the yeast \emph{Saccharomyces cerevisiae} have been modeled as
graphs by \citet{JMBO}, \citet{Wagner}, \citet{MS}, \citet{SPSK}
and \citet{VFMA} using data from two-hybrid assays and other experiments
\citep{Uetzetal,ICOYHS}. Figure \ref{cap:protein} shows the protein
interaction map for \emph{S. cerevisiae}. In the graph each node represents
a protein and an undirected link connects two proteins if they are
known to interact. There are 1870 nodes and 2240 links.
\item \textbf{Internet}: The Internet is a network of interconnected computers.
Typically a few computers at one physical location (say a university
or a company) are connected by a `local area network'. These LANs
interact with each other via `routers' or `gateways'. The topology
of the Internet can be studied at various levels of detail. For example,
\cite{FFF} have studied a graph of the Internet at both the router
level, where each node corresponds to a router, and at a more coarse-grained
level, where each node corresponds to a group of routers. Another
study of the Internet at the router level was done by \citet{GT}.
Figure \ref{cap:routerInternet} shows a graph of the Internet at
the router level. Another representation of the Internet is in terms
of `autonomous systems'; each autonomous system ``approximately maps
to an internet service provider (ISP) and its links are inter-ISP
connections'' \citep{P-SVV}.
\item \textbf{World-Wide Web}: The WWW is a network of interconnected hypertext
documents. The connections are in the form of `hyperlinks' that can
be followed to other documents on the WWW. The structure of the WWW
has been studied at this level of resolution \citep{AJB1,Broder_web,Kumar_web}
as well as at the level of `sites', collections of all the documents
stored on the same web server \citep{HA}.
\item \textbf{Peer to peer networks}: The Gnutella network is one example.
It consists of a set of computers connected to each other for the
purpose of sharing files, with no central coordinating computer \citep{ALH}. 
\item \textbf{Power grid networks}: \cite{WS} have studied a graph of the
electricity transmission grid of western USA. The nodes of the graph
represent generators, transformers and substations, and links represent
transmission lines.
\end{itemize}

\begin{itemize}
\newpage
\item \textbf{Transportation networks}: The Indian railway system, consisting
of a network of stations connected by train services, has been studied
by \cite{SDCSMM}. A similar network of the world's airports has been
reconstructed from data on number of passengers arriving and leaving
each airport \citep{ASBS}.
\item \textbf{Economic networks} \citep{Kirman2}: An economic system can
be thought of as a network of interacting `agents' (individuals, companies
or even nations). Several different types of interactions can be visualized.
For instance, one company may use the product of another as a raw
material. In a market, individuals may interact by barter or monetary
transactions. Stock market traders interact by buying and selling
stocks and shares.
\item \textbf{Word co-occurrence network}: \citet{iCS} and \citet{DM0}
have studied a graph representing the English language. The nodes
represent the different words of the English language and links signify
that two words occur adjacent to each other in at least one sentence
in the database of texts studied. The resulting graph has approximately
470 thousand nodes and 17 million links.
\item \textbf{Friendship networks}: \cite{ASBS} have studied the properties
of a graph representing the friendship network of a group of 417 high
school students in USA. 
\item \textbf{Sexual contact network}: A graph with each node representing
an individual and links representing sexual contacts was constructed
and studied by \cite{LEASA}, based on a Swedish survey of 2810 people
in the age range 18-74 years.
\item \textbf{Film actor/actress collaboration network}: A graph can be
constructed where nodes represent actors and actresses and a link
is put between two if they have acted together in any film. \cite{WS}
have studied such a graph of collaborations for all actors listed
in the Internet Movie Database (\emph{http://us.imdb.com}).
\item \textbf{Scientific collaboration networks}: A similar graph can be
constructed for scientific collaborations, where nodes represent researchers
and a link represents co-authorship of one or more papers. \cite{Newman1,Newman2}
has studied such a graph of collaborations constructed from databases
of papers in physics, biomedical research and computer science.
\item \textbf{Citation networks} \citep{Seglen,Redner}: It is common for
research papers to cite several other papers. Thus a network of mutual
citations interconnects research papers.
\end{itemize}

\newpage
\section{\label{sec:Graph-representation-of}Graph representation of a network}

As is evident from the list above, the starting point of many studies
is to model networks using graphs. A graph representation is useful
for formulating questions about the structure and functioning of a
network more precisely. Graph theory suggests several quantities that
can characterize the structure of a graph. These can be computed for
a real network and used for comparison with, for instance, random
and regular graphs. 

The correlation of these graph theoretic quantities with the functioning
of the network can also be studied, thus exposing, in a more precise
way, the connection between a network's structure and its functioning.
Another possibility a graph representation allows is of a cross comparison
of very different systems -- for instance, the metabolic network of
a cell can be compared with the Internet. 

In many natural systems, the network of interactions is itself a dynamical
variable: The transcription regulatory network of an organism changes
as genes evolve, the World Wide Web changes as web pages get added
and deleted, and ecosystems change as species become extinct and new
species arise. With a graph representation of the network, one can
model processes that alter the graph with time, either in discrete
steps or continuously.

\section{\label{sec:Difficulties-of-creating}Difficulties of creating a graph
representation}

Representing a network as a graph is not as straightforward as it
may seem at first glance. Firstly, it is likely that several different
graphs can be drawn for a given system, each of which focus on different
aspects of the network. Consider, for example, the metabolism of the
bacterium \emph{Escherichia coli}, which is specified by a set of
chemical reactions involving different substrates that are catalyzed
by various enzymes. \cite{FW} construct two different graphs from
this information. The `reaction graph' has a node for each chemical
reaction with a directed link from one reaction to another if a product
of the first reaction is used as a reactant for the second. The `substrate
graph' has one node for each substrate with a directed link from one
substrate to another if there is a reaction in which the first substrate
is a reactant and the second a product. \cite{JTAOB} represent the
metabolic network of each of 43 organisms as a graph consisting of
two types of nodes, one type for each substrate and one type for each
reaction. The graph is bipartite with links from substrate nodes to
reaction nodes and reaction nodes to substrate nodes but no links
connecting two substrate or two reaction nodes. This graph has more
information than the previous two but also leaves out some information,
for example, the stoichiometry of the reactions. Thus, each of these
three types of graphs encodes different information about the metabolic
network, and each of them excludes certain information. 

Another problem is the incompleteness and uncertainty of data \citep{Kohn}.
When the system consists of a complex network of interactions, even
a single missed interaction or an erroneously added interaction, can
drastically affect the dynamics. Food webs tend to suffer from a bias
against including parasites \citep{WM}. Incompleteness of data is
suggested by \citet{WM} and \citet{CGA} as one of the possible reasons
why the Ythan estuary food web appears to differ, in several properties,
from the other food webs they have studied. Protein interaction networks
are another example. The two hybrid experiments that have been used
to build these networks are susceptible to both false positives as
well as false negatives, i.e., some protein interactions may be erroneously
missed and some non-existent protein interactions may be included
\citep{Uetzetal,ICOYHS}.

\section{\label{sec:Structure-of-networks}Structure of networks}

How can one characterize the structure of the Internet, the neural
network of \emph{Caenorhabditis elegans}, or the Indian railway system?
Representing a network as a graph makes it possible to use various
graph theoretic measures to characterize its structure. For instance,
one can compute the length of the shortest path between two nodes
averaged over all pairs of nodes in the graph. This quantity can be
compared with that expected for a `random graph'. Random graphs --
graphs where each pair of nodes has a probability, $p$, to be connected
by a link (see section \ref{sec:Random-graphs} for a more precise
definition) -- were introduced by \cite{ER1}, and have several known
structural characteristics. In the limit where the number of nodes
of the graph, $N$, tends toward infinity, then if $p<1/N$ (equivalently,
if the number of links $l<N/2$) the graph consists of several connected
clusters of nodes, each of which has only a finite number of nodes,
and if $p\ge 1/N$ (or $l\ge N/2$) a `giant' connected cluster exists
which has an infinite number of nodes \citep{Bollo}. The average
shortest path length for an undirected random graph of $N$ nodes,
with enough links to contain a giant cluster, grows as $\ln N$ \citep{Bollo}.
At the other extreme are regular graphs, such as a one-dimensional
chain of $N$ nodes whose average shortest path length scales in proportion
to $N$ (for a $d$-dimensional hyper-cubic lattice it would scale
as $N^{1/d}$).

Graphs of the Indian railway network \citep{SDCSMM}, the western
USA power grid network, the movie actor network, the neural network
of \emph{C. elegans} \citep{WS}, the world airport network and polymer
chain networks \citep{ASBS}, and ecological networks \citep{MoS}
are known to have low average path lengths similar to random graphs
with the same number of nodes and links. 

However they differ from random graphs in other characteristics, like
`clustering'. A set of nodes is `clustered' if in some sense the nodes
are more `strongly connected' to each other than to other nodes of
the graph outside the set. One measure of the amount of clustering
in a graph is the `clustering coefficient', which is the probability
that two neighbours of any node are also neighbours of each other.
Regular graphs have a relatively large clustering coefficient, while
random graphs have a relatively low coefficient. 

The networks mentioned above, which have a low average path length,
have much higher clustering coefficients than random graphs with the
same number of nodes and links. Graphs of this type which have a comparable
average path length but a much higher clustering coefficient than
a similar random graph have been dubbed `small-world' graphs \citep{WS,Watts}.

Another measure is the degree distribution of a graph. The degree
of a node (the total number of outgoing and incoming links to a node,
see section \ref{sec:Degrees-and-dependency}) takes a range of values
over all the nodes of any graph. The distribution of these values
can be analytically calculated for the random graph described above,
and is a binomial distribution. Given a graph of a real network, say
the Internet, one can compute its degree distribution and compare
with the binomial distribution expected for a random graph with the
same number of nodes and links. In contrast, a regular graph, such
as the lattice of a crystal solid, has nodes whose degrees take at
most a few different values. For example, in a body-centred cubic
lattice (such as that formed by iron or CsCl) each node (atom) has
a degree eight. Many networks, such as the metabolic network of \emph{E.
coli} \citep{JTAOB}, the Internet \citep{FFF} and citation networks
\citep{Seglen,Redner} have neither the binomial degree distribution
of a random graph nor the discrete distribution of a regular graph.
Instead their degree distribution has a power law tail. Such networks
have been termed `scale-free' \citep{BA}. \citet{MoS} claim that
ecological networks also have a scale-free degree distribution, a
conclusion that is disputed by \citet{CGA}. Several properties of
scale-free graphs have been elucidated \citep{NSW,BR}. For instance,
scale-free graphs are small-world, having a small average shortest
path length \citep{BA} and a high clustering coefficient \citep{Watts}.
However, the converse is not true: \cite{ASBS} show that not all
small-world networks are scale-free -- among several different networks
that are small-world, the degree distribution of some have an exponential
tail like random graphs, some are scale-free, while others have a
power-law distribution which is truncated by an exponential or a gaussian
tail.

Yet another measure used to characterize a graph is its eigenvalue
spectrum. Any graph with $N$ nodes can be specified by an $N\times N$
`adjacency' matrix. The collection of eigenvalues of this matrix is
the eigenvalue spectrum of the graph. Some structural characteristics
of a graph are reflected in the eigenvalue spectrum. For instance,
in a directed graph, the largest eigenvalue is directly related to
the presence or absence of cycles in the graph (see section \ref{sec:PF}).
Though only some aspects of the connection between the spectrum and
the structure of the graph have been worked out, it nevertheless serves
to classify graphs into different groups. Once again, some properties
of the spectrum of a random graph are known. For an undirected random
graph, with enough links to contain a giant cluster, the value of
the largest eigenvalue scales as $pN$, when $N\rightarrow \infty $
, while the distribution of the rest of the eigenvalues converges
to a semi-circle whose width is proportional to $\sqrt{N}$ \citep{FDBV}.
Therefore, the amount by which the eigenvalue spectrum of a real network
deviates from this `semi-circle law' is a measure of the non-randomness
of the graph. \citet{FDBV} and \citet{GKK} have shown that the eigenvalue
distributions of sparse random graphs and scale-free graphs are very
different from the semi-circle distribution.

The clustering and degree of a node are `local' quantities, in the
sense that they depend only on the immediate links of the node or
at most the links of its neighbours. In contrast, shortest path lengths,
eigenvalues and eigenvectors are `non-local' quantities, in the sense
that the shortest path between two nodes, or the component of an eigenvector
corresponding to a node, depends on the entire graph and not just
on the immediate links of the nodes in question. It is therefore not
surprising that the possibility of missing or extra links is a more
serious problem for the non-local quantities than for the local quantities.
The addition or removal of a small number of links will not significantly
affect the degree distribution or the clustering coefficient. In contrast,
adding a small number of random links to a regular graph reduces the
average path length drastically \citep{WS}. Similarly, the addition
of a single link can create a cycle, where there was none before,
thereby changing the largest eigenvalue. 

Another non-local graph-theoretic measure is the dependency of a node,
which, for a directed graph, is defined to be the number of links
that lie on paths leading to the node in question \citep{JKemerg}.
The average dependency of nodes of a graph is termed the `interdependency'
of the graph because it is a measure of how interdependent are the
nodes of the graph. Section \ref{sec:Degrees-and-dependency} will
discuss more about these measures.

A structural characteristic that is of interest, especially for chemical
networks, is the property of autocatalysis. The concept of an autocatalytic
set of chemical species was introduced by \citet{Eigen}, \citet{Kauffman1}
and \citet{Rossler}, and is defined in that context to be a set of
molecular species that contains, within itself, a catalyst for each
of its member species. This notion is closely related to that of positive
feedback loops and can be generalized to different systems which have
other kinds of `beneficial' interactions -- for instance, symbiosis
in ecological networks. \citet{HFRR} and \citet{LSYG} have each
constructed a pair of different molecules that catalyze one another's
(as well as their own) production and hence form an autocatalytic
set of this kind. \citet{Wacht} and \citet{MKYC} use a slightly
different, but related, definition of autocatalysis to argue that
the reductive citric acid cycle is autocatalytic. A graph-theoretic
definition of autocatalytic sets \citep{JKauto} is discussed in section
\ref{sec:Autocatalytic-sets-(ACSs)}. Autocatalytic sets will play
a major role in the dynamics of the model studied in this thesis.

Several other structural characteristics of graphs that have this
non-local character have been studied: \citet{DBCT} have analyzed
the structure of the shortest path spanning all nodes in a variety
of graphs derived from regular lattices; \citet{JG} define the ``weight-bearing
capacity'' of a branching hierarchical network and analyze which network
structures enhance this capacity; \citet{HHLM} and \citet{Lauff}
argue that cellular networks are `modular', i.e., consisting of distinguishable
clusters of nodes that play some functional role; \citet{RSMOB} suggest
that the metabolic network of \emph{E. coli} may have a hierarchical
structure; \emph{}\citet{MSIKCA}, \citet{SMMA} and \cite{MSA} describe
a way of decomposing the transcription regulatory network of \emph{E.
coli} into commonly occurring `motifs', which are specific patterns
of connections between a small set of nodes.

\section{\label{sec:Dynamical-systems-on}Dynamical systems on networks}

A different set of questions about such networks concerns the dynamics
of variables `living' on the graph -- for example, the concentrations
of different chemical substrates in a cell, or the populations of
species in an ecosystem, or the number of times a document on the
WWW is accessed. The network topology affects the dynamics of these
variables. The food-web structure of an ecosystem affects the dynamics
of its constituent species' populations; the network of human contacts
influences the spread of a contagious disease. 

Such dynamical systems on networks are often modeled as a set of coupled
differential equations in which the couplings are specified by the
network of interactions. An example is a network of coupled (possibly
nonlinear) oscillators where one common question asked is whether
groups of oscillators eventually synchronize despite different natural
frequencies and initial phases. The synchronization properties of
coupled dynamical systems have been studied for network structures
ranging from a fully connected graph \citep{Strogatz1} to regular
lattices \citep{Kaneko,PCDCA}, to trees \citep{GCR}, and a variety
of sparse network structures \citep{JA,Sinha}. \citet{Ramram} has
described the synchronization of coupled strange non-chaotic attractors.
The stability of the synchronization to perturbations of the coupling
strengths has been studied by \citet{CRD} for a large class of coupled
maps and differential equations.

In ecosystem models, the Lotka-Volterra equation, replicator equation
and other more complicated differential equations have been used to
represent the dynamics of the populations of species \citep{DMc}.
The replicator equation, for example, displays a range of different
behaviour depending on the network structure -- fixed point, limit
cycle, heteroclinic cycle and chaotic attractors have been observed
\citep{HSig}. Biochemical signaling networks too exhibit a variety
of dynamics \citep{BI,Bhalla1}.

The dynamics of neural networks has been extensively studied. One
example is a model of a neural network that exhibits a periodic cycling
between different meta-stable states ({}``memories'') as well as
intermittent transitions to a high activity state resembling epileptic
seizures \citep{BD1,BD2}. \citet{SC} review what is known about
synchronization in neural network models. For a comprehensive review
of neural network models see \citep{Arbib}.

Some dynamical systems on networks have the property of displaying
a variety of different dynamics even for the same network topology.
\citet{SCS} describe a simple model of a biochemical pathway that
can show fixed point, periodic, birhythmic and chaotic dynamics as
the rates of the reactions comprising the pathway are changed, while
keeping the topology of the pathway fixed. \citet{BRI} analyze a
biochemical signaling pathway whose dynamics is either monostable
or bistable depending on the parameter values.

In some systems it is more appropriate to represent the dynamics by
difference equations, rather than differential equations, or by cellular
automata. This is especially true for systems where the variables
of interest take discrete values and assuming them to be continuous
variables is a bad approximation. For instance, models of traffic
on transportation networks in which particles (vehicles) are individually
represented often show a different behaviour (e.g., different phase
transitions from a free flowing state to a traffic jam state) from
models in which a continuous traffic flow rate is the dynamical variable.
\cite{Nagel} reviews different traffic models on various types of
networks. Several studies of neural networks model the neurons as
cellular automata \citep{Arbib}. Two studies where the dynamics occurs
in discrete time steps are described by \citet{PA2}, who have studied
the dynamics of random walkers on a class of small-world networks,
and \citet{DBM}, who have studied the effect of different pricing
schemes on the dynamics of queues of people accessing an Internet
web server.

When the dynamics can be assumed to be reaching some steady state
or fixed point, there may be short cuts to finding the steady state
without actually having to solve the equations of motion. For instance,
\cite{Palsson} describe a procedure for determining the rates of
the chemical reactions comprising the intermediary metabolism of \emph{E.
coli}, assuming a steady state. \emph{}Instead of finding the rates
as the attractor of some difference or differential equations, the
procedure is based on linear optimization of a specifically chosen
function under the constraints imposed by the stoichiometry of the
reactions. The structure of the steady state reaction fluxes can be
studied as a function of different network topologies.

Local search strategies are another example of dynamical systems on
a network \citep{ALH}. The Gnutella file sharing system consists
of a network of computers; there is no central coordinating computer
that has information about the entire network, in particular, which
files are available at which nodes. Therefore, each user must use
a local search strategy, an algorithm that uses only local information,
such as the identities and connections of a particular node's neighbours,
its neighbour's neighbours, etc., to find where a particular file
is located on the network. Again, the network topology is crucial
because, for example, the most efficient search algorithms for regular
graphs are quite different from those for random or scale-free graphs
\citep{ALH}.

Another set of studies involving dynamics on networks is in the field
of epidemiology. In the models studied by \citet{WS} and \citet{P-SV}
diseases spread faster in small-world or scale-free graphs than in
regular and random graphs. Similar models are used to study the spread
of computer viruses through the Internet. These models can suggest
efficient immunization techniques by identifying which links or nodes
of the graphs are contributing most to the spread of the virus \citep{PA1,P-SV}. 

Closely related to this are studies of `attacks' on a network by the
removal of nodes of the network. \citet{AJB2}, \citet{CNSW} and
\citet{CEbAH1,CEbAH2} show that scale-free networks are more robust
to random attacks than random graphs but are susceptible to directed
attacks at the nodes with the highest degree. These studies contain
networks that are changing with time, though the dynamics simply involves
nodes getting removed in a specified order. The next section deals
with studies of evolving networks with more complicated dynamics.

\section{\label{sec:Dynamics-of-networks}Evolution of networks}

So far I have focused mainly on the structure of, and the behaviour
of dynamical systems on, fixed networks. However, real networks are
rarely static. An obvious question, therefore, is how (and why) did
a particular network evolve into the specific structure we see? For
instance, has the structure of a network evolved to optimize certain
functionality? Further, one can ask what are the mechanisms that cause
the network to change? Is the evolution driven by Darwinian natural
selection, or Lamarckian selection, or by other self-organizing processes? 

Some of the structural and dynamical studies in the previous two sections
can be used to make some guesses about the evolution of the network.
For instance, \cite{FW} suggest that the small-world structure of
metabolic networks may have evolved to enable a cell to react rapidly
to perturbations. \cite{WS} suggest that the visual cortex may have
evolved into a small-world architecture because that would aid the
synchronization of neuron firing patterns. The robustness of scale-free
networks to random removal of nodes has been suggested as an evolutionary
reason for the prevalence of scale-free networks \citep{AJB2}. 

While studies of static networks do provide some insight into network
evolution, it is natural to address such evolution-related questions
in models where the network is also a dynamical variable. One possibility
is to analyze a variety of simple models of changing networks, with
different rules governing the dynamics of the network, and see which
rules produce structures like small-world or scale-free graphs that
are commonly observed in real networks: \cite{WS} describe a model
that produces a small-world graph by randomly rewiring or adding links
to a regular graph. A similar model (where the number of nodes is
fixed but links are repeatedly added) that produces scale-free networks,
is described by \citet{MM}. Several other `growing network' models
start from an empty network and add nodes and links at discrete time
steps. The `preferential attachment' rule -- new nodes are preferentially
assigned links to nodes with a high degree -- produces scale-free
graphs \citep{BA}. \citet{AB} and \citet{DM} review several different
models of this type, using both deterministic and stochastic rules,
that give rise to scale-free graphs. The growing network model described
by \citet{Manna} adds another level of complexity by making the addition
of new links to the network dependent on the dynamics of several random
walkers on a regular lattice. Variants of these models exist where
the preferential attachment to nodes of high degree competes with
the preferential attachment to nodes of a lesser age, i.e., nodes
that have been added to the network more recently. \citet{ASBS} have
shown that such aging effects can result in a graph whose power-law
degree distribution is truncated by an exponential tail. \cite{DM2}
discuss models of accelerated growth of networks, where the rate of
addition of new nodes increases with time. \cite{WM} describe a simple
set of rules to grow a network, that produces graphs remarkably similar
in structure to many ecological food webs \citep{CGA}.\\

One aspect missing from these growing network models is that the dynamics
of the network is not intertwined with the dynamics of other variables.
The models that include aging are implicitly using a dynamical variable,
the age of a node, but the dynamics of that variable is not coupled
to the structure of the graph. In many natural systems the dynamics
of the network is tightly coupled to the dynamics of the variables
living on it, and vice versa. For example, a food-web influences the
dynamics of the species' populations, and if a species becomes extinct,
the food-web changes. The firing pattern of neurons depends on the
structure of the neural network and the strengths of synaptic connections
are in turn modified by the repeated firing or non-firing of neurons
\citep{Arbib}. There exist a number of models that incorporate this
feature too. Models of learning and memory in neural networks typically
have neuron firing patterns co-evolving with the network connection
strengths \citep{Arbib}. A number of such evolving ecosystem models
have been studied \citep{LN,SMa,CSK} (see \citealp{DMc}, for a review).
Evolving replicator networks have been described by \citet{HS} and
\citet{TY2}. 

In addition to creating evolving network models that produce graphs
having a similar structure to existing networks, one can also use
evolving network models to address a number of other evolution-related
questions. For instance, how does the existing structure of a network
influence its subsequent evolution? This question is best addressed
in a model where the evolution of the network is coupled to the dynamics
of other variables, which, in turn, are affected by the network structure.
Because of this coupling, the different (sub)structures in the network
affect its evolution. For instance, autocatalytic sets in prebiotic
chemical networks might have been more stable to perturbations because
of their ability to self-replicate. If so, an interesting question
is: Can such stable structures grow, or spread through a network and,
if they can, over what timescales? Further, one can ask how the existing
structures determine the short and long term effects of a perturbation
on the evolution of the network; in an ecosystem, whether an invading
species will be able to survive, and for how long, will depend on
the competition and resources provided by the existing species.

A related set of (meta)questions can be asked about the `evolvability'
of networks. \citet{KG} have defined the evolvability of a biological
organism as the capacity to generate heritable, selectable phenotypic
variation. This definition can be extended to other types of networks.
Then one can ask: How evolvable is a network? Is this evolvability
also subject to selection and has it, therefore, itself evolved over
time? 

One of the inspirations for the model I will discuss in this thesis
are the models of Kauffman, Farmer, Fontana and others who have explored
questions about self-organization, the origin of life and some of
the above evolvability issues in work on artificial chemistries. An
artificial chemistry is a system whose components `react' with each
other in a way analogous to molecules participating in chemical reactions.
Thus, \citet{Kauffman2} and \citet{BFF} consider systems composed
of strings of arbitrary length made from a binary alphabet. Strings
can participate in `cleavage' and `ligation reactions' which involve
the splitting or concatenation of strings to form new ones. Just as
with a real chemistry, not all reactions will be allowed and there
are different ways of specifying which strings participate in which
reactions. The simplest way is to randomly decide which reactions
are allowed, resulting in a `random chemistry' of the type studied
by Kauffman. He found that there is a critical number of links, such
that if a random chemistry has more than that number of links, it
is almost certain to contain an autocatalytic set \citep{Kauffman3}. 

\citet{BFF} have explored an artificial chemistry consisting of catalyzed
ligation and cleavage reactions. Their network evolves with time as
new molecules can be created by the ligation of existing ones. They
address the puzzle of how large molecules with highly specialized
functionality, like enzymes and DNA, could have evolved from an initial
condition that contained only small molecules which interacted weakly
and non-specifically. Again, autocatalytic sets play an important
role in the overall dynamics and are suggested as one of the possible
means by which a complex chemical organization, a `metabolism', could
have evolved on the prebiotic Earth. \citet{Fontana} and \citet{FB}
study more abstract artificial chemistries consisting of functions
expressed in the lambda-calculus. Each function can `react' with other
functions, producing a new function by the composition of the `reactant'
functions. They find that in such a chemistry there is a spontaneous
emergence of groups of cooperative reactions such as self-replicators,
self-replicating sets, autocatalytic cycles, symbiotic and parasitic
functions.

If Darwinian natural selection is driving the evolution of the network
one can ask, what were the selective pressures acting on the network
at different times? How do the selective pressures change with the
structure of the network? Whenever the dynamics of the network has
a stochastic component the detailed structure of the network is typically
a result of many chance events. Nevertheless, some patterns may still
be predictable. Thus, it is of interest to ask which patterns are
predictable and which are a result of historical accidents.

Another aspect of the dynamics of evolutionary systems is the destruction
of structures. In evolving systems, non-random structures not only
emerge but also get destroyed in some circumstances. A number of mass
extinctions are documented in the fossil record \citep{NP}. Financial
markets suffer sudden large crashes \citep{Bouchad,JS}. Discovering
markers that can be used to predict an imminent crash would obviously
be very useful in the context of financial markets. It is of interest
to try to identify the possibly multiple causes of such crashes. Many
mechanisms have been suggested to explain mass extinctions in the
fossil record \citep{MSmith,Glen}. The model of \citet{BS} is one
example of attempts to explain the distribution of extinction sizes.
\citet{TY1} have studied various statistical properties of the extinction
events in their model of an ecosystem. In ecosystems there can exist
`keystone' species whose extinctions are likely to trigger a cascade
of further extinctions \citep{Paine,JTM,SMo}. This raises several
interesting questions: Knowing the structure of an ecological network
can one predict which species are keystone? Are there any mechanisms
that exist to `protect' such important species?

In this thesis I will try to address some of these issues within the
context of a specific mathematical model of an evolving network. Many
of the phenomena exhibited by the model are reminiscent of the phenomena
seen in a variety of evolving networks -- the growth of non-random
structure (in the form of autocatalytic sets), the growth of interdependence
and cooperation between nodes of the network, sudden mass extinctions
due to the extinction of keystone species or the creation of certain
new structures in the network (termed `innovations'). The model provides
a simple mathematical framework within which to analyze the mechanisms
that produce such phenomena. The basic model, and its variants, will
be specific instances of a broad framework for modeling evolving networks
that I present in the following section.

\section{\label{sec:Framework-of-a}Framework of a model in which the network
co-evolves with other variables}

Consider a process that alters a network, represented by a graph,
in discrete steps. The series of graphs produced can be denoted $C_{n},n=1,2,\ldots $.
Each step of the process, taking a graph from $C_{n-1}$ to $C_{n}$,
will be called a \emph{}`graph update event\emph{'.} The following
framework defines a class of evolving network models that consist
of a graph evolving, by a series of graph update events, along with
other system variables:

\begin{description}
\item [Variables:]The dynamical variables are a directed graph $C$ and
a variable $x_{i}$ associated with each node $i$ of the graph. For
example, $x_{i}$ could represent:
\end{description}
\begin{itemize}
\item the concentrations of substrates and enzymes in a metabolic network, 
\item the expression level of genes in a genetic regulatory network, 
\item the populations of species in an ecosystem, 
\item the state (firing or not firing) of neurons in the brain, 
\item the number of times each web page on the world wide web has been accessed, 
\item the number of films each actor/actress has worked in, 
\item the profits of each of a set of interacting companies.
\end{itemize}
\begin{description}
\item [Initialization:]To start with, the graph $C$ and the variables
$x_{i}$ are given some initial values. The particular choice of values
will depend on the system being modeled.
\item [Dynamics:]~
\end{description}
\begin{lyxlist}{00.00.0000}
\item [Step~1:]First, keeping the graph ($C_{n-1}$ at step $n-1$) fixed,
the $x_{i}$ are evolved for a specified time $T$ according to a
set of differential equations that can be schematically written in
the form: $\dot{x_{i}}=f_{i}(C_{n-1},x_{1},x_{2},\ldots )$, where
$f_{i}$ are certain functions that depend upon the graph $C_{n-1}$
and on all the $x_{i}$ variables. 
\item [Step~2:]After this, some nodes may be removed from the graph, along
with their links. Some links may also be removed from the graph. Which
nodes and links are removed will, in general, depend on the $x_{i}$
values of the nodes and the graph $C_{n-1}$.
\item [Step~3:]Similarly, some nodes may be added to the graph. These
new nodes will be assigned links with existing nodes. The rules for
specifying which links will be assigned may depend on the $x_{i}$
values of the other nodes and $C_{n-1}$. The new node will be given
its own $x_{i}$ variable, which will be assigned some initial value.
Some new links may be added between existing nodes, which again may
depend on the $x_{i}$ values of the nodes and $C_{n-1}$.
\end{lyxlist}
Steps 2 and 3 produce a new graph which will be the graph at time
step $n$, $C_{n}$. This process, from step 1 onward, is iterated.

The above provides a nonequilibrium statistical mechanics framework
for a model of an evolving network. The graph is generically in a
nonequilibrium state because of the constant flux of nodes and links
into and out of the graph.  The $x_{i}$ variables may also be in
a nonequilibrium state depending on the form of of $f_{i}$. The graph
dynamics in this framework is a specific case of a Markov process
on the space of graphs. For a general Markov process, at time $n-1$,
the graph $C_{n-1}$ determines the transition probability to all
other graphs. The stochastic process picks the new graph for time
$n$, $C_{n}$, using this probability distribution. In the example
here, the transition probability is not specified explicitly. It arises
implicitly as a consequence of the dynamics of the $x_{i}$ variables
(step 1) and the way the specified rules for steps 2 and 3 use the
$x_{i}$ values to determine which nodes and links will be removed
or added. 

There are two timescales built into this framework. On a timescale
much shorter than $T$, the $x_{i}$ variables can evolve while the
graph remains fixed. On a longer timescale, the graph changes in discrete
steps that involve the possible removal and addition of nodes and
links. The operation of the $x_{i}$ dynamics and the graph dynamics
on different timescales is common to many systems. In the brain, over
short times neurons fire or not depending on their fixed synaptic
connections and the firing of other neurons, while over a long timescale
the firing pattern can cause the synaptic connections to strengthen
or weaken. In an ecosystem, over short timescales the species composition
is fixed while the populations change, and over long times the network
changes because of the extinction and mutation of existing species
and the invasion of the system by new species. In a genetic regulatory
network, over short times the genes and regulatory interactions are
fixed and it is the expression level of genes that changes with time,
and on a longer timescale the genes themselves evolve. These examples
also reiterate what was mentioned before -- the $x_{i}$ dynamics
and the graph dynamics are interdependent. This feature is also built
into the above framework: The dependence of the $x_{i}$ dynamics
on the graph lies in the dependence of the functions $f_{i}$ on $C$.
The dependence of the graph dynamics on $x_{i}$ lies in steps 2 and
3. The particular choices of $f_{i}$ as well as the rules determining
which, if any, nodes and links will be removed or added will depend
on the particular system that is being modeled.

\section{\label{sec:Extensions-of-the}Extensions of the framework}

This framework can be extended to include a larger class of evolving
network models. Firstly, as mentioned in section \ref{sec:Dynamical-systems-on},
it may be preferable to model some systems using difference equations
or cellular automata. For instance, if the populations of all species
in an ecosystem can be assumed to be large then treating them as continuous
variables and representing their dynamics by differential equations
may be reasonable. But if the populations are close to zero this could
be a bad approximation. It is easy to alter step 1 to take into account
these possibilities. Secondly, it may be more appropriate to model
some systems using two or more graphs, or by some generalization of
a graph like a hypergraph. For example, in a cell, the metabolic network
is linked to the genetic regulatory network and the dynamics on one
network can affect the dynamics on the other. A number of neural network
models of learning consist of two interacting networks, the `teacher'
and the `student'; \cite{Kinzel} has reviewed several models that
consist of more than one neural network interacting with one another.
The framework described above can be generalized to deal with such
systems: \cite{Kohn} has described a generalized form of a graph
that he uses to represent all the different types of interactions
in the mammalian cell cycle control and DNA repair systems. Thirdly,
as is the case in some neural network models \citep{Arbib}, it may
be worthwhile to consider dynamical rules where the network changes
continuously with time, rather than in discrete steps as in the above
framework.

\section{\label{sec:The-origin-of}The origin of life: evolution of a chemical
network}

In this thesis I will describe a model -- a particular instance of
the above framework -- whose rules attempt to describe some aspects
of the dynamics of a chemical network in a pool on the prebiotic Earth
\citep{JKauto}. Assume that the pool contains many amino acid monomers
or nucleotide bases, as well as some small polypeptide chains or short
RNA molecules which may have a weak catalytic activity \citep{Joyce}.
For instance, some of these polypeptides or catalytic RNA might catalyze
the production of other polypeptides or RNA from the monomers present
in the pool. Such a chemical network can be represented as a graph
in which the nodes are the polypeptides/RNA and the links represent
their catalytic interactions. This would be the graph $C$ in the
model. The $x_{i}$ could represent the concentrations or relative
populations of each type, or species, of polypeptide or RNA. $x_{i}$
would change with time as the possible chemical reactions proceed
to produce or deplete the different molecular species. Focusing only
on the catalyzed reactions which produce different molecular species
from the monomers, one can imagine that for a short timescale, over
which the pool remains undisturbed, the chemical network (i.e., $C$)
will be fixed with no molecular species entering or exiting the pool.
Thus, the $x_{i}$ will evolve according to the chemical rate equations,
with $C$ fixed. In section \ref{sec:The-population-dynamics}, I
derive the form of $f_{i}$ for such chemical rate equations under
the following assumptions:

\begin{itemize}
\item only reactions involving the catalyzed production of each molecular
species are modeled,
\item the catalyzed reactions are described by the Michaelis-Menten theory
of enzyme catalysis,
\item the Michaelis constants for each reaction are very large compared
to the populations,
\item the concentrations of all the required reactants are non-zero and
fixed,
\item all catalysts have the same strength,
\item the pool is well-stirred, i.e., there are no spatial degrees of freedom.
\end{itemize}
Over long timescales imagine that the pool is subject to perturbations
in the form of floods, tides or storms which may flush out part of
the pool and possibly introduce new molecules into the pool. Removing
a node from the graph would correspond to removing all the molecules
of a particular type from the pool. When a perturbation removes a
random lot of molecules, the molecular species with smaller $x_{i}$
are more likely to be completely wiped out from the pool. For step
2, I will choose a rule that implements an extreme version of such
selection: the node with the least $x_{i}$ will be removed along
with all its links. A perturbation can add a new node to the graph,
with a small $x_{i}$ value. An added node corresponds to a new type
of molecule being brought into the pool by the perturbation. As there
is no reason to assume such a node would have any particular relationship
with existing nodes it is reasonable to assign the links of the new
node randomly. Therefore, for step 3, I choose the rule: add one new
node, whose links with existing nodes are assigned randomly. In this
scenario one can assume that the perturbation would not remove existing
links without removing a node, or create a new link between existing
nodes. Therefore, these possibilities can be excluded from the model.
A detailed description of the model rules is given in chapter 5.

Such a model could address some questions concerning the origin of
life on Earth \citep{JKcoop}. The chemical network of a bacterial
cell of today consists of several thousand types of molecules involved
in thousands of different chemical reactions. Each type of molecule
plays a definite functional role and the entire chemical network is
organized to enable the cell to perform the various functions it needs
to survive and reproduce in often extreme conditions. This chemical
organization is highly non-random -- the molecules found in cells
are a small subset in a very large space of possible molecules and
the graph that describes their interactions is a special kind of graph
in the very large space of graphs. The probability of such a structured
chemical organization arising by pure chance is extremely small. After
the oceans condensed about 4 billion years back on the prebiotic Earth
there was no such chemical network of interactions existing anywhere.
However, if we assume that life originated on Earth about 3.5 to 3.8
billion years ago as suggested by the microfossil evidence \citep{Schopf},
then that leaves a few hundred million years for the first living
cells to form. A puzzle of the origin of life on Earth is: how did
such a structured, non-random chemical organization form in such a
short time? This question can be addressed in the above model. One
can start with a random graph, representing the situation just after
the oceans would have condensed on Earth, where the pool would be
likely to contain a random set of molecules with no particular non-random
chemical organization. As I will show, starting from this initial
state, the chemical network in the model described above evolves inevitably
into a highly non-random, autocatalytic network \citep{JKauto,JKemerg}.
The mechanisms that cause this growth might throw light on the above
puzzle.

\section{\label{sec:Catastrophes-and-recoveries}Catastrophes and recoveries
in evolving networks}

The model I will analyze also exhibits sudden catastrophes -- mass
extinctions where the relative populations of many species fall to
zero in a short time -- which are followed by slow recoveries. The
asymmetry between catastrophe and recovery times, and fat tails in
the size distribution of catastrophes in the model \citep{JKcoresh}
are also observed in catastrophes and recoveries seen in the fossil
record \citep{NP} and in financial markets \citep{Bouchad,JS}. The
mechanisms that cause such mass extinctions are a matter of much debate
\citep{MSmith,Glen}. The model of \citet{BS} is an attempt to explain
the frequency distribution of extinction events of different sizes
using the mechanism of self-organized criticality. The model I will
study is similar in some ways to the Bak-Sneppen model, with the addition
of another dynamical variable -- the network structure of the system.
The advantage of explicitly modeling the network structure is that
the mechanisms causing the catastrophes and their dependence on the
network structure can be analyzed in detail. 

I will show that, in the model, the largest extinctions are caused
mainly by three mechanisms \citep{JKlargeext}. One of the mechanisms
involves the removal of specific species. This is reminiscent of the
notion of `keystone species' that was discussed in section \ref{sec:Dynamics-of-networks}.
Thus, in this model, one can give a graph-theoretic definition of
keystone species and explain why their removal causes mass extinctions
\citep{JKlargeext}.

\subsection{Innovations}

Apart from the removal of species, it is interesting that `innovations'
-- the creation of new structures in the network -- can also cause
catastrophes. In the context of the model it is possible to create
a graph-theoretic definition for the term `innovation' \citep{JKlargeext,JKinn}.
This definition allows me to construct a hierarchy of innovations
and correlate the graph-theoretic structure of an innovation to its
short and long term effects on the evolution of the network \citep{JKwiley,JKinn}.

\section{\label{sec:A-map-of}A map of subsequent chapters}

Chapters 2--4 describe different segments of the basic model, that
are put together in chapter 5, and introduce various results from
graph theory that will be useful for analyzing the model. The later
chapters analyze the diverse phenomena observed in the model using
the results from the previous chapters. In more detail:

\begin{description}
\item [Chapter~2]introduces certain elements of graph theory. I set out
the notation that will be used and give definitions of directed graphs,
degree of a node, paths, cycles and connected components of a graph.
I introduce the notions of `dependency' of a node, and `interdependency'
of a graph. I define the Perron-Frobenius eigenvalue and eigenvectors
of a graph and describe some of their properties. Random graphs are
briefly discussed at the end of the chapter. 
\item [Chapter~3]introduces the concept of an `autocatalytic set'. I describe
the relation between algebraic properties of a graph, such as the
Perron-Frobenius eigenvalue and eigenvectors, and topological properties,
such as the presence or absence of autocatalytic sets in a graph.
I then present the eigenvector profile theorem which, for any graph,
specifies the `profile' (which components are zero and which non-zero)
of all possible Perron-Frobenius eigenvectors of the graph. The notion
of a core and periphery of a Perron-Frobenius eigenvector is introduced.
\item [Chapter~4]presents a dynamical system whose variables `live' on
the nodes of the graph. Their dynamics is described by a set of coupled
differential equations, where the couplings are specified by the graph.
The equations are derived from an idealized version of the chemical
rate equations that would describe the dynamics of catalytic molecules
in a well-stirred chemical reactor. I show that the attractors are
fixed points and for generic initial conditions are Perron-Frobenius
eigenvectors of the graph. I present the attractor profile theorem
which identifies the subset of Perron-Frobenius eigenvectors that
are attractors of the system for a given graph. The `core' and `periphery'
of a graph are defined and the notion of a `keystone node' is introduced.
\item [Chapter~5]introduces a model of an evolving graph along the lines
of section \ref{sec:Framework-of-a}. The model uses the dynamical
system discussed in chapter 4 for step 1 of the dynamics. Some specific
rules are chosen for steps 2 and 3 that aim to capture the way storms,
floods or tides would remove molecular species from, and add new molecular
species to, a pool on the prebiotic Earth. I display the evolution
of various quantities -- the total number of links in the graph, the
number of nodes with non-zero relative population in the attractor,
the Perron-Frobenius eigenvalue, and the interdependency of the graph
-- as a function of time for some example runs of the model. I also
exhibit snapshots of the graph at various times for one run. Three
`phases' of behaviour are observed in these runs which are dubbed
the `random', `growth' and `organized' phases.
\item [Chapter~6]discusses the random and growth phases. Autocatalytic
sets are shown to play an important role. I argue that no graph structure
is stable for very long during the random phase, and that this is
because there is no ACS in the graph. I show that the chance formation
of an ACS triggers the growth phase, in which the number of links
of the graph grow exponentially as the ACS expands by accreting more
and more nodes to itself. This continues until the entire graph is
an ACS at which point the organized phase starts. The timescales of
appearance and growth of the ACS are analytically estimated. I show
that the final fully autocatalytic graph is a highly non-random graph.
The degree and dependency distributions for the fully autocatalytic
graphs produced in the runs are discussed.
\item [Chapter~7]discusses the sudden extinctions of large numbers of
molecular species observed occasionally in the organized phase. I
show that the largest extinction events are `core-shifts', a complete
change of the core of the graph. The notion of an `innovation' is
introduced. I show that core-shifts can occur due to the removal of
a keystone species from the graph, or due to a particular type of
innovation (the addition of a species which creates a new `self-sustaining'
structure in the graph), or a specific combination of both. The timescales
of these large extinction events, as well as the recovery of the system
afterward, are discussed. The structure of the graphs just before
each large extinction, in particular their degree and dependency distributions,
are analyzed.
\item [Chapter~8]discusses the robustness of the formation and growth
of autocatalytic sets to changes in the model rules. I list various
simplifications in the model rules that depart from realism but make
the system analytically tractable. Variants of the model that relax
these simplifications are presented and are shown to exhibit the formation
and growth of autocatalytic sets.
\item [Chapter~9]summarizes the interesting features of the model and
its variants, and discusses the limitations and possible extensions
of the model.
\item [Appendix~A]contains detailed proofs of all propositions made in
the thesis, including the eigenvector, and attractor, profile theorems. 
\item [Appendix~B]describes two computer programs that use different methods
to simulate the model and its variants. The source code of the programs
are provided in the attached CD.
\item [Appendix~C]describes the contents of the attached CD. 
\end{description}

\chapter{\label{cha:Graph-Theory}Definitions and Terminology}

\section{Directed graphs and adjacency matrices}

\begin{description}
\item [Definition~2.1:]Directed graph. \\
A \textit{directed graph} $G=G(S,L)$ is defined by a set $S$ of
`nodes' and a set $L$ of `links', where each link is an ordered pair
of nodes \citep{Harary,RF}. 
\end{description}
The set of nodes can be conveniently labeled by integers, $S=\{1,2,\ldots ,s\}$
for a directed graph of $s$ nodes. Henceforth I will use the term
`graph' to refer to a labeled, directed graph.

\begin{figure}
\begin{center}\includegraphics[  width=15cm,
  keepaspectratio]{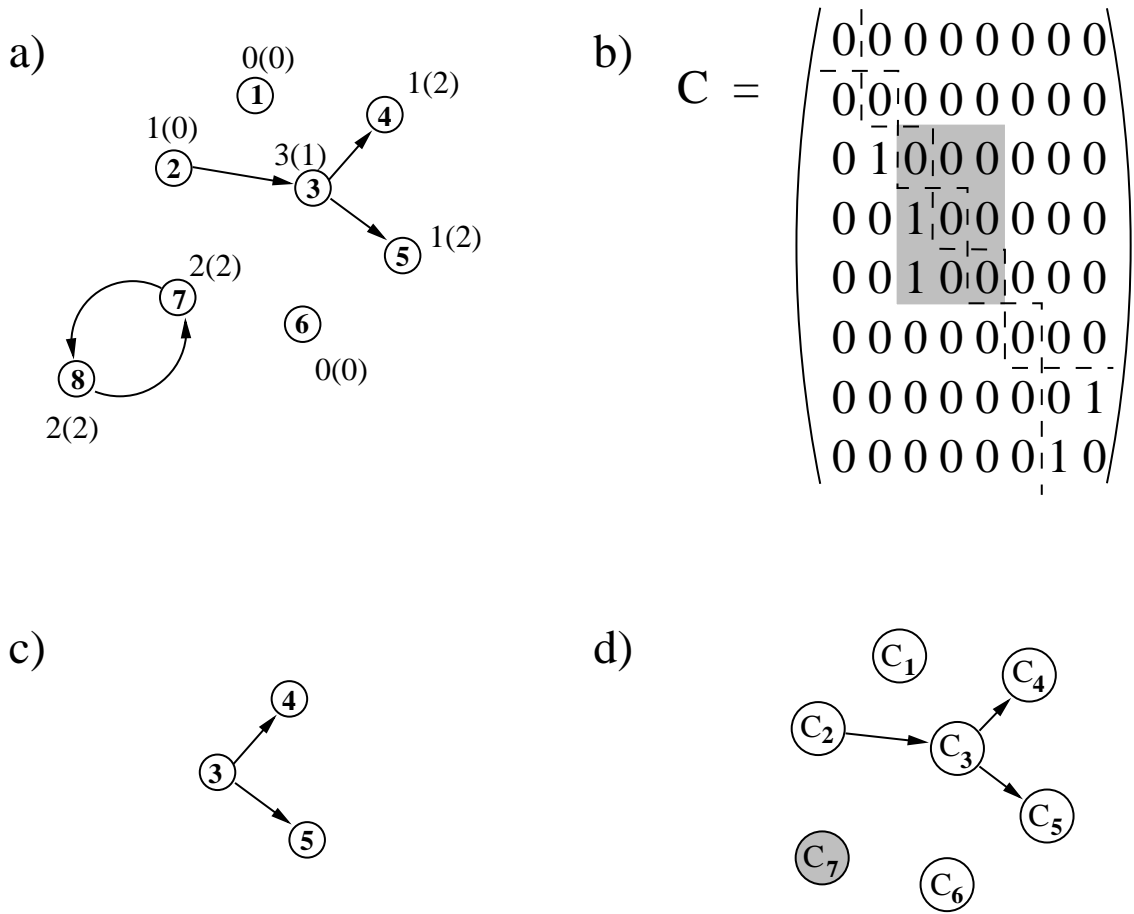}\end{center}

\caption{\textbf{a.} A directed graph with 8 nodes. The unbracketed number
adjacent to each node is the degree of that node, and the bracketed
number is the dependency. \textbf{b.} The adjacency matrix of the
graph in (a). \textbf{c.} A subgraph of the graph in (a) induced by
$S'=\{3,4,5\}$. The adjacency matrix of the subgraph is the shaded
portion of the matrix in (b). \textbf{d.} The condensation of the
graph in (a). The dashed lines in (b) demarcate the portions that
correspond to the strong components $C_{\alpha }$. The only basic
subgraph, $C_{7}$, is shaded.\label{cap:digraph}}
\end{figure}

An example of a graph is given in Figure \ref{cap:digraph}a where
each node is represented by a small labeled circle, and a link $(j,i)$
is represented by an arrow pointing from node $j$ to node $i$. A
graph with $s$ nodes is completely specified by an $s\times s$ matrix,
$C=(c_{ij})$, called the \textit{\emph{`adjacency matrix}}' of the
graph, and vice versa. 

\begin{description}
\item [Definition~2.2:]Adjacency matrix.\\
The \emph{adjacency matrix} of a graph $G=G(S,L)$ with $s$ nodes
is an $s\times s$ matrix, denoted $C=(c_{ij})$, where $c_{ij}=1$
if $L$ contains a directed link $(j,i)$ (arrow pointing from node
$j$ to node $i$), and $c_{ij}=0$ otherwise. 
\end{description}
This convention differs from the usual one \citep{Harary,RF,Bollobas}
where $c_{ij}=1$ if and only if there is a link from node $i$ to
node $j$; the transpose of the adjacency matrix defined above. This
convention has been chosen because it is more convenient in the context
of the dynamical system to be discussed in chapter \ref{cha:Population-Dynamics}.
Figure \ref{cap:digraph}b shows the adjacency matrix corresponding
to the graph in Figure \ref{cap:digraph}a. I will use the terms `graph'
and `adjacency matrix' interchangeably: the phrase `a graph with adjacency
matrix $C$' will often be abbreviated to `a graph $C$'. 

Undirected graphs are special cases of directed graphs whose adjacency
matrices are symmetric \citep{Harary,RF}. A single (undirected) link
of an undirected graph between, say, nodes $j$ and $i$, can be viewed
as two directed links of a directed graph, one from $j$ to $i$ and
the other from $i$ to $j$. See \cite{Bollobas} for many results
concerning undirected graphs.

\begin{description}
\item [Definition~2.3:]Subgraph. \\
A graph $G'=G'(S',L')$ is called a \textit{subgraph} of $G(S,L)$
if $S'\subset S$ and $L'\subset L$ \\
\citep{Harary,Bollobas}.
\item [Definition~2.4:]Induced subgraph.\\
A graph $G'=G'(S',L')$ is called an \textit{induced subgraph} of
$G(S,L)$, or the \emph{subgraph of $G(S,L)$ induced by} $S'$, if
$S'\subset S$ and $L'$ contains all (and only) those links in $L$
with both endpoints in $S'$ \citep{Harary,Bollobas}. 
\end{description}
The graph in Figure \ref{cap:digraph}c is thus an induced subgraph
of the graph in Figure \ref{cap:digraph}a (induced by $S'=\{3,4,5\}$).
For a subgraph it is often more convenient to label the nodes not
by integers starting from 1, but by the same labels the corresponding
nodes had in the parent graph. The adjacency matrix of an induced
subgraph can be obtained by deleting all the rows and columns from
the full adjacency matrix that correspond to the nodes outside the
subgraph. The highlighted portion of the matrix in Figure \ref{cap:digraph}b
is the adjacency matrix of the induced subgraph in Figure \ref{cap:digraph}c.

\section{\label{sec:Degrees-and-dependency}Degrees and dependency}

\subsection{Degree of a node and degree distribution of a graph}

\begin{description}
\item [Definition~2.5:]Degree, in-degree and out-degree.\\
The \emph{degree}, or \emph{total degree}, of a node is the total
number of incoming plus outgoing links from that node, i.e., the degree
of node $i$ is $\sum _{j=1}^{s}(c_{ji}+c_{ji})$.\\
The \emph{in-degree} of a node is the total number of incoming links
to that node, i.e., the in-degree of node $i$ is $\sum _{j=1}^{s}c_{ij}$.\\
The \emph{out-degree} of a node is the total number of outgoing links
from that node, i.e., the out-degree of node $i$ is $\sum _{j=1}^{s}c_{ji}$
\citep{Harary,Bollobas}.
\end{description}
The unbracketed numbers adjacent to each node in Figure \ref{cap:digraph}a
show the degree of that node.

\begin{description}
\item [Definition~2.6:]Degree distribution.\\
The \emph{degree distribution} of a graph, denoted $P(k)$, is the
fraction of nodes with degree $k$.\\
The \emph{out-degree distribution}, $P_{out}(k)$, and the \emph{in-degree
distribution}, $P_{in}(k)$, are similarly defined \citep{Harary,Bollobas}.
\end{description}

\subsection{Dependency and interdependency}

\begin{description}
\item [Definition~2.7:]Dependency.\\
The \emph{dependency}, denoted $d_{i}$, of a node $i$ is the total
number of links in all paths that terminate at that node, each link
counted only once \citep{JKemerg}.
\item [Definition~2.8:]Dependency distribution.\\
The \emph{dependency distribution} of a graph, denoted $D(d)$, is
the fraction of nodes with \\
dependency $d$.
\item [Definition~2.9:]Interdependency.\\
The \emph{interdependenc}y of a graph is defined to be the average
dependency: \\
$\overline{d}\equiv (1/s)\sum _{i=1}^{s}d_{i}=\sum _{d=0}^{\infty }d\times D(d)$
\\
\citep{JKemerg}.
\end{description}
The bracketed numbers adjacent to each node in Figure \ref{cap:digraph}a
show the dependency of that node. The interdependency of the graph
in Figure \ref{cap:digraph}a is 9/8.

Because $d_{i}$ counts how many links ultimately `feed into' the
node $i$, it is a measure of how `dependent' node $i$ is on other
nodes. Thus $\overline{d}$ is a measure of how interdependent are
the nodes in the graph. While the degree of a node is a `local' measure
as it depends only on the immediate connections of a node, the dependency
is more `non-local' in character because links far away from the node
contribute to it.

\section{Walks, paths and cycles}

\begin{description}
\item [Definition~2.10:]Walk, closed walk.\\
A \textit{walk} of length $n$ (from node $i_{1}$ to node $i_{n+1}$)
is an alternating sequence of nodes and links $i_{1}l_{1}i_{2}l_{2}\ldots i_{n}l_{n}i_{n+1}$
such that link $l_{1}$ points from node $i_{1}$ to node $i_{2}$
(i.e. $l_{1}=(i_{1},i_{2})$), $l_{2}$ points from $i_{2}$ to $i_{3}$
and so on. If the first and last nodes $i_{1}$ and $i_{n+1}$ of
a walk are the same, it will be referred to as a \textit{closed}\emph{~walk}
\emph{}\\
\citep{Harary,RF,Bollobas}.
\end{description}
The existence of even one closed walk in the graph implies the existence
of an infinite number of distinct walks in the graph. In the graph
of Figure \ref{cap:digraph}a, there are an infinite number of walks
from node 7 to node 8 (e.g., $7\rightarrow 8$, $7\rightarrow 8\rightarrow 7\rightarrow 8$,
$\ldots $) but no walk from node 7 to node 2. An undirected graph
trivially has closed walks if it has any undirected links at all. 

If $C$ is the adjacency matrix of a graph then it is easy to see
that $(C^{n})_{ij}$ equals the number of distinct walks of length
$n$ from node $j$ to node $i$. E.g., $(C^{2})_{ij}=\sum _{k=1}^{s}c_{ik}c_{kj}$;
each term in the sum is unity if and only if there exists a link from
$j$ to $k$ and from $k$ to $i$, hence the sum counts the number
of walks from $j$ to $i$ of length 2. 

\begin{description}
\item [Definition~2.11:]Path.\\
A walk with all nodes distinct is a \textit{path} \\
\citep{Harary,RF,Bollobas}. 
\end{description}
In a directed graph $C$, I will say node $j$ `has access to' node
$i$, or node $i$ `has access from' node $j$, if there is a path
from node $j$ to node $i$, i.e., for some $n\ge 0$ $(C^{n})_{ij}>0$
\citep{Rothblum}. I refer to a node $i$ as being `downstream' from
a node $j$ if $j$ has access to $i$, but $i$ does not have access
to $j$. Similarly $i$ is `upstream' from $j$ if $i$ has access
to $j$, but $j$ does not have access to $i$. Thus in Figure \ref{cap:digraph}a,
node 5 is downstream from node 2, or equivalently node 2 is upstream
from node 5 because node 2 has access to 5 but not vice versa. Node
2 is neither upstream nor downstream from node 7 as neither have access
to the other. Node 8 is also neither upstream nor downstream from
7 because each has access to the other along some directed path.

\begin{description}
\item [Definition~2.12:]Cycle, $n$-cycle.\\
An $n$-\emph{cycle} is a closed walk with $n$ links, and all intermediate
nodes distinct \\
\citep{Harary,RF,Bollobas}. 
\end{description}
I will also use the term `cycle' to refer to the subgraph consisting
of the nodes and links that form the cycle. Thus any subgraph with
$n\geq 1$ nodes that contains exactly $n$ links and also contains
a closed walk that covers all $n$ nodes is an \textit{}\textit{\emph{$n$-cycle}}.
E.g., the subgraph induced by nodes 7 and 8 in Figure \ref{cap:digraph}a
is a 2-cycle. Clearly any graph that has a closed walk contains a
cycle.

\section{Connected components of a graph}

Given a directed graph $C$, its \textit{\emph{`associated undirected
graph'}} (or `symmetrized version') $C^{(s)}$ can be obtained by
adding additional links as follows: for every link $(j,i)$ in $L$,
add another link $(i,j)$ if the latter is not already in $L$. 

\begin{description}
\item [Definition~2.13:]Strongly, weakly and unilaterally connected nodes.\\
Two nodes $i$ and $j$ of a directed graph $C$ will be said to be
\textit{}\\
\textit{\emph{(i)}} \textit{strongly connected} \textit{\emph{if $i$
has access to $j$ and $j$ also has access to $i$}}\textit{,}\\
\textit{\emph{(ii)}} \textit{unilaterally connected} \textit{\emph{if
either $i$ has access to $j$, or $j$ has access to to $i$}}\textit{,}\\
\textit{\emph{(iii)}} \textit{weakly connected} if there exists a
path between them in the associated \\
undirected graph $C^{(s)}$,\\
(iv) \emph{disconnected} if none of the above is true \citep{Harary,RF}.
\end{description}
It is evident that any strongly connected nodes are also unilaterally
connected, and any unilaterally connected nodes are weakly connected,
but the converse need not be true. A graph will be termed strongly,
unilaterally, or weakly connected if \emph{all} pairs of its nodes
are, respectively, strongly, unilaterally or weakly connected. Any
directed graph can be decomposed into strongly-, unilaterally- and
weakly-connected components which are subgraphs induced by maximal
sets of strongly, unilaterally and weakly connected nodes \citep{Harary,RF}
(e.g., the graph of Figure \ref{cap:digraph}a has five weakly connected
components induced, respectively, by the nodes $\{1\},\{2,3,4,5\},\{6\}$
and $\{7,8\}$). By convention a single node is considered strongly-connected
to itself even if there is no self-link \citep{Harary,RF}. Therefore,
if a graph contains a single node with no links the single node is
also considered a (trivial) strong component of the graph.

\section{Partitioning a graph into its strong components}

The nodes of any graph can be grouped into a unique set of strong
components as follows: 

\begin{enumerate}
\item Pick any node, say $i$. Find all the nodes having access from $i$.
Denote this set by $S_{1}$; it may include $i$ itself. Similarly
find all the nodes having access to $i$. Denote this set by $S_{2}$.
Denote the subgraph induced by the set of nodes $\{i\}\cup (S_{1}\cap S_{2})$
as $C_{1}$. $C_{1}$ is one strongly connected component of the graph.
\item Pick another node that is not in $C_{1}$ and repeat the procedure
with that node to get another subgraph, $C_{2}$. The sets of nodes
comprising the two subgraphs will be disjoint.
\item Repeat this process until all nodes have been placed in some $C_{\alpha }$,
$\alpha =1,2,\ldots ,M$. Each $C_{\alpha }$ is a strong component
of the graph.
\end{enumerate}
Irrespective of which nodes are picked and in which order, this procedure
will produce for any graph a unique (upto labeling of the $C_{\alpha }$)
set of disjoint subgraphs, each of which is strongly connected, encompassing
all the nodes of the graph \citep{Harary,RF}. The graph in Figure
\ref{cap:digraph}a will decompose into 7 such subgraphs (comprising
nodes $\{1\},\{2\},\{3\},\{4\},\{5\},\{6\}$ and $\{7,8\}$).

One says there is a path from a strong component $C_{1}$ to another
strong component $C_{2}$ if there is a path in $C$ from any node
of $C_{1}$ to any node of $C_{2}$. The terms `access', `downstream'
and `upstream' can thus be used unambiguously for the $C_{\alpha }$.

\section{\noindent \label{sec:Condensation-of-a}Condensation of a graph}

\begin{description}
\item [Definition~2.14:]Condensation of a graph.\\
Determine all the strong components $C_{1},C_{2},\ldots ,C_{M}$ of
the graph as described above. Construct a new graph of $M$ nodes,
one node for each $C_{\alpha }$, $\alpha =1,\ldots ,M$. The new
graph has a directed link from $C_{\beta }$ to $C_{\alpha }$ if,
in the original graph, any node of $C_{\beta }$ has a link to any
node of $C_{\alpha }$. This new graph is called the \emph{condensation}
\emph{of the graph} $C$ \\
\citep{Harary,RF}. 
\end{description}
\noindent Figure \ref{cap:digraph}d illustrates the condensation
of Figure \ref{cap:digraph}a. Clearly the condensed graph cannot
have any closed walks. For if it were to have a closed walk then the
$C_{\alpha }$ subgraphs comprising the closed walk would together
have formed a larger strong component in the first place. Therefore
one can renumber the $C_{\alpha }$ such that if $\alpha >\beta $,
$C_{\beta }$ is never downstream from $C_{\alpha }$. Now one can
renumber the nodes of the original graph such that nodes belonging
to a given $C_{\alpha }$ are labeled by contiguous node numbers,
and whenever a pair of nodes $i$ and $j$ belong to different subgraphs
$C_{\alpha }$ and $C_{\beta }$ respectively, then $\alpha >\beta $
implies $i>j$. Such a renumbering is in general not unique, but with
any such renumbering the adjacency matrix takes the following canonical
form:

\[
C=\left(\begin{array}{cccccc}
 C_{1} &  &  &  &  & 0\\
  & C_{2} &  &  &  & \\
  &  & . &  &  & \\
  &  &  & . &  & \\
  &  &  &  & . & \\
 R &  &  &  &  & C_{M}\end{array}\right),\]
 where $0$ indicates that the upper block triangular part of the
matrix contains only zeroes while the lower block triangular part,
$R$, is not equal to zero in general. The matrix in Figure \ref{cap:digraph}b
is already in this canonical form; the dashed lines demarcate the
portions that correspond to the $C_{\alpha }$.

\section{Irreducible graphs and matrices}

\begin{description}
\item [Definition~2.15:]Irreducible graph.\\
A graph is termed \textit{irreducible} if each node in the graph has
access to every other node \citep{Harary,Seneta}. 
\end{description}
\noindent The simplest irreducible subgraph is a 1-cycle. In Figure
\ref{cap:digraph}a the subgraph induced by nodes 7 and 8 is irreducible,
but the subgraph induced by nodes 2, 3, 4, and 5 is not irreducible
as there is, for example, no path from node 5 to node 2.

\begin{description}
\item [Definition~2.16:]Irreducible matrix.\\
A matrix $C$ is \textit{irreducible} if for every ordered pair of
nodes $i$ and $j$ there exists a positive integer $k$ such that
$(C^{k})_{ij}>0$ \citep{Seneta}. 
\end{description}
Thus, if a graph is irreducible then its adjacency matrix is also
\textit{\emph{irreducible}}, and vice versa. Irreducible graphs are,
by definition, strongly connected graphs. In fact all strongly connected
graphs are irreducible, except the graph consisting of a single node
with no self-link which is strongly connected (by definition) but
not irreducible. 

\section{\label{sec:PF}Perron-Frobenius eigenvectors (PFEs)}

\begin{description}
\item [Definition~2.17:]Eigenvector, eigenvalue.\\
A column vector ${\textbf {x}}=(x_{1},x_{2},\ldots ,x_{s})^{T}$ is
said to be a \emph{right eigenvector} of an $s\times s$ matrix $C$
with an eigenvalue $\lambda $ if for each $i$, $\sum _{j=1}^{s}c_{ij}x_{j}=\lambda x_{i}$.
The \emph{eigenvalues} of a matrix $C$ are roots of the \textit{\emph{characteristic}}
\textit{}\textit{\emph{equation}} of the matrix: $|C-\lambda I|=0$,
where $I$ is the identity matrix of the same dimensionality as $C$
and $|A|$ denotes the determinant of the matrix $A$ \citep{Seneta}.
\end{description}
\noindent A `left eigenvector' \emph{}is similarly defined to be a
row vector ${\textbf {x}}=(x_{1},x_{2},\ldots ,x_{s})$ such that
for each $i$, $\sum _{j=1}^{s}c_{ji}x_{j}=\lambda x_{i}$; the left
eigenvectors of $C$ are simply the transpose of the right eigenvectors
of $C^{T}$, the transpose of $C$. In this thesis I will only use
right eigenvectors and will therefore refer to them simply as eigenvectors.
Note that the set of eigenvalues corresponding to all right eigenvectors
is identical to the set of eigenvalues corresponding to the left eigenvectors.
In general a matrix will have complex eigenvalues and eigenvectors,
but an adjacency matrix of a graph has special properties, because
it is a `non-negative' matrix, i.e., it has no negative entries.

\subsection{The Perron-Frobenius theorem }

The Perron-Frobenius theorem for irreducible non-negative matrices
states \citep{Seneta}:\\
 Suppose $T$ is an $s\times s$ non-negative irreducible matrix.
Then there exists an eigenvalue $r$ such that:\\
 a) $r$ is real, $>0$;\\
 b) with $r$ can be associated strictly positive left and right eigenvectors;\\
 c) $r\geq |\lambda |$ for any eigenvalue $\lambda \ne r$;\\
 d) the eigenvectors associated with $r$ are unique to constant multiples;\\
 e) Let $B$ be any $s\times s$ non-negative matrix. If $0\leq B\leq T$
and $\beta $ is an eigenvalue of $B$, then $|\beta |\leq r$. Moreover,
$|\beta |=r$ implies $B=T$;\\
 f) $r$ is a simple root of the characteristic equation of $T$.\\

If the matrix is not irreducible a weaker form of the above theorem
holds. The Perron-Frobenius theorem for a general non-negative matrix
states \citep{Seneta}:\\
 Suppose $T$ is an $s\times s$ non-negative matrix. Then there exists
an eigenvalue $r$ such that:\\
 a') $r$ is real, $\geq 0$;\\
 b') with $r$ can be associated non-negative left and right eigenvectors;\\
 c') $r\geq |\lambda |$ for any eigenvalue $\lambda \ne r$;\\
 e') Let $B$ be any $s\times s$ non-negative matrix. If $0\leq B\leq T$
and $\beta $ is an eigenvalue of $B$, then $|\beta |\leq r$.\\

\begin{description}
\item [Definition~2.18:]Perron-Frobenius eigenvalue.\\
For any graph $C$, the eigenvalue of $C$ that is real and larger
than or equal to all other eigenvalues in magnitude will be called
the \emph{Perron-Frobenius eigenvalue} of the graph and denoted by
$\lambda _{1}(C)$. The Perron-Frobenius theorem guarantees the existence
of such an eigenvalue.
\item [Definition~2.19:]Perron-Frobenius eigenvector (PFE).\\
For any graph $C$, each eigenvector corresponding to $\lambda _{1}(C)$
consisting only of real and non-negative components will be referred
to as \emph{Perron-Frobenius eigenvector (PFE}). The Perron-Frobenius
theorem guarantees the existence of at least one such eigenvector. 
\end{description}
The Perron-Frobenius eigenvalue of the graph in Figure \ref{cap:digraph}a
is 1 and $\mathbf{x}=(0,0,0,0,0,0,1,1)^{T}$ is a Perron-Frobenius
eigenvector. 

The presence or absence of closed walks in a graph can be determined
from the Perron-Frobenius eigenvalue of its adjacency matrix:

\begin{description}
\item [Proposition~2.1:]\textit{\emph{If a graph, $C$,}}\emph{}\\
 \emph{}(i) \emph{}\textit{\emph{has no closed walk then $\lambda _{1}(C)=0$}}\emph{,}\\
 \emph{}(ii) \emph{}\textit{\emph{has a closed walk then $\lambda _{1}(C)\geq 1$.}}
\end{description}
The proof of this and all subsequent propositions can be found in
appendix A. 

Note that $\lambda _{1}$ cannot take values between zero and one
because of the discreteness of the entries of $C$, which are either
zero or unity.

\subsection{Basic subgraphs}

In section \ref{sec:Condensation-of-a} it was shown that the adjacency
matrix of any graph can always be written in the following canonical
form by an appropriate renumbering of the nodes:

\[
C=\left(\begin{array}{cccccc}
 C_{1} &  &  &  &  & 0\\
  & C_{2} &  &  &  & \\
  &  & . &  &  & \\
  &  &  & . &  & \\
  &  &  &  & . & \\
 R &  &  &  &  & C_{M}\end{array}\right).\]

From the above form of $C$ it follows that \[
|C-\lambda I|=|C_{1}-\lambda I|\times |C_{2}-\lambda I|\times \ldots \times |C_{M}-\lambda I|.\]
 Therefore the set of eigenvalues of $C$ is the union of the sets
of eigenvalues of $C_{1},\ldots ,C_{M}$, which implies that $\lambda _{1}(C)={\textrm{max}}_{\alpha }\{\lambda _{1}(C_{\alpha })\}$.
Thus, if a given graph $C$ has a Perron-Frobenius eigenvalue $\lambda _{1}$
then it contains at least one strong component with Perron-Frobenius
eigenvalue $\lambda _{1}$. 

\begin{description}
\item [Definition~2.20:]Basic subgraph.\\
Each strong components of $C$ with Perron-Frobenius eigenvalue equal
to $\lambda _{1}(C)$ is referred to as a \textit{basic} \emph{subgraph}
\citep{Rothblum}. 
\end{description}
The shaded node in Figure \ref{cap:digraph}d corresponds to the only
basic subgraph of Figure \ref{cap:digraph}a. 

\begin{description}
\item [Proposition~2.2:]\textit{\emph{If all the basic subgraphs of a
graph $C$ are cycles then $\lambda _{1}(C)=1$}}, and vice versa. \emph{}
\end{description}
\begin{figure}
\includegraphics[  width=12cm,
  keepaspectratio]{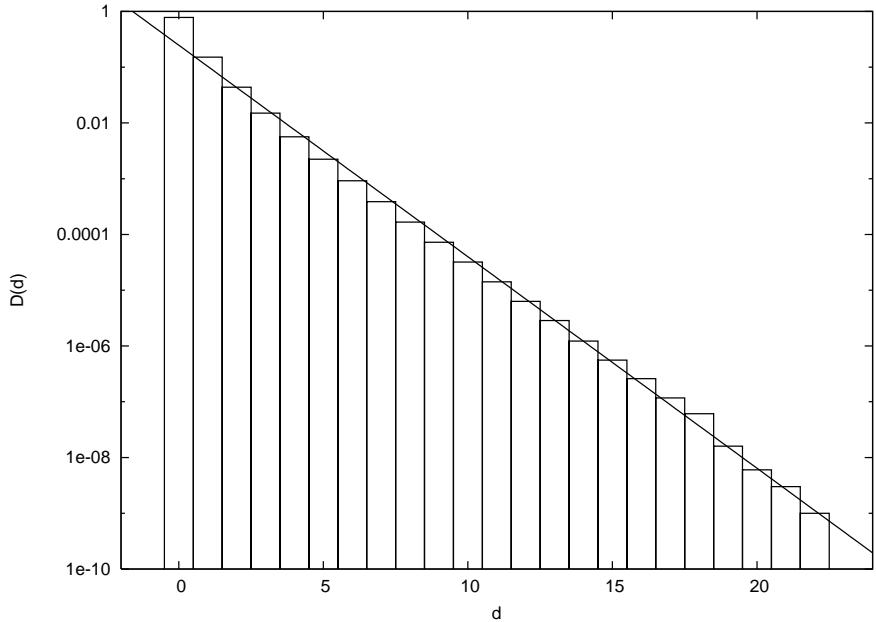}

\caption{Dependency distribution of nodes of $10^{7}$ random graphs from
the ensemble $G_{s}^{p}$ with $s=100,p=0.0025$. The average dependency
is 0.33 and the standard deviation of dependency values is 0.76. The
straight line plots the function $A\exp (-\alpha d)$ with $A=0.25,\alpha =-0.87$.
This is the best straight line fit to the dependency distribution
on a semi-log plot.\label{cap:depdistrnd}}\lyxline{\normalsize}

\end{figure}

\section{\label{sec:Random-graphs}Random graphs}

A random graph is an ensemble of graphs with each member of the ensemble
having an associated probability (introduced by \citealp{ER1}). Two
ensembles have been extensively studied. One, denoted $G_{s}^{l}$,
is the ensemble of graphs with $s$ vertices and $l$ links (and no
self-links) with a uniform associated probability. The second is the
ensemble of graphs, denoted $G_{s}^{p}$, having $s$ vertices, with
each possible link (except self-links) present with probability $p$;
in other words the ensemble of all graphs with the probability associated
with a graph being $p^{l}(1-p)^{s(s-1)-l}$ if it has $l$ links.
These two ensembles have the same properties in the $s\rightarrow \infty $
limit if $l=ps(s-1)$ \citep{Bollobas,Bollo}. For the random graph
$G_{s}^{p}$ the average number of links is $ps(s-1)$, while the
(total) degree distribution is binomial: $P(k)=\: ^{2s-2}C_{k}\, p^{k}(1-p)^{2s-2-k}$
\citep{Bollobas,Bollo}. Figure \ref{cap:depdistrnd} shows the dependency
distribution of nodes of $10^{7}$ graphs picked from the ensemble
$G_{s}^{p}$ with $s=100,p=0.0025$ (the program \texttt{rndgraph.cpp}
that is included in the attached CD, see appendix C, was used to generate
the random graphs). The dependency averaged over all nodes was 0.33,
and the standard deviation of dependency values was 0.76. The distribution
appears to decline exponentially as a function of dependency value,
but much slower than would be expected for a Poisson distribution
with the same mean. See \citep{Bollobas,Bollo} for further results
concerning random graphs.

\chapter{\label{cha:Autocatalytic-Sets}Autocatalytic Sets}

This chapter discusses the notion of an `autocatalytic set', which
will play an important role in the dynamics of the model studied in
this thesis. After first providing a graph-theoretic definition of
an autocatalytic set, I discuss its relationship with the Perron-Frobenius
eigenvectors of a graph. The rest of the chapter analyzes the structure
of PFEs of different graphs, both ones which contains autocatalytic
sets and ones which do not. The results derived here will prove useful
in the analysis of the dynamical system discussed in chapter \ref{cha:Population-Dynamics}.

\section{\label{sec:Autocatalytic-sets-(ACSs)}Autocatalytic sets (ACSs)}

\begin{description}
\item [Definition~3.1:]Autocatalytic set (ACS).\\
An \textit{autocatalytic set} is a graph, each of whose nodes has
at least one incoming link from a node belonging to the same graph
\citep{JKauto}.
\end{description}
The concept of an ACS was introduced in the context of a set of catalytically
interacting molecules where it was defined to be a set of molecular
species that contains, within itself, a catalyst for each of its member
species \citep{Eigen,Kauffman1,Rossler}. Such a set of molecular
species can collectively self-replicate under certain circumstances
even if none of its component molecular species can individually self-replicate.
This definition matches the one above if you imagine a node in a directed
graph to represent a molecular species and a link from $j$ to $i$
to signify that $j$ is a catalyst for $i$. 

\begin{figure}
\begin{center}\includegraphics{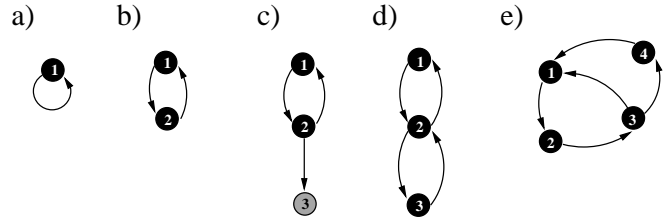}\end{center}

\caption{Examples of autocatalytic sets (ACSs). \textbf{a.} A 1-cycle, the
simplest ACS. \textbf{b.} A 2-cycle. \textbf{c.} An ACS that is not
an irreducible graph. \textbf{d,e} Examples of ACSs that are irreducible
graphs but not cycles.\label{cap:ACSegs}}\lyxline{\normalsize}

\end{figure}

Figure \ref{cap:ACSegs} shows various ACSs. The simplest is a 1-cycle;
Figure \ref{cap:ACSegs}a. Figures \ref{cap:ACSegs}a and \ref{cap:ACSegs}b
are graphs that are irreducible as well as cycles, \ref{cap:ACSegs}c
is an ACS that is not an irreducible graph and hence not a cycle,
while \ref{cap:ACSegs}d and \ref{cap:ACSegs}e are irreducible graphs
that are not cycles. 

\begin{description}
\item [Proposition~3.1:](i) All cycles are irreducible graphs and all
irreducible graphs are ACSs.\\
(ii) Not all ACSs are irreducible graphs and not all irreducible graphs
are cycles.
\end{description}

\section{Relationship between Perron-Frobenius eigenvectors and autocatalytic
sets}

The ACS is a useful graph-theoretic construct in part because of its
connection with the PFE. Firstly, the Perron-Frobenius eigenvalue
of a graph is an indicator of the existence of an ACS in the graph
\citep{JKauto,JKemerg}:

\begin{description}
\item [Proposition~3.2:](i) \textit{\emph{An ACS must contain a}} closed
walk\emph{.} Consequently,\\
 (ii) \textit{\emph{If a graph $C$ has no ACS then}} \textit{$\lambda _{1}(C)=0$}.\\
 (iii) \textit{\emph{If a graph $C$ has an ACS then}} \textit{$\lambda _{1}(C)\geq 1$}.
\end{description}
\begin{figure}
\begin{center}\includegraphics{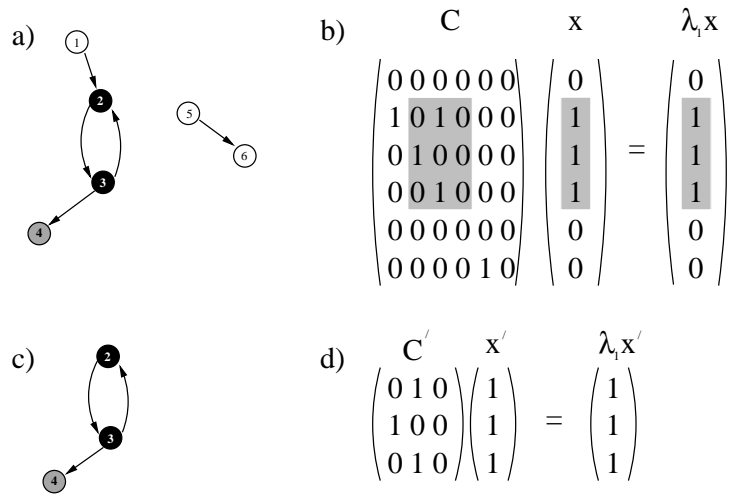}\end{center}

\caption{\textbf{a.} A graph with 6 nodes. \textbf{b.} \textbf{x} is an eigenvector
of its adjacency matrix $C$ with eigenvalue $\lambda _{1}=1$, which
is the Perron-Frobenius eigenvalue of the graph. The non-zero components
of ${\textbf {x}}$ and the corresponding rows and columns of $C$
are shaded. \textbf{c.} The subgraph of the PFE ${\textbf {x}}$.
The subgraph is an ACS and contains one of the basic subgraphs of
the graph in (a). \textbf{d.} The vector \textbf{x}' constructed by
removing the zero components of ${\textbf {x}}$ is an eigenvector
of the adjacency matrix, $C'$, of the PFE subgraph. Its corresponding
eigenvalue is unity, which is also the Perron-Frobenius eigenvalue
of the PFE subgraph.\label{cap:PFEsubgraph}}\lyxline{\normalsize}

\end{figure}

\noindent Let ${\textbf {x}}$ be a PFE of a graph. Consider the set
of all nodes $i$ for which $x_{i}$ is non-zero. I will call the
subgraph induced by these nodes the `subgraph of the PFE \textbf{x}'.
For example consider the graph in Figure \ref{cap:PFEsubgraph}a.
For this graph $\lambda _{1}=1$. Figure \ref{cap:PFEsubgraph}b shows
a PFE of the graph and how it satisfies the eigenvalue equation. For
this PFE, nodes 1, 5 and 6 have $x_{i}=0$. Removing these nodes produces
the PFE subgraph shown in Figure \ref{cap:PFEsubgraph}c. If all the
components of the PFE are non-zero then the subgraph of the PFE is
the entire graph. One can show that \citep{JKauto,JKemerg}:

\begin{description}
\item [Proposition~3.3:]\textit{\emph{If $\lambda _{1}(C)\ge 1$, then
the subgraph of any PFE of $C$ is an ACS}}\emph{.}
\end{description}
For the PFE of Figure \ref{cap:PFEsubgraph}b this is immediately
verified by inspection. Note that this result relates an algebraic
property of a graph, its PFE, to a topological structure, an ACS.
Further, this result is not true if we considered irreducible graphs
instead of ACSs. 

\begin{description}
\item [Proposition~3.4:]Let ${\textbf {x}}$ be a PFE of a graph $C$,
and let $C'$ denote the adjacency matrix of the subgraph of ${\textbf {x}}$.
Let $\lambda _{1}(C')$ denote the Perron-Frobenius eigenvalue of
$C'$. Then $\lambda _{1}(C')=\lambda _{1}(C)$ and $C'$ must contain
at least one of the basic subgraphs of $C$.
\end{description}
Figure \ref{cap:PFEsubgraph} illustrates this point. The adjacency
matrix of the PFE subgraph, $C'$, is obtained by removing rows 1,
5, 6 and columns 1, 5, 6 from the original matrix. Figure \ref{cap:PFEsubgraph}d
illustrates that the vector constructed by removing the zero components
of the PFE is an eigenvector of $C'$ with eigenvalue 1. $C'$ contains
one basic subgraph of $C$ -- the one induced by the nodes 2 and 3. 

\begin{description}
\item [Definition~3.2:]Simple PFE.\\
If the subgraph of a PFE contains only one basic subgraph then the
PFE is termed a \emph{simple PFE} \citep{JKwiley}.
\end{description}
The PFE in Figure \ref{cap:PFEsubgraph} is simple.

\section{Eigenvector profile theorem}

I now present a series of propositions that describe some aspects
of the `profile' of possible PFEs of a graph, i.e., which components
of the PFE are zero and which are non-zero. These propositions lead
toward the eigenvector profile theorem which provides an algorithm
for finding the profile of all the simple PFEs of a graph, and consequently
the profile of the possible non-simple PFEs also. This theorem is
completely general, applying to any given graph. After I had generated
the proofs for the theorem and the propositions it depends upon, I
found that these results had already been derived by \cite{Rothblum}
as part of a more general theorem concerning the profile of generalized
eigenvectors of a non-negative matrix. The proofs provided in appendix
A (Krishna and Jain, forthcoming) are, however, different from his;
they make no use of the principle of mathematical induction, which
Rothblum uses extensively.

For convenience I will often abbreviate statements like `there exists
a PFE in which the components corresponding to the nodes in subgraph
$C_{1}$are non-zero/zero' to `there exists a PFE in which the nodes
in subgraph $C_{1}$ are non-zero/zero'. 

\begin{description}
\item [Proposition~3.5:]\textit{\emph{If a node is non-zero in a PFE then
all nodes it has access to are non-zero in that PFE.}}
\end{description}
It follows that if any node of an irreducible subgraph is non-zero
in a PFE then all the other nodes in that subgraph must be non-zero
in that PFE. A situation where some nodes of an irreducible subgraph
are non-zero and some are zero in the same PFE cannot happen.

\begin{description}
\item [Proposition~3.6:]\textit{\emph{If a node is upstream from a basic
subgraph then that node is zero in any PFE.}} \textit{}
\item [Proposition\textit{~}\textit{\emph{3.7:}}]\textit{\emph{If a node
does not have access from a basic subgraph then it is zero in any
PFE.}}
\end{description}
Propositions 3.6 and 3.7 describe which nodes will be zero in the
PFEs, while proposition 3.4 and 3.5 describe which nodes will be non-zero
-- at least one of the basic subgraphs will be non-zero in every PFE
along with all nodes it has access to. In other words the subgraph
of any PFE consists of one or more of the basic subgraphs and those
nodes that they have access to, i.e., the subgraph of any PFE is necessarily
an ACS (recall proposition 3.3, this is an alternate proof of that
proposition).

Propositions 3.4--3.7 can be used to prove the following theorem about
the profile of PFEs of any graph:

\begin{description}
\item [Theorem~3.1:]Eigenvector profile theorem\\
For any graph $C$, determine all the basic subgraphs of $C$. Denote
them by $D_{1},\ldots ,D_{K}$. Determine which of these does not
have any other $D_{i}$ downstream from it. Denote these by $E_{1},\ldots ,E_{N}$.
Then: \\
\textit{\emph{(i) For each $i=1,\ldots ,N$ there exists a unique
(upto constant multiples) PFE in which the nodes of $E_{i}$ and all
nodes having access from them are non-zero and all other nodes are
zero. It is evident that these PFEs are simple. Moreover these are
the only simple PFEs of the graph $C$. }}\\
\textit{\emph{(ii) Any PFE is a linear combination of these $N$ simple
PFEs.}}
\end{description}

\section{\noindent \label{sec:Core-and-periphery}Core and periphery of a
simple PFE}

\begin{description}
\item [Definition~3.3:]Core and periphery of a simple PFE.\\
If $C$ is the subgraph of a simple PFE, the basic subgraph contained
in $C$ will be called the \textit{core} of the simple PFE (or equivalently,
the `core of $C$'), and denoted $Q$. The set of the remaining nodes
and links of $C$ that are not in its core will together be said to
constitute the \textit{periphery} of the simple PFE \citep{JKwiley}. 
\end{description}
For example, for the PFE in Figure \ref{cap:PFEsubgraph}b the core
is the 2-cycle comprising nodes 2 and 3. In Figure \ref{cap:PFEsubgraph}
the core nodes are black and the periphery nodes grey. Note that the
periphery is not a subgraph because it contains links not just between
periphery nodes but also from nodes outside the periphery (like the
link from node 3 to 4 in Figure \ref{cap:PFEsubgraph}c). 

\begin{figure}
\begin{center}\includegraphics{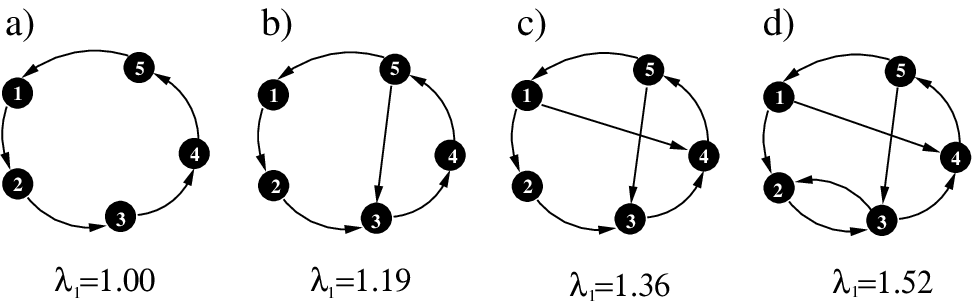}\end{center}

\caption{$\lambda _{1}$ is a measure of the multiplicity of internal pathways
in the core of a simple PFE. Four irreducible graphs are shown. An
irreducible graph always has a unique PFE that is simple and whose
core is the entire graph. The Perron-Frobenius theorem ensures that
adding a link to the core of a simple PFE necessarily increases its
Perron-Frobenius eigenvalue $\lambda _{1}$. \label{cap:lambda_core}}\lyxline{\normalsize}

\end{figure}

It follows from the Perron-Frobenius theorem for irreducible graphs
that $\lambda _{1}(Q)$ will necessarily increase if any link is added
to the core. Similarly removing any link will decrease $\lambda _{1}(Q)$.
Thus $\lambda _{1}$ measures the multiplicity of internal pathways
in the core. Figure \ref{cap:lambda_core} illustrates this point.

The core and periphery can be shown to have the following topological
property \citep{JKwiley}:

\begin{description}
\item [Proposition~3.8:]\textit{\emph{Every node in the core of a simple
PFE has access to every other node of the PFE subgraph. No periphery
node has access to any core node.}} 
\end{description}
\noindent Thus starting from the core one can reach the periphery
but not vice versa.

\section{\noindent Core and periphery of a non-simple PFE }

\noindent Because any PFE of a graph can be written as a linear combination
of the simple PFEs (which, upto constant multiples, are unique for
any graph), the definitions of core and periphery can be readily extended
to any PFE as follows: 

\begin{description}
\item [Definition~3.4:]Core and periphery of a (non-simple) PFE.\\
The \textit{core of a PFE}, denoted $Q$, is the union of the cores
of those simple PFEs whose linear combination forms the given PFE.
The rest of the nodes and links of the PFE subgraph constitute its
\emph{periphery} \citep{JKwiley}. 
\end{description}
\noindent It follows from the above discussion that $\lambda _{1}(Q)=\lambda _{1}(C)$.
When the core is a union of disjoint cycles then $\lambda _{1}(Q)=1$,
and vice versa (recall proposition 2.2).

\begin{figure}
\begin{center}\includegraphics{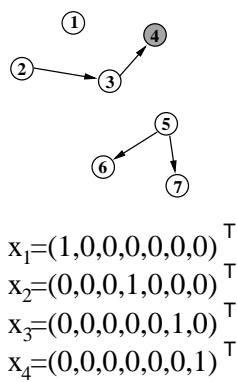}\end{center}

\caption{A graph that has no ACS. The graph has four simple PFEs (upto constant
multiples).\label{cap:noACSeg}}
\end{figure}

\begin{figure}
\begin{center}\includegraphics{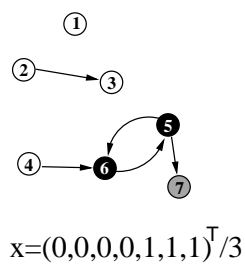}\end{center}

\caption{A graph that has an ACS, but only one basic subgraph, along with
various non-ACS structures. The only simple PFE of the graph is also
displayed.\label{cap:simpleACSeg}}
\end{figure}

\begin{figure}
\begin{center}\includegraphics{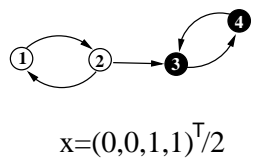}\end{center}

\caption{One 2-cycle downstream from another. Both 2-cycles are basic subgraphs.
The graph has one simple PFE in which only the downstream cycle is
non-zero.\label{cap:two2cycles}}
\end{figure}

\section{The profile of PFEs when there is no ACS}

\noindent If $\lambda _{1}(C)=0$, the graph has no ACS, then the
profile of PFEs is as follows: for every node that has no outgoing
links, there is a PFE with that node non-zero and all other nodes
zero. For example the graph in Figure \ref{cap:noACSeg} has four
(upto constant multiples) simple PFEs that are displayed in the same
figure. This follows from theorem 3.1 because for a graph with $\lambda _{1}=0$
each node is a basic subgraph. A general PFE is a linear combination
of all such simple PFEs. Because there is no closed walk there is
no core (or periphery) for any PFE of the graph. The core of all PFEs
of such a graph may be defined to be the null set, $Q=\Phi $.

\section{\noindent The profile of PFEs when there is an ACS but only one basic
subgraph}

Now consider a graph that has $\lambda _{1}\ge 1$, i.e., it contains
an ACS, but has only one basic subgraph. In addition to the ACS it
could possibly have various non-ACS structures such as chains, trees
and isolated nodes. An example is the graph in Figure \ref{cap:simpleACSeg}.
For such a graph theorem 3.1 says that there is only one simple PFE
in which the nodes in the basic subgraph and all nodes having access
from it are non-zero, while the other nodes are zero (displayed in
Figure \ref{cap:simpleACSeg}). In particular the nodes in all non-ACS
structures are zero in the (unique) PFE of the graph because they
are don't have access from the basic subgraph. The core nodes (nodes
of the single basic subgraph) are black and the periphery nodes are
grey in Figure \ref{cap:simpleACSeg}. The white nodes are the nodes
that are zero in the PFE of the graph.

\section{The profile of PFEs when there is an ACS and many basic subgraphs}

For any graph containing an ACS, i.e., one that has $\lambda _{1}\ge 1$,
there are possibly several simple PFEs as specified by theorem 3.1.
However it follows from the theorem that in all the simple PFEs the
nodes of non-ACS structures will be zero because they are not downstream
from any basic subgraph. 

While the nodes of non-ACS structures are guaranteed to be zero, it
can also happen that the nodes of some basic subgraphs are zero in
every PFE. Figure \ref{cap:two2cycles} illustrates such a case where
the graph consists of one 2-cycle downstream from another. The Perron-Frobenius
eigenvalue of this graph is $\lambda _{1}=1$, and each 2-cycle is
a basic subgraph. From theorem 3.1 it follows that this graph has
only one (simple) PFE (displayed in Figure \ref{cap:two2cycles})
in which the upstream 2-cycle is zero.

\chapter{\label{cha:Population-Dynamics}Population Dynamics}

\section{\label{sec:The-population-dynamics}The population dynamics equation}

In this chapter, I will analyze the attractors of the dynamical system
described by the equation: \begin{equation}
\dot{x}_{i}=\sum _{j=1}^{s}c_{ij}x_{j}-x_{i}\sum _{k,j=1}^{s}c_{kj}x_{j},\label{xdot}\end{equation}

where $\mathbf{x}$ lies on the simplex $J=\{{\textbf {x}}\equiv (x_{1},x_{2},\ldots ,x_{s})^{T}\in {\textbf {R}}^{s}|0\leq x_{i}\leq 1,\sum _{i=1}^{s}x_{i}=1\}$
of normalized non-negative vectors in $s$ dimensions.

As discussed in section \ref{sec:Framework-of-a}, the set of variables
$x_{i}$ can be thought of as `living' on the nodes of a graph whose
adjacency matrix is $C=(c_{ij})$. The links of the graph represent
the interactions between the variables $x_{i}$. It is useful to see
how equation (\ref{xdot}) arises in a chemical context.

Let $i\in \{1,\ldots ,s\}$ denote a chemical (or molecular) species
in a well-stirred chemical reactor. Molecules can react with one another
in various ways; I focus on only one aspect of their interactions:
catalysis. The catalytic interactions can be described by a directed
graph with $s$ nodes. The nodes represent the $s$ species and the
existence of a link from node $j$ to node $i$ means that species
$j$ is a catalyst for the production of species $i$. In terms of
the adjacency matrix, $C=(c_{ij})$ of this graph, $c_{ij}$ is set
to unity if $j$ is a catalyst of $i$ and is set to zero otherwise.
The operational meaning of catalysis is as follows:

Each species $i$ will have an associated non-negative population
$y_{i}$ in the reactor that changes with time. Let species $j$ catalyze
the ligation of reactants $A$ and $B$ to form the species $i$,
$A+B\stackrel{j}{\rightarrow }i$. Assuming that the rate of this
catalyzed reaction is given by the Michaelis-Menten theory of enzyme
catalysis, $\dot{y}_{i}=V_{max}ab\frac{y_{j}}{K_{M}+y_{j}}$ \citep{Gutfreund},
where $a,b$ are the reactant concentrations, and $V_{max}$ and $K_{M}$
are constants that characterize the reaction. If the Michaelis constant
$K_{M}$ is very large this can be approximated as $\dot{y}_{i}\propto y_{j}ab$.
Combining the rates of the spontaneous and catalyzed reactions and
also putting in a dilution flux $\phi $, the rate of growth of species
$i$ is given by $\dot{y}_{i}=k(1+\nu y_{j})ab-\phi y_{i}$, where
$k$ is the rate constant for the spontaneous reaction, and $\nu $
is the catalytic efficiency. Assuming the catalyzed reaction is much
faster than the spontaneous reaction, and that the concentrations
of the reactants are non-zero and fixed, the rate equation becomes
$\dot{y_{i}}=Ky_{j}-\phi y_{i}$, where $K$ is a constant. In general
because species $i$ can have multiple catalysts, $\dot{y}_{i}=\sum _{j=1}^{s}K_{ij}y_{j}-\phi y_{i}$,
with $K_{ij}\sim c_{ij}$. I make the further idealization $K_{ij}=c_{ij}$
giving:\begin{equation}
\dot{y}_{i}=\sum _{j=1}^{s}c_{ij}y_{j}-\phi y_{i}.\label{ydot}\end{equation}

The relative population of species $i$ is by definition $x_{i}\equiv y_{i}/\sum _{j=1}^{s}y_{j}$.
As $0\leq x_{i}\leq 1,\sum _{i=1}^{s}x_{i}=1$, ${\textbf {x}}\equiv (x_{1},\ldots ,x_{s})^{T}\in J$.
Taking the time derivative of $x_{i}$ and using (\ref{ydot}) it
is easy to see that $\dot{x}_{i}$ is given by (\ref{xdot}). Note
that the $\phi $ term, present in (\ref{ydot}), cancels out and
is absent in (\ref{xdot}).

Because we ultimately ignore the concentrations of the reactants we
would get the same rate equations even if the reaction scheme were
different, e.g., if only one reactant was required as in the reaction
$A\stackrel{j}{\rightarrow }i$. Such dynamics might also be relevant
in an economic context where the presence of one commodity increases
the rate of production of another commodity.

Equation (\ref{xdot}) has been used by \citet{EMS} to describe the
dynamics of the relative populations of self-replicating strings.
Each node of $C$ represents one string. Two strings are connected
by an undirected link if mutations of one string can produce the other;
the strength of the link being a decreasing function of the number
of mutations required. Eigen et al. have determined the attractor
of equation (\ref{xdot}) for the specific form this interpretation
imposes on $C$. Below, I analyze the attractors of equation (\ref{xdot})
for an arbitrary graph $C$, with no restrictions placed on its structure.

\section{Attractors of the population dynamics equation }

An important motivation for the choice of the population dynamics
(\ref{xdot}) is it's analytical tractability. The following properties
of the attractor can be derived (\citealp{JKwiley}; Krishna and Jain,
forthcoming):

\begin{description}
\item [Proposition~4.1:]\textit{\emph{For any graph $C$,}} \emph{}\\
 \emph{}(i) \emph{}\textit{\emph{Every eigenvector of $C$ that belongs
to the simplex $J$ is a fixed point of equation (\ref{xdot}), and
vice versa.}}\emph{}\\
 (ii) \emph{}\textit{\emph{Starting from any initial condition in
the simplex $J$, the trajectory converges to some fixed point (generically
denoted ${\textbf {X}}$) in $J$.}}\emph{}\\
 (iii) \emph{}\textit{\emph{For generic initial conditions in $J$,
${\textbf {X}}$ is a Perron-Frobenius eigenvector (PFE) of $C$.}}\emph{}\\
(iv) \emph{}\textit{\emph{If $C$ has a unique (upto constant multiples)
PFE, it is the unique stable attractor of equation (\ref{xdot}). }}
\newpage
\item [~]~\\
(v) \textit{\emph{If $C$ has more than one linearly independent PFE,
then ${\textbf {X}}$ can depend upon the initial conditions. The
set of allowed ${\textbf {X}}$ is a linear combination of a subset
of the PFEs.}} The interior of this set in $J$ may then be said to
be the `attractor' of (\ref{xdot}), in the sense that for generic
initial conditions all trajectories converge to a point in this set. 
\end{description}
Statement (i) is easy to see: let ${\textbf {x}}^{\lambda }\in J$
be an eigenvector of $C$, $\sum _{j}c_{ij}x_{j}^{\lambda }=\lambda x_{i}^{\lambda }$.
Substituting this on the r.h.s. of (\ref{xdot}), one gets zero. Conversely,
if the r.h.s. of (\ref{xdot}) is zero, ${\textbf {x}}={\textbf {x}}^{\lambda }$,
with $\lambda =\sum _{k,j}c_{kj}x_{j}$.

Appendix A provides proof of the above propositions but they can be
motivated by considering the underlying dynamics (\ref{ydot}) from
which (\ref{xdot}) is derived. Because (\ref{xdot}) is independent
of $\phi $, we can set $\phi =0$ in (\ref{ydot}) without any loss
of generality. With $\phi =0$ the general solution of (\ref{ydot}),
which is a linear system, can be schematically written as: \[
{\textbf {y}}(t)=e^{Ct}{\textbf {y}}(0),\]
 where ${\textbf {y}}(0)$ and ${\textbf {y}}(t)$ are viewed as column
vectors. Suppose ${\textbf {y}}(0)$ is a right eigenvector of $C$
with eigenvalue $\lambda $, denoted ${\textbf {y}}^{\lambda }$.
Then \[
{\textbf {y}}(t)=e^{\lambda t}{\textbf {y}}^{\lambda }.\]
 As this time dependence is merely a rescaling of the eigenvector,
this is an alternative way of showing that ${\textbf {x}}^{\lambda }={\textbf {y}}^{\lambda }/\sum _{j=1}^{s}y_{j}^{\lambda }$
is a fixed point of (\ref{xdot}). If the eigenvectors of $C$ form
a basis in $R^{s}$, ${\textbf {y}}(0)$ is a linear combination:
${\textbf {y}}(0)=\sum _{\lambda }a_{\lambda }{\textbf {y}}^{\lambda }$.
In that case, for large $t$ it is clear that the term with the largest
value of $\mathrm{Re}(\lambda )$ will grow fastest, hence, \[
{\textbf {y}}(t)\stackrel{t\rightarrow \infty }{\sim }e^{\lambda _{1}t}{\textbf {y}}^{\lambda _{1}},\]
 where $\lambda _{1}$ is the eigenvalue of $C$ with the largest
real part (which is the same as its Perron-Frobenius eigenvalue) and
${\textbf {y}}^{\lambda _{1}}$ an associated eigenvector. Therefore,
for generic initial conditions the trajectory of (\ref{xdot}) will
converge to ${\textbf {X}}={\textbf {x}}^{\lambda _{1}}$, a PFE of
$C$. If the eigenvectors of $C$ do not form a basis in $R^{s}$,
the above result is still true (see appendix A).

Note that $\lambda _{1}$ can be interpreted as the `population growth
rate' at large $t$, because $\dot{\textbf {y}}(t)\stackrel{t\rightarrow \infty }{\sim }\lambda _{1}{\textbf {y}}$.
In section \ref{sec:Core-and-periphery}, I had mentioned that $\lambda _{1}$
measures a topological property of the graph, the multiplicity of
internal pathways in the core of the graph. Thus, in the present model,
$\lambda _{1}$ has both a topological and dynamical significance,
which relates two distinct properties of the system, one structural
(multiplicity of pathways in the core of the graph), and the other
dynamical (population growth rate). The higher the multiplicity of
pathways in the core, the greater is the population growth rate of
the dominant ACS.

\newpage
\section{Attractor profile theorem}

While any eigenvector of $C$ is a fixed point it need not be stable.
In other words, only for special initial conditions, forming a space
of measure zero in $J$, can ${\textbf {X}}$ be some other eigenvector
of $C$, not a PFE. Henceforth, I ignore such special initial conditions.
However, not all PFEs are attractors either. The following theorem
indicates which subset of PFEs form the attractor set (Krishna and
Jain, forthcoming):

\begin{description}
\item [Theorem~4.1:]Attractor profile theorem\\
 1) Determine all the strong components of the given graph $C$. Denote
these by $C_{1},\ldots ,C_{M}$. \\
 2) Determine which of these are basic subgraphs. Denote them by $D_{1},\ldots ,D_{K}$.\\
3) Construct a graph, denoted $D^{*}$, with $K$ nodes, representing
the $D_{i}$, and a link from node $j$ to node $i$ if there is a
path from any node of $D_{j}$ to any node of $D_{i}$ that does not
contain a node of any other basic subgraph.\\
 4) Determine which of the $D_{i}$ are at the ends of the longest
paths in the above graph $D^{*}$. Denote these by $F_{i},i=1,\ldots ,N$.\textit{}\\
\textit{\emph{For each $i=1,\ldots ,N$ there exists a unique (upto
constant multiples) PFE that has only nodes of $F_{i}$ and all nodes
having access from them non-zero, and all other nodes zero. The attractor
set consists of all linear combinations of these PFEs that lie on
the simplex $J$.}}
\end{description}

\section{\label{sec:The-attractor-of}The attractor for a graph with no ACS}

\begin{figure}
\begin{center}\includegraphics{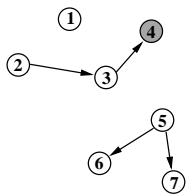}\end{center}

\caption{A graph that has no ACS. Only node 4, coloured grey, is non-zero
in the attractor.\label{cap:noACSeg2}}\lyxline{\normalsize}

\end{figure}

For any graph that consists of only chains, trees and isolated nodes,
i.e. any graph with $\lambda _{1}=0$, the theorem reduces to the
following \citep{JKauto,JKemerg}:

\begin{description}
\item [Proposition~4.2:]For any graph with $\lambda _{1}(C)=0$, in the
attractor only the nodes at the ends of the longest paths are non-zero.
All other nodes are zero.
\end{description}
This follows from theorem 4.1 because for a graph with $\lambda _{1}=0$
each node is a basic subgraph. Appendix A has an alternate proof based
directly on the underlying $y_{i}$ dynamics. Figure \ref{cap:noACSeg2}
shows a graph that has $\lambda _{1}=0$. The white nodes are those
that are zero in the attractor and the grey node is the only non-zero
node in the attractor. This graph is the same as the one in Figure
\ref{cap:noACSeg}. Notice that of the four PFEs of the graph only
one, $(0,0,0,1,0,0,0)^{T}$, is the attractor.

\section{The dominant ACS of a graph}

For graphs with an ACS, i.e., with $\lambda _{1}\ge 1$, recall that
proposition 3.3 showed a close relationship between the PFEs of the
graph and ACSs. A similar result holds for the attractor \citep{JKauto,JKemerg}:

\begin{description}
\item [Proposition~4.3:]\textit{\emph{For any graph $C$,}} \emph{}\\
 \emph{}(i) \emph{}\textit{\emph{For every ${\textbf {X}}$ belonging
to the attractor set, the set of nodes $i$ for which $X_{i}>0$ is
the same and is uniquely determined by $C$. The subgraph formed by
this set of nodes will be called the `subgraph of the attractor' of
(\ref{xdot}) for the graph $C$.}} \\
 (ii) \textit{\emph{If $\lambda _{1}(C)\ge 1$, the subgraph of the
attractor is an ACS.}} \emph{}
\end{description}

\begin{description}
\item [Definition~4.1:]Dominant ACS.\\
For any graph with $\lambda _{1}\ge 1$ the subgraph of the attractor,
which is an ACS, will be called the \textit{dominant ACS} of the graph
\citep{JKauto,JKemerg}. 
\end{description}
\noindent The dominant ACS is independent of (generic) initial conditions
and is \textit{\emph{the subgraph induced by the set of nodes belonging
to all the $F_{i}$ and all nodes having access from them in $C$.
The following sections use theorem 4.1 to determine the structure
of the dominant ACS of several different types of graphs.}}

\section{\label{sec:Examples-of-the}Examples of the attractor for specific
graphs}

It is instructive to consider examples of graphs and see how the trajectory
converges to a PFE, and what the dominant ACS looks like in each case.\\

\smallskip{}

\begin{figure}
\begin{center}\includegraphics[  width=8cm,
  keepaspectratio]{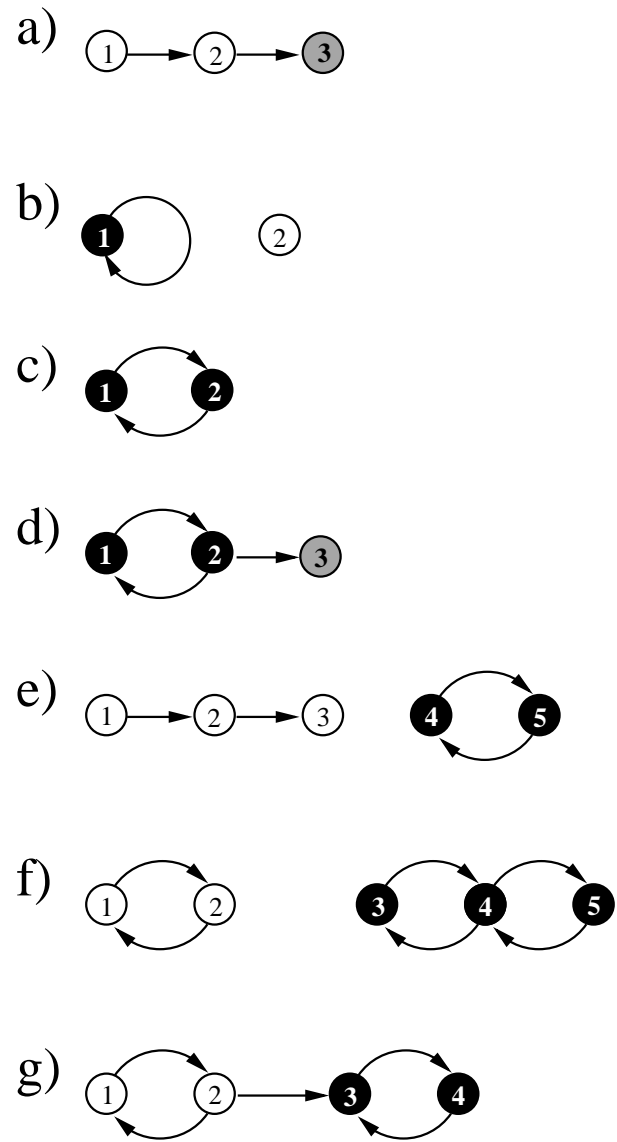}\end{center}

\caption{Examples of graphs with a unique PFE (upto constant multiples), which
is, therefore, the unique stable attractor. The coloured nodes show
the subgraph of the PFE, i.e., the nodes that are non-zero in the
attractor. The black nodes are core nodes, the grey nodes are periphery
nodes and the white nodes are the nodes that are zero in the attractor.\label{cap:uniquePFEfig}}
\end{figure}

\noindent \textbf{Example 1}. A simple chain, Figure \ref{cap:uniquePFEfig}a:\\
 The adjacency matrix of this graph has all eigenvalues zero; $\lambda _{1}=0$.
The unique PFE (upto constant multiples) of this graph is ${\textbf {e}}=(0,0,1)^{T}$.
Because node 1 has no catalyst, its rate equation is (henceforth taking
$\phi =0$) $\dot{y_{1}}=0$. Therefore $y_{1}(t)=y_{1}(0)$, which
is a constant independent of $t$. The rate equation for node 2 is
$\dot{y_{2}}=y_{1}=y_{1}(0)$. Thus $y_{2}(t)=y_{2}(0)+y_{1}(0)t$.
Similarly $\dot{y_{3}}=y_{2}$ implies that $y_{3}(t)=(1/2)y_{1}(0)t^{2}+y_{2}(0)t+y_{3}(0)$.
At large $t$, $y_{1}={\textrm{constant}}$, $y_{2}\sim t$, $y_{3}\sim t^{2}$;
hence $y_{3}$ dominates. Therefore, $X_{i}=\lim _{t\rightarrow \infty }x_{i}(t)$
is given by $X_{1}=0,X_{2}=0,X_{3}=1$. Thus ${\textbf {X}}$ equals
the unique PFE ${\textbf {e}}$, independent of initial conditions.
\\

\newpage
\noindent \textbf{Example 2}. A 1-cycle, Figure \ref{cap:uniquePFEfig}b:\\
 This graph has two eigenvalues, $\lambda _{1}=1$, $\lambda _{2}=0$.
The unique PFE is ${\textbf {e}}=(1,0)^{T}$. The rate equations are
$\dot{y_{1}}=y_{1},\dot{y_{2}}=0$, with the solutions $y_{1}(t)=y_{1}(0)e^{t},y_{2}(t)=y_{2}(0)$.
At large $t$ node 1 dominates, hence ${\textbf {X}}=(1,0)^{T}={\textbf {e}}$.
The exponentially growing population of 1 is a consequence of the
fact that 1 is a self-replicator, as embodied in the equation $\dot{y_{1}}=y_{1}$.
\\

\noindent \textbf{Example 3}. A 2-cycle, Figure \ref{cap:uniquePFEfig}c:\\
 The corresponding adjacency matrix has eigenvalues $\lambda _{1}=1,\lambda _{2}=-1$.
The unique normalized PFE is ${\textbf {e}}=(1,1)^{T}/2$. The population
dynamics equations are $\dot{y_{1}}=y_{2},\dot{y_{2}}=y_{1}.$ \\

\noindent The general solution to these is (note $\ddot{y_{1}}=y_{1}$)
\[
y_{1}(t)=Ae^{t}+Be^{-t},\quad \quad y_{2}(t)=Ae^{t}-Be^{-t}.\]
 Therefore at large $t$, $y_{1}\rightarrow Ae^{t},y_{2}\rightarrow Ae^{t}$,
hence ${\textbf {X}}=(1,1)^{T}/2={\textbf {e}}$. Neither 1 nor 2
is individually a self-replicating node, but collectively they function
as a self-replicating entity. \\

\noindent \textbf{Example 4}. A 2-cycle with a periphery, Figure \ref{cap:uniquePFEfig}d:\\
 This graph has $\lambda _{1}=1$ and a unique normalized PFE ${\textbf {e}}=(1,1,1)^{T}/3$.
The population equations for $y_{1}$ and $y_{2}$ and consequently
their general solutions are the same as Example 3, but now in addition
$\dot{y_{3}}=y_{2}$, yielding $y_{3}(t)=Ae^{t}+Be^{-t}+{\textrm{constant}}$.
Again for large $t$, $y_{1},y_{2},y_{3}$ grow as $\sim Ae^{t}$,
hence ${\textbf {X}}=(1,1,1)^{T}/3={\textbf {e}}$. The dominant ACS
includes all the three nodes.

This example shows how a parasitic periphery (which does not feed
back into the core) is supported by an autocatalytic core. This is
also an example of the following general result: when a subgraph $C'$,
with largest eigenvalue $\lambda _{1}'$, is \textit{\emph{downstream}}
from another subgraph $C''$ with largest eigenvalue $\lambda _{1}''>\lambda _{1}'$,
then the populations of the former also grow at the rate at which
the populations of $C''$ are growing. Therefore if $C''$ is non-zero
in the attractor, so is $C'$. In this example $C'$ is the single
node 3 with $\lambda _{1}'=0$ and $C''$ is the 2-cycle of nodes
1 and 2 with $\lambda _{1}''=1$. \\

\noindent \textbf{Example 5}. A 2-cycle and a chain, Figure \ref{cap:uniquePFEfig}e:
\\
 The graph in Figure \ref{cap:uniquePFEfig}e combines the graphs
of Figures \ref{cap:uniquePFEfig}a and c. Following the analysis
of those two examples it is evident that for large $t$, $y_{1}\sim t^{0},y_{2}\sim t^{1},y_{3}\sim t^{2},y_{4}\sim e^{t},y_{5}\sim e^{t}$.
Because the populations of the 2-cycle are growing exponentially they
will eventually completely overshadow the populations of the chain
which are growing only as powers of $t$. Therefore the attractor
will be ${\textbf {X}}=(0,0,0,1,1)^{T}/2$ which, it can be verified,
is a PFE of the graph (it is an eigenvector with eigenvalue 1). In
general when a graph consists of one or more ACSs and other nodes
that are not part of any ACS, the populations of the ACS nodes grow
exponentially while the populations of the latter nodes grow at best
as powers of $t$. Hence ACSs always outperform non-ACS structures
in the population dynamics (see also Example 2). 

\noindent \textbf{Example 6}. A 2-cycle and another irreducible graph
disconnected from it, Figure \ref{cap:uniquePFEfig}f:\\
 One can ask, when there is more than one ACS in the graph which is
the dominant ACS? Figure \ref{cap:uniquePFEfig}f shows a graph containing
two strong components. The 2-cycle subgraph has a Perron-Frobenius
eigenvalue 1, while the other strong component has a Perron-Frobenius
eigenvalue $\sqrt{2}$. The unique PFE of the entire graph is ${\textbf {e}}=(0,0,1,\sqrt{2},1)^{T}/(2+\sqrt{2})$
with eigenvalue $\sqrt{2}$. The population dynamics equations are
$\dot{y_{1}}=y_{2},\dot{y_{2}}=y_{1},\dot{y_{3}}=y_{4},\dot{y_{4}}=y_{3}+y_{5},\dot{y_{5}}=y_{4}$.
The first two equations are completely decoupled from the last three
and the solutions for $y_{1}$ and $y_{2}$ are the same as for Example
3. For the other irreducible graph the solution is (because $\ddot{y_{4}}=\dot{y_{3}}+\dot{y_{5}}=2y_{4}$)
\[
y_{4}(t)=Ae^{\sqrt{2}t}+Be^{-\sqrt{2}t},\quad \quad y_{3}(t)={\frac{1}{\sqrt{2}}}(Ae^{\sqrt{2}t}+Be^{-\sqrt{2}t})+C,\]
 \[
y_{5}(t)={\frac{1}{\sqrt{2}}}(Ae^{\sqrt{2}t}+Be^{-\sqrt{2}t})-C.\]
 Thus, the populations of nodes 3, 4 and 5 also grow exponentially
but at a faster rate, reflecting the higher Perron-Frobenius eigenvalue
of the subgraph comprising those nodes. Therefore, this structure
eventually overshadows the 2-cycle, and the attractor is ${\textbf {X}}={\textbf {e}}$.
The dominant ACS in this case is the irreducible subgraph formed by
nodes 3, 4 and 5. More generally, when a graph consists of several
disconnected ACSs with different individual $\lambda _{1}$, only
the ACSs whose $\lambda _{1}$ are the largest (and equal to $\lambda _{1}(C)$)
end up with non-zero relative populations in the attractor. \\

\noindent \textbf{Example 7}. A 2-cycle downstream from another 2-cycle,
Figure \ref{cap:uniquePFEfig}g:\\
 What happens when the graph contains two ACSs whose individual $\lambda _{1}$
equals $\lambda _{1}(C)$, and one of those ACSs is downstream of
another? In Figure \ref{cap:uniquePFEfig}g nodes 3 and 4 form a 2-cycle
that is downstream from another 2-cycle comprising nodes 1 and 2.
The unique PFE of this graph, with $\lambda _{1}=1$, is ${\textbf {e}}=(0,0,1,1)^{T}/2$.
The population dynamics equations are $\dot{y_{1}}=y_{2},\dot{y_{2}}=y_{1},\dot{y_{3}}=y_{4}+y_{2},\dot{y_{4}}=y_{3}$.
Their general solution is: \[
y_{1}(t)=Ae^{t}+Be^{-t},\quad \quad y_{2}(t)=Ae^{t}-Be^{-t},\]
 \[
y_{3}(t)={\frac{t}{2}}(Ae^{t}-Be^{-t})+Ce^{t}+De^{-t},\]
 \[
y_{4}(t)={\frac{t}{2}}(Ae^{t}+Be^{-t})+(C-{\frac{A}{2}})e^{t}+({\frac{B}{2}}-D)e^{-t}.\]
 It is clear that for large $t$, $y_{1}\sim e^{t},y_{2}\sim e^{t},y_{3}\sim te^{t},y_{4}\sim te^{t}$.
While all four grow exponentially with the same rate $\lambda _{1}$,
as $t\rightarrow \infty $ $y_{3}$ and $y_{4}$ will overshadow $y_{1}$
and $y_{2}$. The attractor will be therefore be ${\textbf {X}}=(0,0,1,1)^{T}/2={\textbf {e}}$.
Here the dominant ACS is the 2-cycle of nodes 3 and 4. This result
generalizes to other kinds of ACSs: if one strong component is downstream
of another with the same Perron-Frobenius eigenvalue, the latter will
have zero relative population in the attractor.\\

The above examples displayed graphs with a unique PFE, and illustrated
proposition 4.1(iv). The stability of the attractor follows from the
fact that the constants $A,B,C,D$, etc., in the above examples, that
can be traded for the initial conditions of the populations, do not
appear anywhere in the attractor ${\textbf {X}}$. Now I consider
examples where the PFE is not unique.\\

\begin{figure}
\begin{center}\includegraphics[  height=6.3cm,
  keepaspectratio]{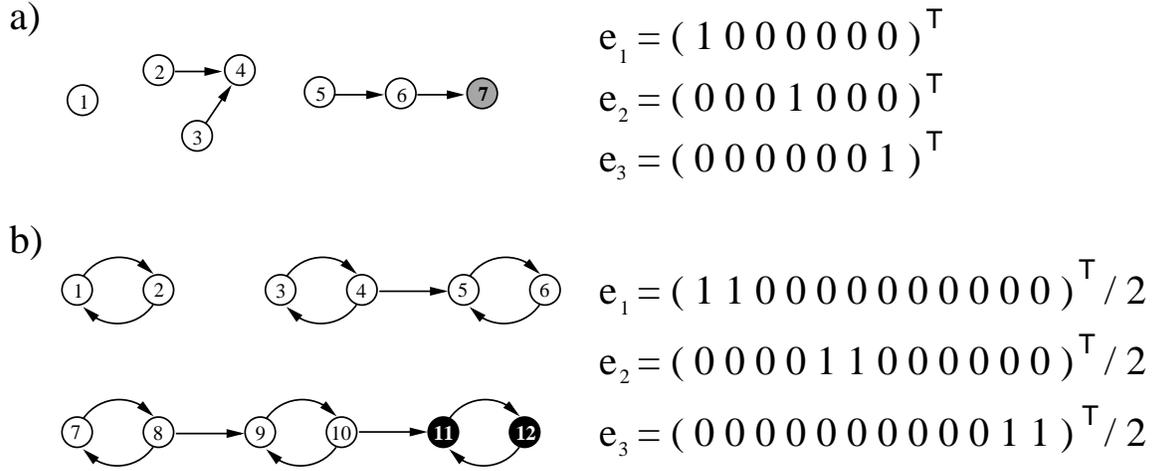}\end{center}

\caption{Examples of graphs with multiple PFEs. \textbf{a.} ${\textbf {e}}_{1},{\textbf {e}}_{2},{\textbf {e}}_{3}$
are all eigenvectors with eigenvalue $\lambda _{1}=0$. Only ${\textbf {e}}_{3}$
is the attractor. Thus for generic initial conditions, only node 7,
which sits at the end point of the longest chain of nodes is non-zero
in the attractor. \textbf{b.} ${\textbf {e}}_{1},{\textbf {e}}_{2},{\textbf {e}}_{3}$
are all eigenvectors with eigenvalue $\lambda _{1}=1$, but only ${\textbf {e}}_{3}$
is the attractor. Only the 2-cycle, of nodes 11 and 12, which sits
at the end of the longest chain of cycles, is non-zero in the attractor.\label{cap:nonuniquePFEfig}}\lyxline{\normalsize}

\end{figure}

\noindent \textbf{Example 8}. Graph with $\lambda _{1}=0$ and three
weakly connected components, Figure \ref{cap:nonuniquePFEfig}a:\\
This graph has three independent PFEs, displayed in Figure \ref{cap:nonuniquePFEfig}a.
The attractor is ${\textbf {X}}={\textbf {e}}_{3}$. This is an immediate
generalization of Example 1 above. Using the same argument as for
Example 1, we can see that $y_{i}\sim t^{k}$ if the longest path
ending at node $i$ is of length $k$. Thus, the populations of nodes
1, 2, 3 and 5 are constant, those of 4 and 6 increase $\sim t$ for
large $t$, and of 7 as $\sim t^{2}$. Therefore, only nodes at the
ends of the longest paths will be non-zero in the attractor.\\

\noindent \textbf{Example 9}. Several weakly connected components
containing 2-cycles, Figure \ref{cap:nonuniquePFEfig}b:\\
 Here again there are three PFEs, one for each weakly connected component.
The population of nodes in 2-cycles that are not downstream of other
2-cycles (nodes 1,2,3,4,7 and 8) will grow as $e^{t}$. As in Example
7, Figure \ref{cap:uniquePFEfig}g, the nodes of 2-cycles that are
downstream of one 2-cycle (nodes 5, 6, 9 and 10) will grow as $te^{t}$.
It can be verified that the populations of nodes in 2-cycles downstream
from two other 2-cycles (nodes 11 and 12) will grow as $t^{2}e^{t}$.
The pattern is clear: in the attractor only the 2-cycles at the ends
of the longest chains of 2-cycles will have non-zero relative populations.
Therefore, the attractor is ${\textbf {X}}={\textbf {e}}_{3}$.

\section{\noindent Timescale for reaching the attractor}

\noindent How long does it take the system to reach the attractor?
This depends on the structure of the graph $C$. For instance in Example
2, the attractor is approached as the population of node 1, $y_{1}$,
overwhelms the population $y_{2}$. Because $y_{1}$ grows exponentially
as $e^{t}$, the attractor is reached on a timescale $\lambda _{1}^{-1}=1$.
(In general, when I say that `the timescale for the system to reach
the attractor is $\tau $', I mean that for $t\gg \tau $, ${\textbf {x}}(t)$
is `exponentially close' to its final destination ${\textbf {X}}\equiv \lim _{t\rightarrow \infty }{\textbf {x}}(t)$,
i.e. for all $i$, $|x_{i}(t)-X_{i}|\sim e^{-t/\tau }t^{\alpha }$,
with some finite $\alpha $.) In contrast, in Example 1, the attractor
is approached as $y_{3}$ overwhelms $y_{1}$ and $y_{2}$. Because
in this case all the populations are growing as powers of $t$, the
timescale for reaching the attractor is infinite. When the populations
of different nodes are growing at different rates, this timescale
depends on the difference in growth rate between the fastest growing
population and the next fastest growing population.

For graphs with $\lambda _{1}=0$ like those in Example 1 and 8, all
populations grow as powers of $t$, hence the timescale for reaching
the attractor is infinite.

For graphs that have $\lambda _{1}\geq 1$ but all the basic subgraphs
are in different weakly connected components, such as Examples 2-6,
the timescale for reaching the attractor is given by $\left(\lambda _{1}-{\textrm{Re}}\lambda _{2}\right)^{-1}$,
where $\lambda _{2}$ is the eigenvalue of $C$ with the next largest
real part, compared to $\lambda _{1}$.

For graphs having $\lambda _{1}\ge 1$ with at least one basic subgraph
downstream from another basic subgraph, the ratio of the fastest growing
population to the next fastest growing one will always be a power
of $t$ (as in Examples 7 and 9), therefore, the timescale for reaching
the attractor is again infinite.

\section{\noindent Core and periphery of a graph}

\noindent As the dominant ACS is specified by a particular PFE, I
will define the core of the dominant ACS to be the core of the corresponding
PFE. If the PFE is simple, the core of the dominant ACS consists of
just one basic subgraph. If the PFE is non-simple the core of the
dominant ACS will be a union of more than one basic subgraph. The
dominant ACS is uniquely determined by the graph. This motivates the
definition of the core and periphery of a graph: 

\begin{description}
\item [Definition~4.2:]Core and periphery of a graph.\\
The \textit{core} of a graph $C$, denoted $Q(C)$, is the core of
the dominant ACS of $C$, if an ACS exists in the graph. The \textit{periphery}
of $C$ is the periphery of the dominant ACS of $C$. If no ACS exists
in the graph, i.e., $\lambda _{1}(C)=0$ then $Q(C)=\Phi $ \citep{JKwiley}.
\end{description}
\noindent In all cases $\lambda _{1}(Q(C))=\lambda _{1}(C)$. Henceforth
all graphs will be depicted with the following colour scheme: black
for core nodes, grey for periphery nodes (and for nodes with non-zero
$X_{i}$ when there is no ACS), and white for nodes that are not in
any of the PFE subgraphs.

\section{\noindent \label{sec:Keystone-nodes}Keystone nodes}

\noindent %
\begin{figure}
\begin{center}\includegraphics[  keepaspectratio]{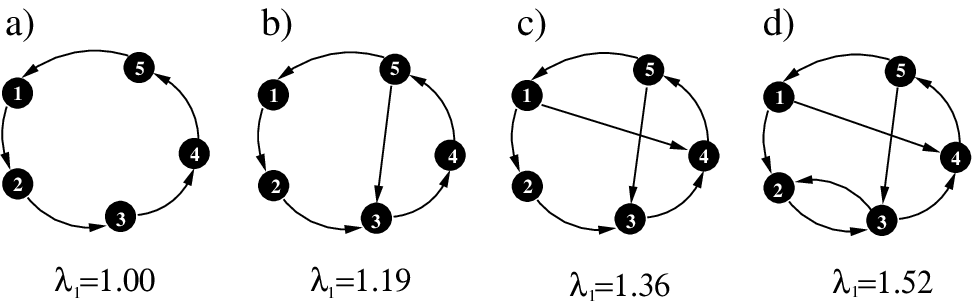}\end{center}

\caption{Keystone nodes. \textbf{a.} All nodes of the graph are keystone.
\textbf{b.} Nodes 3, 4 and 5 are keystone. \textbf{c.} Nodes 4 and
5 are keystone. \textbf{d.} None of the nodes are keystone.\label{cap:keystone1}}\lyxline{\normalsize}

\end{figure}

\noindent In ecology certain species are referred to as keystone species
-- those whose extinction or removal would seriously disturb the ecosystem
\citep{Paine,JTM,SMo}. One might similarly look for the notion of
a keystone node in a directed graph that captures some important organizational
role played by a node. Consider the impact of the hypothetical removal
of any node $i$ from a graph $C$. One can, for example, study the
structure of the core of the graph $C-i$ that would result if node
$i$ (along with all its links) were removed from $C$. 

\begin{description}
\item [Definition~4.3:]Core overlap.\\
Given any two graphs $C$ and $C'$ whose nodes are labeled, the \textit{core
overlap} between them, denoted $Ov(C,C')$, is the number of common
links in the cores of $C$ and $C'$, i.e., the number of ordered
pairs $(j,i)$ for which $Q_{ij}$ and $Q'_{ij}$ are both non-zero.
If either $C$ or $C'$ does not have a core, $Ov(C,C')$ is defined
to be zero \citep{JKcoresh}.
\item [Definition~4.4:]Keystone node.\\
I will refer to a node $i$ of a graph $C$ as a \textit{keystone
node} if $C$ has a non-vanishing core and $Ov(C,C-i)=0$ \citep{JKlargeext,JKwiley}. 
\end{description}
Thus a keystone node is one whose removal modifies the organizational
structure of the graph (as represented by its core) drastically. In
each of Figures \ref{cap:keystone1}a-d, for example, the core is
the entire graph. In Figure \ref{cap:keystone1}a, all the nodes are
keystone, because the removal of any one of them would leave the graph
without an ACS (and hence without a core). In general when the core
of a graph is a single $n$-cycle, for any $n$, all the core nodes
are keystone. In Figure \ref{cap:keystone1}b, nodes 3, 4 and 5 are
keystone but the other nodes are not, and in Figure \ref{cap:keystone1}c
only nodes 4 and 5 are keystone. In Figure \ref{cap:keystone1}d,
there are no keystone nodes. These examples show that the more internal
pathways a core has (generally, this implies a higher value of $\lambda _{1}$),
the less likely it is to have keystone nodes, and hence the more robust
its structure is to removal of nodes.

\begin{figure}
\begin{center}\includegraphics[  keepaspectratio]{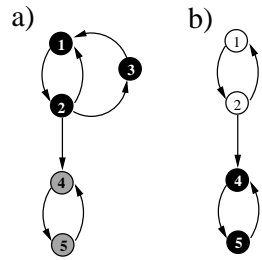}\end{center}

\caption{Another example of a keystone node. Node 3 is a keystone node of
the graph (a) because its removal produces the graph (b) which has
a zero core overlap with the graph (a). The core nodes of both graphs
are coloured black. \label{cap:keystone2}}\lyxline{\normalsize}

\end{figure}

Figure \ref{cap:keystone2} illustrates another type of graph structure
that has a keystone node. The graph in Figure \ref{cap:keystone2}a
consists of two strong components -- a 2-cycle (nodes 4 and 5) downstream
from an irreducible subgraph consisting of nodes 1, 2 and 3. The core
of this graph is the latter irreducible subgraph. Figure \ref{cap:keystone2}b
shows the graph that results if node 3 is removed with all its links.
This consists of one 2-cycle downstream from another. Though both
2-cycles are basic subgraphs of the graph, as discussed in Example
7, Figure \ref{cap:uniquePFEfig}g, this graph has a unique (upto
constant multiples) PFE, whose subgraph consists of the downstream
cycle (nodes 4 and 5) only. Thus the 2-cycle 4-5 is the core of the
graph in Figure \ref{cap:keystone2}b. Clearly $Ov(C,C-3)=0$. Therefore,
node 3 in Figure \ref{cap:keystone2}a is a keystone node.

The above purely graph theoretic definition of a keystone node turns
out to be useful in the dynamical system discussed in this and the
following chapters. For other dynamical systems, alternate definitions
of keystone might be more useful.

\chapter{\label{cha:Graph-Dynamics}Graph Dynamics}

In this chapter, I present a model of an evolving network that is
a specific instance of the framework described in section \ref{sec:Framework-of-a}.
The dynamical system described in the previous chapter is used for
step 1 of the dynamics, and certain rules are chosen for steps 2 and
3 that aim to capture some features of the evolution of chemical networks
on the prebiotic Earth.

\section{Graph dynamics rules}

The initial graph is constructed as follows: For every ordered pair
$(i,j)$ with $i\ne j$ (where $i,j\in S=\{1,2,\ldots ,s\}$), $c_{ij}$
is independently chosen to be unity with a probability $p$ and zero
with a probability $1-p$. $c_{ii}$ is set to zero for all $i\in S$.
Thus the initial graph is a random graph, a member of the ensemble
$G_{s}^{p}$ described in section \ref{sec:Random-graphs}. Each $x_{i}$
is chosen randomly, with uniform probability, in $[0,1]$ and all
$x_{i}$ are rescaled so that $\sum _{i=1}^{s}x_{i}=1$.

\begin{figure}
\includegraphics[  width=15cm,
  keepaspectratio]{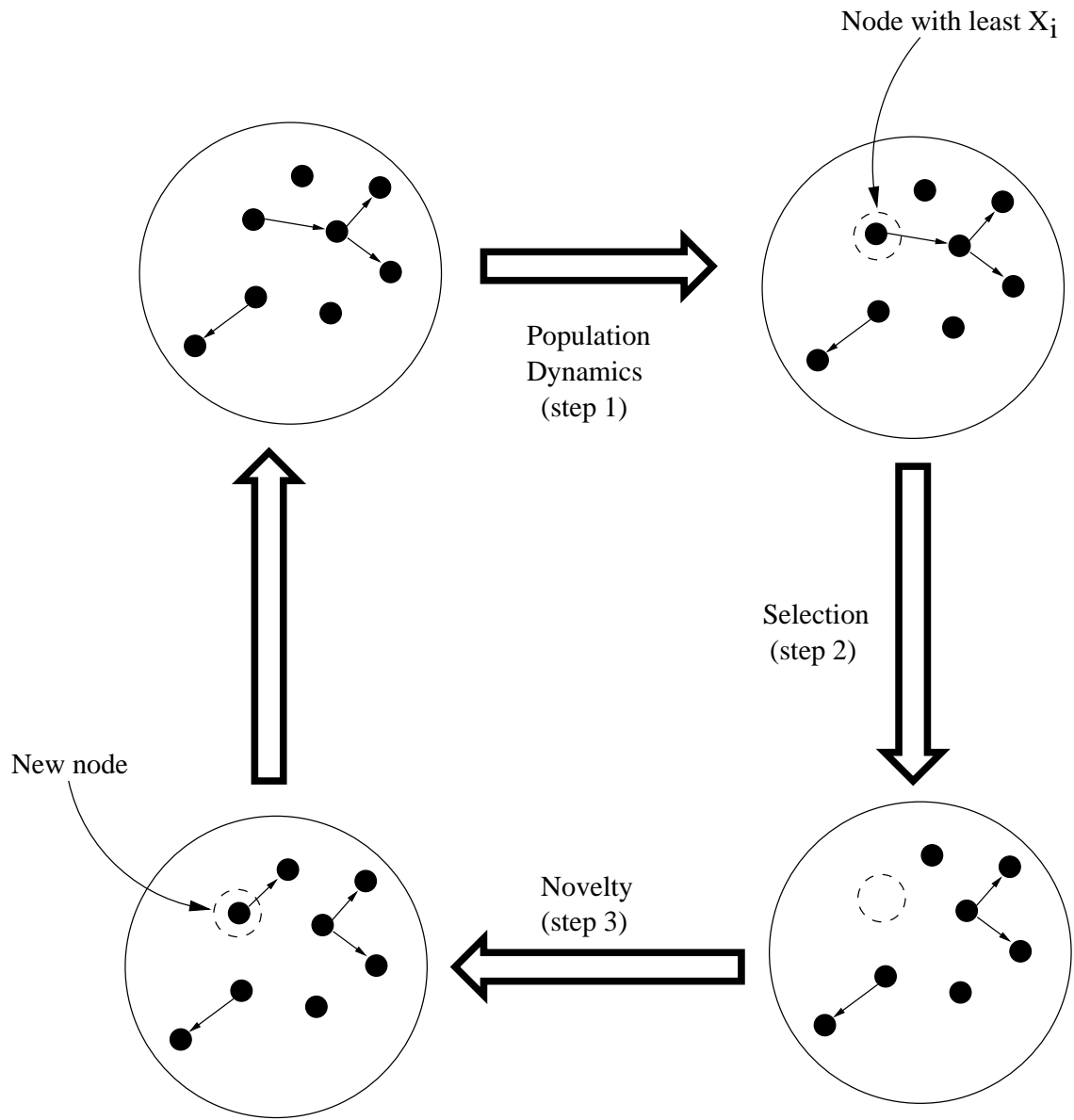}

\caption{Schematic depiction of the graph dynamics. Starting with a random
graph, first the population dynamics is allowed to run until the relative
populations reach their attractor (step 1). Then one of the nodes
with the least relative population is removed from the graph (selection,
step 2). It is replaced by a new node which is assigned random links
to existing nodes (novelty, step 3). This process is then iterated.\label{cap:dynamics}}
\end{figure}

\begin{lyxlist}{00.00.0000}
\item [Step~1.]With $C$ fixed, ${\textbf {x}}$ is evolved according
to (\ref{xdot}) until it converges to a fixed point, \\
denoted ${\textbf {X}}$. 
\item [Step~2.]The set $\mathcal{L}$ of nodes with the least $X_{i}$
is determined, i.e, $\mathcal{L}=\{i\in S|X_{i}=\min _{j\in S}X_{j}\}$.
A node, say node $k$, is picked randomly from $\mathcal{L}$ and
is removed from the graph along with all its links.
\item [Step~3.]A new node (also denoted $k$) is added to the graph. Links
to and from $k$ to other nodes are assigned randomly according to
the same rule, i.e, for every $i\ne k$ $c_{ik}$ and $c_{ki}$ are
independently reassigned to unity with probability $p$ and zero with
probability $1-p$, irrespective of their earlier values, and $c_{kk}$
is set to zero. All other matrix elements of $C$ remain unchanged.
$x_{k}$ is set to a small constant $x_{0}$, all other $x_{i}$ are
perturbed by a small amount from their existing value $X_{i}$, and
all $x_{i}$ are rescaled so that $\sum _{i=1}^{s}x_{i}=1$.
\end{lyxlist}
This process, from step 1 onward, is iterated many times. Figure \ref{cap:dynamics}
depicts the graph dynamics schematically.

\section{Features of the graph dynamics}

\subsection{Evolution in a prebiotic pool}

These graph dynamics rules are motivated by the puzzle of how a complex
chemical organization might have emerged from an initial `random soup'
of chemicals. As discussed in section \ref{sec:The-origin-of}, the
rules model the imagined dynamics of an evolving chemical network
such as might have existed in a pool on the prebiotic Earth. Several
assumptions and idealizations underlie such a chemical interpretation
of the model rules. Some of these have been mentioned in sections
\ref{sec:The-origin-of} and \ref{sec:The-population-dynamics}, however
I will defer a more detailed discussion to section \ref{sec:Limitations-of-the}.

The model has two main sources of inspiration. One is the set of models
studied by Farmer, Kauffman, Packard and others \citep{Kauffman2,Kauffman3,FKP,BF,BFF},
and by \citet{Fontana}, and \citet{FB} (see also \citealp{SFM,Dyson}).
Like these models, the present one employs an artificial chemistry
of catalyzed reactions, albeit a much simpler one, in which populations
of species evolve over time. To this I add the feature, inspired by
the model of \cite{BS}, that the `least fit species mutates' with
`fitness' here being equated to the relative population. Unlike the
Bak-Sneppen model, however, the `mutation' of a species also changes
its links to other species.

\subsection{Coupling of population and graph dynamics: two timescales}

As discussed in section \ref{sec:Framework-of-a}, the coupling of
the population dynamics and the graph dynamics is built into the framework
of the model: the evolution of the $x_{i}$ depends on the graph $C$
in step 1, and the evolution of $C$ in turn depends on the $x_{i}$
through the choice of which node to remove in step 2. There are two
timescales in the dynamics, a short timescale over which the graph
is fixed while the $x_{i}$ evolve, and a longer timescale over which
the graph is changed.

\subsection{\label{sub:Absence-of-self-replicators}Absence of self-replicators}

While in previous sections I have considered graphs with 1-cycles,
the requirement $c_{ii}=0$ in the present section forbids 1-cycles
in the graph. The motivation is the following: 1-cycles represent
self-replicating species (see previous section, Example 2). Such species,
e.g., RNA molecules, are difficult to produce and maintain in a prebiotic
scenario and it is possible that it requires a self-supporting molecular
organization to be in place \textit{before} an RNA world, for example,
can take off \citep{JSMO,Joyce}. Thus, I wish to address the question:
can one get molecular organizations that can collectively self-replicate
without introducing self-replicating species `by hand' into the system? 

\subsection{Selection and novelty}

The rules for changing the graph implement \textit{selection} and
\textit{novelty}, two important features of natural evolution. Selection
is implemented by removing the node that is `performing the worst',
with `performance' in this case being equated to a node's relative
population (step 2). Adding a new node introduces novelty into the
system (step 3). Note that although the actual connections of a new
node with other nodes are created randomly, the new node has the same
average connectivity as the initial set of nodes. Thus, the new node
is not biased in any way toward increasing the complexity of the chemical
organization. Steps 2 and 3 implement the interaction of the system
with the external environment. The phenomena to be described in the
following sections are all consequences of the interplay between selection,
novelty and the population dynamics.

\begin{figure}
\includegraphics[  width=15cm,
  keepaspectratio]{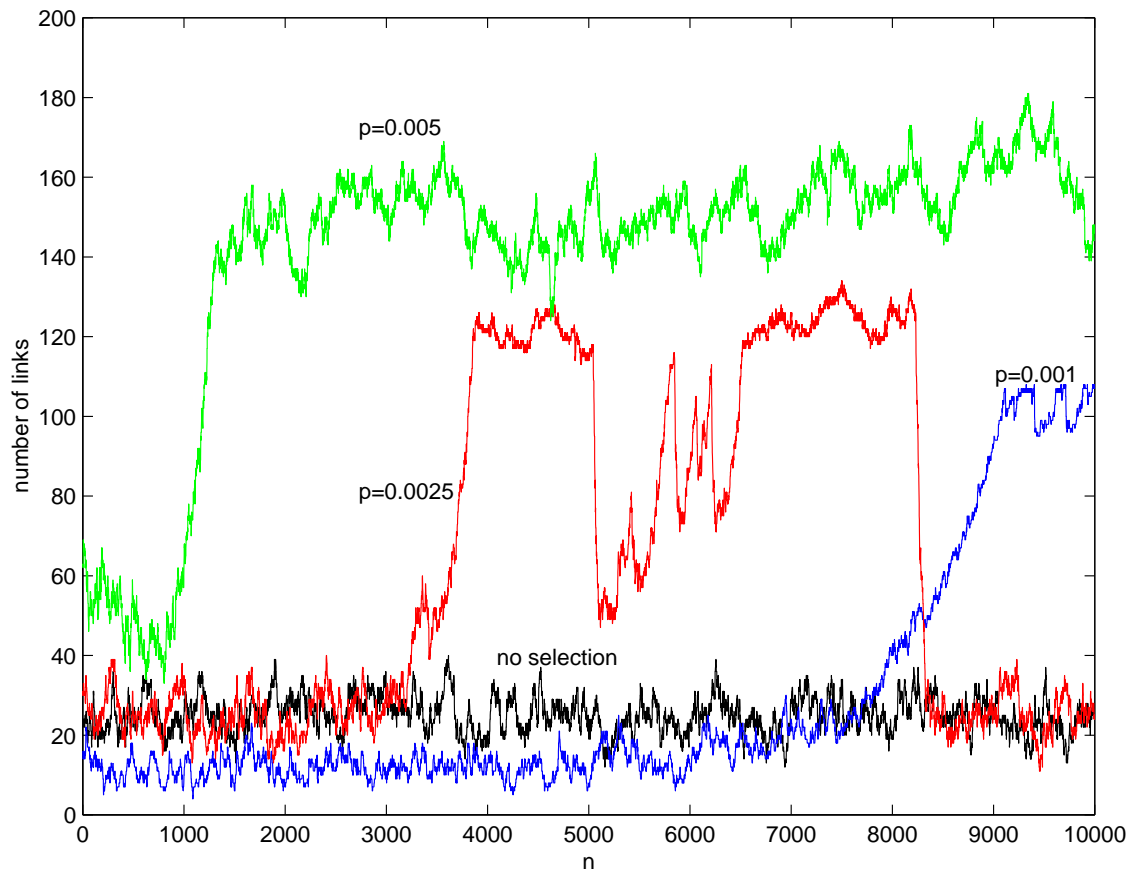}

\caption{The number of links versus time, $n$, for various runs. Each run
had $s=100$. The black curve is a run with no selection: a random
node is picked for removal at each graph update. The other curves
show runs with selection and with different $p$ values: blue, $p=0.001$;
red, $p=0.0025$; green, $p=0.005$.\label{cap:links}}\lyxline{\normalsize}

\end{figure}

\section{Implementation}

Two programs are described in appendix B, and are included in the
attached CD, that use different methods to implement the graph dynamics.
The programs differ in the way the attractor is determined at each
time step. The first program uses theorem 4.1. The second numerically
integrates equation (\ref{xdot}) to find its attractor. The programs
are written in C++, with Matlab 5.2 being used to determine eigenvalues
and eigenvectors where required.

\section{\label{sec:Results-of-graph}Results of graph evolution}

\begin{figure}
\textbf{a)}

\includegraphics[  width=8cm,
  keepaspectratio]{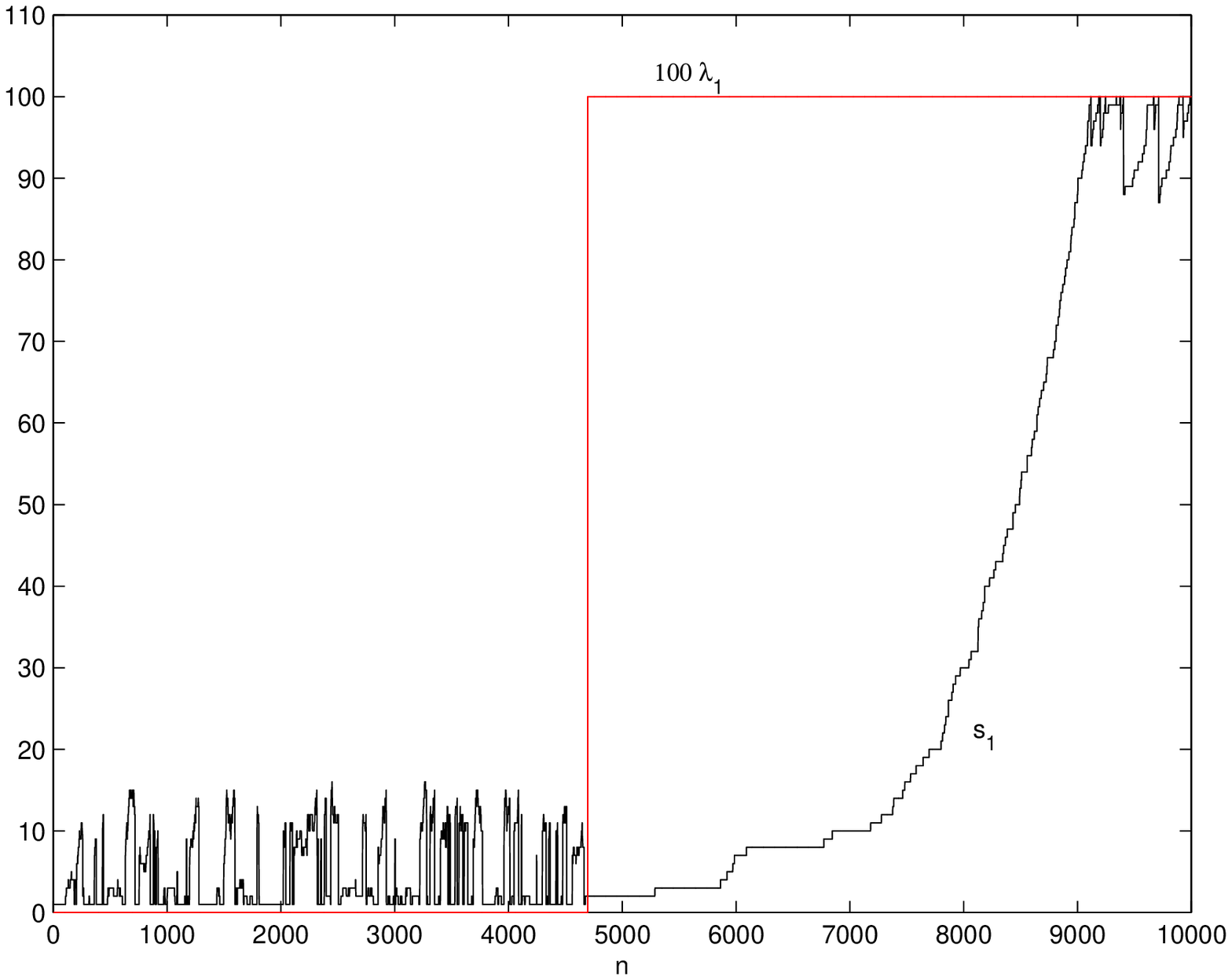}

\textbf{b)}

\includegraphics[  width=8cm,
  keepaspectratio]{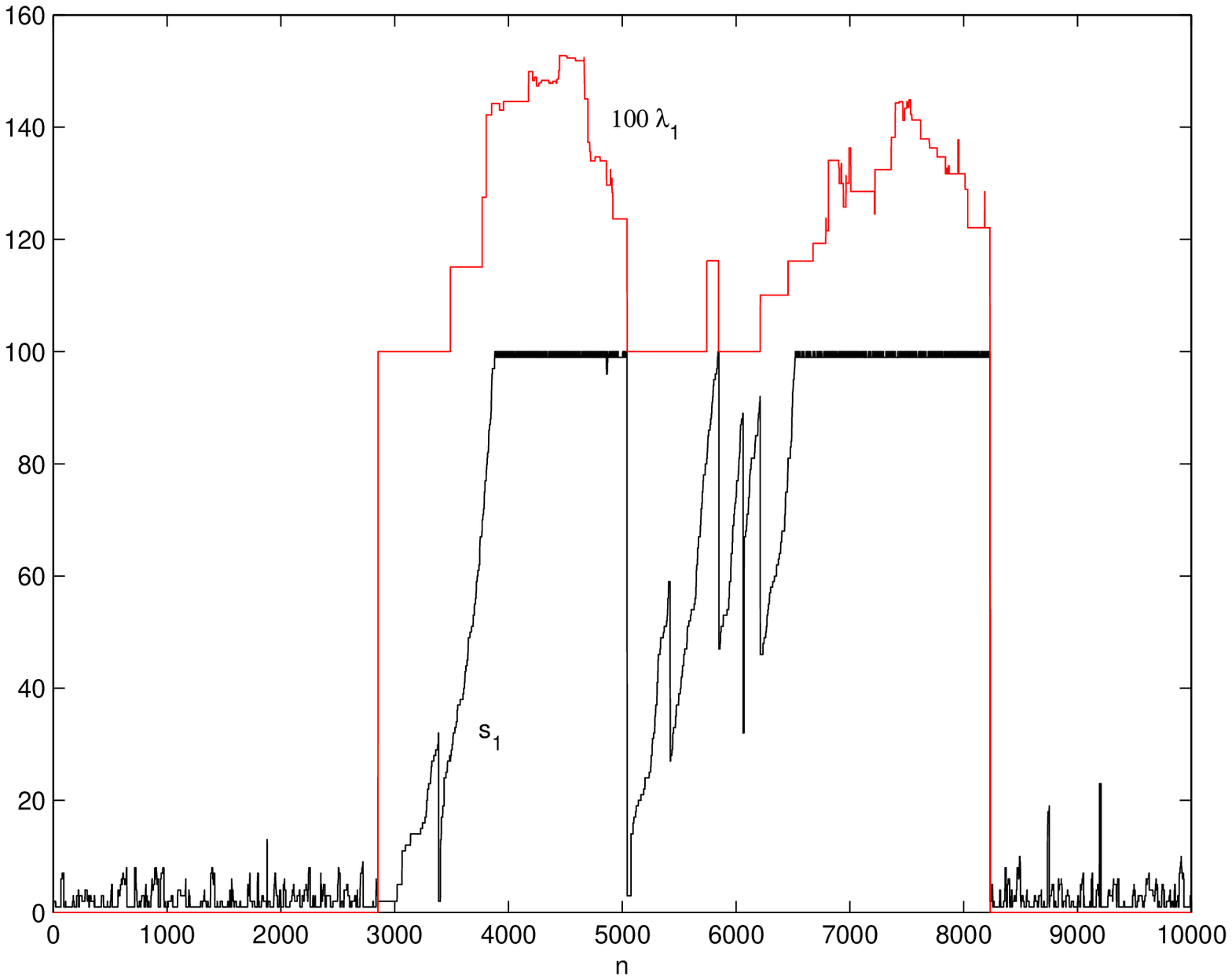}

\textbf{c)}

\includegraphics[  width=8cm,
  keepaspectratio]{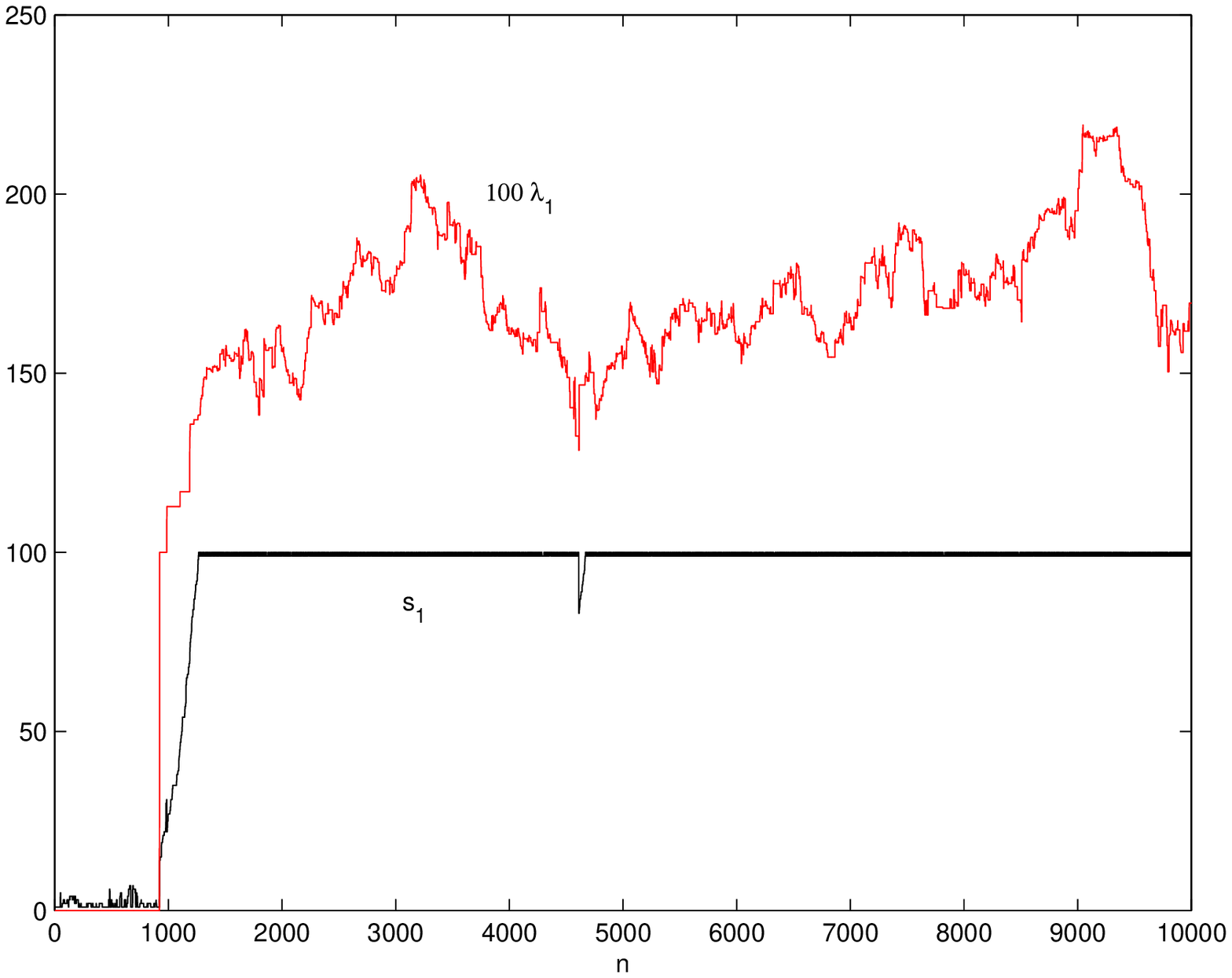}

\caption{Number of nodes with $X_{i}>0$, $s_{1}$, (black curve) and the
Perron-Frobenius eigenvalue of the graph, $\lambda _{1}$, (red curve)
versus time, $n$, for the same three runs shown in Figure \ref{cap:links}.
Each run has $s=100$ and, \textbf{a.} $p=0.001$, \textbf{b. $p=0.0025$}
and \textbf{c.} $p=0.005$, respectively. The $\lambda _{1}$ values
shown have been multiplied by 100 to ease comparison with the $s_{1}$
curves.\label{cap:s1lambda}}
\end{figure}

\begin{figure}
\includegraphics[  width=11.75cm,
  keepaspectratio]{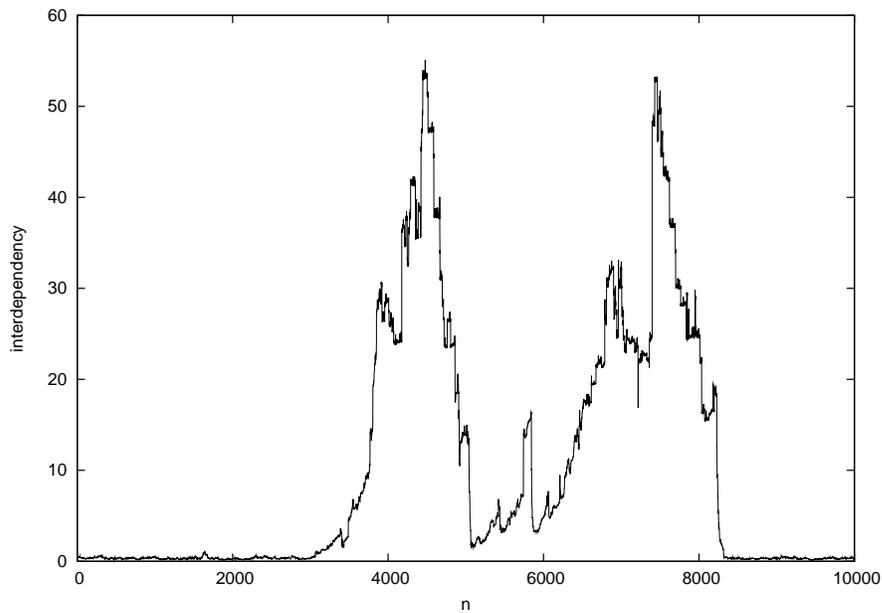}

\caption{Interdependency vs. $n$ for the run with $s=100$ and $p=0.0025$
that is displayed in Figure \ref{cap:s1lambda}b. \label{cap:interdependency}}
\end{figure}

\begin{figure}
\includegraphics[  width=12cm,
  keepaspectratio]{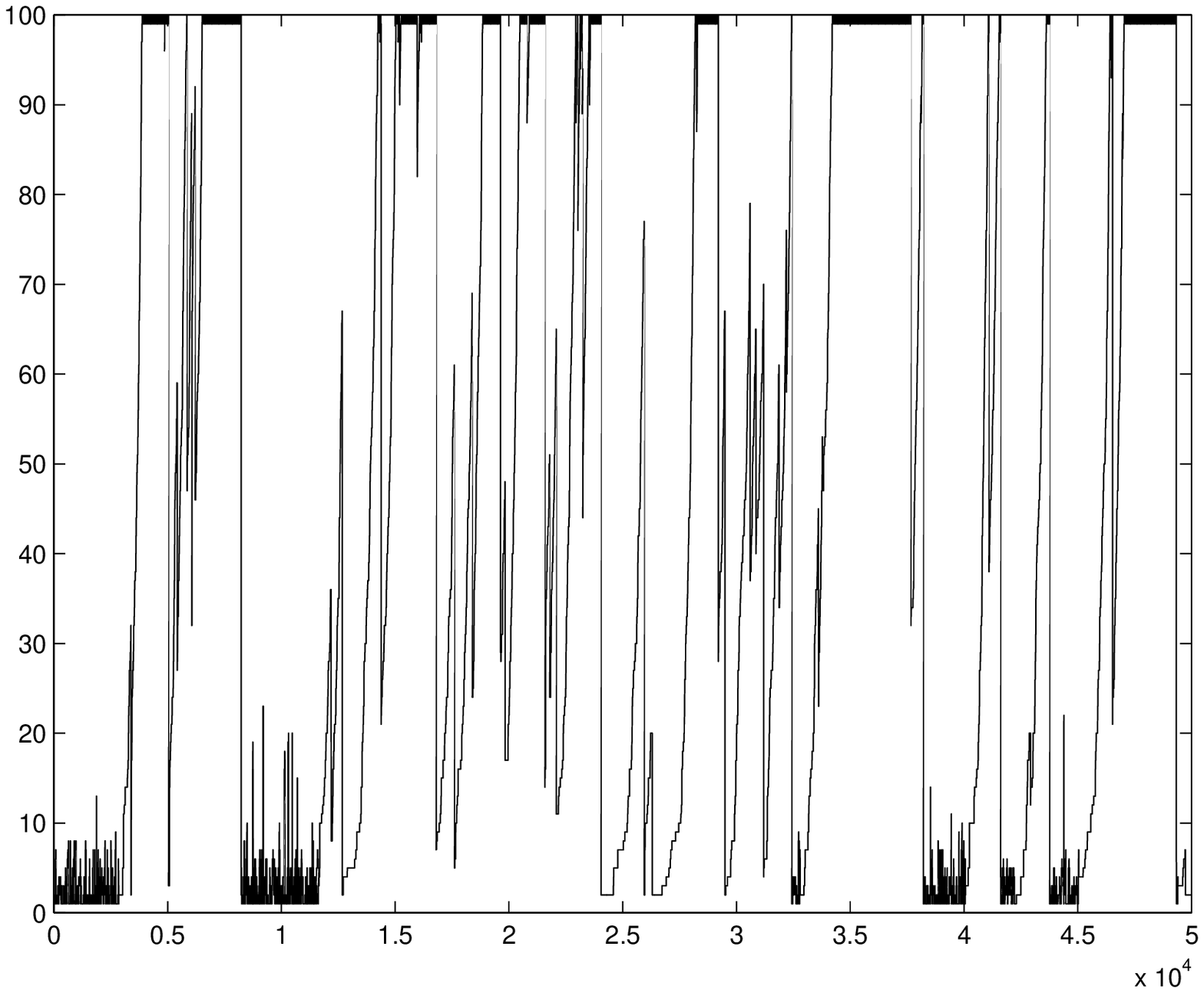}

\caption{$s_{1}$ vs. $n$ for the same run displayed in Figure \ref{cap:s1lambda}b
($s=100,p=0.0025$) for a longer time, till $n=50000$. Repeated rounds
of catastrophes and recoveries are seen.\label{cap:longtimefig}}
\end{figure}

Figure \ref{cap:links} shows the total number of links in the graph
versus time ($n$, the number of graph updates). Three runs of the
model described in the previous section, each with $s=100$ and different
values of $p$ are exhibited. Also exhibited is a run where there
was \textit{no selection} (in which step 2 is modified: instead of
picking one of the nodes of $\mathcal{L}$, any one of the $s$ nodes
is picked randomly and removed from the graph along with all its links.
The rest of the procedure remains the same). Figure \ref{cap:s1lambda}
shows the time evolution of two more quantities for the same three
runs with selection displayed in Figure \ref{cap:links}. The quantities
plotted are $s_{1}$, the number of nodes with $X_{i}>0$, and $\lambda _{1}$,
the Perron-Frobenius eigenvalue of the graph. Figure \ref{cap:interdependency}
shows the evolution of another graph theoretic measure, the interdependency
of the graph, $\bar{d}$, for the run of Figure \ref{cap:s1lambda}b
(the data files for these runs are included in the attached CD, see
appendix C). 

For all the displayed runs the size of the network has been taken
to be $s=100$. Computational constraints have prevented a detailed
exploration of larger values of $s$, though the few runs I have studied
with values of $s$ upto 250 and $ps<1$ have shown a similar behaviour.
The values of the parameters $p$ and $s$ for the displayed runs
were chosen to lie in the regime $ps<1$, and much of the analytical
work described in later chapters, such as estimation of various timescales,
assumes that $ps\ll 1$. This range is the one of interest from the
point of view of the origin of life problem because it is likely that
the catalytic networks existing in prebiotic pools would have been
quite sparse. However, it is worthwhile to extend the analysis and
simulations to other parameter ranges, such as large $s$ or large
$p$ values, and compare with the $ps<1$ behaviour.

Coming back to the runs displayed in Figure \ref{cap:links}, it is
clear why the number of links fluctuates about its random graph value
$\approx ps^{2}$ in the run without selection. This is because each
graph update replaces a randomly chosen node with another that has
on average the same connectivity and, therefore, the graph remains
random like the starting graph. As soon as selection is added the
behaviour becomes more interesting. Three regimes can be observed.
First, the `random phase' where the number of links fluctuates around
$ps^{2}$, while $s_{1}$ and $\bar{d}$ are small. Second, the `growth
phase' where $l$, $s_{1}$ and $\bar{d}$ show a clear rising tendency.
Finally, the `organized phase' where $l$ and $\bar{d}$ hover (with
large fluctuations) about a value much higher than the initial random
graph value, and $s_{1}$ remains close to its maximum value $s$
(with some fluctuations). The time spent in each phase depends on
$p$ and $s$. I analyze the structure of the graph in the three phases
in chapter \ref{cha:Formation-and-Growth}.

Figure \ref{cap:longtimefig} shows $s_{1}$ for the same run as that
of Figure \ref{cap:s1lambda}b for a longer time, from $n=1$ to $n=50,000$.
In this long run one can see several sudden, large drops in $s_{1}$:
\textit{\emph{`catastrophes'}} in which a large fraction of the $s$
nodes become extinct, i.e., the $X_{i}$ values of the nodes fell
to zero. Some of the drops seem to take the system back into the random
phase, others are followed by \textit{\emph{`recoveries'}} in which
$s_{1}$ rises back towards its maximum value $s$. The recoveries
are comparatively slower than the catastrophes, which in fact occur
in a single time step. An analysis of crashes and recoveries is the
subject of chapter \ref{cha:Destruction-of-ACSs}.

\begin{figure}
\textbf{y) n=8500\hfill{}z) n=10000}

\includegraphics[  width=6.5cm,
  keepaspectratio]{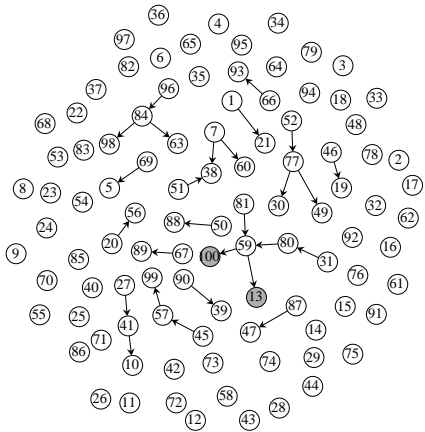}\hfill{}\textbf{\includegraphics[  width=6.5cm,
  keepaspectratio]{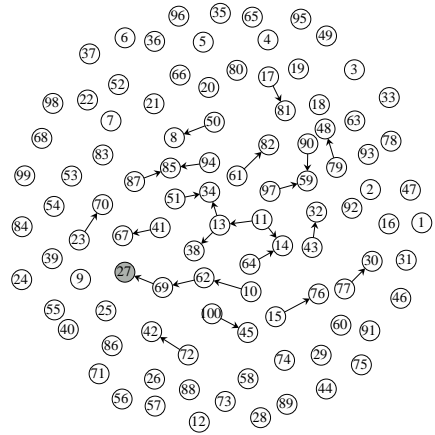}}

\caption{(see previous pages also) Snapshots of the graph at various times
for the run shown in Figure \ref{cap:s1lambda}b with $s=100$ and
$p=0.0025$. See the text for a description of the major events. In
all graphs, white nodes are those with $X_{i}=0$. All shaded nodes
have $X_{i}>0$. If there is an ACS in the graph, the core nodes are
coloured black and the periphery nodes grey. The graphs have been
drawn using LEDA. Data files for these graphs, in LEDA format, and
files containing their adjacency matrices can be found in the attached
CD (see appendix C). \label{cap:snapshots}}\lyxline{\normalsize}

\end{figure}

It may be useful for the reader to trace the steps through one example
run. Figure \ref{cap:snapshots} shows snapshots of the graph at various
times for the run shown in Figure \ref{cap:s1lambda}b, which has
$p=0.0025$. The initial random graph ($n=1$, Figure \ref{cap:snapshots}a)
has 31 links. 55 nodes are isolated with no incoming or outgoing links,
and there are no cycles. In the attractor only node 13 is non-zero,
all others are zero. Figures \ref{cap:snapshots}b-d show further
snapshots of the graph in the random phase.

The first cycle comprising nodes 26 and 90 forms at $n=2854$ (Figure
\ref{cap:snapshots}e), and these nodes are the only non-zero ones
in the attractor. This marks the beginning of the growth phase. At
$n=3022$ (Figure \ref{cap:snapshots}f) more nodes get attached to
the cycle and their relative populations also become non-zero. By
$n=3386$ (Figure \ref{cap:snapshots}g) 28 nodes have added onto
the 2-cycle. All these 30 nodes are non-zero in the attractor, while
the other nodes are zero. At $n=3387$ (Figure \ref{cap:snapshots}h)
there is a sudden drop in $s_{1}$ as a new 2-cycle comprising nodes
41 and 98 is formed. This 2-cycle is downstream from the earlier 2-cycle,
and for such a graph, as explained in section \ref{sec:Examples-of-the}
(Example 7), only the downstream cycle will be non-zero. By $n=3402$
(Figure \ref{cap:snapshots}i) the situation has not altered much,
the old 2-cycle still survives by chance. But then node 38 gets removed,
which breaks the path connecting the cycles. So at $n=3403$ (Figure
\ref{cap:snapshots}j) the graph consists of two disconnected 2-cycles.
Both these cycles and all nodes having access from them are non-zero
in the attractor. By $n=3488$ (Figure \ref{cap:snapshots}k) several
nodes have joined the 26-90 cycle. By chance none have joined the
41-98 cycle. At $n=3489$ (Figure \ref{cap:snapshots}l) the 26-90
cycle gets strengthened by a second cycle. This subgraph now has a
larger Perron-Frobenius eigenvalue and so the other cycle is zero
in the attractor. After this the dominant ACS keeps accreting nodes
until it spans the entire graph for the first time at $n=3880$ (Figure
\ref{cap:snapshots}m) signaling the start of the organized phase.
At $n=4448$ (Figure \ref{cap:snapshots}n) the core is at its largest.
Most subsequent graph updates involve the removal of a node with no
more drastic effect on the dominant ACS. This removed node will keep
getting repeatedly replaced until it rejoins the dominant ACS. At
$n=4695$ (Figure \ref{cap:snapshots}o) the node 36 which is outside
the ACS and therefore zero in the attractor, gets replaced and the
new node rejoins the ACS resulting in the graph at $n=4696$ (Figure
\ref{cap:snapshots}p). Notice that node 36 forms a 2-cycle with node
74. Thus, this graph update has created a new irreducible subgraph
in the periphery. At the moment this structure does not affect the
dominant ACS because the core has a larger $\lambda _{1}$. However,
a few hundred time steps later this 2-cycle plays a significant role.

\begin{figure}
\textbf{a) n=1\hfill{}b) n=78}

\includegraphics[  width=6.5cm,
  keepaspectratio]{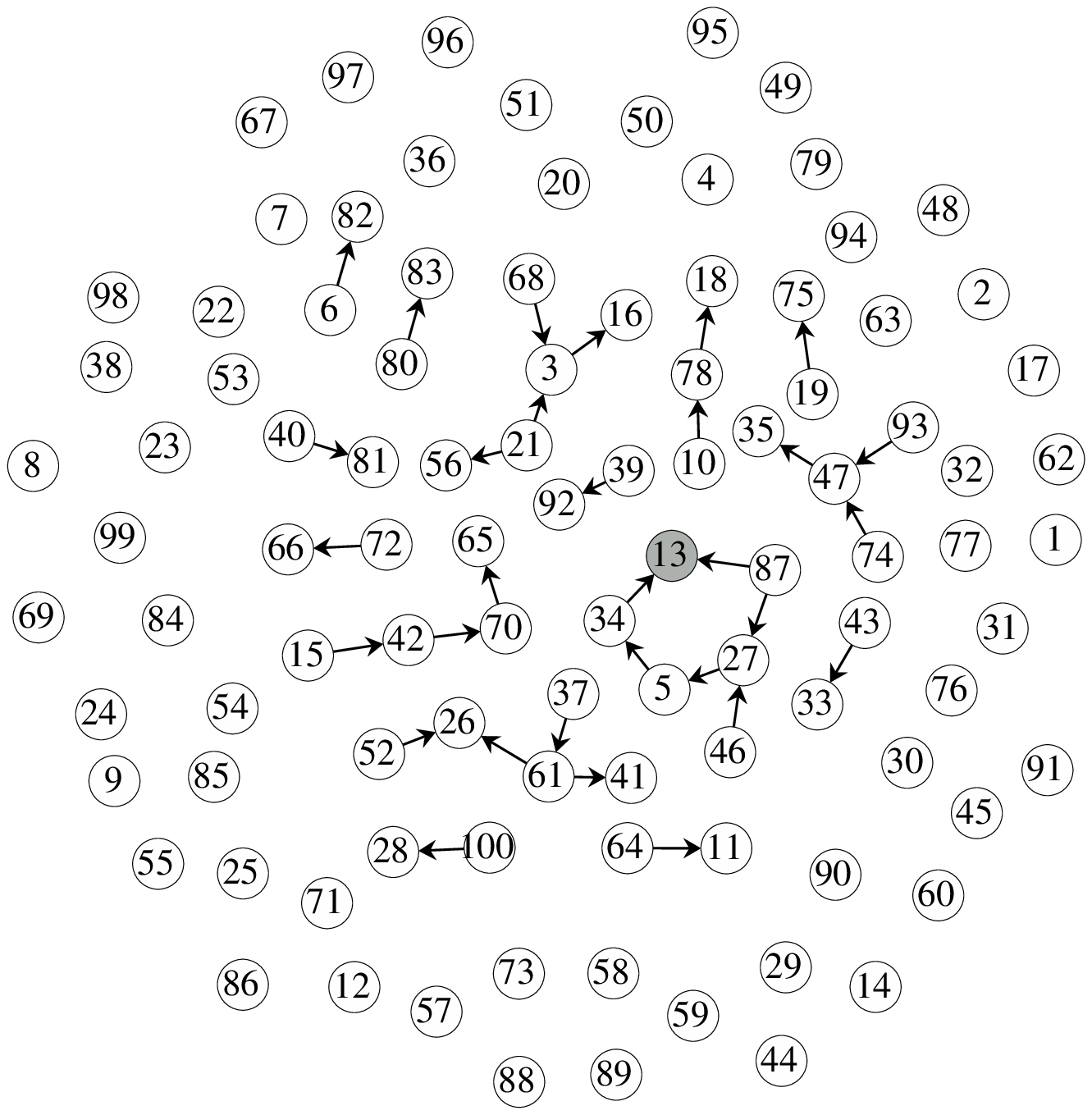}\hfill{}\includegraphics[  width=6.5cm,
  keepaspectratio]{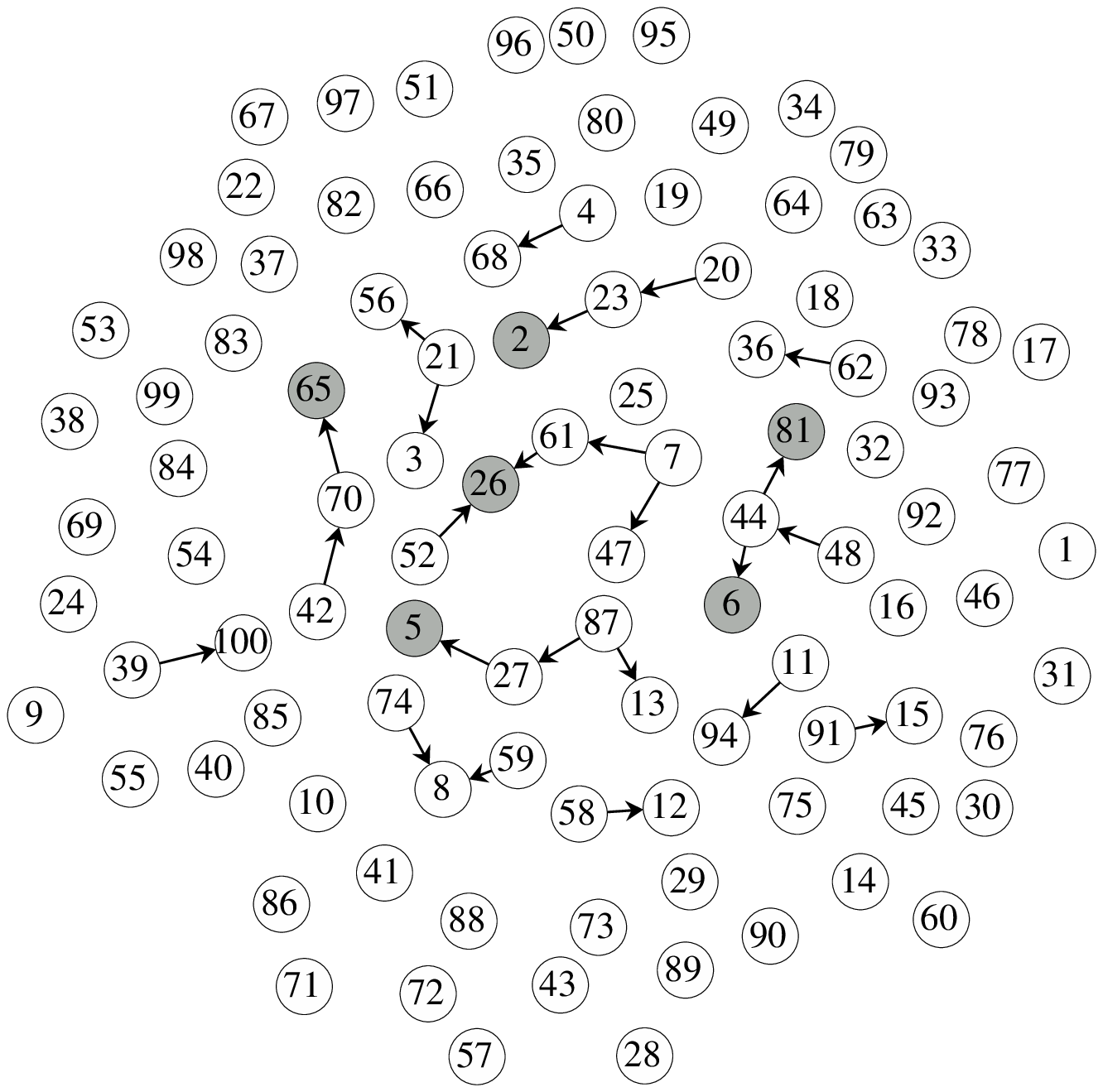}

\textbf{c) n=79\hfill{}d) n=2853}

\includegraphics[  width=6.5cm,
  keepaspectratio]{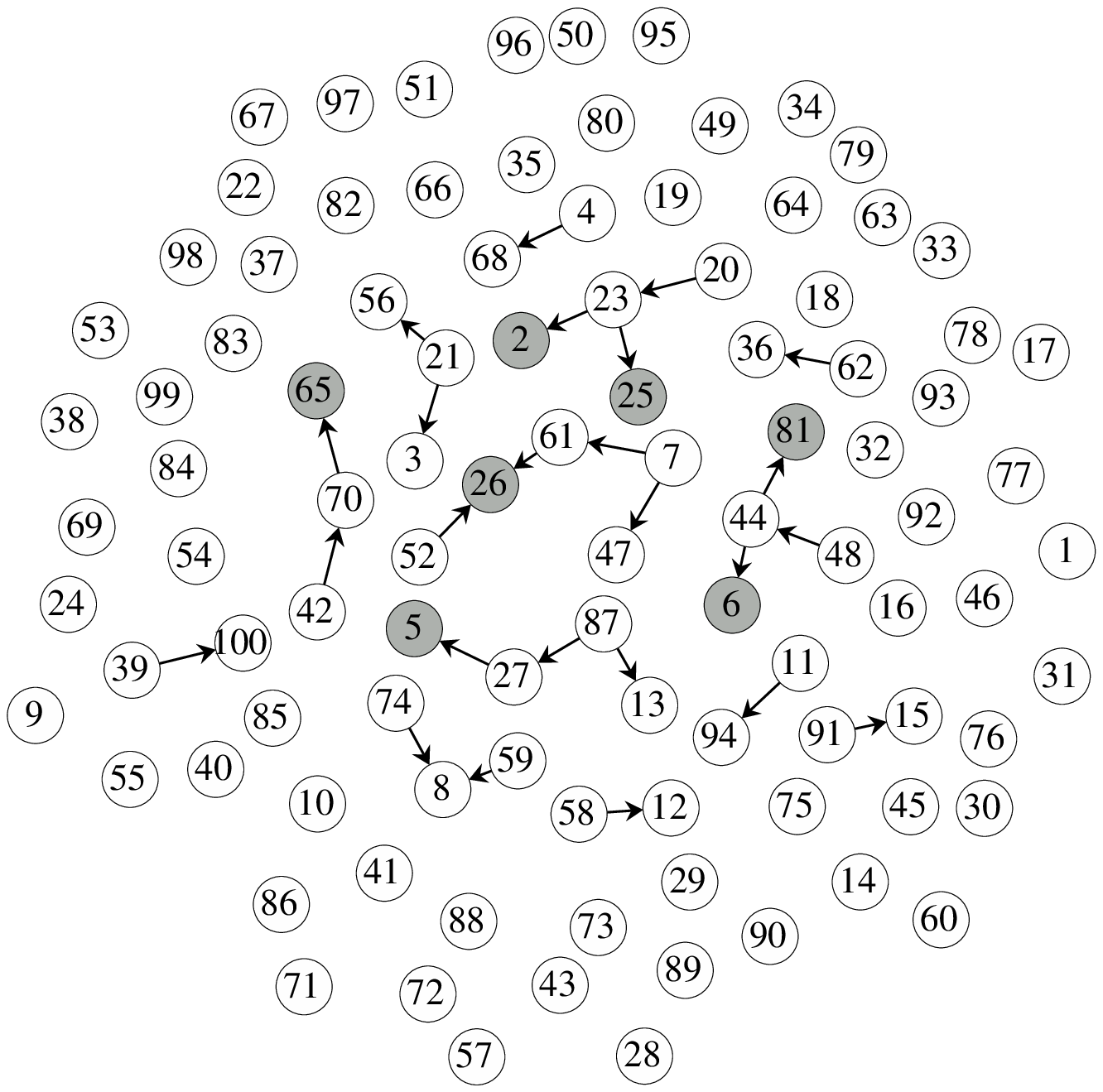}\hfill{}\includegraphics[  width=6.5cm,
  keepaspectratio]{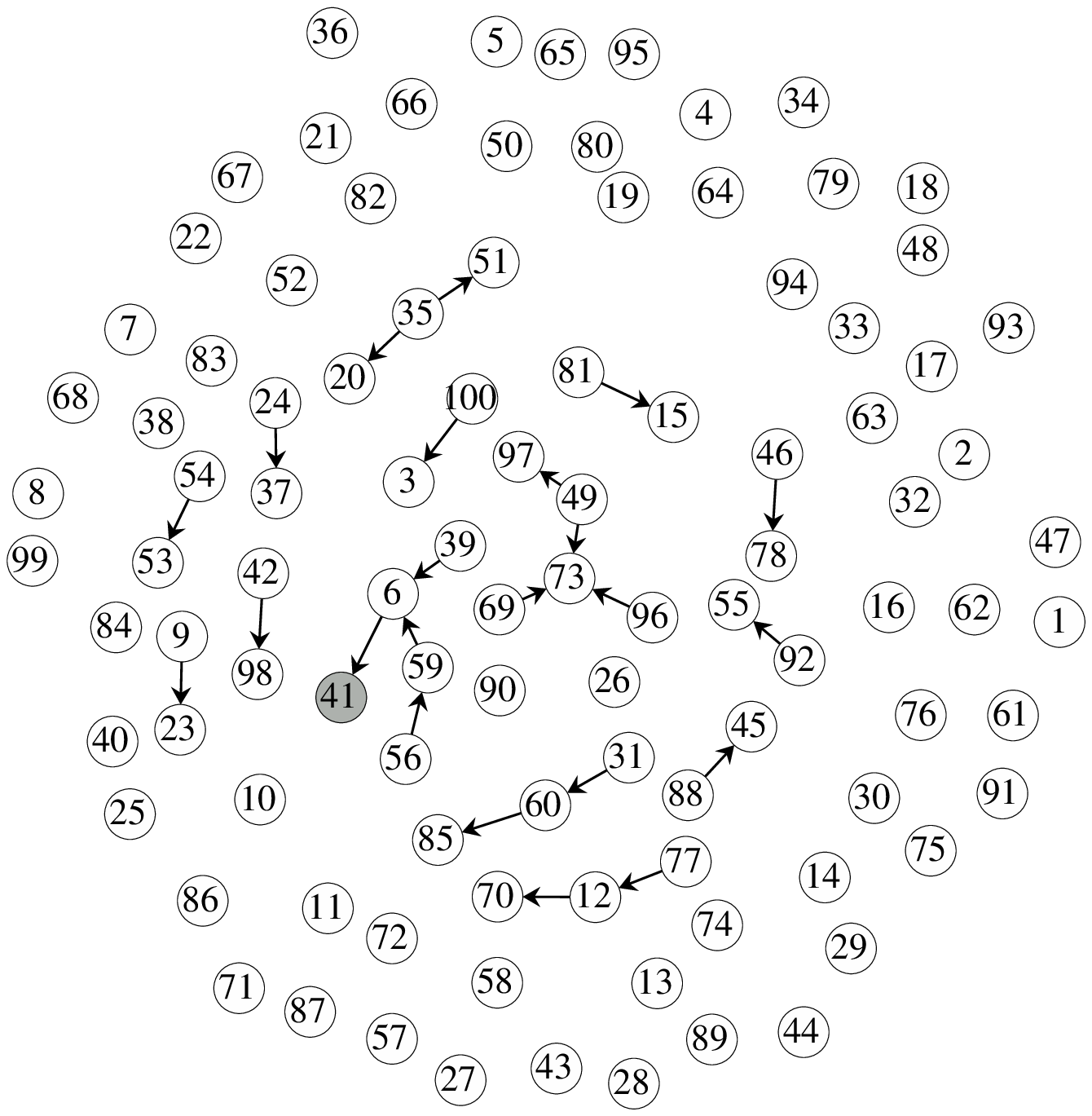}

\textbf{e) n=2854\hfill{}f) n=3022}

\includegraphics[  width=6.5cm,
  keepaspectratio]{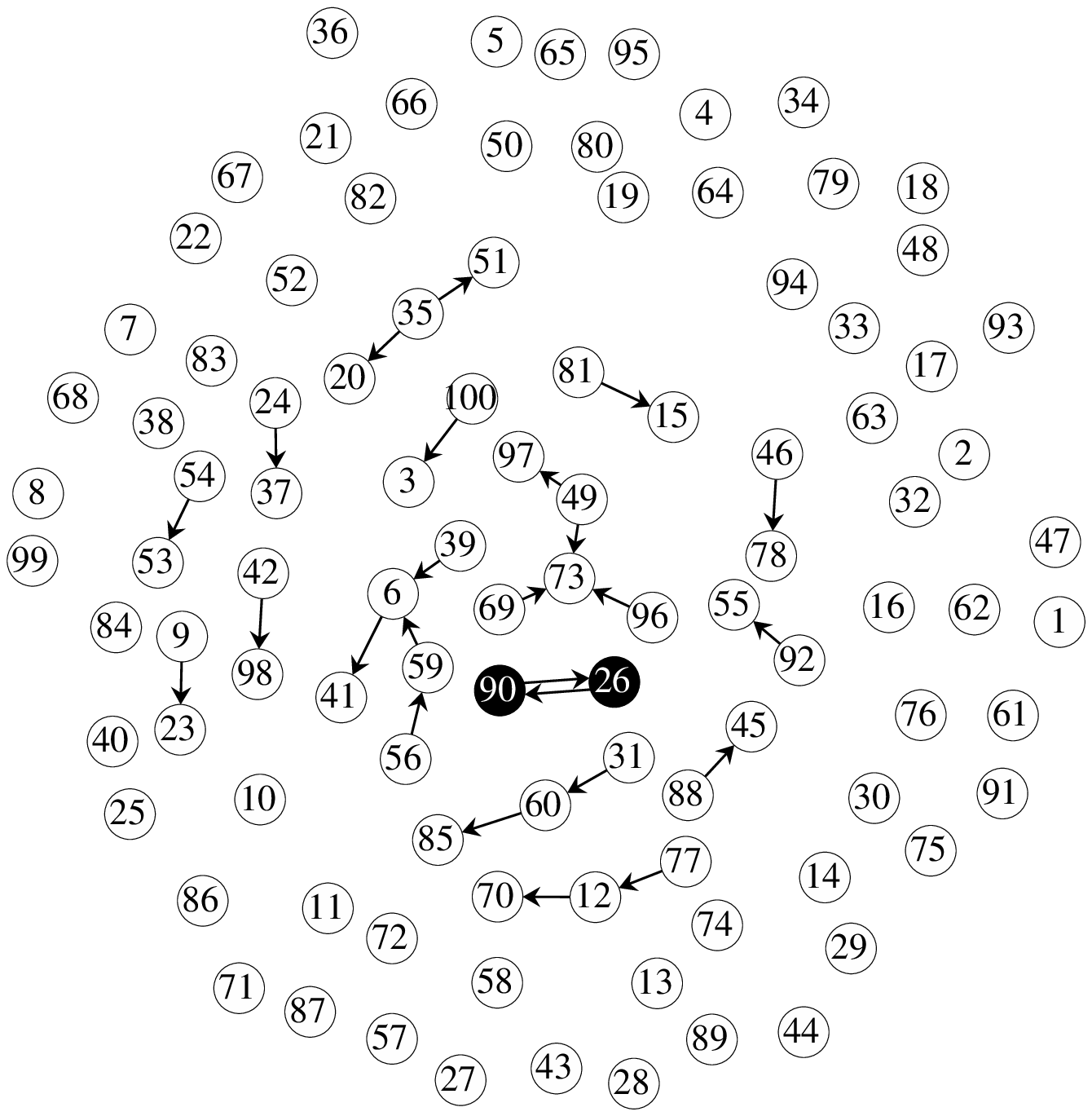}\hfill{}\includegraphics[  width=6.5cm,
  keepaspectratio]{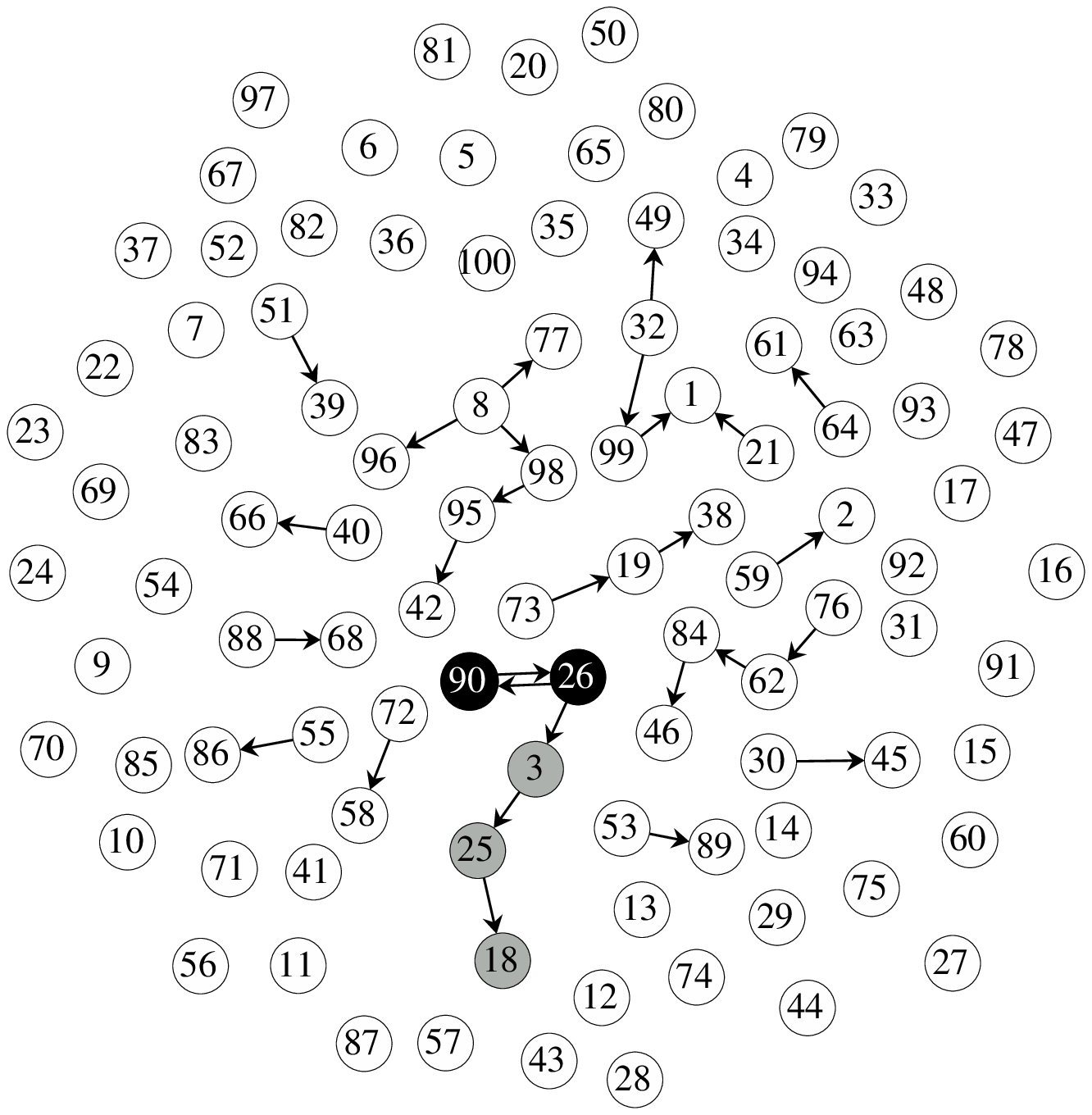}

\begin{center}Figure \ref{cap:snapshots}a--f, continued on next page.\end{center}
\end{figure}
\begin{figure}
\textbf{g) n=3386\hfill{}h) n=3387}

\includegraphics[  width=6.5cm,
  keepaspectratio]{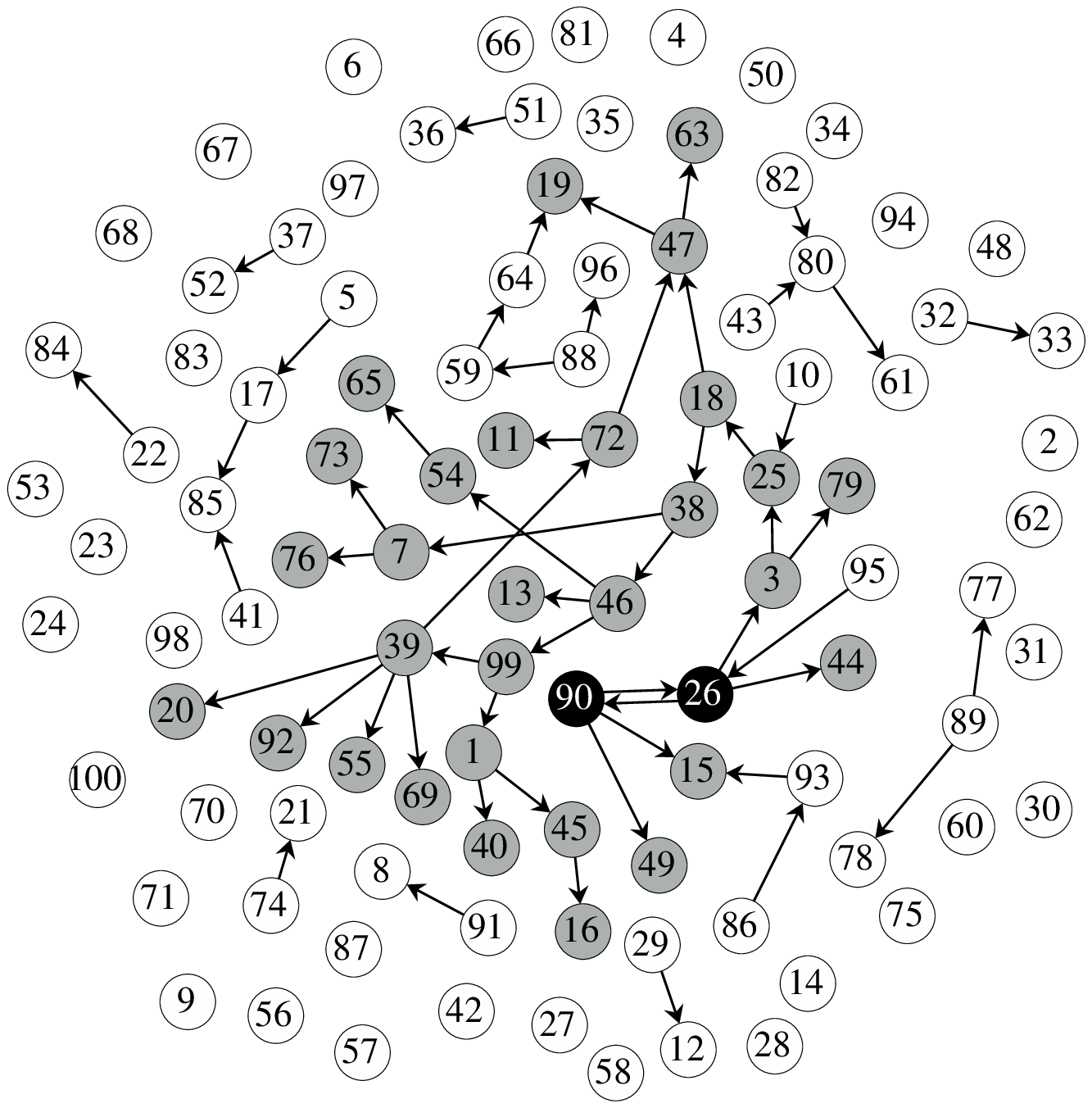}\hfill{}\includegraphics[  width=6.5cm,
  keepaspectratio]{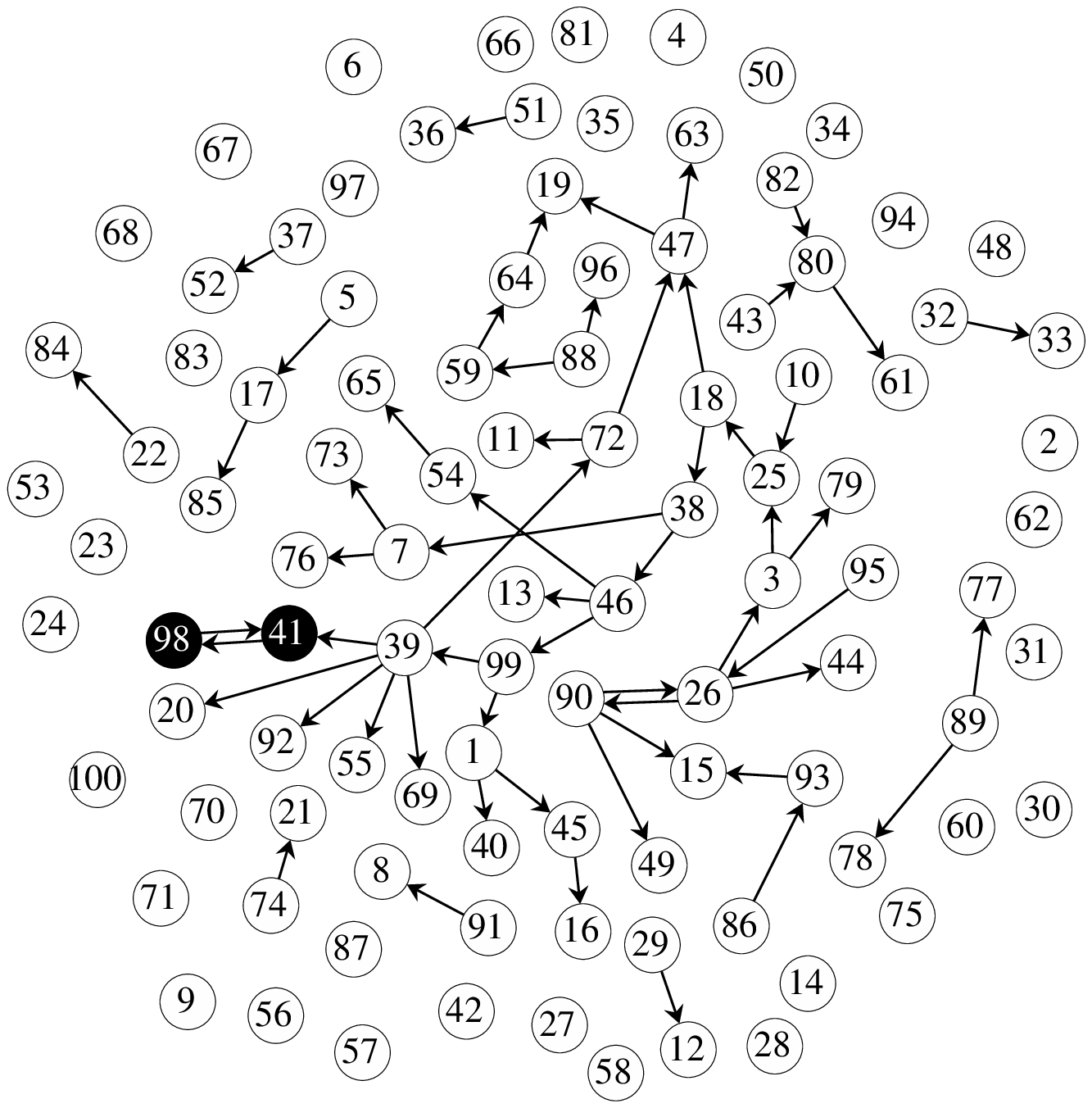}

\textbf{i) n=3402\hfill{}j) n=3403}

\includegraphics[  width=6.5cm,
  keepaspectratio]{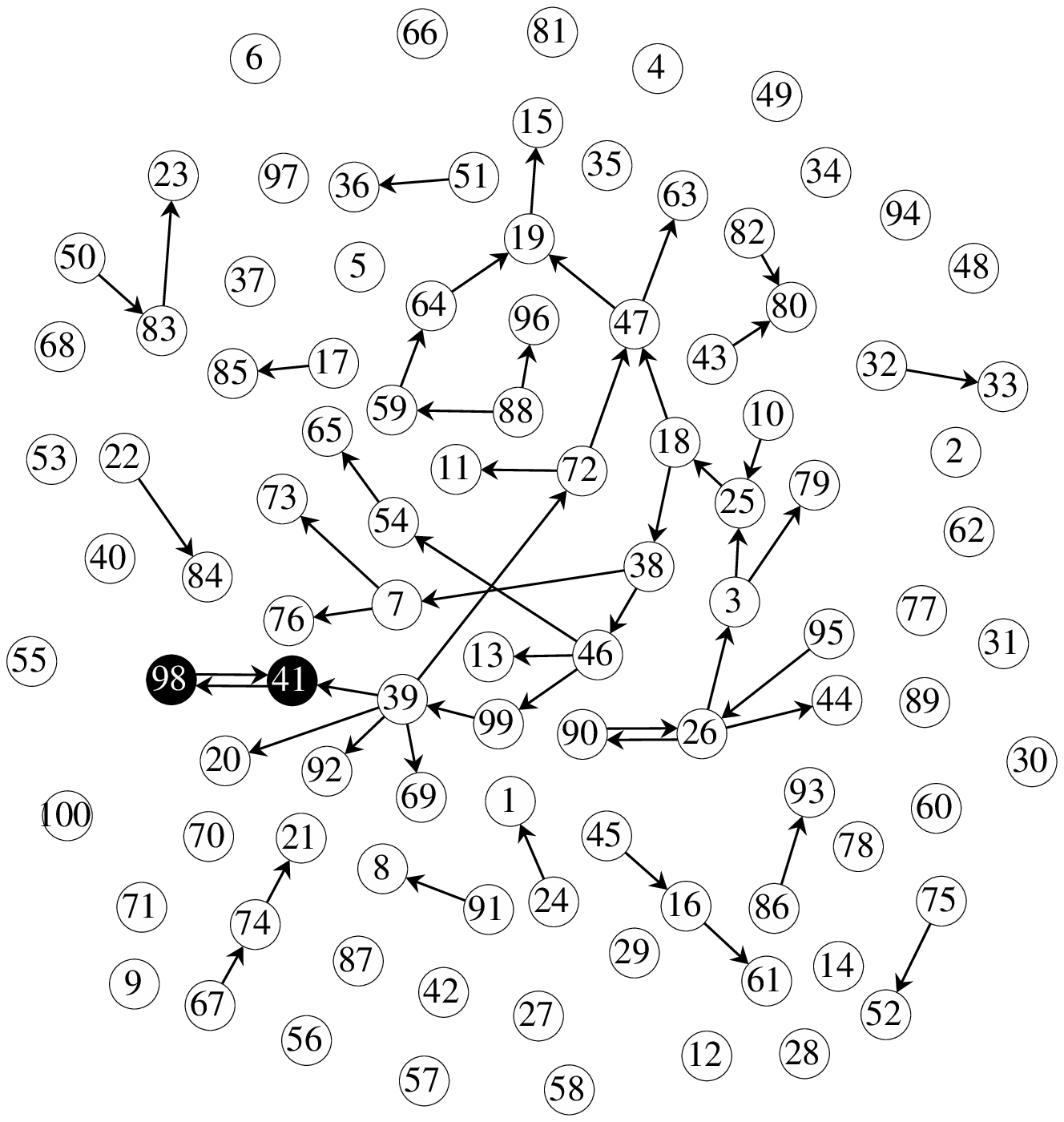}\hfill{}\includegraphics[  width=6.5cm,
  keepaspectratio]{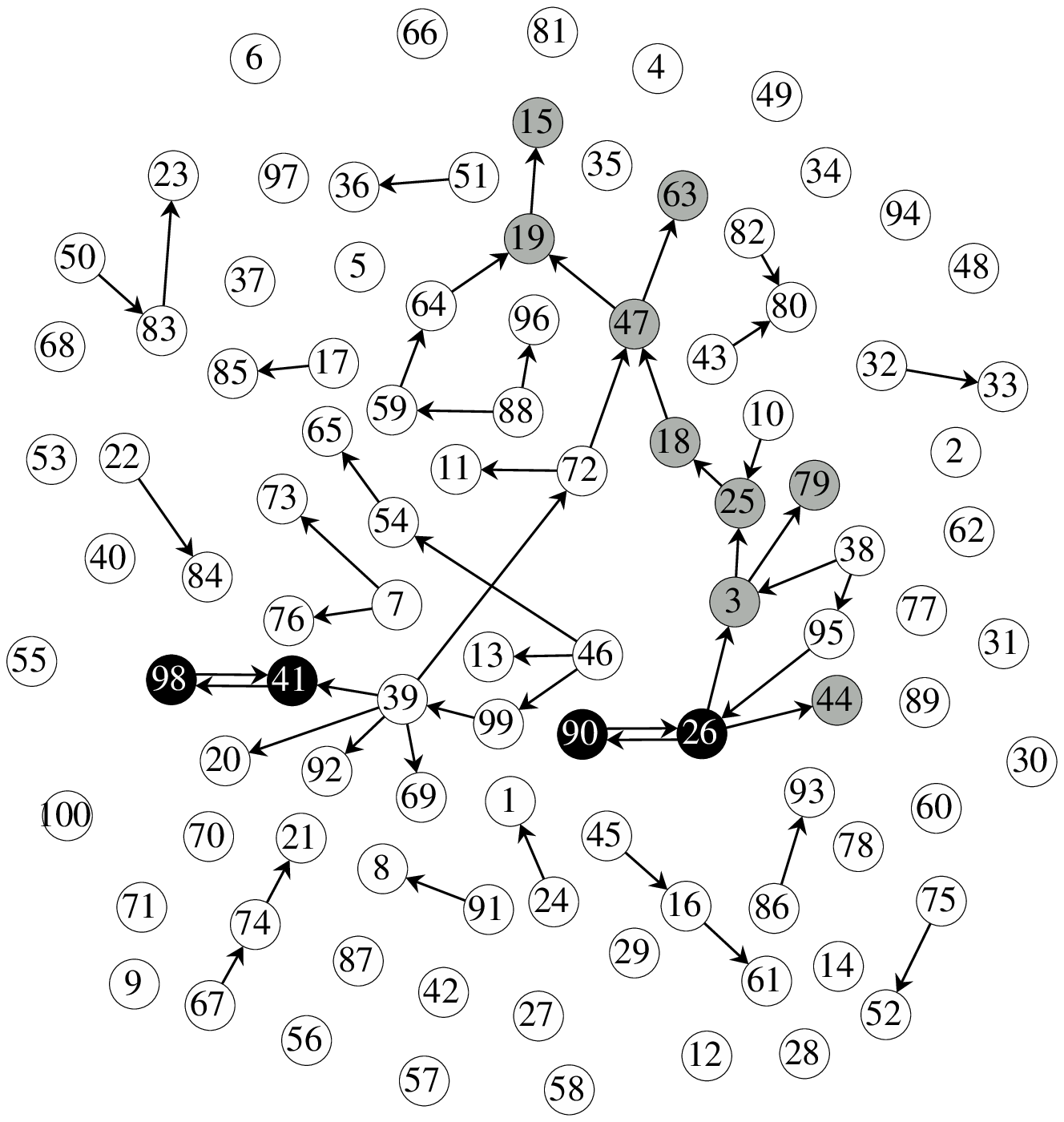}

\textbf{k) n=3488\hfill{}l) n=3489}

\includegraphics[  width=6.5cm,
  keepaspectratio]{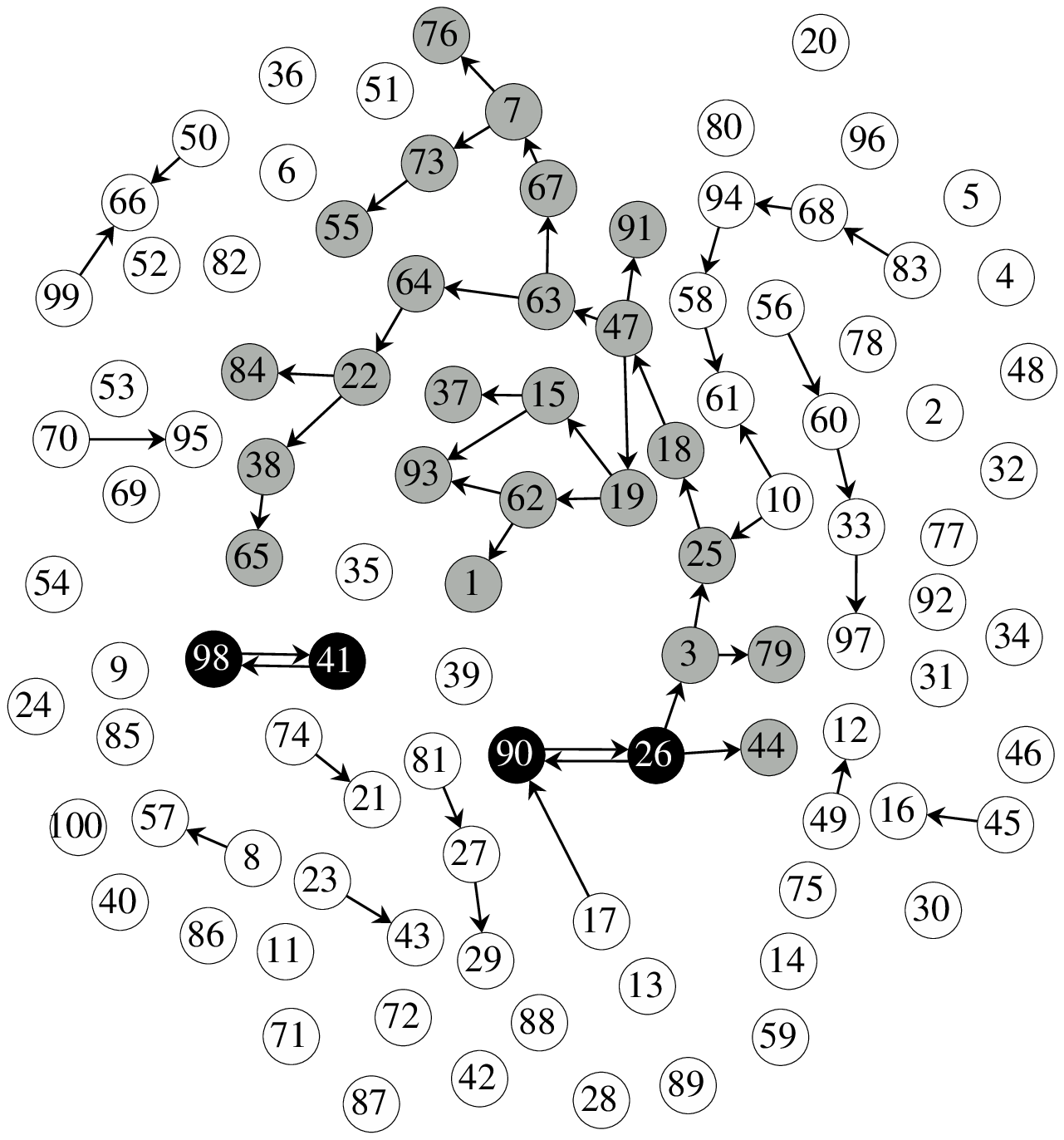}\hfill{}\includegraphics[  width=6.5cm,
  keepaspectratio]{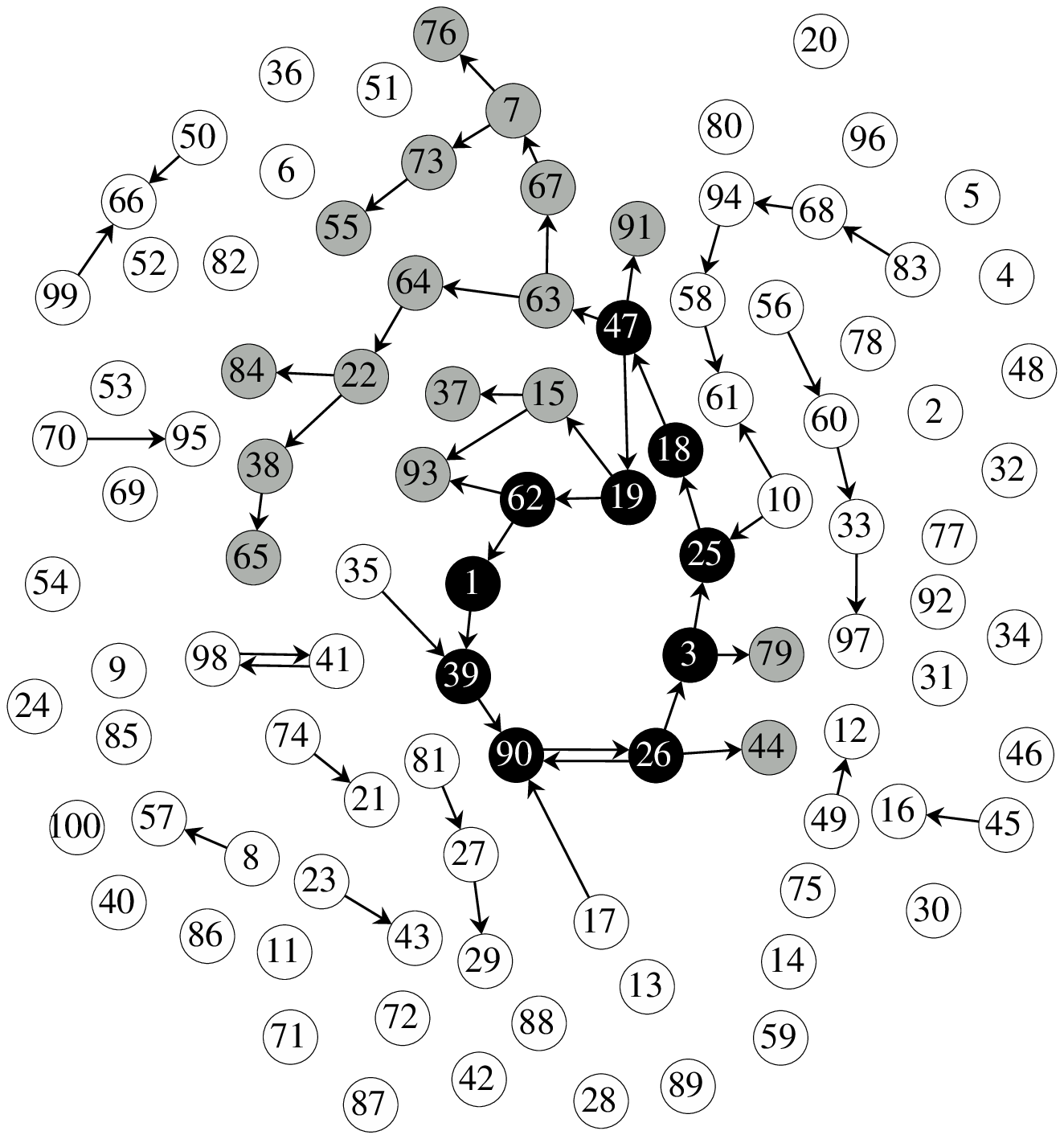}

\begin{center}Figure \ref{cap:snapshots}g--l, continued on next page.\end{center}
\end{figure}
\begin{figure}
\textbf{m) n=3880\hfill{}n) n=4448}

\includegraphics[  width=6.5cm,
  keepaspectratio]{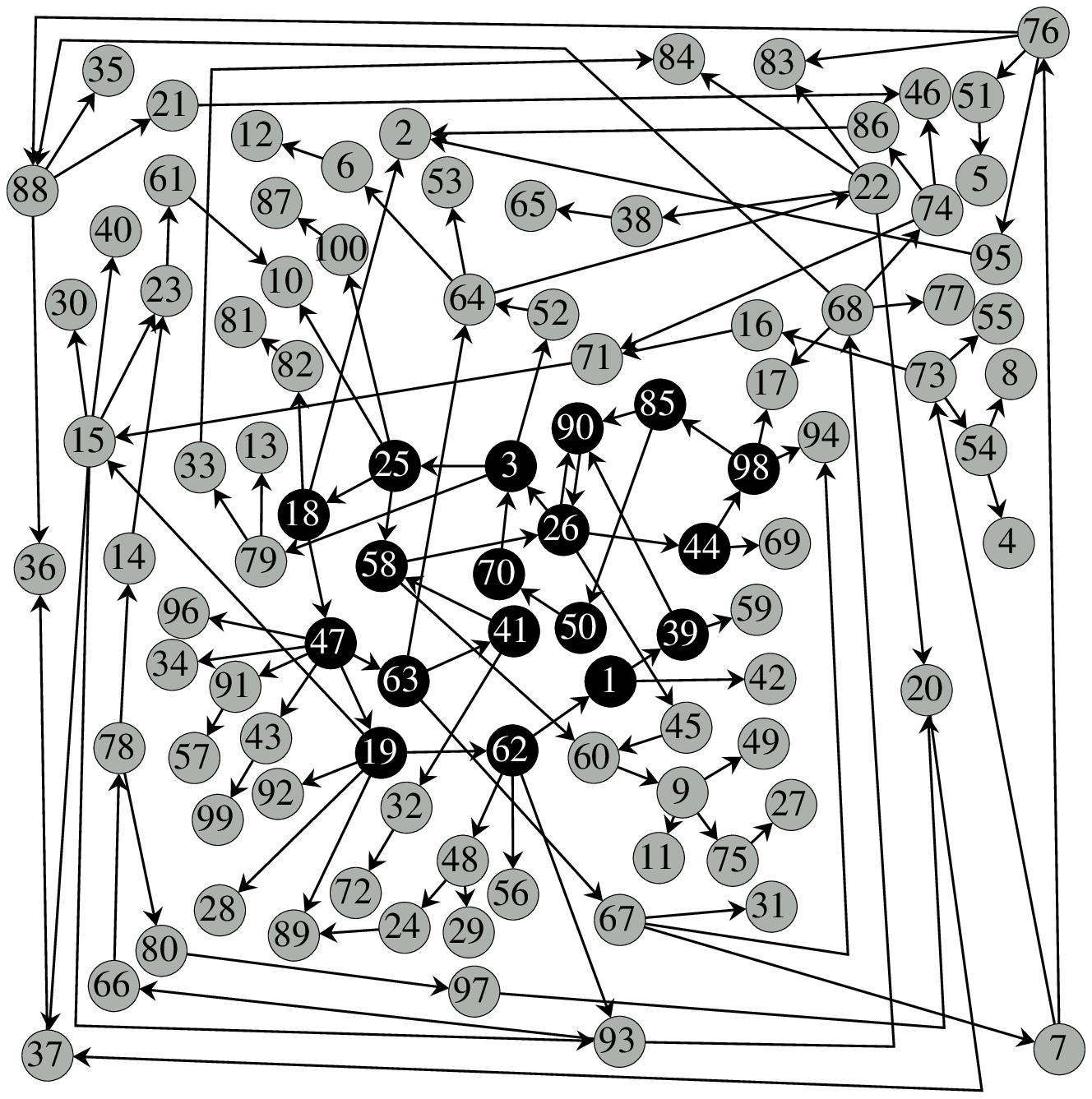}\hfill{}\includegraphics[  width=6.5cm,
  keepaspectratio]{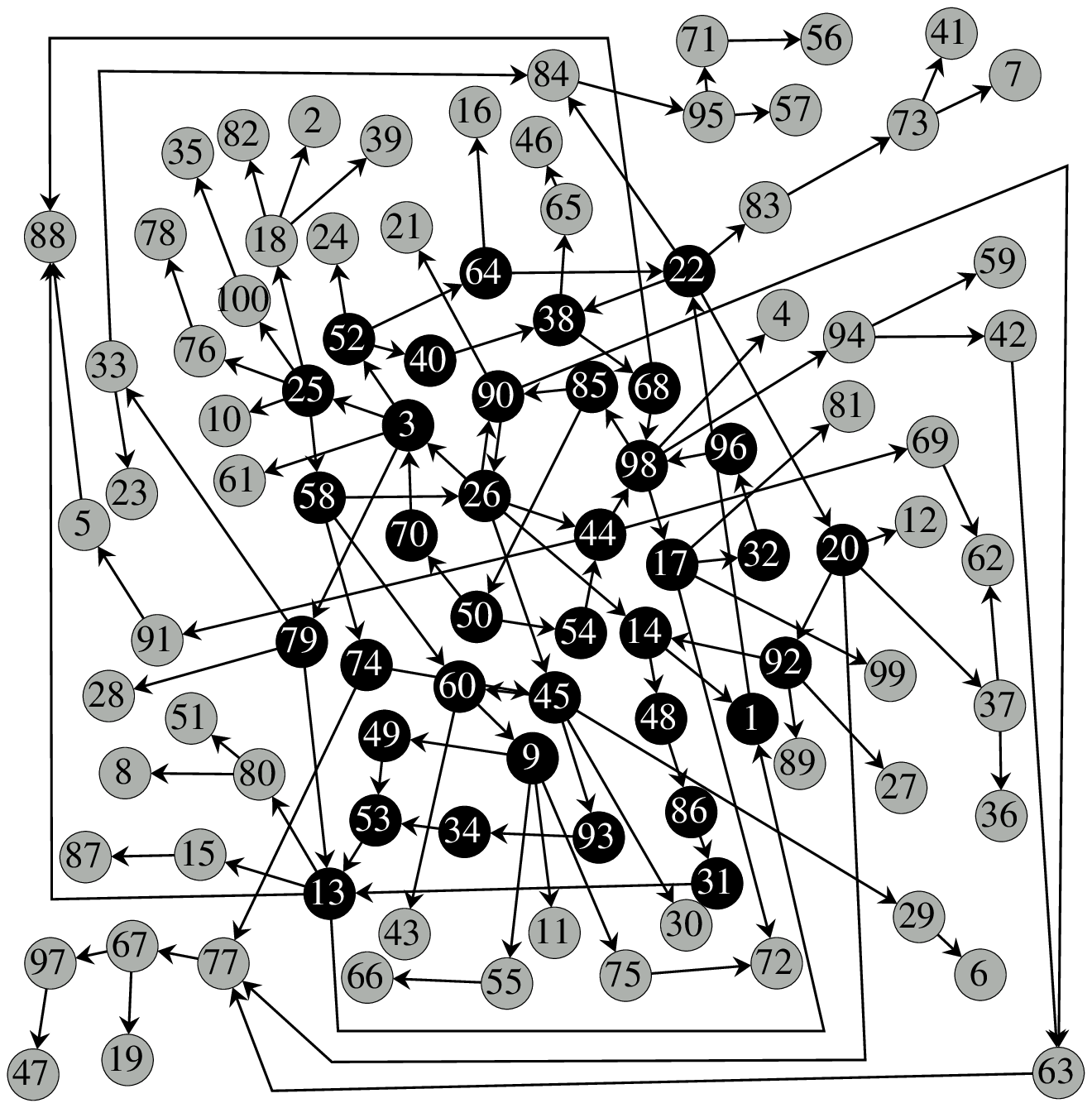}

\textbf{o) n=4695\hfill{}p) n=4696}

\includegraphics[  width=6.5cm,
  keepaspectratio]{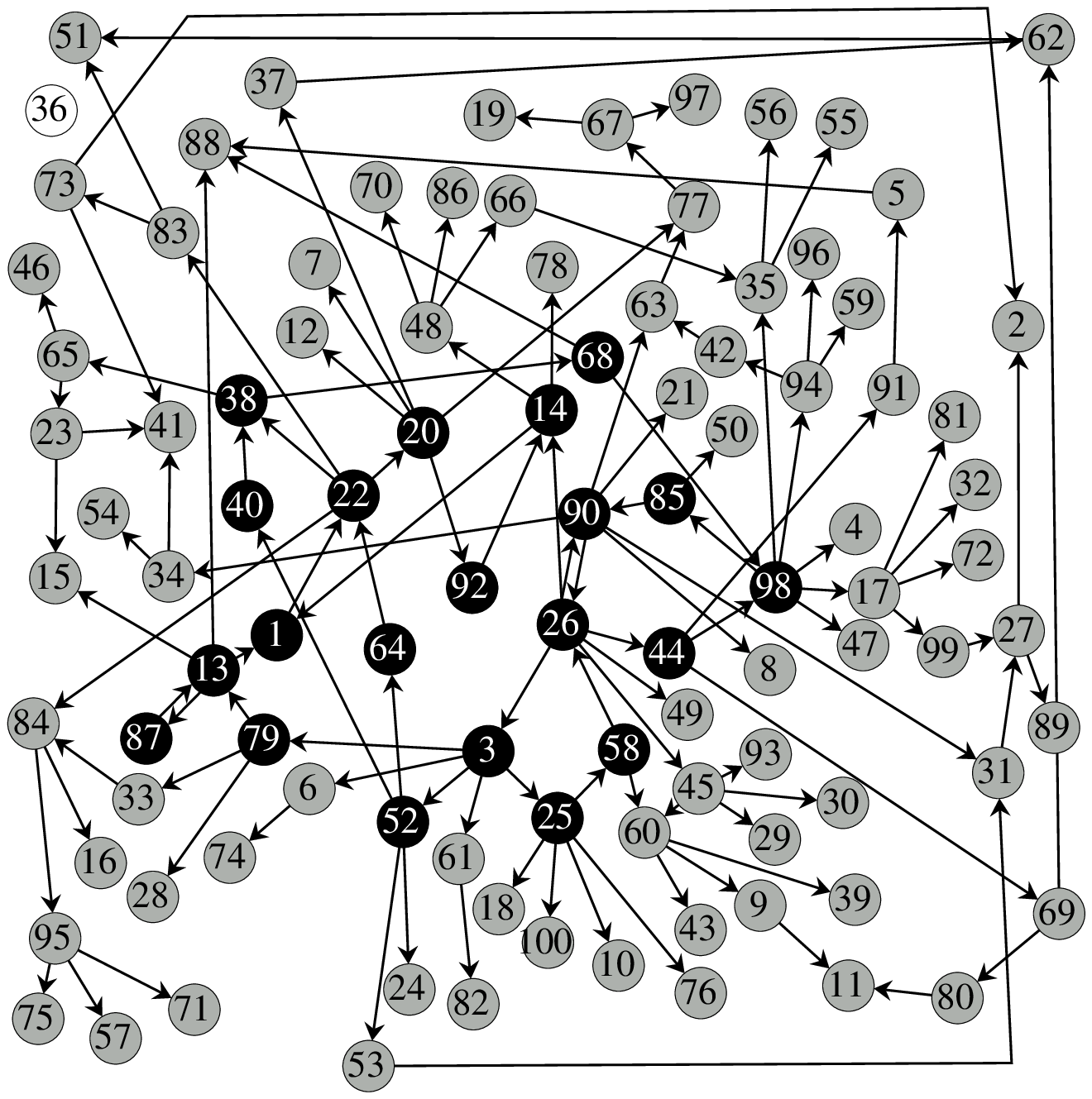}\hfill{}\includegraphics[  width=6.5cm,
  keepaspectratio]{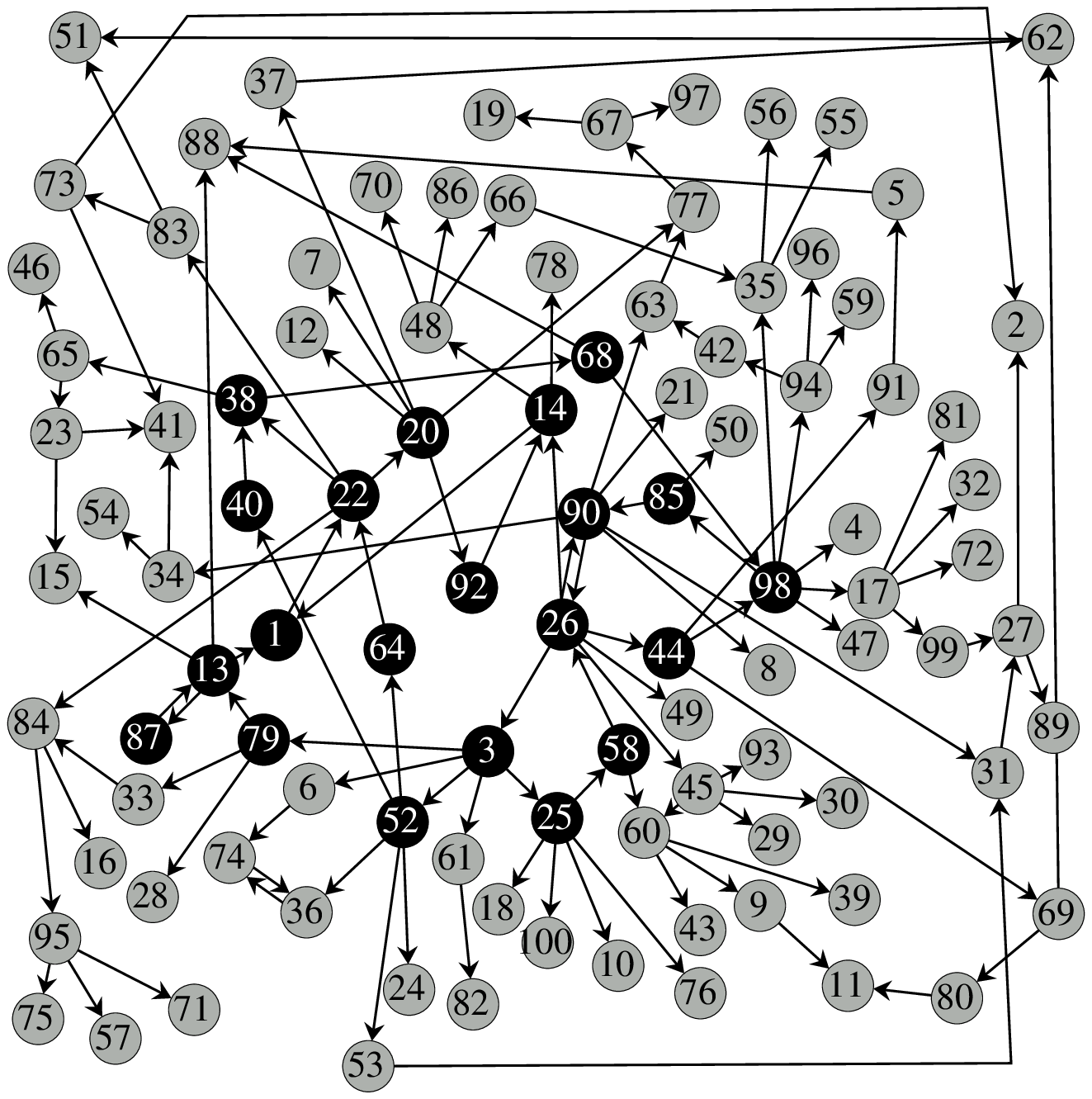}

\textbf{q) n=5041\hfill{}r) n=5042}

\includegraphics[  width=6.5cm,
  keepaspectratio]{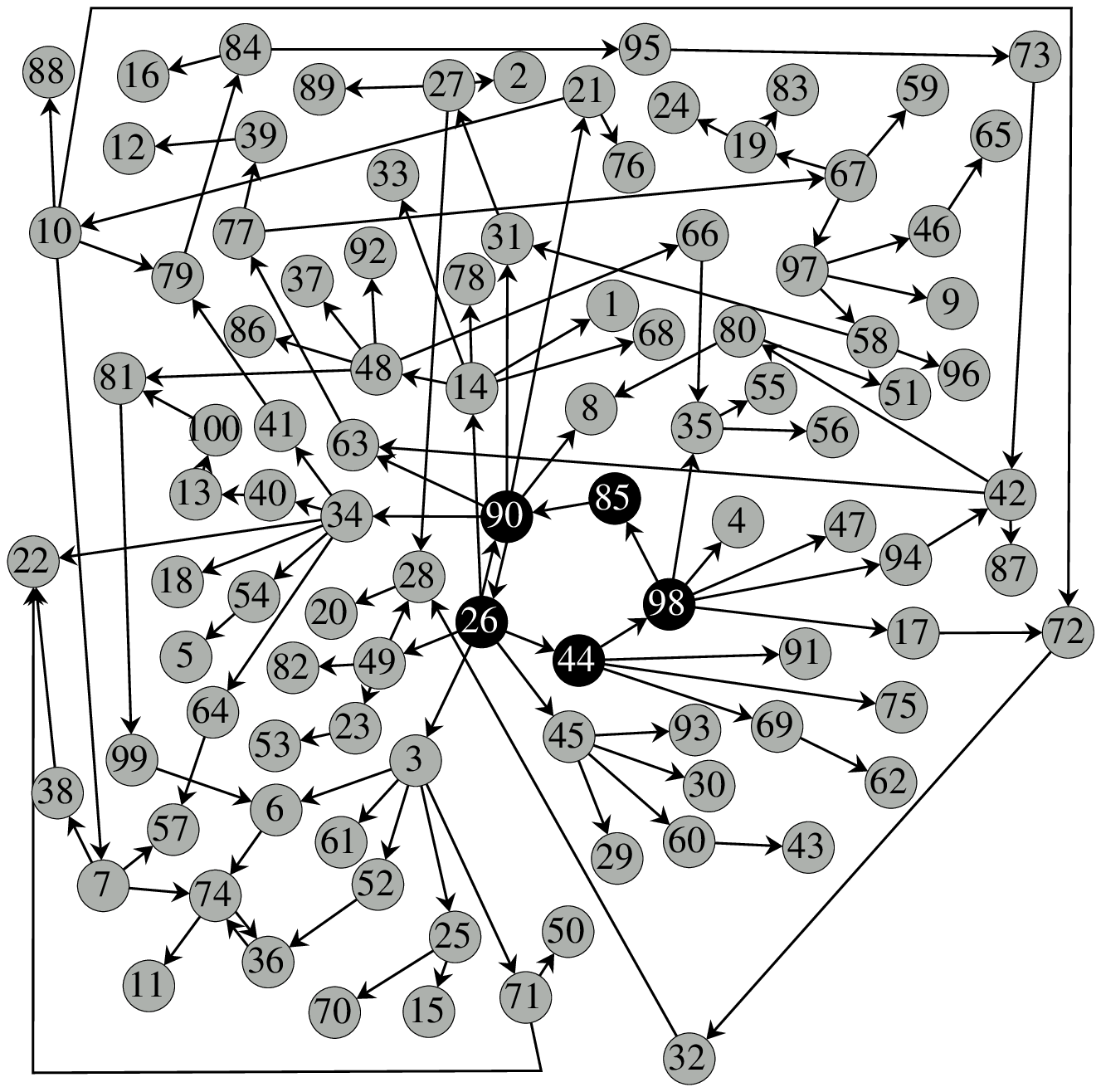}\hfill{}\includegraphics[  width=6.5cm,
  keepaspectratio]{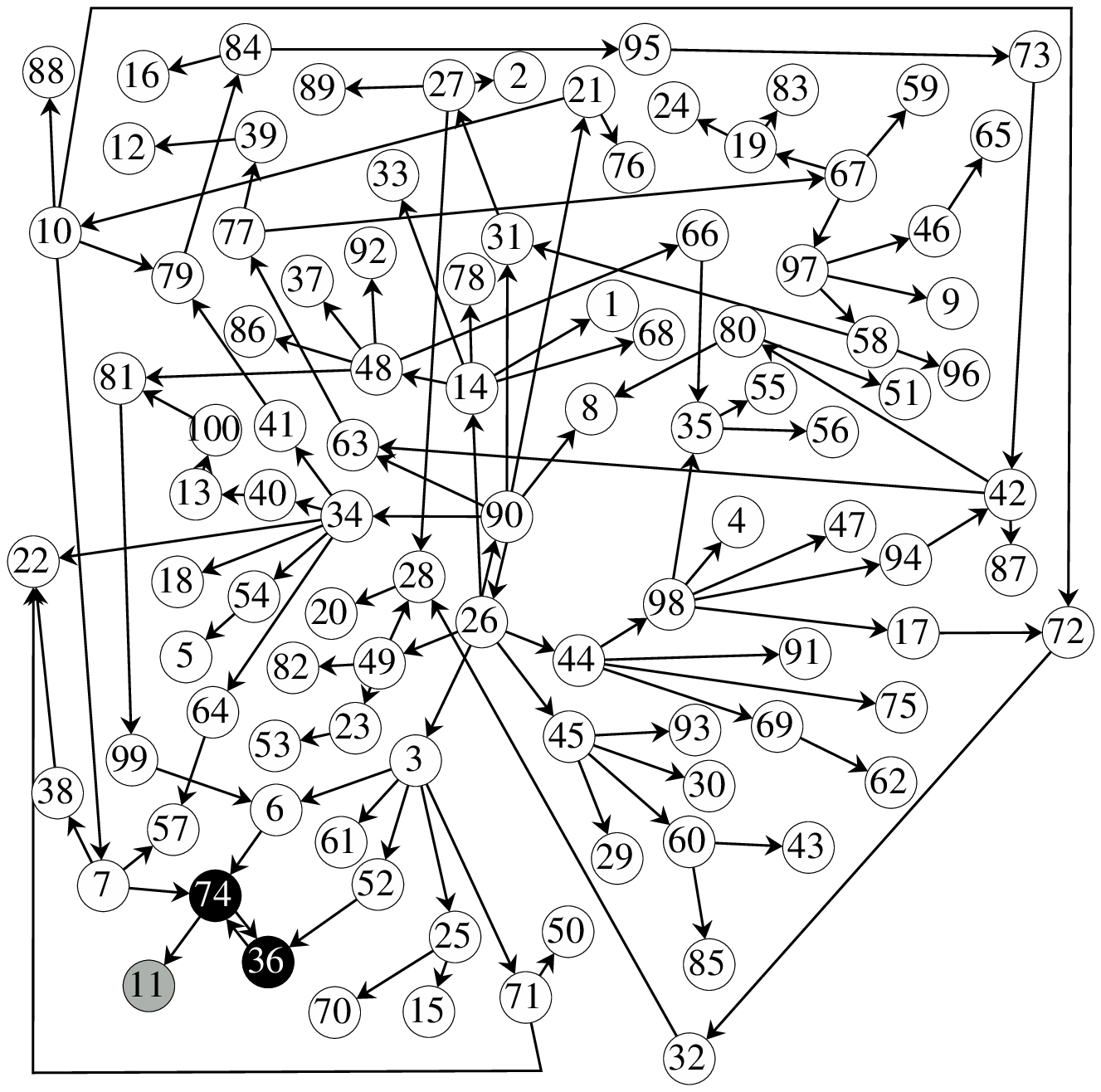}

\begin{center}Figure \ref{cap:snapshots}m--r, continued on next page.\end{center}
\end{figure}
\begin{figure}
\textbf{s) n=6061\hfill{}t) n=6062}

\includegraphics[  width=6.5cm,
  keepaspectratio]{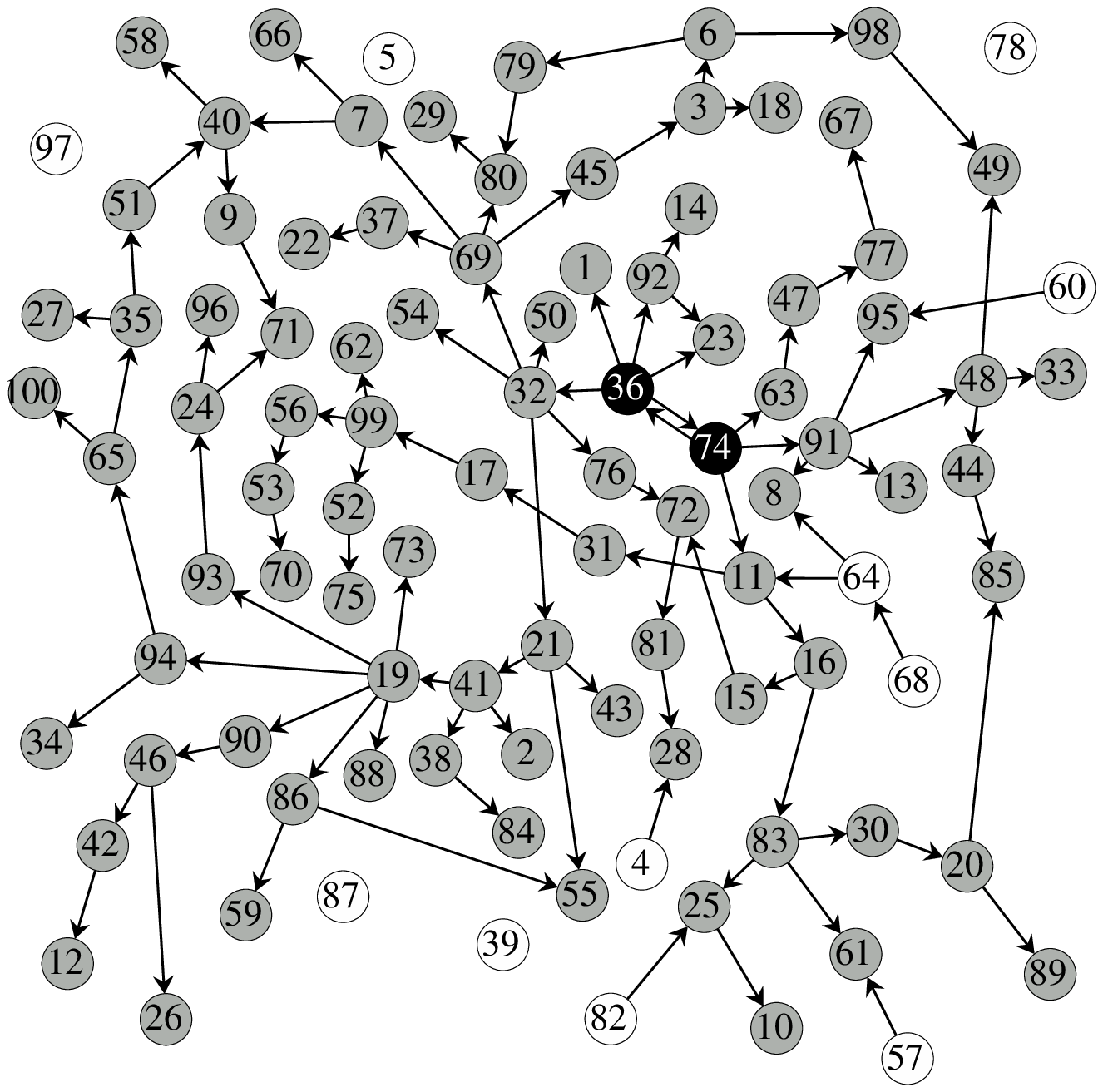}\textbf{\hfill{}}\includegraphics[  width=6.5cm,
  keepaspectratio]{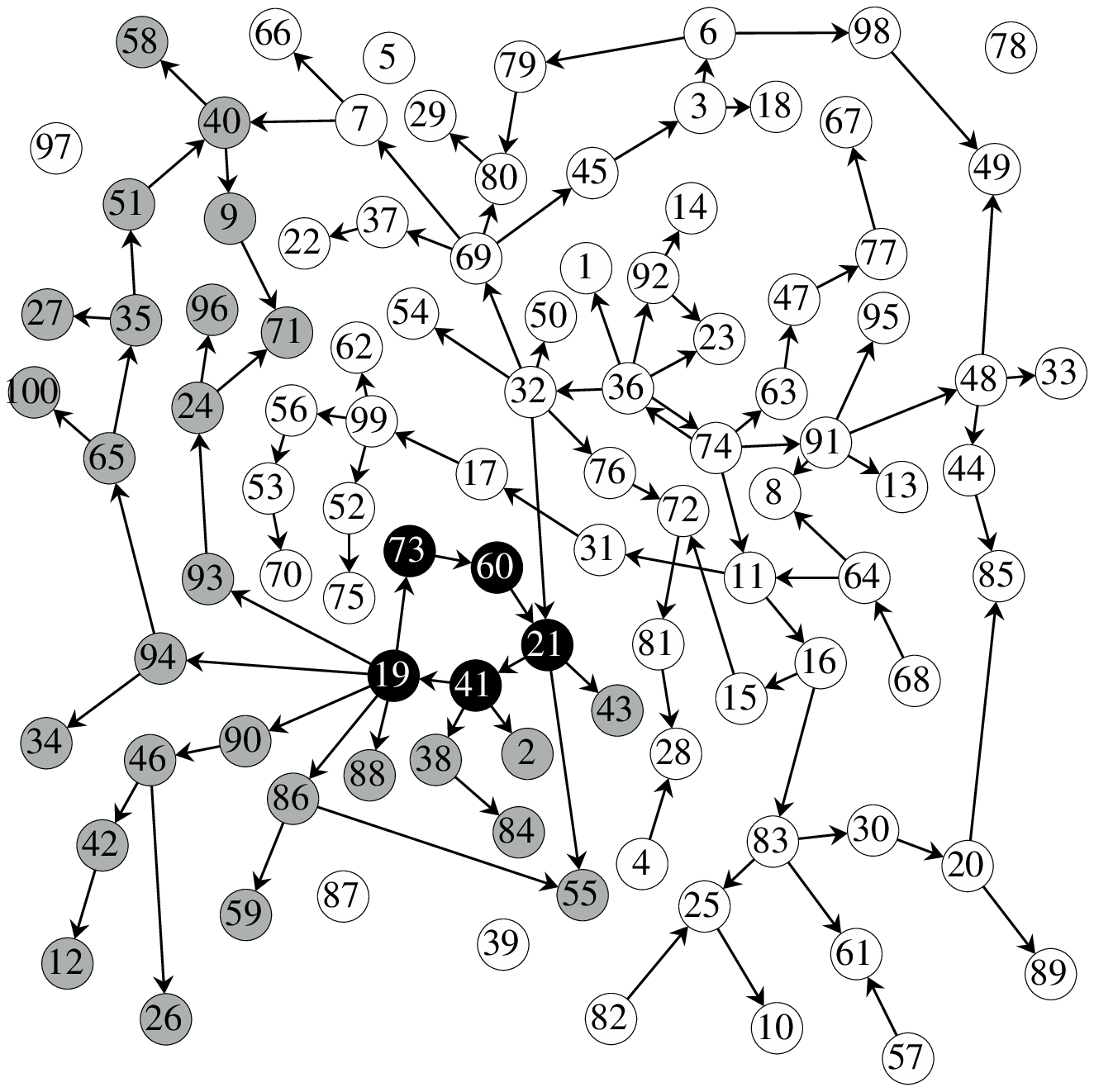}

\textbf{u) n=6070\hfill{}v) n=6212}

\includegraphics[  width=6.5cm,
  keepaspectratio]{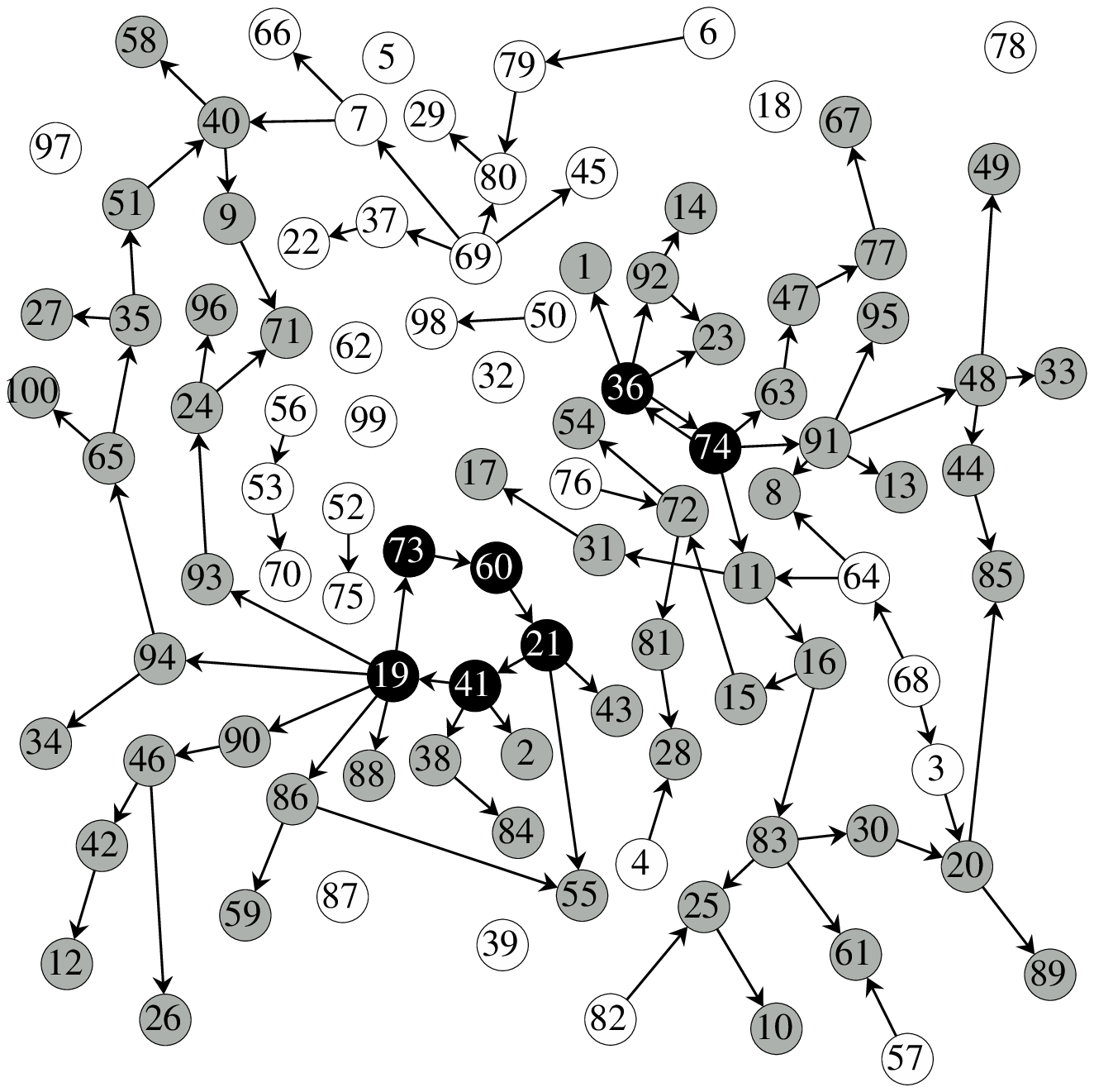}\textbf{\hfill{}}\includegraphics[  width=6.5cm,
  keepaspectratio]{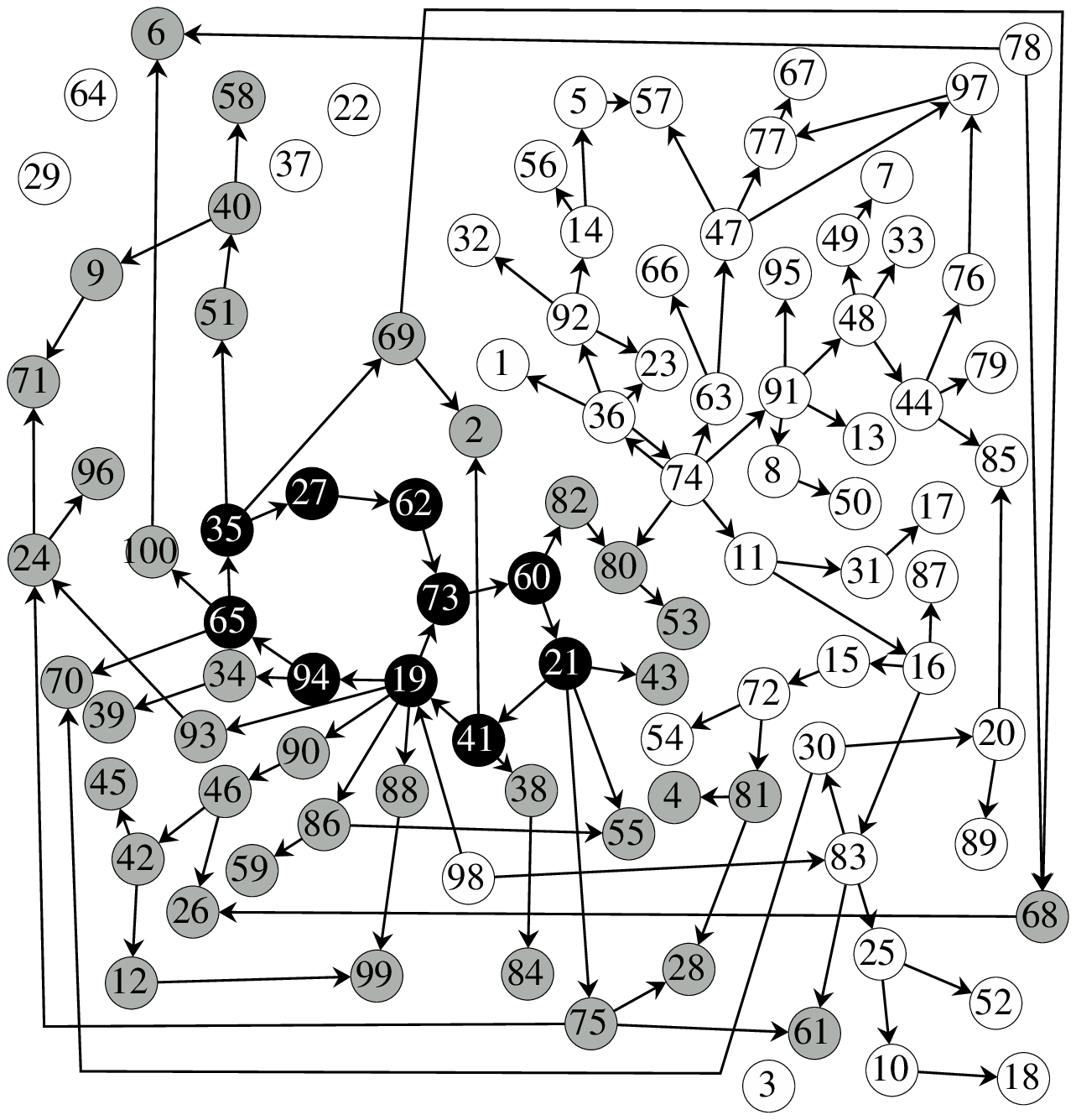}

\textbf{w) n=8232\hfill{}x) n=8233}

\includegraphics[  width=6.5cm,
  keepaspectratio]{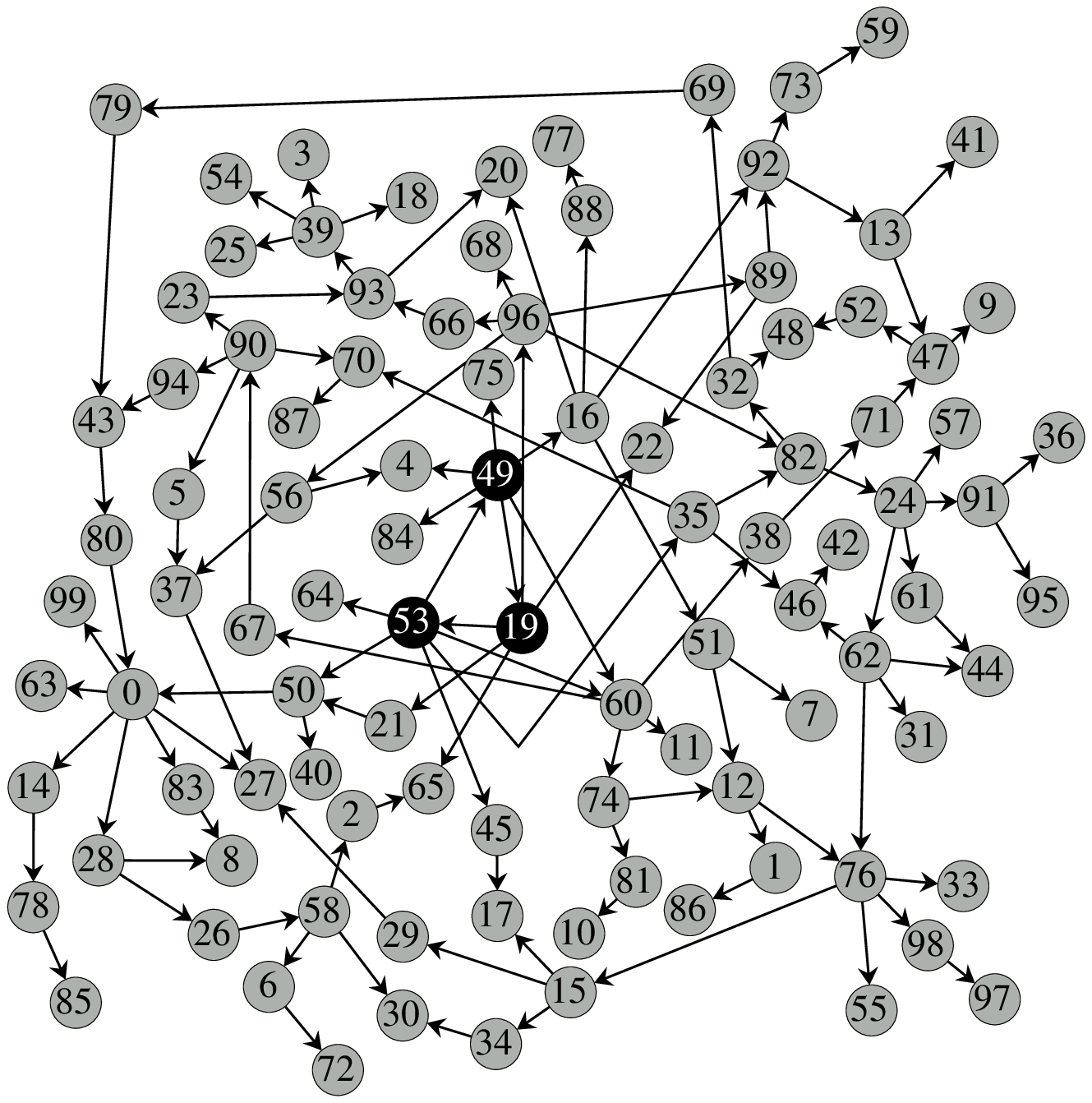}\textbf{\hfill{}}\includegraphics[  width=6.5cm,
  keepaspectratio]{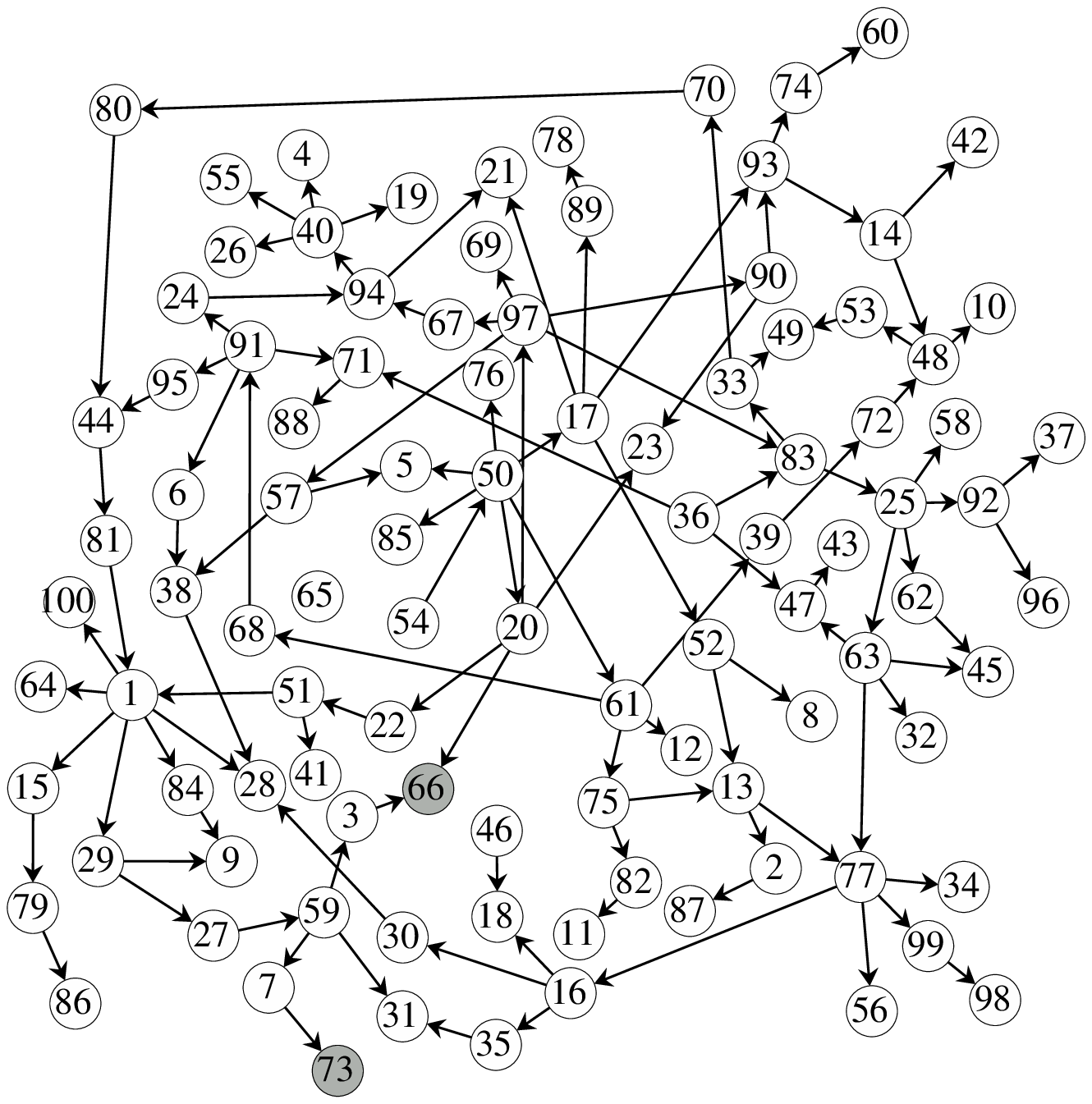}

\begin{center}Figure \ref{cap:snapshots}s--x, continued on next page.\end{center}
\end{figure}

\newpage
By $n=5041$ (Figure \ref{cap:snapshots}q) the core has shrunk to
5 nodes. Node 85 gets removed which leaves the graph at $n=5042$
(Figure \ref{cap:snapshots}r) with one 2-cycle (nodes 36 and 74)
downstream from another (26 and 90). Only nodes 36, 74 and node 11
(downstream from 74) are non-zero in the new attractor. All other
nodes have become extinct, triggering a drop in $s_{1}$ from 100
to 3. Over the next thousand time steps the dominant ACS rebuilds
itself around the new core comprising nodes 36 and 74 ($n=6061$,
Figure \ref{cap:snapshots}s). Then node 60 comes in, completing a
cycle of 5 nodes that is downstream from the 36-74 cycle ($n=6062$,
Figure \ref{cap:snapshots}t) and again there is a large drop in $s_{1}$
as 36, 74 and many nodes having access from them become extinct. At
$n=6070$ (Figure \ref{cap:snapshots}u), however, the 36-74 cycle
is resurrected as the path connecting it to the 5-cycle is broken.
It does not last for too long as the 5-cycle is strengthened by another
cycle at $n=6212$ (Figure \ref{cap:snapshots}v) and the 36-74 cycle
again becomes extinct. This time it does not revive, as the dominant
ACS builds around the strengthened 5-cycle. About two thousand time
steps later, at $n=8232$ (Figure \ref{cap:snapshots}w), the graph
is again fully autocatalytic but is on the verge of a major collapse.
The core is just a 3-cycle and node 54 gets removed, thus destroying
the only cycle in the graph. The graph at $n=8233$ (Figure \ref{cap:snapshots}x)
is left without an ACS; the system is now once again in the random
phase. Within a few hundred time steps all of the structure is destroyed
and the graph at $n=8500$ (Figure \ref{cap:snapshots}z) and $n=10000$
(Figure \ref{cap:snapshots}z) are similar to the initial graph at
$n=1$ (Figure \ref{cap:snapshots}a). At this point, I will stop
following the run but eventually a new ACS arises, grows and spans
the entire graph, then gets destroyed and another round starts (see
Figure \ref{cap:longtimefig}).

\chapter{\label{cha:Formation-and-Growth}Formation and Growth of Autocatalytic
Sets}

\section{The random phase}

\begin{figure}
\begin{center}\includegraphics[  width=15cm,
  keepaspectratio]{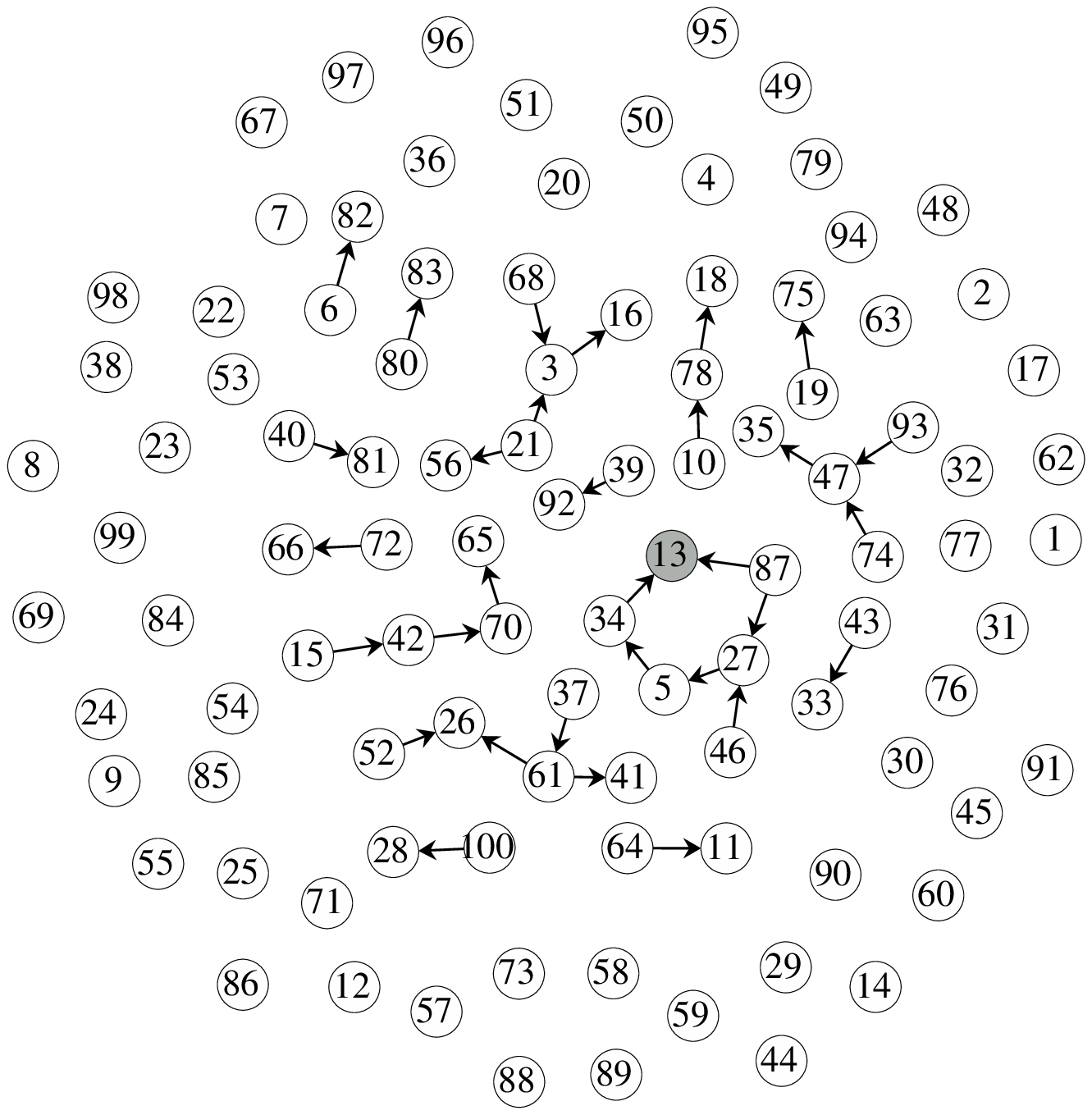}\end{center}

\caption{The initial random graph, at $n=1$, for the run shown Figure \ref{cap:s1lambda}b.
Node 13 is the only node with non-zero $X_{i}$ because it is at the
end of the longest path.\label{cap:n=3D1}}
\end{figure}

For the run with $s=100,p=0.0025$ displayed in Figure \ref{cap:s1lambda}b,
the initial random graph at $n=1$ contains no cycles, and hence no
ACSs, and its Perron-Frobenius eigenvalue is $\lambda _{1}=0$. Figure
\ref{cap:n=3D1} shows this initial graph. 

For such a graph $X_{i}>0$ for all nodes that are at the ends of
the longest paths of nodes, and $X_{i}=0$ for every other node (see
section \ref{sec:The-attractor-of}). In Figure \ref{cap:n=3D1},
there are two paths of length 4, which are the longest paths in the
graph. Both end at node 13, which is therefore the only non-zero node
in the attractor for this graph. This node, then, is the only node
protected from removal during the graph update. Node 13 has a high
relative population but its supporting nodes do not have high relative
populations. Inevitably within a few graph updates a supporting node
will be removed from the graph. When that happens node 13 which presently
has non-zero $X_{i}$ will no longer be at the end of the longest
path and hence will get $X_{i}=0$. In the run node 34 is replaced
at the 8th time step. After that node 13 joins the set $\mathcal{L}$
of nodes with least $X_{i}$. Thus no structure is stable when there
is no ACS. Eventually, all nodes are removed and replaced.

\begin{figure}
\textbf{a)}

\includegraphics[  width=12cm,
  keepaspectratio]{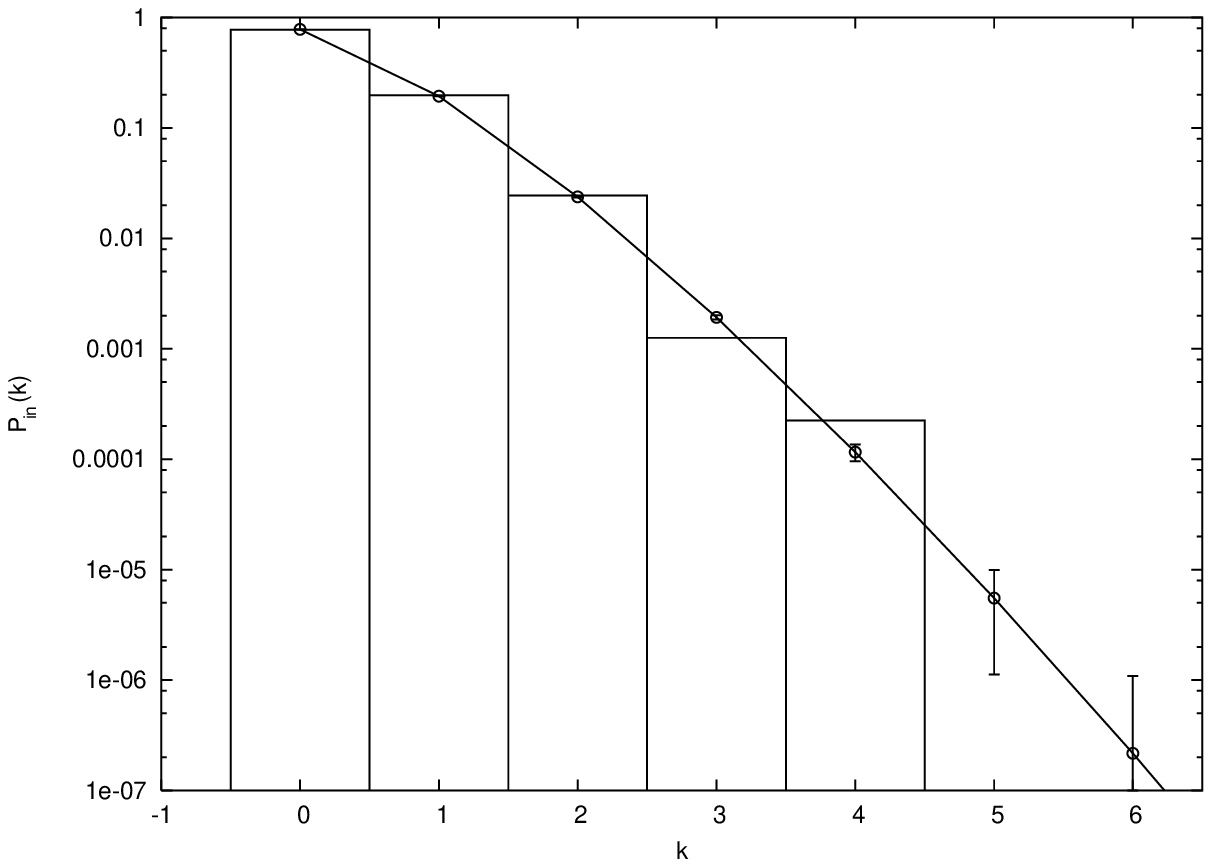}

\textbf{b)}

\includegraphics[  width=12cm,
  keepaspectratio]{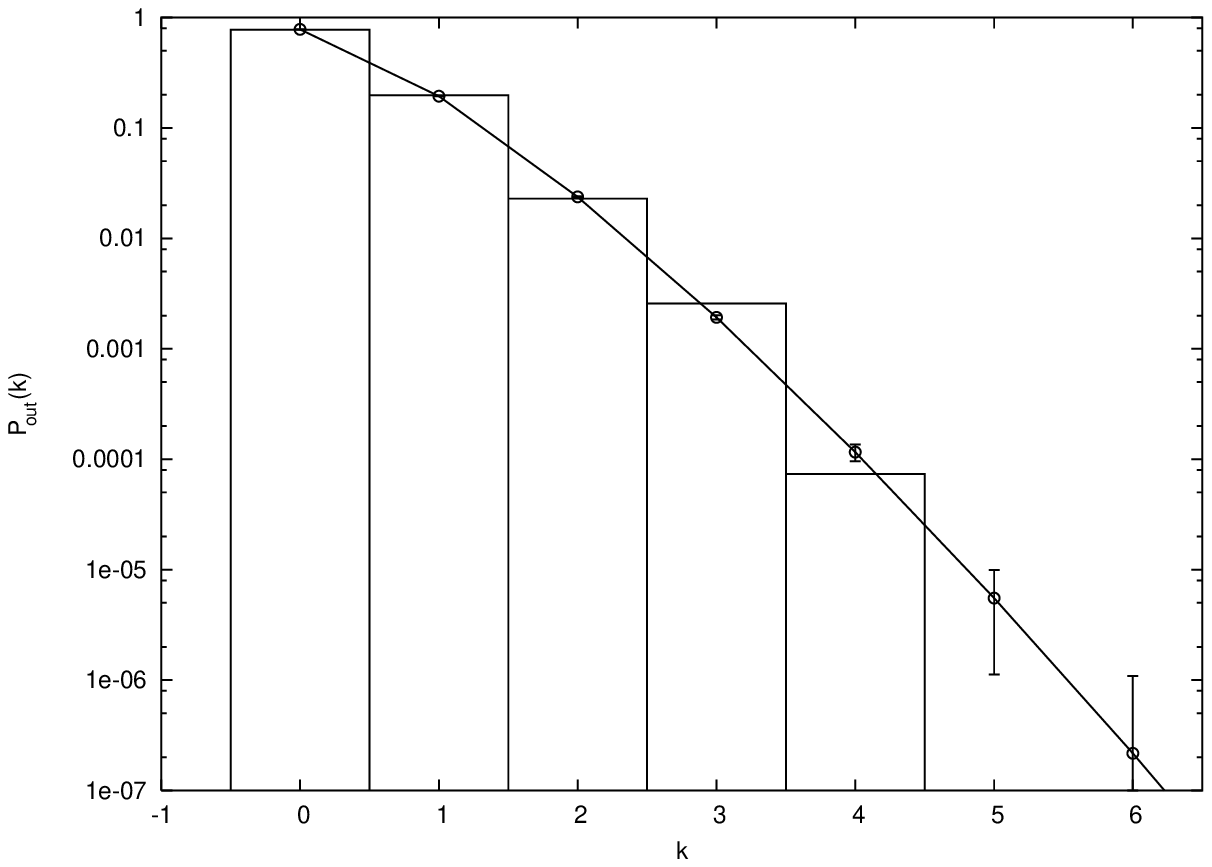}

\caption{Comparison of the \textbf{a.} in-, and \textbf{b.} out-degree distributions
(bars) of 2853 graphs from the random phase ($n=1$ to $n=2853$ of
the run in Figure \ref{cap:s1lambda}b) with the binomial degree distribution
$B_{p}^{s-1}(k)\equiv \: ^{s-1}C_{k}\, p^{k}(1-p)^{s-1-k}$ (lines
with circles) expected for a random graph with $s=100,p=0.0025$.
The error bars show the expected standard deviation, $\sqrt{\frac{B_{p}^{s-1}(k)\left[1-B_{p}^{s-1}(k)\right]}{285300}}$,
of the distribution from the binomial for a finite sample of 2853
random graphs each with 100 nodes. \label{cap:degdistcmp}}
\end{figure}

\begin{figure}
\includegraphics[  width=13cm,
  keepaspectratio]{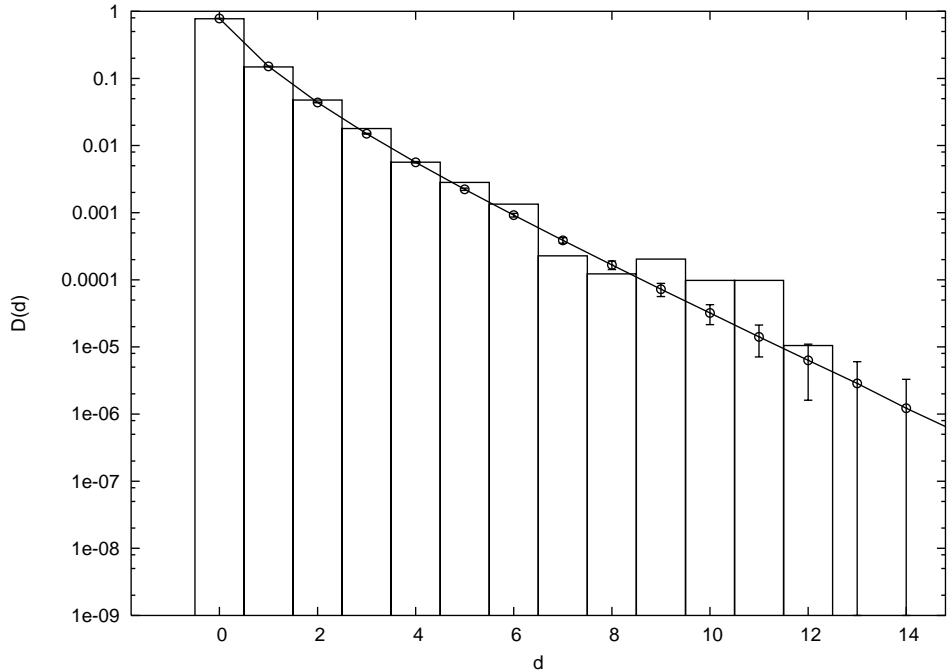}

\caption{Comparison of the dependency distribution of 2853 graphs (bars) from
the random phase ($n=1$ to $n=2853$ of the run in Figure \ref{cap:s1lambda}b)
with the dependency distribution (shown in figure \ref{cap:depdistrnd})
of $10^{7}$ random graphs with $s=100,p=0.0025$ (line with circles).
The error bars show the expected standard deviation, $\sqrt{\frac{D(d)\left[1-D(d)\right]}{285300}}$,
of the distribution for a finite sample of 2853 random graphs each
with 100 nodes (this estimate of the error assumes that the dependencies
of each node is independently picked from the distribution $D(d)$
and does not take into account any correlations that would exist,
even for random graphs, between dependencies of nodes of the same
graph).\label{cap:depdeistcmp}}\lyxline{\normalsize}

\end{figure}

Figures \ref{cap:degdistcmp} and \ref{cap:depdeistcmp} compare properties
of the graphs from the first random phase of the run (from $n=1$
to $n=2853$, inclusive), with properties of the random graph ensemble,
$G_{s}^{p}$, described in section \ref{sec:Random-graphs}. The average
number of links for the 2853 graphs in the random phase is 25.2, with
standard deviation 5.3. For the ensemble $G_{s}^{p}$, with $s=100,p=0.0025$,
the average number of links is $ps(s-1)=24.75$. Figure \ref{cap:degdistcmp}
shows that the in and out degree distributions of graphs taken from
the random phase are close to the binomial distribution expected for
$G_{s}^{p}$ for smaller values of degree but depart from the binomial
for larger degrees. Figure \ref{cap:depdeistcmp} compares the dependency
distribution of the random phase with that of $10^{7}$ graphs taken
from the ensemble $G_{s}^{p}$. Again, the distributions are similar
for smaller dependency values, and differ for higher values. \\

Although the properties of the random phase graphs do not exactly
match the properties of graphs from the ensemble $G_{s}^{p}$, the
main point about the graph dynamics holds nevertheless: In the random
phase no graph structure is stable for very long and eventually all
nodes are removed and replaced.

Note that the initial random graph is likely to contain no cycles
when $p$ is small ($ps\ll 1$). If larger values of $p$ are chosen,
it becomes more likely that the initial graph will contain a cycle.
If it does, there is no random phase; the system is then in the growth
phase, discussed below, right from the initial time step.

\section{\label{sec:The-growth-phase}The growth phase}

At some graph update an ACS is formed by pure chance. The probability
of this happening can be closely approximated by the probability of
a 2-cycle (the simplest ACS with 1-cycles being disallowed) forming
by chance, which is $p^{2}s$ (= the probability that in the row and
column corresponding to the replaced node in $C$, any matrix element
and its transpose are both assigned unity). Thus, the `average time
of appearance' of an ACS is $\tau _{a}=1/p^{2}s$, and the distribution
of times of appearance is $P(n_{a})=p^{2}s(1-p^{2}s)^{n_{a}-1}$.
This approximation is better for small $p$. In the run displayed
in Figure \ref{cap:s1lambda}b, a 2-cycle between nodes 26 and 90
formed at $n=2854$. This is a graph that consists of a 2-cycle and
several other chains and trees. For such a graph, I have shown in
Example 3 in section \ref{sec:Examples-of-the} that the attractor
has non-zero $X_{i}$ for nodes 26 and 90 and zero for all other nodes.
The dominant ACS consists of nodes 26 and 90. Therefore these nodes
cannot be picked for removal at the graph update and, hence, a graph
update cannot destroy the nodes and links of the dominant ACS. \textit{The
autocatalytic property is guaranteed to be preserved until the dominant
ACS spans the whole graph}. In fact, an even stronger statement can
be made \citep{JKauto,JKemerg}:

\begin{description}
\item [Proposition~6.1:]\textit{\emph{$\lambda _{1}$ is a non-decreasing
function of $n$ as long as $s_{1}<s$}}\emph{.}
\end{description}
When a new node is added to the graph at a graph update, one of three
things will happen:

\begin{enumerate}
\item The new node will not have any links from the dominant ACS and will
not form a new ACS. In this case the dominant ACS will remain unchanged,
the new node will have zero relative population and will be part of
$\mathcal{L}$. For small $p$ this is the most likely possibility.
\item The new node gets an incoming link from the dominant ACS and hence
becomes a part of it. In this case the dominant ACS grows to include
the new node. For small $p$, this is less likely than the first possibility,
but such events do happen and in fact are the ones responsible for
the growth of the dominant ACS.
\item The new node forms another ACS. This new ACS competes with the existing
dominant ACS. Whether it now becomes dominant, overshadowing the previous
dominant ACS or it gets overshadowed, or both ACSs coexist depends
on the Perron-Frobenius eigenvalues of their respective subgraphs
and how they are connected. Again, for small $p$, this is a rare
event compared with possibilities 1 and 2.
\end{enumerate}
Typically the dominant ACS keeps growing by accreting new nodes, usually
one at a time, until the entire graph is an ACS. At this point the
growth phase stops and the organized phase begins.

\subsection{\noindent Timescale for growth of the dominant ACS}

\begin{figure}
\includegraphics[  width=13cm,
  keepaspectratio]{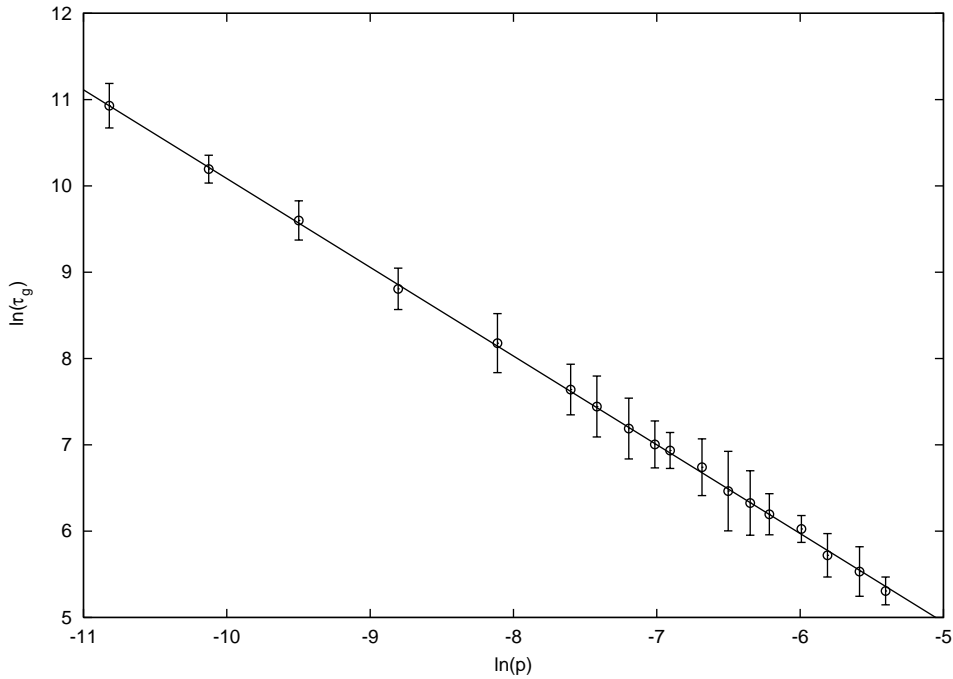}

\caption{\textit{\emph{Dependence of $\tau _{g}$ (the growth timescale of
the dominant ACS) on $p$. Each data point shows the average of $\ln \tau _{g}$
over 5 different runs with $s=100$ and the given $p$ value. The
error bars correspond to one standard deviation of the $\ln \tau _{g}$
values for each $p$. The straight line is the best linear fit to
the data points on a log-log plot. It has slope $-1.03\pm 0.03$ and
intercept $-0.20\pm 0.25$ which is consistent with the expected slope
$-1$ and intercept $0$.}}\label{cap:taug}}\lyxline{\normalsize}

\end{figure}

\noindent Assuming that possibility 3 above is rare enough to neglect,
and that the dominant ACS grows by adding a single node at a time,
one can estimate the time required for it to span the entire graph.
Let the dominant ACS consist of $s_{1}(n)$ nodes at time $n$. The
probability that the new node gets an incoming link from the dominant
ACS and hence joins it is $ps_{1}$. Thus in $\Delta n$ graph updates,
the dominant ACS will grow, on average, by $\Delta s_{1}=ps_{1}\Delta n$
nodes. Therefore $s_{1}(n)=s_{1}(n_{a})\mathrm{exp}((n-n_{a})/\tau _{g})$,
where $\tau _{g}=1/p$, $n_{a}$ is the time of appearance of the
first ACS and $s_{1}(n_{a})$ is the size of the first ACS (=2 for
the run shown in Figure \ref{cap:s1lambda}b). Thus $s_{1}$ is expected
to grow exponentially with a characteristic timescale $\tau _{g}=1/p$.
The time taken from the appearance of the ACS to its spanning is $\tau _{g}\ln (s/s_{1}(n_{a}))$.
This analytical result is confirmed by simulations (see Figure \ref{cap:taug}).
In the displayed run, after the first ACS (a 2-cycle) is formed at
$n=2854$, it takes 1026 time steps, until $n=3880$ for the dominant
ACS to span the entire graph. This explains how an autocatalytic network
structure and the positive feedback processes inherent in it can bootstrap
themselves into existence from a small seed. The small seed, in turn,
is guaranteed to appear on a certain timescale ($1/p^{2}s$ in the
present model) just by random processes. 

\section{\noindent Fully autocatalytic graphs }

\begin{figure}
\begin{center}\includegraphics[  width=15cm,
  keepaspectratio]{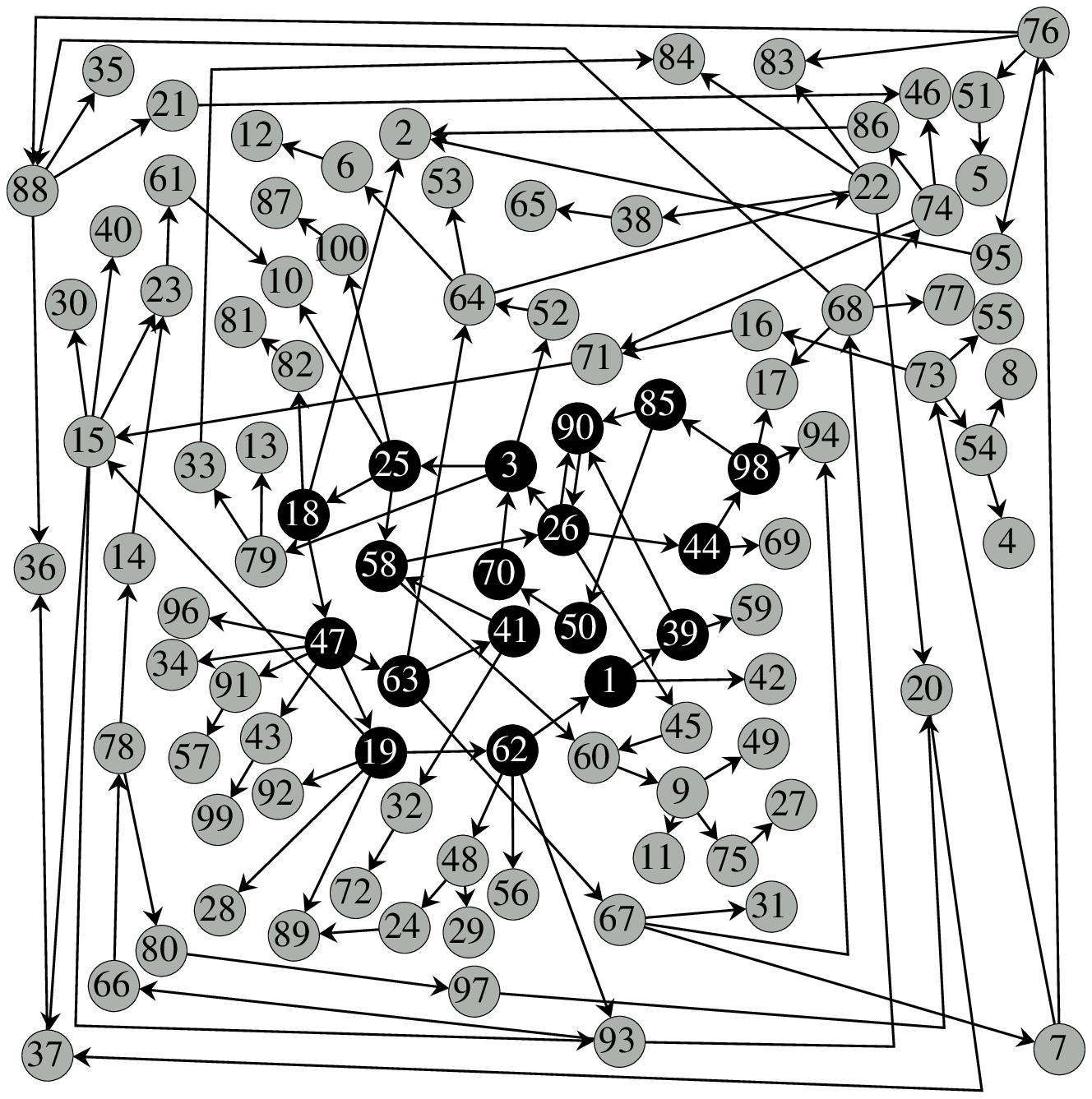}\end{center}

\caption{The graph, at $n=3880$, for the run shown Figure \ref{cap:s1lambda}b.
For the first time in this run the graph is fully autocatalytic. All
nodes are part of the dominant ACS and are non-zero in the attractor.
\label{cap:n=3D3880}}
\end{figure}

The growth phase continues until the dominant ACS spans the entire
graph. In the displayed run this happened at $n=3880$. Figure \ref{cap:n=3D3880}
shows the graph at this time -- it is a fully autocatalytic graph,
all nodes are part of the dominant ACS.

\subsection{\label{sub:Probability-of-a}Probability of a random graph being
fully autocatalytic}

A fully autocatalytic graph such as this is a highly improbable structure
for a random graph. Consider a graph of $s$ nodes and let the probability
of a positive link existing between any pair of nodes (except self-links)
be $p^{*}$. Such a graph has on average $m^{*}=p^{*}(s-1)$ incoming
or outgoing positive links per node. For the entire graph to be an
ACS, each node must have at least one incoming link, i.e. each row
of the matrix $C$ must contain at least one positive element. Hence
the probability, $P$, for the entire graph to be an ACS is\\

\begin{tabular}{ccl}
 $P$&
 $=$&
 probability that every row has at least one positive entry\\
&
 $=$&
 {[}probability that a row has at least one positive entry$]^{s}$\\
&
 $=$&
 $[1-($probability that every entry of a row is zero$)]^{s}$\\
&
 $=$&
 $[1-(1-p^{*})^{s-1}]^{s}$\\
&
 $=$&
 $[1-(1-m^{*}/(s-1))^{s-1}]^{s}$.\\
\end{tabular}\\

Note from Figure \ref{cap:links} that at spanning the number of links
is 124, i.e., $O(s)$. Thus the average degree $m^{*}$ at spanning
is $O(1)$. I have found this to be true in all the runs I have done
where $ps$ was $O(1)$ or less.

For large $s$ and $m^{*}\sim O(1)$, $P\approx (1-e^{-m^{*}})^{s}\sim e^{-\alpha s}$,
where $\alpha $ is positive, and $O(1)$. Thus, a fully autocatalytic
graph is exponentially unlikely to form if it were being assembled
randomly (or picked at random from the random graph ensemble $G_{s}^{p^{*}}$).
In the present model nodes are being added completely randomly but
the underlying population dynamics and the selection imposed at each
graph update result in the inevitable arrival of an ACS (in, on average,
$\tau _{a}=1/p^{2}s$ time steps) and its inevitable growth into a
fully autocatalytic graph in (on average) an additional $\sim \tau _{g}\ln s$
time steps.

For the displayed run if we take $m^{*}=ps\approx 0.25$ we get $P\approx 3\times 10^{-66}$.
Even if we take $m^{*}$ to be the average degree of the graph at
$n=3880$, i.e., $m^{*}=1.24,$ we get $P\approx 3\times 10^{-15}$;
one would expect such a fully autocatalytic graph to take $\sim 10^{15}$
time steps to form if it were being assembled randomly, in contrast
to the 3880 time steps it actually took in this run.

As mentioned in section \ref{sec:The-origin-of}, one of the puzzles
of the origin of life is: how could a highly structured chemical network
form in such a short time after the oceans condensed on the Earth?
In this model, a highly non-random structure, a fully autocatalytic
set, inevitably forms. Further, the time taken for it to form is extremely
short, growing only as a logarithm of the size of the network. This
model, therefore, suggests a mechanism -- the growth of ACSs by the
accretion of nodes around a small initial ACS -- by which highly structured
chemical networks (that are exponentially unlikely to form by chance)
can form in a very short time. This mechanism, or its analogue in
a more realistic chemical model, might explain the emergence of a
structured chemical organization on the prebiotic Earth. This is discussed
further in section \ref{sec:Implications-for-the}.

\subsection{\label{sub:Clustering-coefficient}Clustering coefficient}

In several runs with $s=100,p=0.0025$ (totaling 1.55 million iterations)
160659 graphs were fully autocatalytic. The mean clustering coefficient
of these 160659 graphs was 0.025, with standard deviation 0.016, which
is comparable to the clustering coefficient for random graphs with
the same connectivity ($p^{*}=1.27/(s-1)$, see below). The graph
in Figure \ref{cap:n=3D3880}, in fact, has a clustering coefficient
of zero: none of the neighbours of any node are connected to each
other, as can be verified from the figure. Thus, the fully autocatalytic
graphs produced in these runs do not have a small-world structure.

\subsection{\label{sub:Degree-distribution}Degree and dependency distributions}

\begin{figure}
\includegraphics[  width=15.5cm,
  keepaspectratio]{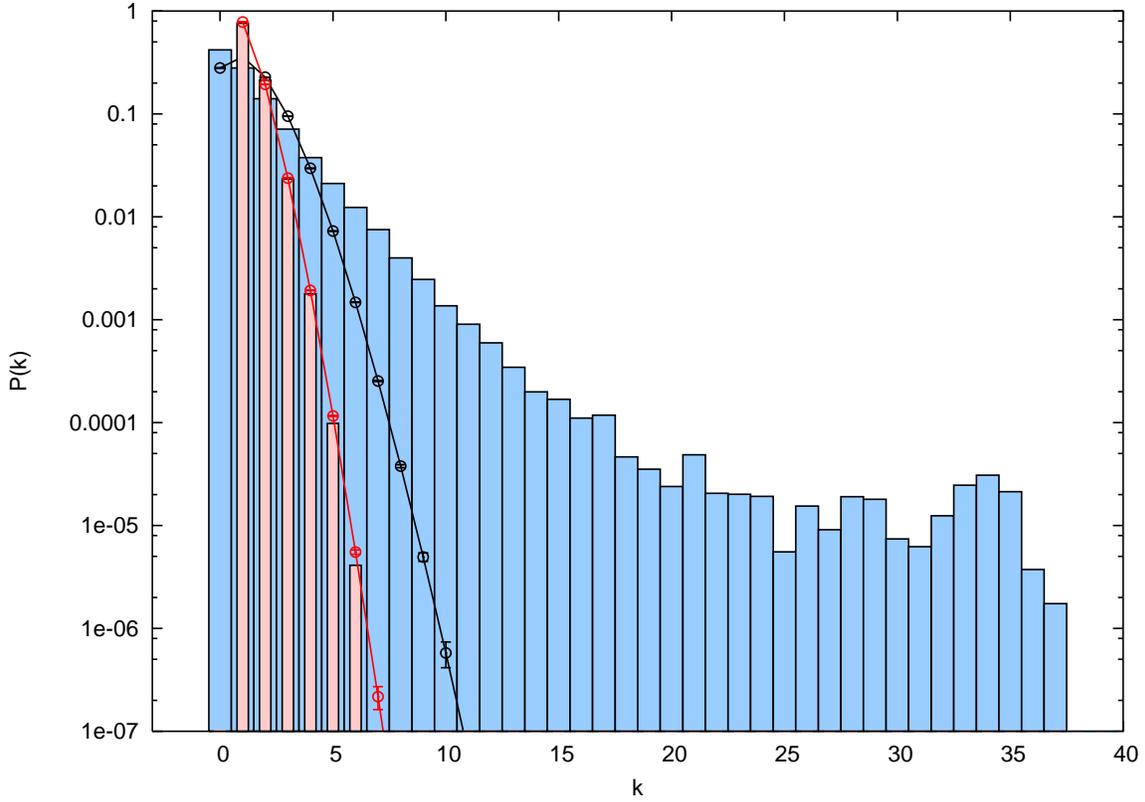}

\caption{Light red bars: in-degree distribution of nodes of 160659 fully autocatalytic
graphs from several runs with $s=100,p=0.0025$ (totaling 1.55 million
iterations). Mean in-degree is 1.27. Light blue bars: out-degree distribution
for the same graphs. Mean out-degree is 1.27. Black curve: the binomial
distribution $B_{p^{*}}^{s-1}(k)$ expected for a random graph $G_{s}^{p^{*}},p^{*}=1.27/(s-1)$;
error bars are the expected standard deviation, $\sqrt{\frac{B_{p^{*}}^{s-1}(k)\left[1-B_{p^{*}}^{s-1}(k)\right]}{16065900}}$,
for a finite sample of 160659 graphs. Red curve: the shifted binomial
distribution $B_{p}^{s-1}(k-1)$ expected for a random fully autocatalytic
set; error bars are the expected standard deviation, $\sqrt{\frac{B_{p}^{s-1}(k-1)\left[1-B_{p}^{s-1}(k-1)\right]}{16065900}}$.
\label{cap:facsdeg}}
\end{figure}

\begin{figure}
\includegraphics[  width=11cm,
  keepaspectratio]{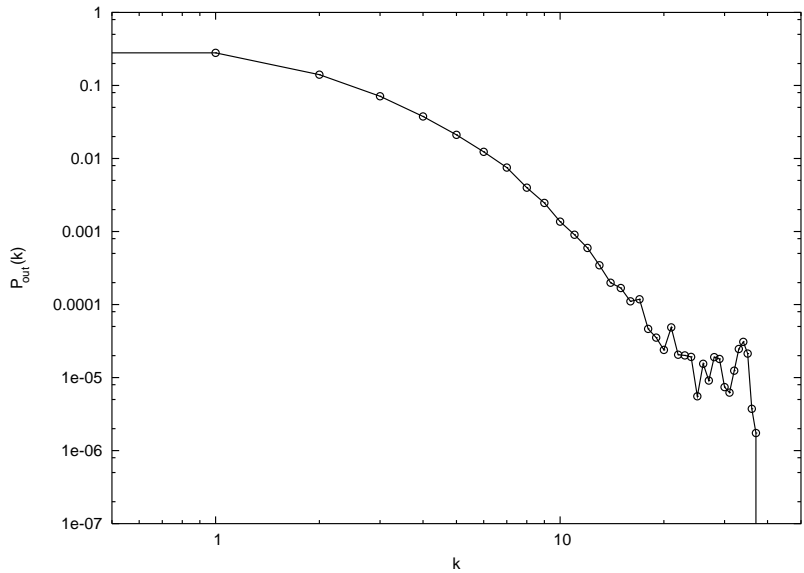}

\caption{The out-degree distribution of fully autocatalytic sets in several
runs with $s=100,p=0.0025$ (light blue bars in Figure \ref{cap:facsdeg})
on a log-log plot. A scale-free distribution would be a straight line.
\label{cap:facsdegout}}
\end{figure}

The in-degree and out-degree distributions of the nodes of all the
160659 fully autocatalytic graphs produced in these runs are shown
in Figure \ref{cap:facsdeg}. The mean in-degree for nodes of these
graphs was 1.27, and the standard deviation was 0.50. The mean out-degree
was 1.27, with standard deviation 1.75. The distributions are compared
with the binomial distribution $B_{p^{*}}^{s-1}(k)$ (as expected
for random graphs from the ensemble $G_{s}^{p^{*}}$; $p^{*}=1.27/(s-1)$)
and the shifted binomial $B_{p}^{s-1}(k-1)$ (as would be expected
for a \emph{random} autocatalytic graph). The in-distribution is quite
similar in functional form to the shifted binomial for small in-degrees
but departs from it for higher values. In contrast, the out-distribution
falls off much more slowly, though it is not scale-free either because
it is not a straight line in a log-log plot (Figure \ref{cap:facsdegout}).
However, as $s=100$, the allowed values for the degree of a node
cover a range of only two decades. If a portion of the out-degree
distribution is a power-law then it is more likely to show up for
graphs with higher $s$, where the degree distributions would span
more decades. For such networks, with larger $s$, whether the fully
autocatalytic graphs produced are scale-free or not is an open question.

\begin{figure}
\includegraphics[  width=11cm,
  keepaspectratio]{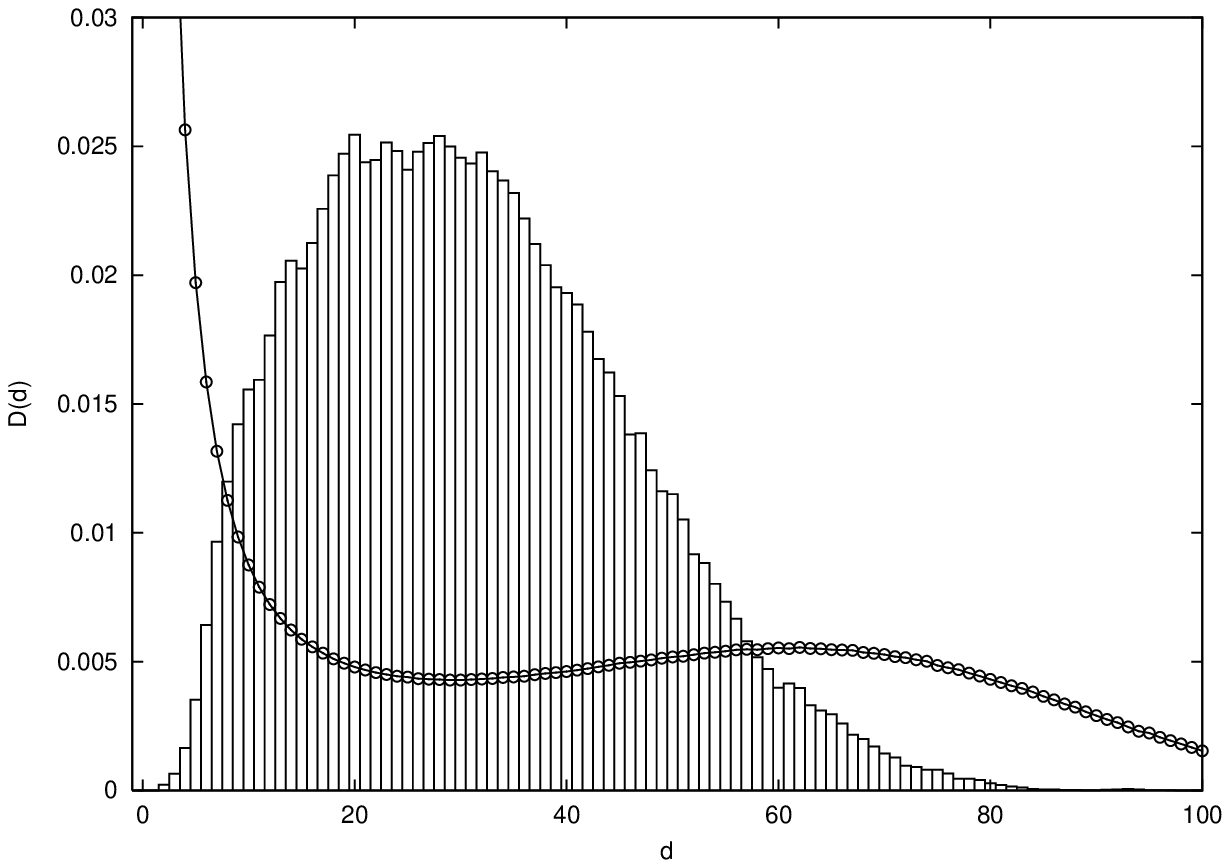}

\caption{Bars: The dependency distribution of nodes of 160659 fully autocatalytic
graphs from several runs with $s=100,p=0.0025$ (totaling 1.55 million
iterations). Mean dependency is 30.93 with standard deviation 14.71.
Circles: The dependency distribution of $1.5\times 10^{6}$ graphs
from the ensemble $G_{s}^{p^{*}}$, with $s=100,p^{*}=1.27/(s-1)$.
The mean dependency is 23.70, with standard deviation 30.82. \label{cap:facsdep}}
\end{figure}

The dependency distribution of the fully autocatalytic graphs produced
in the runs (Figure \ref{cap:facsdep}) has a substantially larger
mean than the distribution for graphs of the random phase (compare
with Figure \ref{cap:depdeistcmp}). This is consistent with the growth
of interdependency observed in Figure \ref{cap:interdependency}.
Further, as evident from Figure \ref{cap:facsdep}, the dependency
distribution is also clearly different from, and much narrower than,
the distribution for random graphs with a comparable connectivity
$p^{*}=1.27/(s-1)$. The mean dependency of nodes of the fully autocatalytic
graphs (30.93, with standard deviation 14.71) is slightly larger than
the mean dependency for $1.5\times 10^{6}$ random graphs, $G_{s}^{p^{*}}$,
with $p^{*}=127/(s-1),s=100$ (23.70, with standard deviation 30.82),
but the maximum dependency observed for the fully autocatalytic graphs
(100) is smaller than the maximum for the random graphs (163). In
all the fully autocatalytic graphs, the minimum dependency was two,
and the dependency distribution is small for small values of dependency.
In contrast, the dependency distribution for the random graphs is
maximum at zero dependency.

\chapter{\label{cha:Destruction-of-ACSs}Destruction of Autocatalytic Sets}

\section{Catastrophes and recoveries in the organized and growth phases}

Once an ACS spans the entire graph the effective dynamics again changes
although the microscopic dynamical rules are unchanged. At spanning,
for the first time since the formation of the initial ACS, a node
of the dominant ACS will be picked for removal. This is because at
spanning all nodes belong to the dominant ACS and have non-zero relative
populations; one node nevertheless has to be picked for removal. Most
of the time the removal of the node with the least $X_{i}$ will result
in minimal damage to the ACS. The rest of the ACS will remain with
non-zero relative populations, and the new node will repeatedly be
removed and replaced until it once again joins the ACS. Thus $s_{1}$
will fluctuate between $s$ and $s-1$ most of the time. However,
once in a while, the node that is removed happens to be playing a
crucial role in the graph structure despite its low relative population.
Then its removal can trigger large changes in the structure and catastrophic
drops in $s_{1}$ and $l$. Alternatively, it can sometimes happen
that the new node added can trigger a catastrophe because of the new
graph structure it creates. 

\begin{figure}
\includegraphics[  width=15cm,
  keepaspectratio]{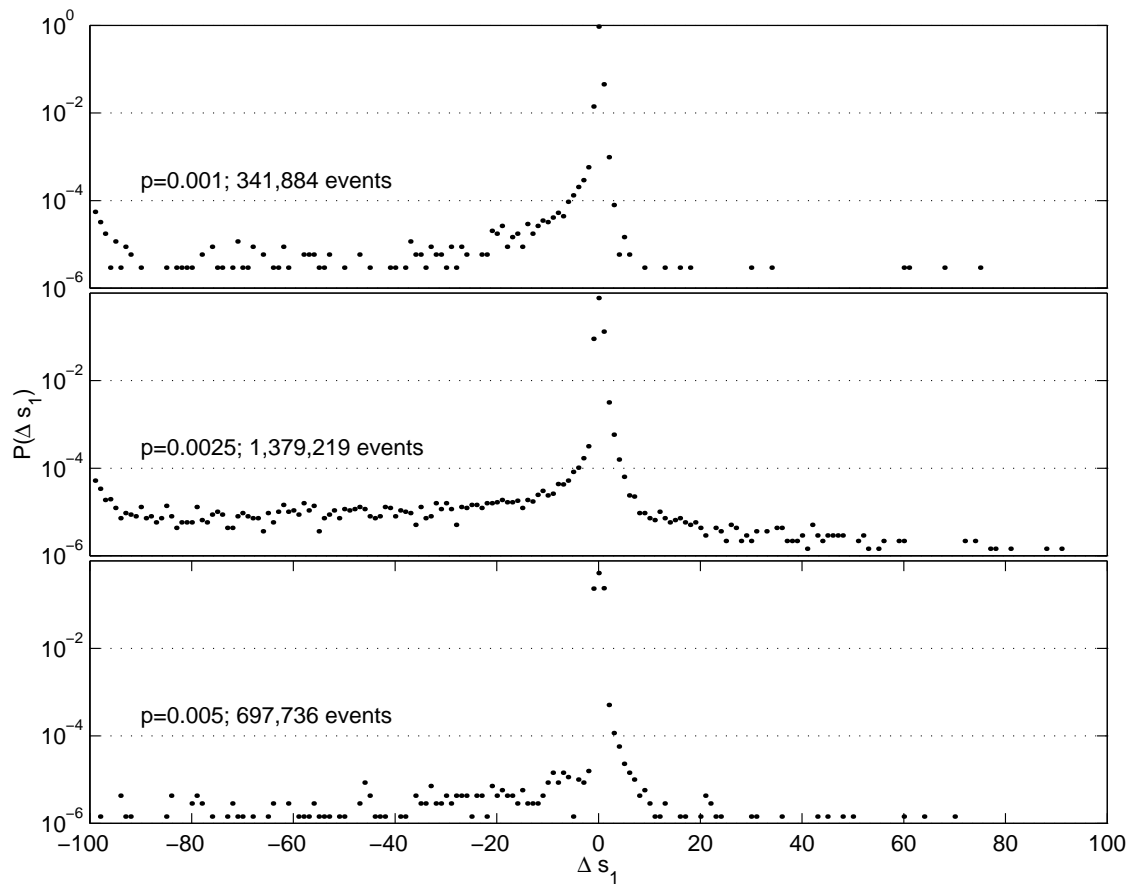}

\caption{Probability distribution of changes in $s_{1}$, the number of nodes
with $X_{i}>0$. $P(\Delta s_{1})$ is the fraction of time steps
in which $s_{1}$ changes by an amount $\Delta s_{1}$ in one time
step in an ensemble of runs with $s=100$ and $p=0.001,0.0025,0.005$.
Only time steps where an ACS initially exists are counted.\label{cap:dropdist}}
\end{figure}

In Figure \ref{cap:longtimefig}, which shows $s_{1}$ for the same
run as that in Figure \ref{cap:s1lambda}b but for a much longer time,
one can observe several of the large and sudden drops in $s_{1}$
that will be discussed in this chapter. These `catastrophes' in the
organized and growth phases are followed by `recoveries', in which
$s_{1}$ rises on a certain timescale. Figure \ref{cap:dropdist}
shows the probability distribution $P(\Delta s_{1})$ of changes in
the number of nodes with $X_{i}>0$; $\Delta s_{1}(n)\equiv s_{1}(n)-s_{1}(n-1)$.
The asymmetry between rises and drops as well as fat tails in the
distribution are evident. For low $p$ the probability of large drops
is an order of magnitude greater than intermediate size drops (also
see Figure \ref{cap:s1lambda}).

\begin{figure}
\includegraphics[  width=15cm,
  keepaspectratio]{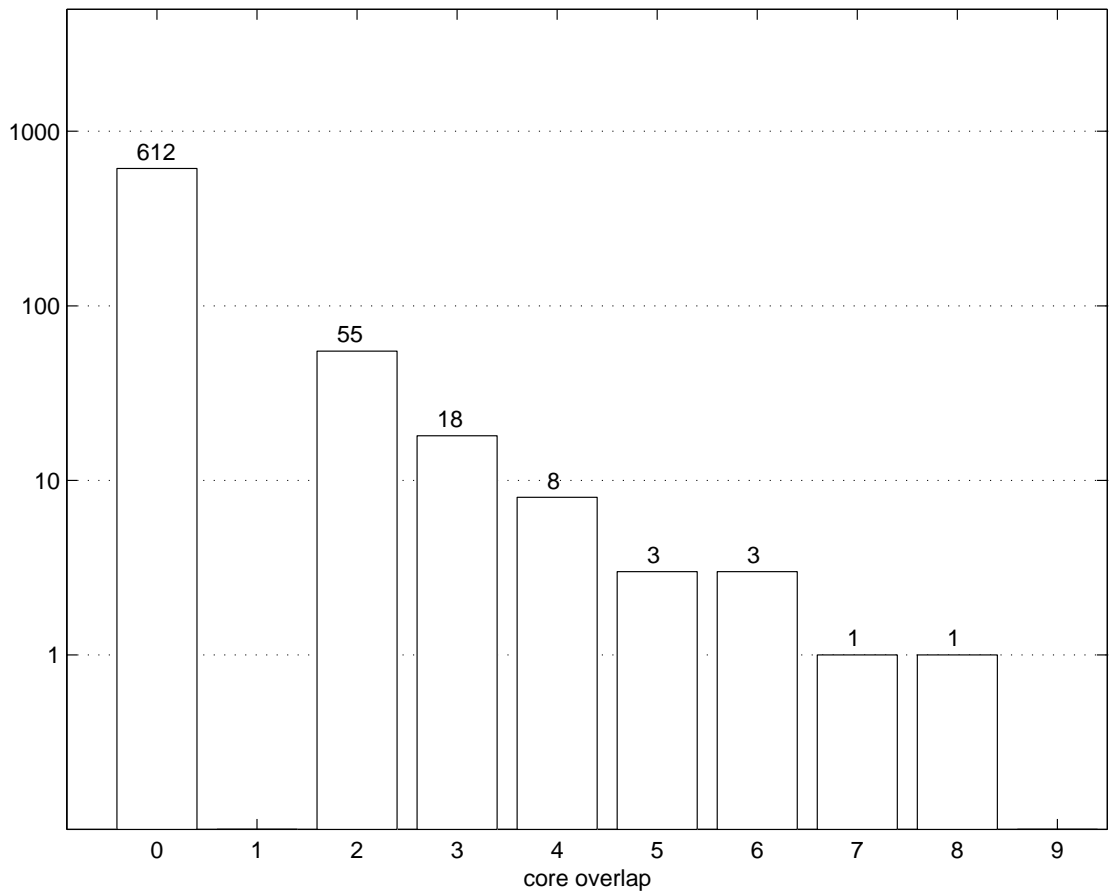}

\caption{Crashes are predominantly core-shifts. A histogram of core overlaps
for the 701 crashes, in which $s_{1}$ dropped by more than $s/2$,
observed in several runs with $s=100$ and $p=0.0025$, totaling 1.55
million iterations. The attached CD contains data files for all these
runs, as well as information about the 701 crashes (see appendix C).
\label{cap:histogram}}
\end{figure}

\section{\label{sec:Crashes-and-core-shifts}Crashes and core-shifts}

\begin{description}
\item [Definition~7.1:]Crash\textit{.}\\
\textit{\emph{A}} \textit{crash} is a graph update event $n$ for
which $\Delta s_{1}(n)<-s/2$ \\
\citep{JKcoresh,JKlargeext}.
\end{description}
A crash is an event in which a significant fraction (arbitrarily chosen
as 50\%) of the nodes become extinct. In several runs with $s=100,p=0.0025$
totaling 1.55 million iterations there were 701 crashes. The first
task is to see if these large drops in $s_{1}$ are correlated to
specific changes in the structure of the graph. 

\begin{description}
\item [Definition~7.2:]Core-shift\textit{.}\\
\textit{\emph{A}} \textit{core-shift} is a graph update event $n$
for which $Ov(C_{n-1},C_{n})=0$\\
\citep{JKcoresh,JKlargeext}.
\end{description}
Figure \ref{cap:histogram} shows a histogram of core overlaps $Ov(C_{n-1},C_{n})$
for these 701 crashes. 612 of these have zero core overlap, i.e.,
they are core-shifts. (If one looks at only those events in which
more than 90\% of the nodes went extinct, i.e., $\Delta s_{1}(n)<-0.9s$,
then one finds 235 such events in the same runs, out of which 226
are core-shifts.) Of the remaining 89 crashes 10 were events in which
the core remained unchanged and 79 were events in which the core was
affected but some overlap remained between the new and old cores.
Most of the crashes happen when there is a core-shift -- a drastic
change in the structure of the dominant ACS.

\begin{figure}
\begin{center}\includegraphics[  width=7cm,
  keepaspectratio]{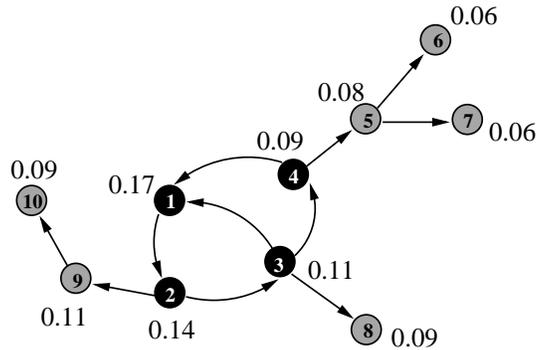}\end{center}

\caption{A graph with $\lambda _{1}=1.22$. The $X_{i}$ values (rounded off
to two decimal places) are shown adjacent to each node. Notice that
if a node has only one incoming link then its $X_{i}$ value is $1/\lambda _{1}$
times the $X_{i}$ value of the node it is getting the link from.
E.g. $X_{6}=X_{5}/\lambda _{1}$. Thus, nodes further down a chain
of single links have lower $X_{i}$. Nodes 6 and 7 have the lowest
$X_{i}$.\label{cap:attenutation}}\lyxline{\normalsize}

\end{figure}

In the 612 core-shifts, the average number of incoming plus outgoing
links is 2.27 for all nodes in the graph, 2.25 for the node that is
removed and 1.25 for the new node. Thus the nodes whose removal or
addition causes the crash are not excessively rich in links. `Nondescript'
nodes such as these cause system wide crashes because of their critical
location in a small core (the average core size at the 612 core-shifts
is 6.3 nodes). Core-shifts in which the ACS is completely destroyed
typically cause the largest damage (of 612 core-shifts these are 136
in number, with $|\Delta s_{1}|=98.2\pm 1.2$). The remaining 476
core-shifts in which an ACS exists after the core-shift have $|\Delta s_{1}|=75.0\pm 14.2$.
The former constitute an increasing fraction of the crashes at smaller
$p$ values, causing the upturn in $P(\Delta s_{1})$ at large negative
$\Delta s_{1}$ for small $p$ (Figure \ref{cap:dropdist}).

\section{Addition and deletion of nodes from a graph}

In order to understand what is happening during the crashes and subsequent
recoveries I begin by examining the possible changes that an addition
or a deletion of a node can make to the core of the dominant ACS.

\subsection{\noindent Deletion of a node: keystone nodes }

\noindent We have already seen how the deletion of a node can change
the core -- recall the discussion of keystone nodes in section \ref{sec:Keystone-nodes}:
the removal of a keystone node results in a core-shift, i.e., a zero
overlap between the cores of the dominant ACS before and after the
removal. In an actual run a keystone node can only be removed if it
happens to be one of the nodes with the least $X_{i}$. However, the
core nodes are often `protected' by having higher $X_{i}$. The reason
for this is:

${\textbf {X}}$ is an eigenvector of $C$ with eigenvalue $\lambda _{1}$.
Therefore, when $\lambda _{1}\ne 0$ it follows that for nodes of
the dominant ACS, $X_{i}=(1/{\lambda _{1}})\sum _{j}c_{ij}X_{j}$.
If node $i$ of the dominant ACS has only one incoming link (from
the node $j$, say) then $X_{i}=X_{j}/\lambda _{1}$; $X_{i}$ is
`attenuated' with respect to $X_{j}$ by a factor $\lambda _{1}$.
The periphery of an ACS is a tree-like structure emanating from the
core, and for small $p$ most periphery nodes have a single incoming
link. For instance, the graph in Figure \ref{cap:attenutation}, whose
$\lambda _{1}=1.22$, has a chain of nodes $4\rightarrow 5\rightarrow 6$.
The farther down such a chain a periphery node is, the lower is its
$X_{i}$ because of the cumulative attenuation. For such an ACS the
`leaves' of the periphery tree (such as node 6) will have the least
$X_{i}$ while core nodes will have larger $X_{i}$.

However, when $\lambda _{1}=1$ there is no attenuation. At $\lambda _{1}=1$
the core must be a cycle or a set of disjoint cycles (proposition
2.2), hence each core node has only one incoming link within the dominant
ACS. All core nodes have the same value of $X_{i}$. As one moves
out toward the periphery, $\lambda _{1}=1$ implies there is no attenuation,
hence each node in the periphery that receives a single link from
one of the core nodes will also have the same $X_{i}$. Some periphery
nodes may have higher $X_{i}$ if they have more than one incoming
link from the core. Iterating this argument as one moves further outward
from the core, it is clear that at $\lambda _{1}=1$ the core is not
protected and in fact will always belong to $\mathcal{L}$ (the set
of nodes with the least $X_{i}$) if the dominant ACS spans the graph.
We have already seen in section \ref{sec:Keystone-nodes} that when
$\lambda _{1}=1$ and the core is a single cycle every core node is
a keystone node. Thus, when $\lambda _{1}=1$ the organization is
fragile and susceptible to core-shifts caused by the removal of a
keystone node. 

\begin{figure}
\includegraphics[  width=15cm,
  keepaspectratio,
  angle=180,
  origin=lB]{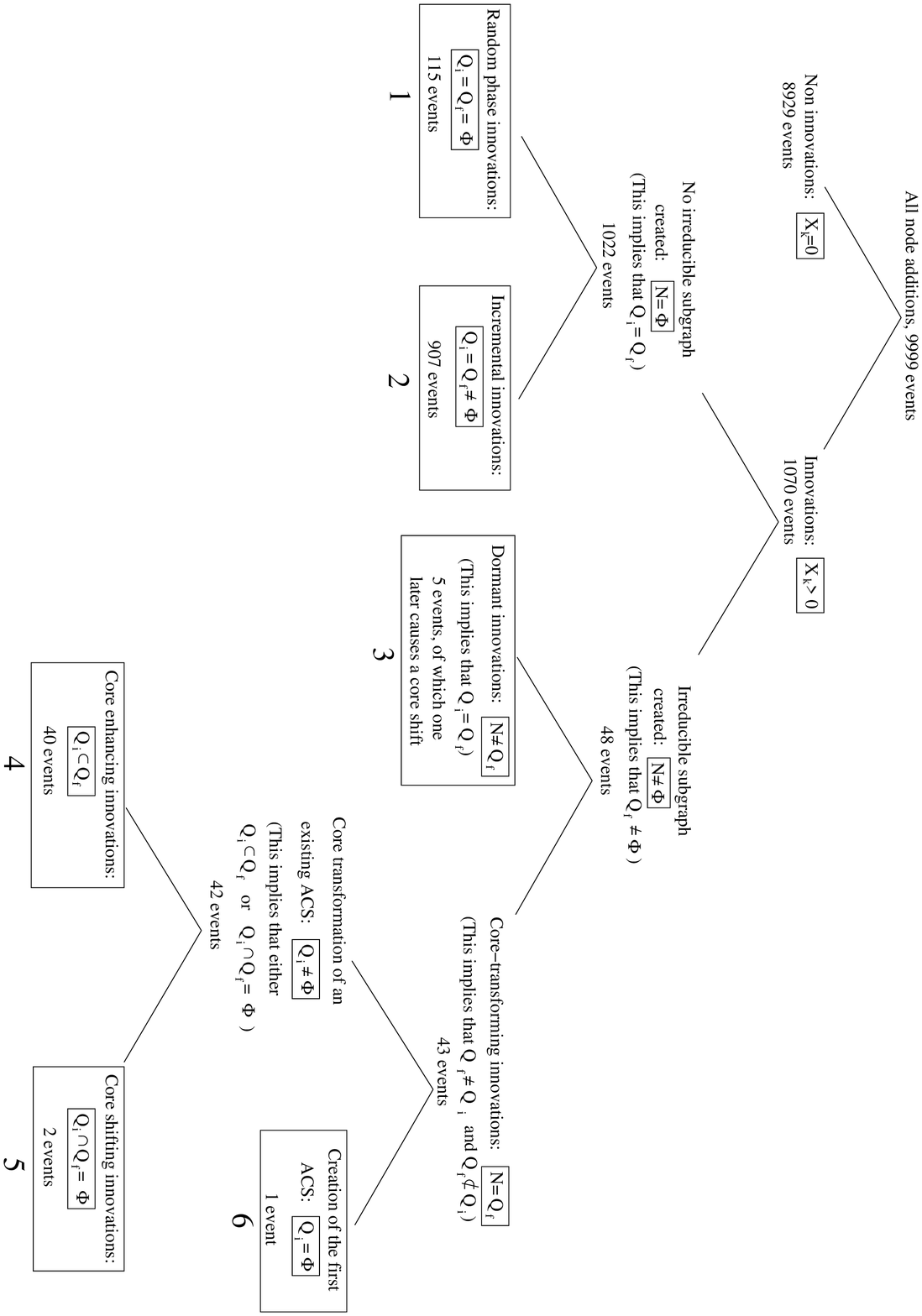}
\end{figure}

\begin{figure}

\caption{(opposite page) A hierarchy of innovations. Each node in this binary
tree represents a class of node addition events. Each class has a
name; the small box contains the mathematical definition of the class.
All classes of events except the leaves of the tree are subdivided
into two exhaustive and mutually exclusive subclasses (represented
by the two branches emanating downward from the class). The number
of events in each class pertain to the run of Figure \ref{cap:s1lambda}b
with a total of 9999 graph updates, between $n=1$ (the initial graph)
and $n=10000$. In that run, out of 9999 node addition events, most
(8929 events) are not innovations. The rest (1070 events), which are
innovations, are classified according to their graph structure. $X_{k}$
is the relative population of the new node in the attractor of equation
(\ref{xdot}) that is reached in step 1 of the graph dynamics immediately
following the addition of that node. If the new node causes a new
irreducible subgraph to be created, $N$ is the \textit{maximal} irreducible
subgraph that includes the new node. If not, $N=\Phi $ (the empty
set). $Q_{i}$ is the core of the graph just before the addition of
the node (just before step 3 of the graph dynamics) and $Q_{f}$ the
core just after the addition of the node. The six leaves of the innovation
subtree are numbered from 1 to 6 and correspond to the classes discussed
in the text. Some classes of events happen rarely (e.g., classes numbered
5 and 6) but have a major impact on the dynamics of the system. The
precise impact of all these classes of innovations on the system over
a short timescale (before the next graph update) as well as their
probable impact over the medium term (upto a few thousand graph updates)
can be predicted from the structure of $N$ and the rest of the graph
at the moment these innovations appear in a run.\label{cap:hierarchy}}\lyxline{\normalsize}

\end{figure}

\subsection{\noindent \label{sub:Addition-of-a}Addition of a node: innovations}

\noindent I now turn to the effects of the addition of a node. The
removal and subsequent addition of a new node in a graph update creates
a new graph that is likely to have a different attractor from the
previous graph. In the new attractor the new node, denoted $k$, may
become extinct, i.e., $X_{k}$ may be zero, or it may survive, i.e.,
$X_{k}$ is non-zero. If the new node becomes extinct then it remains
in the set $\mathcal{L}$ and there is no change to the dominant ACS.
So I will focus on events in which the new node survives in the new
attractor.

\begin{description}
\item [Definition~7.3:]Innovation.\\
An \textit{innovation} is a graph update event in which the relative
population of the new node in the new attractor is non-zero, i.e.
an event in which the new node survives till the next graph update
\citep{JKlargeext,JKinn}. 
\end{description}

I will also use the term `innovation' to refer to the new structures
created -- the new node and the new links brought in by the graph
update event. Figure \ref{cap:hierarchy} shows a graph-theoretic
classification of innovations in terms of a hierarchy. The innovations
that have the least impact on the relative populations of the nodes
and the evolution of the graph on a short timescale (of a few graph
updates) are ones that do not affect the core of the dominant ACS,
if it exists. Such innovations are of three types (see boxes 1--3
in Figure \ref{cap:hierarchy}): 

\newpage
\textbf{1. Random phase innovations.} These are innovations that occur
in the random phase when no ACS exists in the graph, and they do not
create any new ACSs. These innovations are typically short lived and
have little short or long term impact on the structure of the graph.
In the displayed run the first random phase innovation appeared at
$n=79$ (node 25 in Figure \ref{cap:snapshots}c).

\textbf{2. Incremental innovations.} These are innovations that occur
in the growth and organized phases, which add new nodes to the periphery
of the dominant ACS without creating any new irreducible subgraph.
See Figure \ref{cap:snapshots}f, $n=3022$, for the first incremental
innovation of the displayed run. In the short term these innovations
only affect the periphery and are responsible for the growth of the
dominant ACS. In a longer term they can also affect the core as chains
of nodes from the periphery join the core of the dominant ACS. 

\textbf{3. Dormant innovations.} These are innovations that occur
in the growth and organized phases, which create new irreducible subgraphs
in the periphery of the dominant ACS (e.g. the innovation at $n=4696$,
Figure \ref{cap:snapshots}p, that created the 2-cycle 36-74 in the
periphery). These innovations too affect only the periphery in the
short term. But they have the potential to cause core-shifts later
if the right conditions occur (see section \ref{sub:Takeovers-by-dormant}).

Innovations that do immediately affect the core of the existing dominant
ACS are always ones which create a new irreducible subgraph. They
are also of three types (boxes 4--6 in Figure \ref{cap:hierarchy}): 

\textbf{4. Core enhancing innovations.} These innovations result in
the expansion of the existing core by the addition of new links and
nodes from the periphery or outside the dominant ACS. They result
in an increase of $\lambda _{1}$ of the graph. The innovation at
$n=3489$ (Figure \ref{cap:snapshots}l) is an example.

\textbf{5. Core-shifting innovations.} These are innovations that
cause an immediate core-shift often accompanied by the extinction
of a large number of nodes. The innovation at $n=6062$ (Figure \ref{cap:snapshots}t)
is an example (also see section \ref{sub:Takeovers-by-core}).

\textbf{6. Creation of the first ACS.} This is an innovation that
creates the first ACS in a graph which till then had none. In the
displayed run, the first ACS was created at $n=2853$ (Figure \ref{cap:snapshots}d,e).
The innovation moves the system from the random phase to the growth
phase.

\begin{description}
\item [Definition~7.4:]Core transforming innovation.\\
An innovations of type 4, 5 or 6 which immediately affects the core
of the dominant ACS will be called a \textit{core transforming innovation}
\citep{JKwiley}. 
\end{description}
The following proposition makes precise the conditions under which
a core transforming innovation can occur \citep{JKwiley}. Let $C'_{n}\equiv C_{n-1}-k$
denote the graph of $s-1$ nodes just after the node $k$ is removed
from $C_{n-1}$. $Q'_{n}$ will stand for the core of $C'_{n}$.

\begin{description}
\item [Proposition~7.1:]Let $N_{n}$ denote the maximal new irreducible
subgraph that includes the new node at time step $n$. $N_{n}$ will
become the new core of the graph, replacing the old core $Q_{n-1}$,
whenever either of the following conditions are true: \\
 (a) $\lambda _{1}(N_{n})>\lambda _{1}(Q'_{n})$ or, \\
 (b) $\lambda _{1}(N_{n})=\lambda _{1}(Q'_{n})$ and $N_{n}$ is downstream
of $Q'_{n}$.
\end{description}
Such a core transforming innovation will fall into category 4 above
if $Q_{n-1}\subset N_{n}$. However, if $Q_{n-1}$ and $N_{n}$ are
disjoint, we get a core-shift and the innovation is of type 5 if $Q_{n-1}$
is non-empty or type 6 otherwise.

\section{\noindent \label{sec:Classification-of-core-shifts}Classification
of core-shifts }

\begin{figure}
\begin{center}\includegraphics[  width=14cm,
  keepaspectratio]{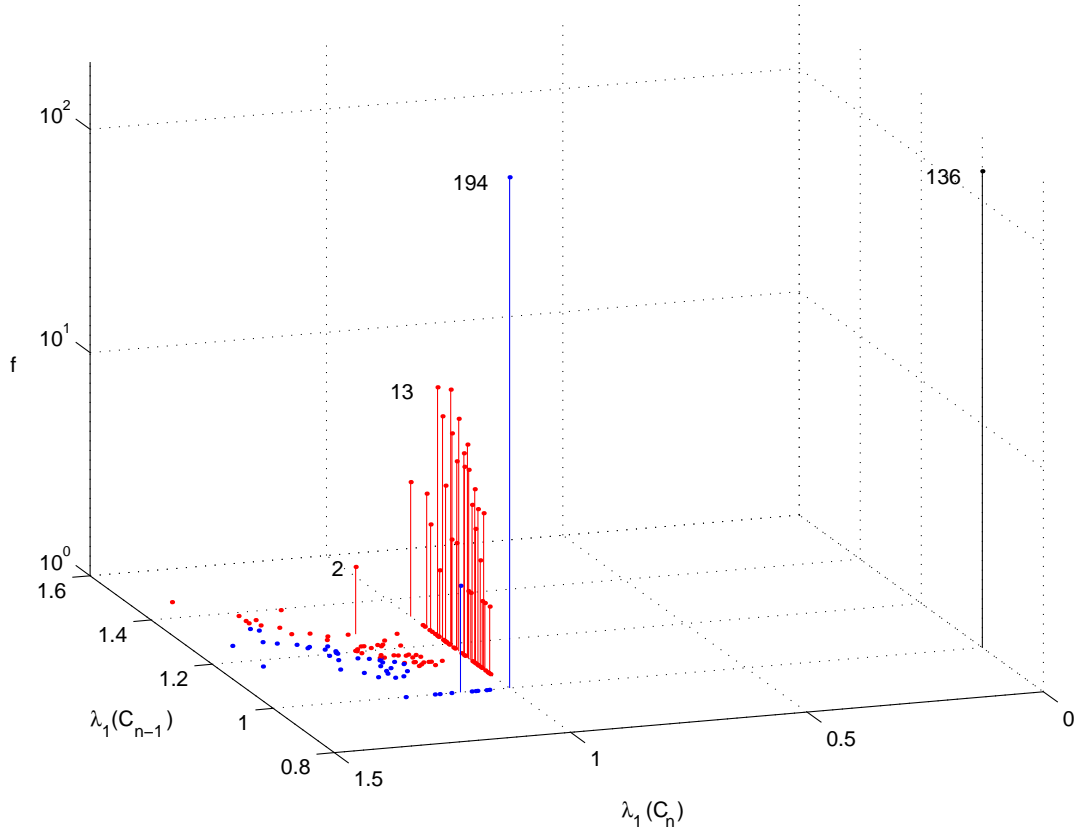}\end{center}

\caption{Classification of core-shifts into three categories. The graph shows
the frequency, $f$, of the $612$ core-shifts observed (see Figure
\ref{cap:histogram}) in a set of runs with $s=100$ and $p=0.0025$
vs. the $\lambda _{1}$ values before, $\lambda _{1}(C_{n-1})$, and
after, $\lambda _{1}(C_{n})$, the core-shift. Complete crashes (black:
$\lambda _{1}(C_{n-1})=1$, $\lambda _{1}(C_{n})=0$), takeovers by
core-transforming innovations (blue: $\lambda _{1}(C_{n})\geq \lambda _{1}(C_{n-1})\geq 1$)
and takeovers by dormant innovations (red: $\lambda _{1}(C_{n-1})>\lambda _{1}(C_{n})\geq 1$)
are distinguished. Numbers alongside vertical lines represent the
corresponding $f$ value.\label{cap:classify}}\lyxline{\normalsize}

\end{figure}

\noindent Using the insights from the above discussion of the effects
of deletion or addition of a node, one can classify the different
mechanisms that cause core-shifts. Figure \ref{cap:classify} differentiates
the 612 core-shifts observed among the 701 crashes. They fall into
three categories: (i) complete crashes ($136$ events), (ii) takeovers
by core-transforming innovations ($241$ events), and (iii) takeovers
by dormant innovations ($235$ events).

\subsection{\noindent \label{sub:Complete-crashes}Complete crashes }

\begin{figure}
\textbf{n=8232}\hfill{}\textbf{n=8233}

\includegraphics[  width=6.5cm,
  keepaspectratio]{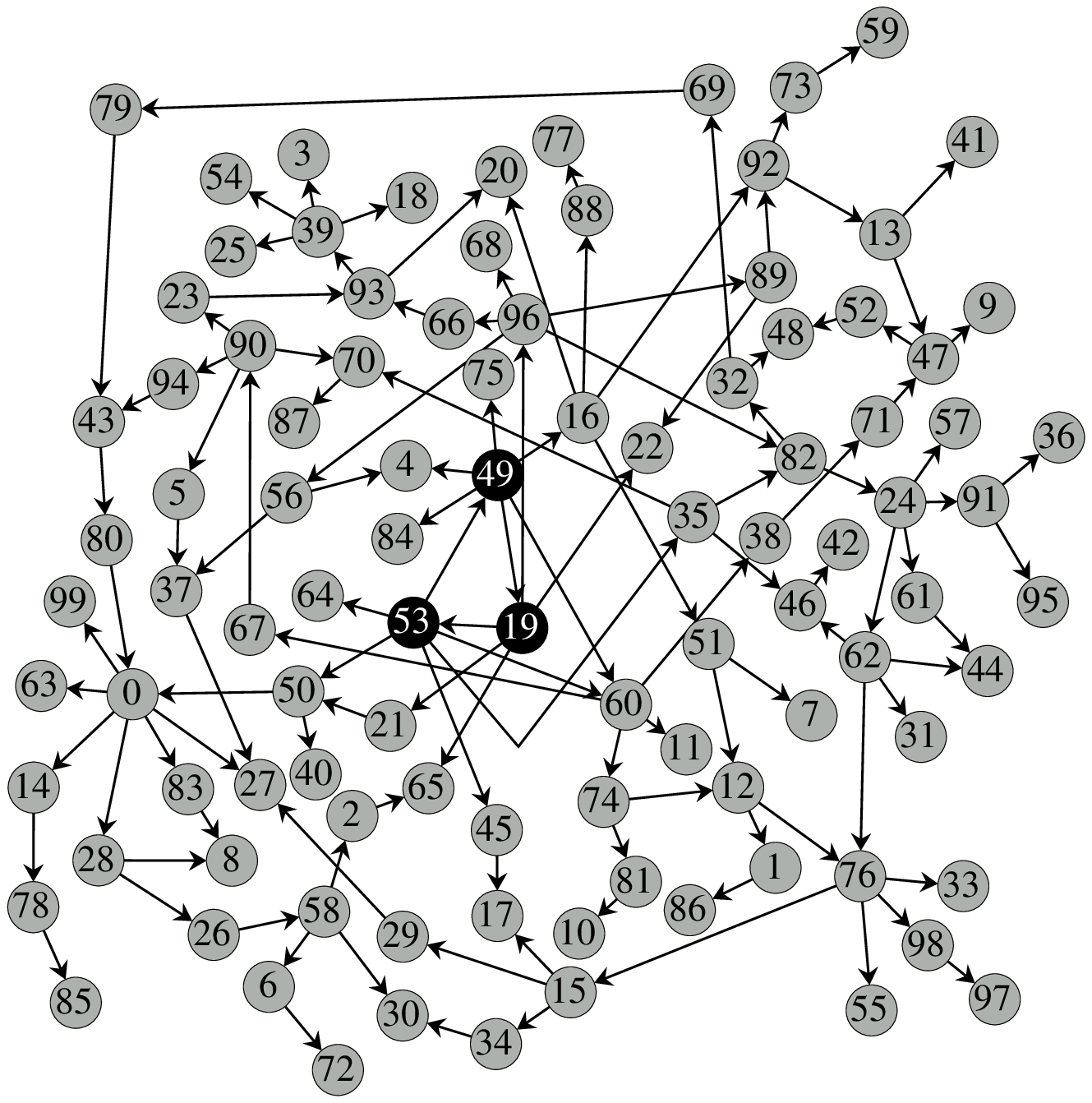}\hfill{}\includegraphics[  width=6.5cm,
  keepaspectratio]{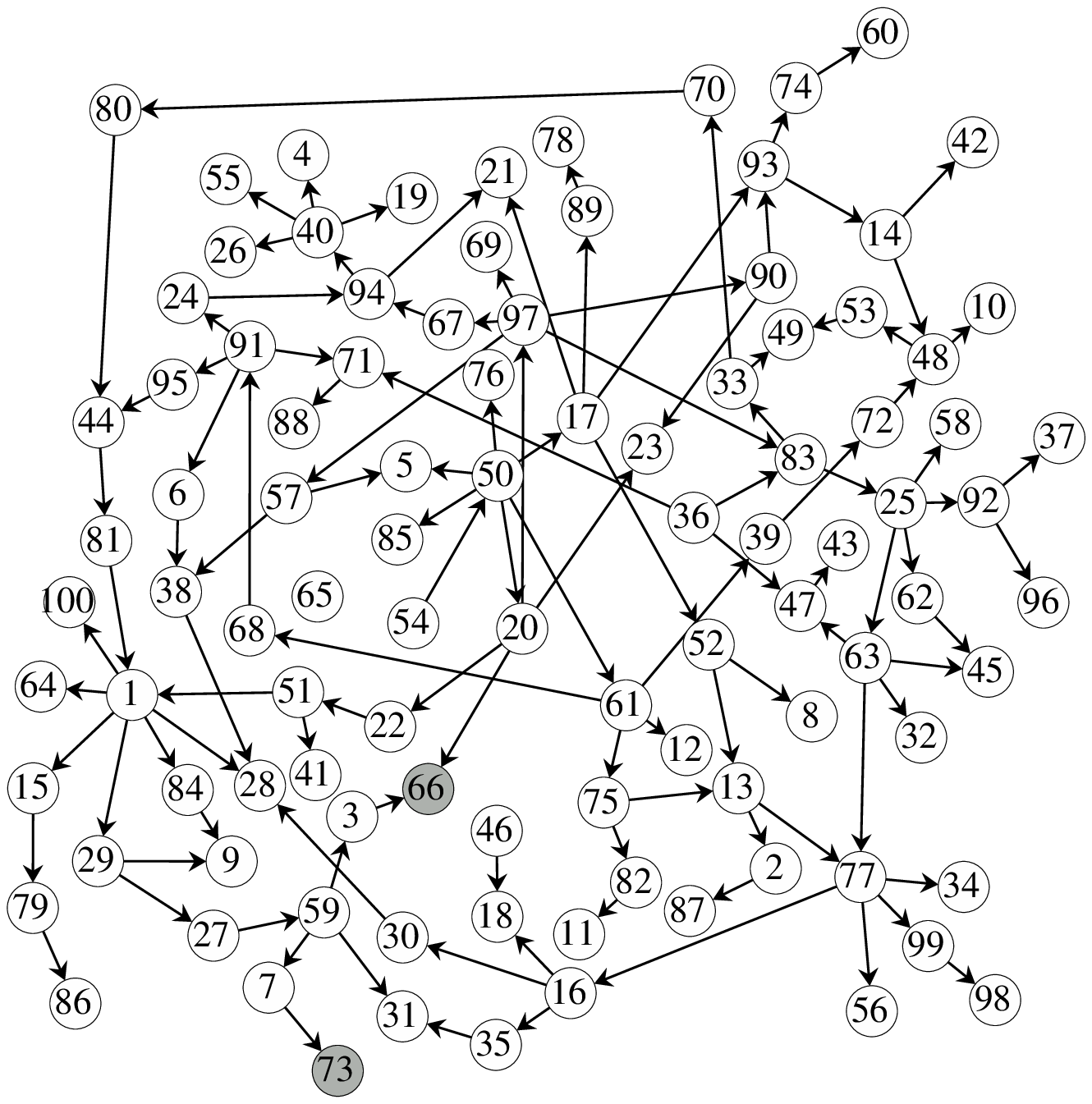}

\caption{A complete crash. In the run displayed in Figure \ref{cap:s1lambda}b
at $n=8232$ the core is a 3-cycle comprising nodes 20, 50 and 54.
Node 54 is the node with the least $X_{i}$ that is removed from the
graph. This breaks the only cycle in the graph resulting in a graph,
at $n=8233$, that has no ACS. The system goes from the organized
phase to the random phase, $\lambda _{1}$ drops from 1 to 0, and
$s_{1}$ drops form 100 to 2.\label{cap:8232,8233}}\lyxline{\normalsize}

\end{figure}

\noindent A \textit{\emph{complete crash}} is an event in which an
ACS exists before but not after the graph update. Such an event takes
the system into the random phase. A complete crash can occur when
a keystone node is removed from the graph. For example, in the run
displayed in Figure \ref{cap:s1lambda}b and \ref{cap:snapshots},
at $n=8232$ (see Figure \ref{cap:8232,8233}), the graph had $\lambda _{1}=1$
and its core was the simple 3-cycle of nodes 20, 50 and 54. As we
have seen above, when the core is a single cycle every core node is
a keystone node and is also in the set $\mathcal{L}$ of nodes with
the least $X_{i}$. Node 54 was removed, thus disrupting the 3-cycle.
The resulting graph, at $n=8233$, had no ACS and $\lambda _{1}$
dropped to zero. As discussed earlier, graphs with $\lambda _{1}=1$
are the ones that are most susceptible to complete crashes. This can
be seen in Figure \ref{cap:classify}: every complete crash occurred
from a graph with $\lambda _{1}(C_{n-1})=1$.

\subsection{\noindent \label{sub:Takeovers-by-core}Takeovers by core transforming
innovations }

\begin{figure}
\textbf{n}=\textbf{6061}\hfill{}\textbf{n=6062}

\includegraphics[  width=6.5cm,
  keepaspectratio]{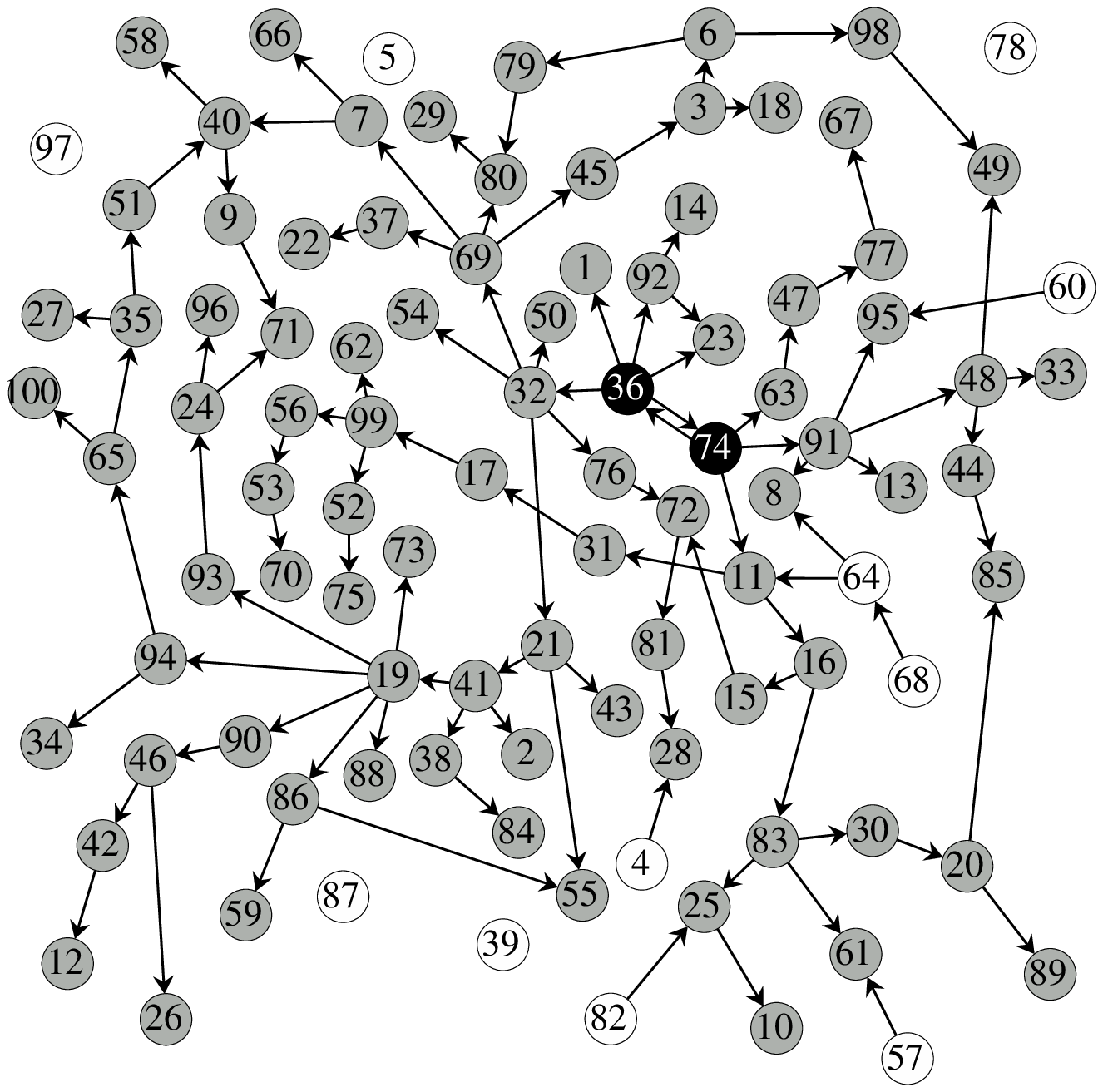}\hfill{}\includegraphics[  width=6.5cm,
  keepaspectratio]{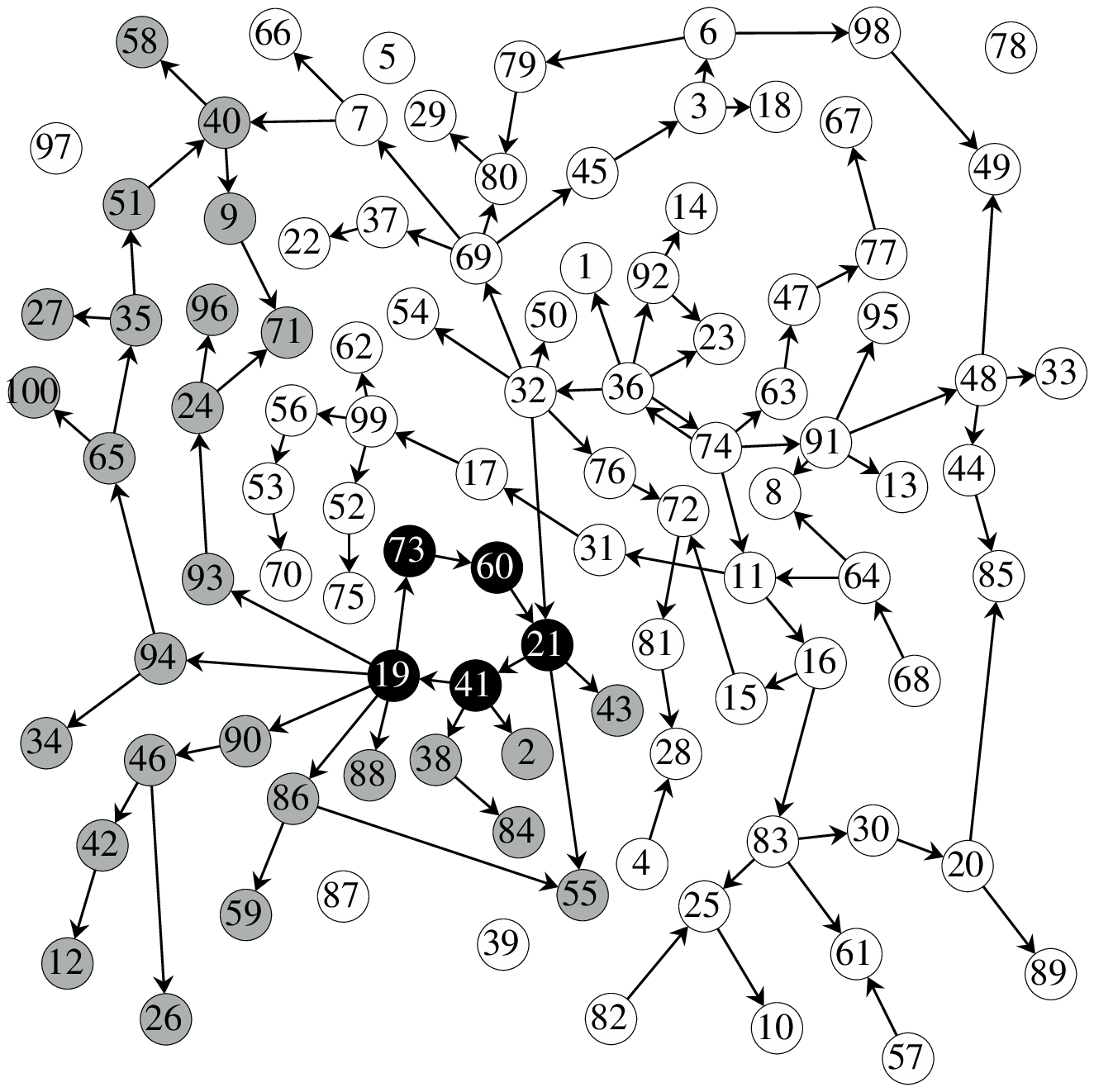}

\caption{Takeover by a core-transforming innovation. In the run displayed
in Figure \ref{cap:s1lambda}b at $n=6061$ the core is a 2-cycle
comprising nodes 36 and 74. The graph update creates a core-transforming
innovation -- a new 5-cycle downstream from the 2-cycle 36-74. As
a result the 2-cycle and all nodes dependent on it become extinct
at $n=6062$ and $s_{1}$ drops from 100 to 32. Because an ACS survives
this event, the system is in the growth phase at $n=6062$.\label{cap:6061,6062}}
\end{figure}

\begin{figure}
\textbf{n=5041}\hfill{}\textbf{n=5042}

\includegraphics[  width=6.5cm,
  keepaspectratio]{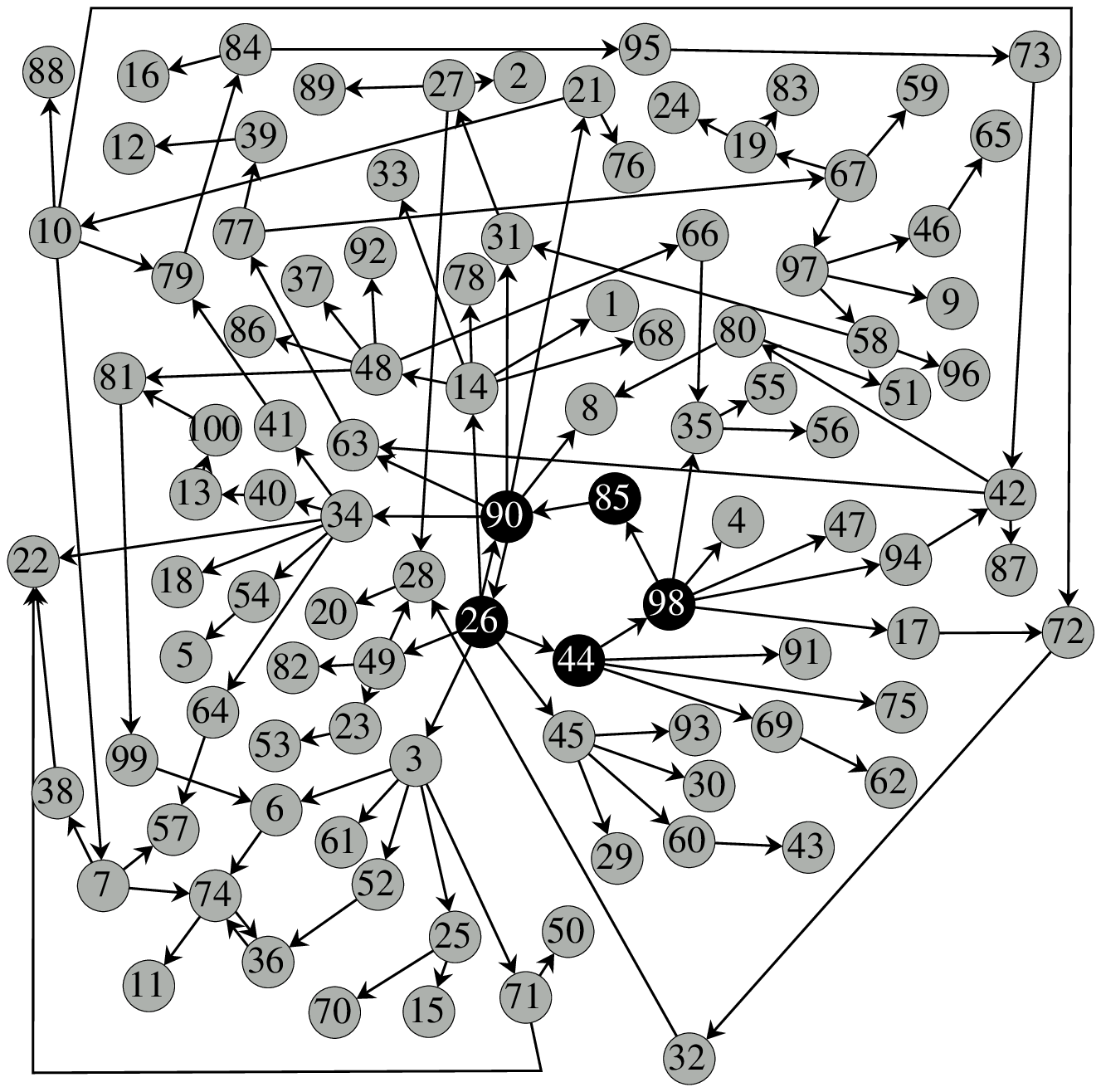}\hfill{}\includegraphics[  width=6.5cm,
  keepaspectratio]{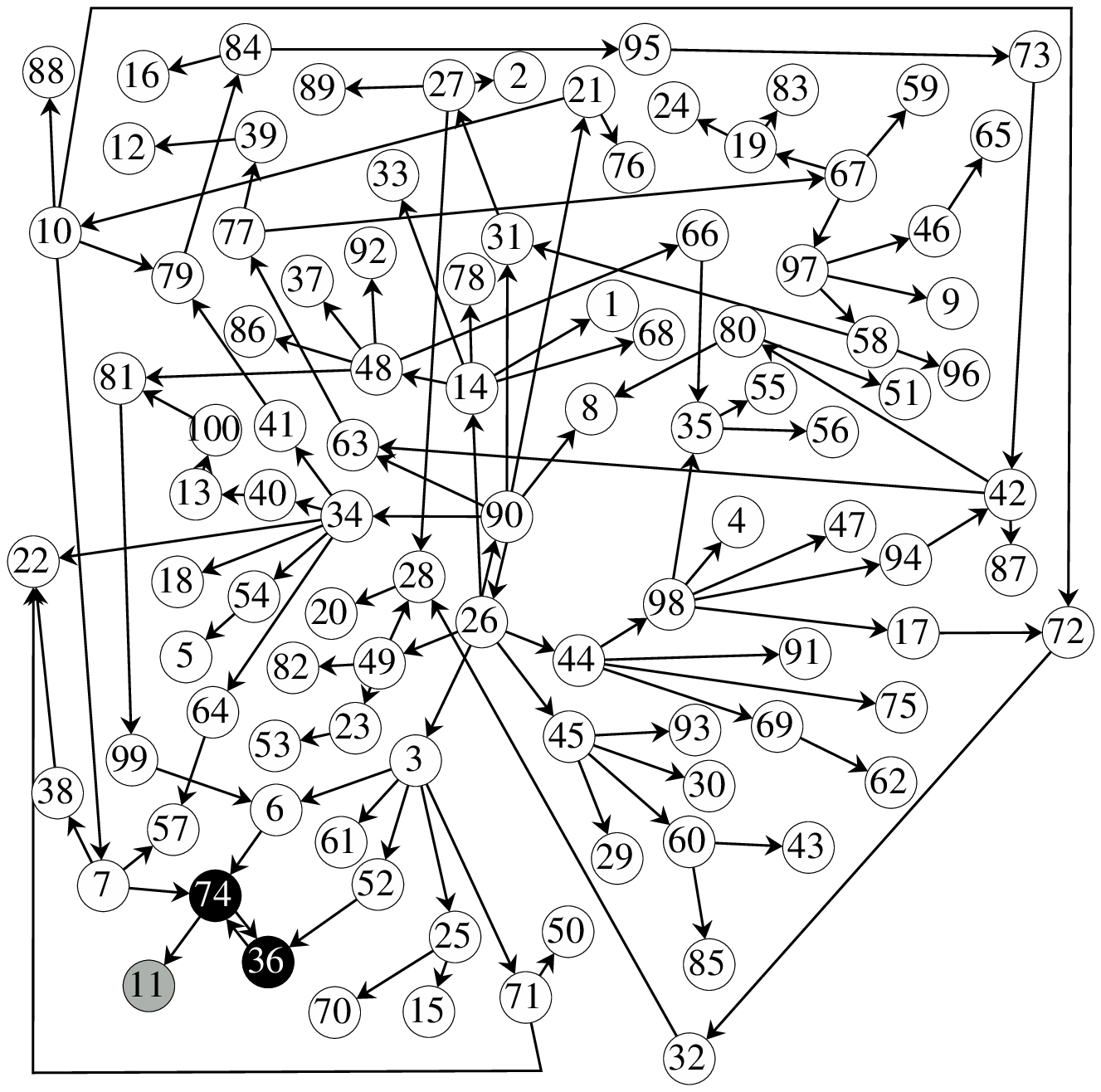}

\caption{Takeover by a dormant innovation. In the run displayed in Figure
\ref{cap:s1lambda}b at $n=5041$ the core has two overlapping cycles.
Downstream from the core is a 2-cycle comprising nodes 36 and 74 which
formed earlier at $n=4696$. Node 85 happens to be one of the nodes
with the least $X_{i}$ and is removed from the graph. Thus, at $n=5042$,
the graph consists of one 2-cycle, 36-74, downstream from another
2-cycle, 26-90. As a result, the upstream cycle, 26-90, and all nodes
dependent on it become extinct and $s_{1}$ drops from 100 to 3. Because
an ACS survives this event, the system is in the growth phase at $n=5042$.\label{cap:5041,5042}}
\end{figure}

\noindent An example of a takeover by a core-transforming innovation
is given in Figure \ref{cap:6061,6062}, taken from the example run
of Figure \ref{cap:s1lambda}b. At $n=6061$ the core was a single
cycle comprising nodes $36$ and $74$. Node $60$ was replaced at
$n=6062$ creating a 5-cycle comprising nodes $60,21,41,19$ and $73$,
downstream from the old core. The graph at $n=6062$ has one cycle
feeding into a second cycle that is downstream from it. We have already
seen in section \ref{sec:Examples-of-the} (see the discussion of
Example 4) that for such a graph only the downstream cycle is non-zero
and the upstream cycle and all nodes dependent on it become extinct.
Thus, the new cycle becomes the new core and the old core becomes
extinct resulting in a core-shift. This is an example of condition
(b) for a core-transforming innovation. For all such events in Figure
\ref{cap:classify}, $\lambda _{1}(Q_{n}')=\lambda _{1}(C_{n-1})$
because $k$ happened not to be a core node of $C_{n-1}$. Thus, these
core-shifts satisfy $\lambda _{1}(C_{n})=\lambda _{1}(N_{n})\geq \lambda _{1}(Q_{n}')=\lambda _{1}(C_{n-1})\geq 1$
in Figure \ref{cap:classify}.

\subsection{\noindent \label{sub:Takeovers-by-dormant}Takeovers by dormant innovations }

\noindent Dormant innovations, which create an irreducible structure
in the periphery of the dominant ACS that does not affect its core
at that time, were discussed in section \ref{sub:Addition-of-a}.
An example is the 2-cycle comprising nodes 36 and 74 formed at $n=4696$.
At a later time such a dormant innovation can result in a core-shift
if the old core gets sufficiently weakened.

In this case the core has become weakened by $n=5041$, when it has
$\lambda _{1}=1.24$ (Figure \ref{cap:5041,5042}). The structure
of the graph at this time is very similar to the graph in Figure \ref{cap:keystone2}a.
Just as node 3 in Figure \ref{cap:keystone2}a was a keystone node,
here nodes 44, 85, and 98 are keystone nodes because removing any
of them results in a graph like Figure \ref{cap:keystone2}b, consisting
of two 2-cycles, one downstream from the other. 

Indeed, at $n=5041$, node $85$ is replaced and the resulting graph
at $n=5042$ has a 2-cycle ($36$ and $74$) downstream from another
cycle ($26$ and $90$). Thus, at $n=5042$, nodes $36$ and $74$
form the new core with only one other downstream node, $11$, being
non-zero. All other nodes become extinct resulting in a drop in $s_{1}$
by $97$. A dormant innovation can takeover as the new core only following
the removal of a keystone node that weakens the old core. In such
an event the new core necessarily has a lower (but non-zero) $\lambda _{1}$
than the old core, i.e., $\lambda _{1}(C_{n-1})>\lambda _{1}(C_{n})\geq 1$
(see Figure \ref{cap:classify}).

Note that $85$ is a keystone node and the graph is susceptible to
a core-shift \textit{because} of the innovation that created the cycle
36-74 earlier. If the cycle between $36$ and $74$ were absent $85$
would \textit{\emph{not}} be a keystone node because its removal would
still leave part of the core intact (nodes $26$ and $90$).

\section{\label{sec:Timescale-of-crashes}Timescale of crashes}

\begin{figure}
\textbf{a)}

\includegraphics[  width=12cm,
  keepaspectratio]{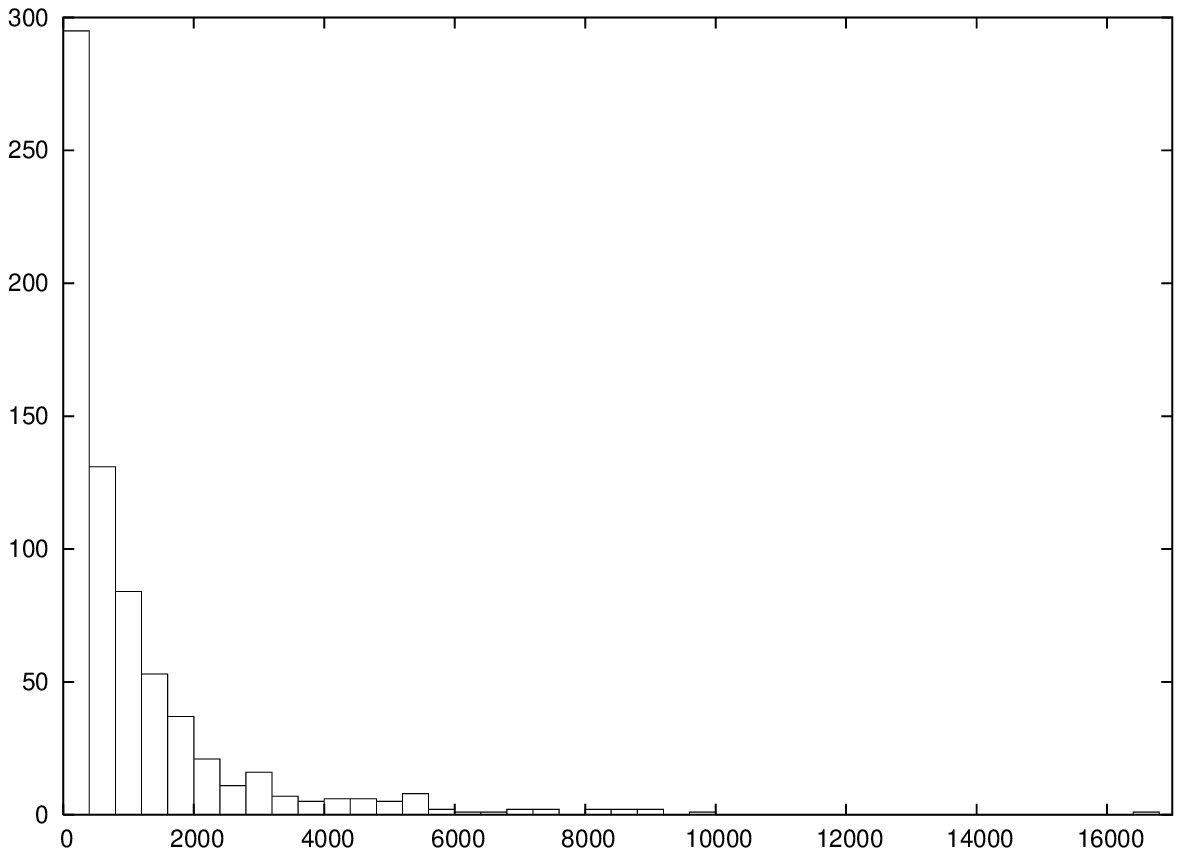}

\textbf{b)}

\includegraphics[  width=12cm,
  keepaspectratio]{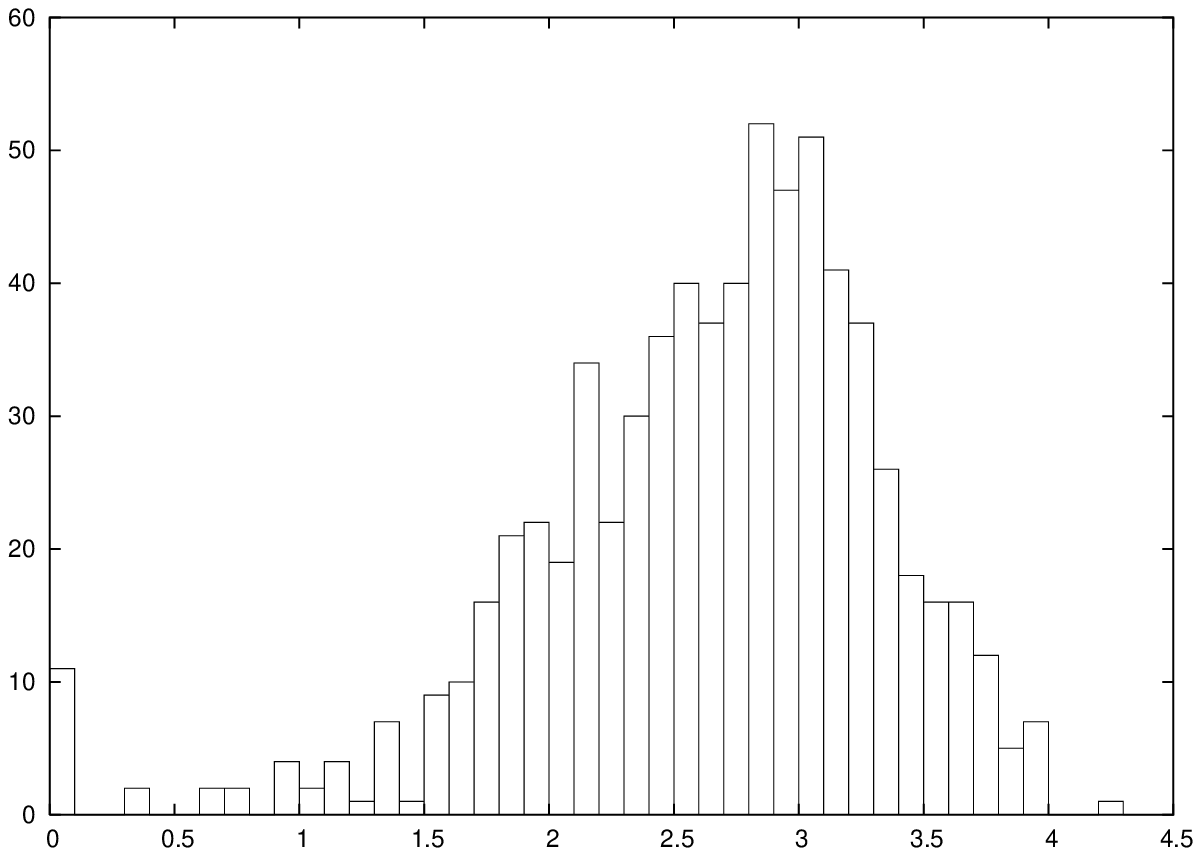}

\caption{\textbf{a.} Distribution of times elapsed between crashes (not counting
the recovery time from the previous crash or beginning of the run),
$n_{s}$, for the 701 crashes observed in runs with $s=100,p=0.0025$.
$\mathrm{min}\{n_{s}\}=1,\mathrm{max}\{n_{s}\}=16625,\tau _{s}\equiv \overline{n_{s}}\approx 1088.3$,
and the standard deviation of the 701 $n_{s}$ values is $\approx 1581.0$.
The bin size is 400. \textbf{b.} Distribution of $\mathrm{log}_{10}n_{s}$
values for the same 701 crashes. The bin size is 0.1. \label{cap:taushist}}
\end{figure}

I will denote by $n_{s}$ the time elapsed between successive crashes,
not counting the time spent in the recovery from the previous crash
(i.e., the time from the previous crash to the first subsequent spanning
of the graph by the dominant ACS). Figure \ref{cap:taushist}a shows
the distribution of $n_{s}$ values for the 701 crashes, discussed
above, observed in runs with $s=100,p=0.0025$. The average $\tau _{s}\equiv \overline{n_{s}}\approx 1088.3$
and the standard deviation of $n_{s}$ values is $\approx 1581.0$
for these 701 crashes. Figure \ref{cap:taushist}b shows the distribution
of $\mathrm{log}_{10}n_{s}$ values. The distribution and the $\tau _{s}$
value must be taken with caution because the distribution appears
to be a very broad one (the standard deviation of $n_{s}$ values
is larger than $\tau _{s}$) and 701 instances may not be enough to
specify it accurately.

\begin{figure}
\textbf{a)}

\includegraphics[  width=12cm,
  keepaspectratio]{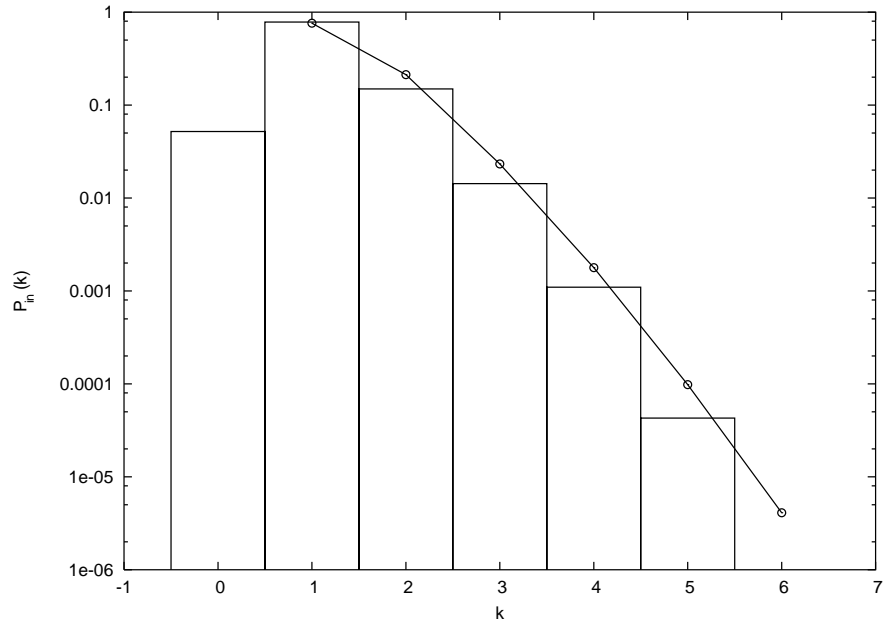}

\textbf{b)}

\includegraphics[  width=12cm,
  keepaspectratio]{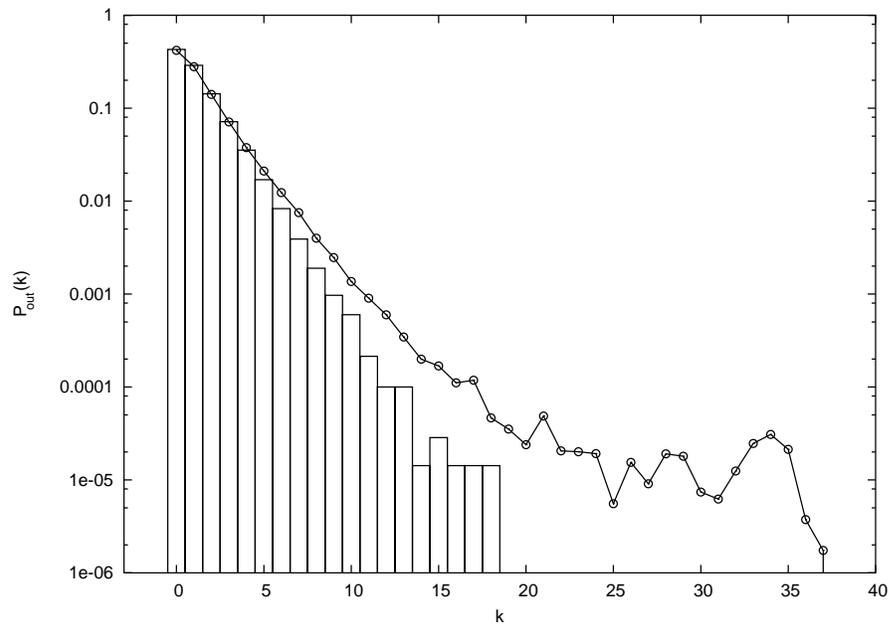}

\caption{\textbf{a.} Comparison of the in-degree distribution of nodes of
the 701 graphs just before each crash (bars) with the in-degree distribution
of all fully autocatalytic graphs in the same runs (line with circles).
\textbf{b.} Comparison of the out-degree distributions for the same
graphs as in (a).\label{cap:crashdeg}}
\end{figure}

\section{Recoveries}

After a complete crash the system is back in the random phase. In
$O(s)$ graph updates each node is removed and replaced by a randomly
connected node, resulting in a graph as random as the initial graph.
Then the process starts again, with a new ACS being formed after an
average of $1/p^{2}s$ time steps and then growing to span the entire
graph after, on average, $(1/p)\ln (s/s_{0})$ time steps, where $s_{0}$
is the size of the initial ACS that forms in this round (typically
$s_{0}=2$ for small $p$).

After other catastrophes, an ACS always survives. In that case the
system is in the growth phase and immediately begins to recover, with
$s_{1}$ growing exponentially on a timescale $1/p$. Note that these
recoveries happen because of innovations (mainly of type 2 and 4,
and some of type 3).

\section{\label{sec:Structure-of-the}Structure of the graph just before a
crash}

\begin{figure}
\includegraphics[  width=14.5cm,
  keepaspectratio]{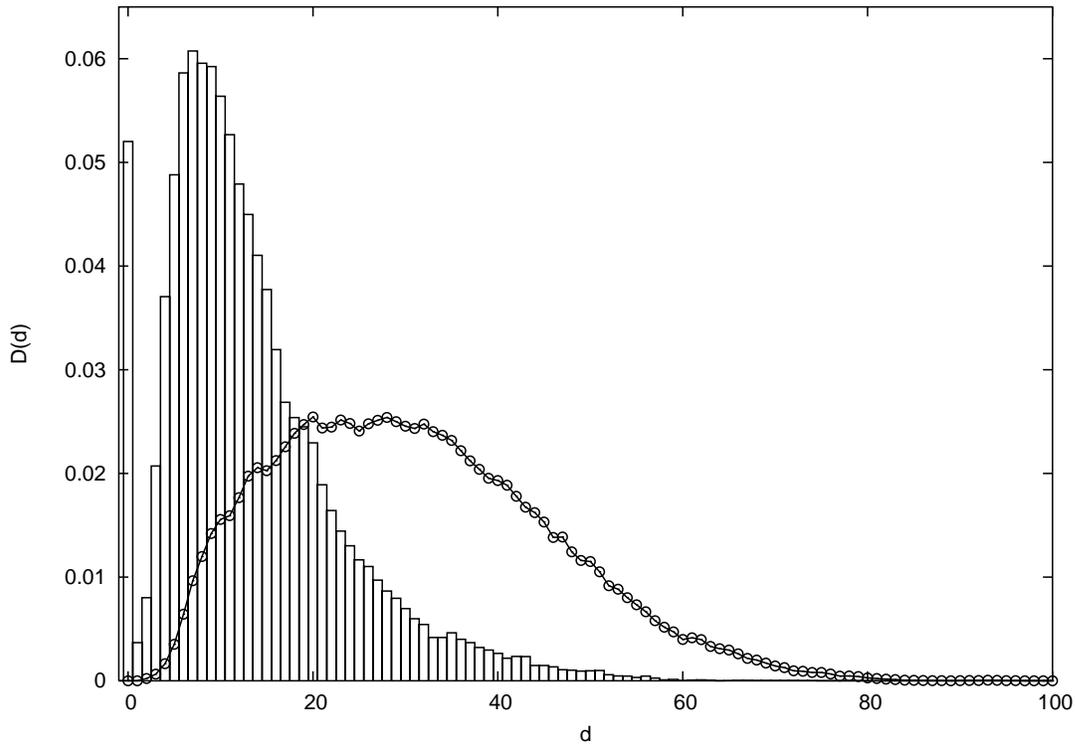}

\caption{Comparison of the dependency distribution of nodes of the 701 graphs
just before each crash (bars) with the distribution for all fully
autocatalytic graphs in the same runs (circles). \label{cap:crashdep}}\lyxline{\normalsize}

\end{figure}

The in-degree and out-degree distributions of the nodes of the 701
graphs just before each crash are shown in Figure \ref{cap:crashdeg}.
The distributions are compared with the distributions for all fully
autocatalytic graphs produced in the same runs. Both the distributions
fall off faster than the distributions for all fully autocatalytic
graphs. 

Another interesting difference between the graphs just before a crash
and all fully autocatalytic graphs lies in their dependency distributions:
The dependency distribution of the graphs before crashes (Figure \ref{cap:crashdep})
is bimodal, in contrast to the distribution for all fully autocatalytic
graphs which is unimodal and has a larger mean.

\begin{figure}
\includegraphics[  width=14.5cm,
  keepaspectratio]{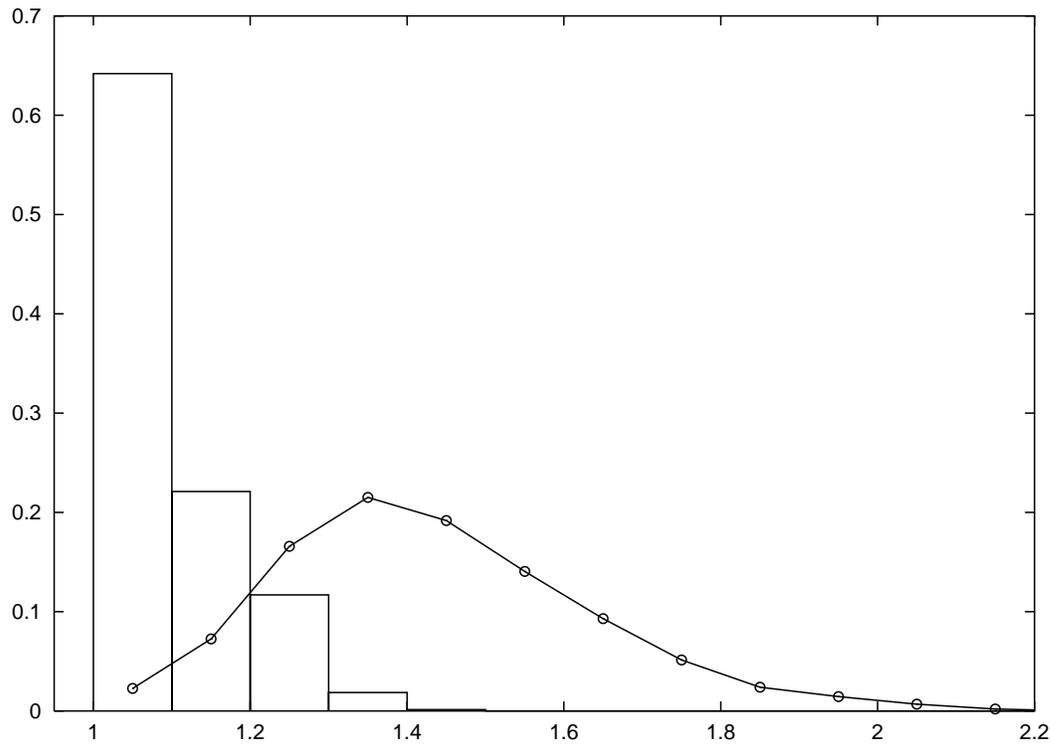}

\caption{Comparison of the distribution of Perron-Frobenius eigenvalues of
the 701 graphs just before each crash (bars) with the distribution
for all fully autocatalytic graphs in the same runs (circles). \label{cap:crashlm}}\lyxline{\normalsize}

\end{figure}

Figure \ref{cap:crashlm} compares the distribution of Perron-Frobenius
eigenvalues for the 701 graphs with the distribution for all the fully
autocatalytic graphs. For the graphs just before each of the 701 crashes
the mean Perron-Frobenius eigenvalue is 1.07, with standard deviation
0.10 -- clearly lower than the mean for the fully autocatalytic graphs,
which is 1.43, with standard deviation 0.20. The three mechanisms
that cause core-shifts, discussed previously, are all more likely
to happen with graphs whose cores are cycles. Therefore, it is not
surprising that the graph just before a crash tends to have a Perron-Frobenius
value close to unity.

\chapter{\label{cha:Variants-of-model}Robustness of the ACS Growth Mechanism}

\section{Variants of the model}

In this chapter, I examine the effect of various modifications to
the model rules on the formation and growth of ACSs. These modifications
relax several of the simplifications that have been made, which depart
from realism but make the model simpler to analyze:

\begin{enumerate}
\item $c_{ij}$ take values 0 and 1 only; all catalysts have the same catalytic
efficiency.
\item $c_{ij}$ take positive values only; negative links -- inhibitors
-- are not present in the model.
\item Diagonal elements $c_{ii}$ are zero; no self-replicators are allowed.
\item The node with the \emph{least} $X_{i}$ is removed at each graph update;
extremal selection. 
\item If several nodes have the same least $X_{i}$ only one is removed.
\item The total number of nodes is fixed; the rate of node removals and
node additions is exactly the same.
\item The population dynamics is linear in the $y_{i}$ variables.
\end{enumerate}

I consider the following variants of the model, that relax these simplifications:

\begin{enumerate}
\item \textbf{Variable link strengths.} $c_{ij}$, when non-zero, are allowed
to take any value in the interval $[0,1]$. This relaxes simplification
(1).
\item \textbf{Negative links.} $c_{ij}$, when non-zero, are allowed to
take negative values also. This relaxes simplifications (1) and (2).
\item \textbf{Self-replicators.} The diagonal elements, $c_{ii}$, are allowed
to take non-zero values. This relaxes simplification (3).
\item \textbf{Non-extremal selection.} Simplification (4) is relaxed in
three variants, in which the dynamics is made non-extremal in different
ways:

\begin{enumerate}
\item A fixed fraction $f$ of the nodes with the least $X_{i}$ are replaced
at each graph update.
\item A single node is picked for removal, with a probability proportional
to $1/X_{i}$, at each graph update. When some nodes have $X_{i}=0$
then one of them is picked randomly.
\item The selection depends on a parameter, denoted $q$. At each graph
update, the node to be removed is selected from the set of nodes with
least $X_{i}$ with a probability $q$, or randomly from all nodes
with a probability $1-q$.
\end{enumerate}
\item \textbf{Variable number of nodes.} A fixed threshold $x_{t}$ is chosen
and at each graph update all nodes with $X_{i}<x_{t}$ are removed
and one new node is brought in. This relaxes simplifications (5) and
(6).
\item \textbf{Different population dynamics.} An equation for $\dot{x}_{i}$
in which the first term is quadratic in the relative population variables
is used instead of equation (\ref{xdot}). This relaxes simplification
(7).
\end{enumerate}

\section{Variable link strengths}

The first modification I examine is allowing $c_{ij}$ to take values
in the interval $[0,1]$, instead of just the values 0 or 1. The initial
graph is constructed as follows: for each ordered pair $(i,j)$ with
$i\ne j$, $c_{ij}$ is non-zero with probability $p$ and $c_{ij}=0$
with probability $1-p$. If non-zero, the value of $c_{ij}$ is randomly
picked from the interval $[0,1]$ with uniform probability. As before
$c_{ii}$ is set to zero for all $i$ to disallow self-replicators.
Step 3 of the graph dynamics -- addition of a new node -- is similarly
modified. Apart from these alterations the model remains unchanged.

\begin{figure}
\textbf{a)}

\includegraphics[  width=13cm,
  keepaspectratio]{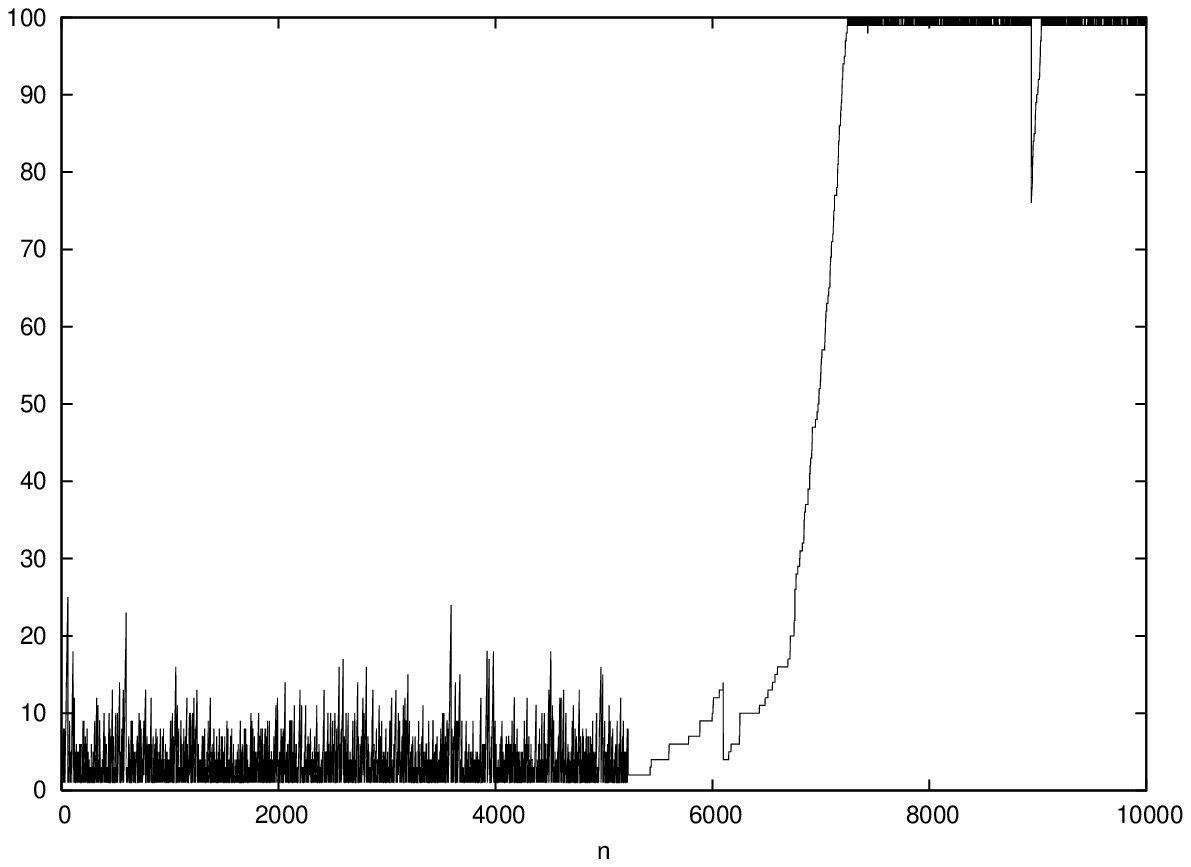} \textbf{}

\textbf{b)}

\includegraphics[  width=13cm,
  keepaspectratio]{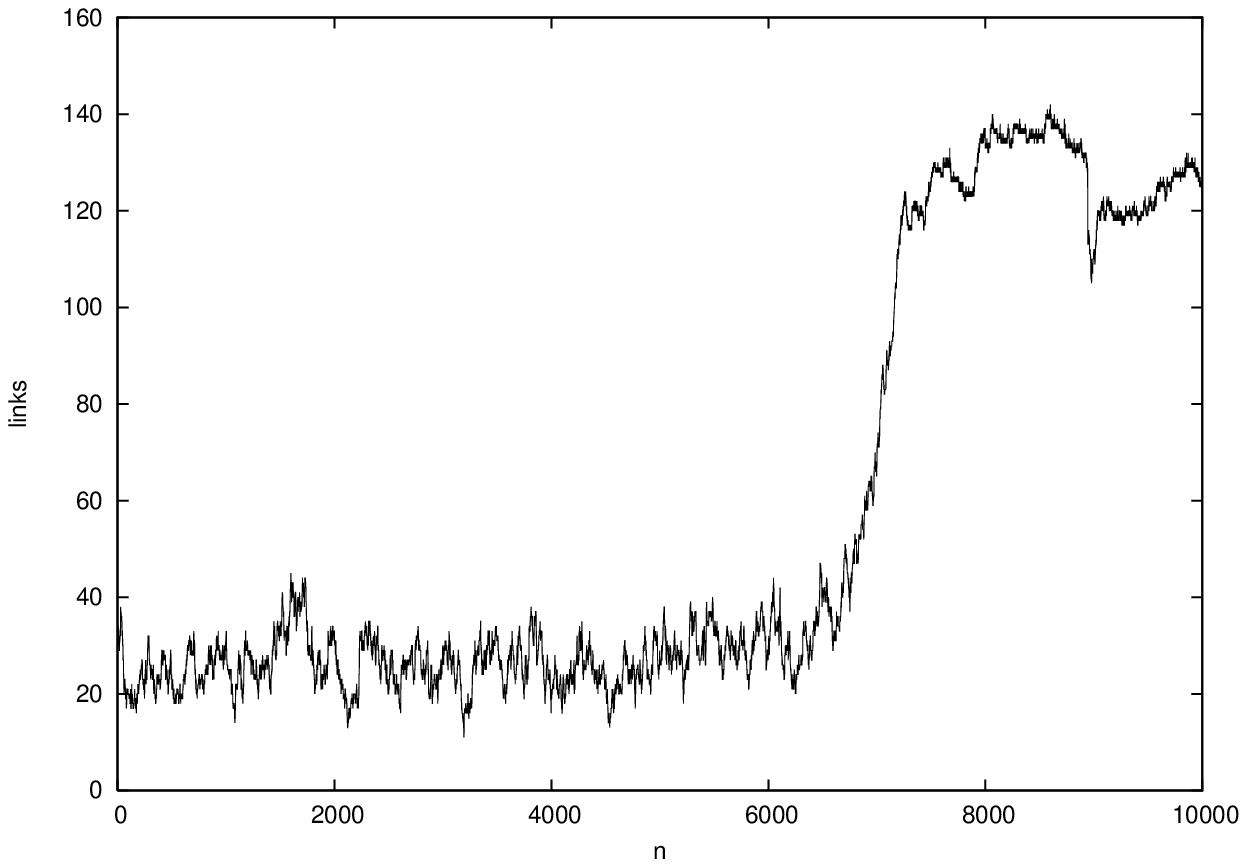}

\caption{A run with $s=100,p=0.0025$ with variable strength links: $c_{ij}$
when non-zero can take any value in the interval $[0,1]$. \textbf{a.}
number of nodes with $X_{i}>0$, $s_{1}$, versus time, $n$. \textbf{b.}
number of links versus $n$.\label{cap:varylinkfig}}
\end{figure}

Figure \ref{cap:varylinkfig} shows a run with these modifications
in the rules. The same behaviour is seen. There is a `random phase'
where there is no ACS and the graph remains random. The chance formation
of an ACS on an average timescale $\tau _{a}=1/p^{2}s$ triggers the
`growth phase' in which the number of non-zero nodes grows exponentially
with a characteristic timescale $\tau _{g}=1/p$, until the ACS spans
the entire graph.

\begin{figure}
\begin{center}\includegraphics{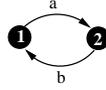}\end{center}

\caption{A two cycle with link strengths $a,b\in [0,1]$. The Perron-Frobenius
eigenvalue of this graph is $\sqrt{ab}$.\label{cap:twocycleab}}\lyxline{\normalsize}

\end{figure}

This is expected because most of the analytical results hold, with
minor modifications, for this variant too. The Perron-Frobenius theorem
holds for the adjacency matrices of the graphs produced because they
are still non-negative. The only significant modification in the analytical
results is that now $\lambda _{1}$ can take values between 0 and
1 also. For example, consider the 2-cycle shown in Figure \ref{cap:twocycleab}.
The strengths of the two links are $a$ and $b$, any real numbers
in the interval $[0,1]$. For such a graph $\lambda _{1}=\sqrt{ab}$. 

Thus, the condition $\lambda _{1}\ge 1$ is changed to $\lambda _{1}>0$
in all propositions that contain it. Proposition 2.1(ii) is modified
to read: If a graph $C$ has a closed walk then $\lambda _{1}>0$.
Proposition 3.2(iii) is similarly modified: If a graph $C$ has an
ACS then $\lambda _{1}>0$.

The only difference this makes to the dynamics of the system is that
in this variant it is much less likely for two disjoint ACSs with
the same $\lambda _{1}$ value to exist in some graph. In the original
model two or more ACSs, typically cycles, often coexisted in runs
with small $p$ (see Figures \ref{cap:snapshots}j,u). With variable
link strengths even if two disjoint cycles form the chance of their
having exactly the same $\lambda _{1}$ value is negligible.

\section{\label{sec:Negative-links}Negative links: emergence of cooperation}

Now I generalize the model to include inhibitory reactions (where
a species can decrease the rate of production of another species)
along with variable link strengths. I implement this by modifying
the way links are assigned in the initial random graph and to the
new node at each graph update: if $c_{ij}$ is chosen to be non-zero
then it is assigned randomly a value in the interval $[-1,1]$ if
$i\ne j$ or $[-1,0]$ if $i=j$ (self-replicators continue to be
disallowed but self-inhibitors are possible). With negative links
the populations can become negative. To prevent this a cutoff must
be put on $\dot{y}_{i}$ when $y_{i}$ becomes zero: \begin{equation}
\dot{y}_{i}=\left\{ \begin{array}{l}
 r_{i}\quad \mathrm{if}\; y_{i}>0\; \mathrm{or}\; r_{i}\ge 0\\
 0\quad \mathrm{if}\; y_{i}=0\; \mathrm{and}\; r_{i}<0\end{array}\right.\label{ydot-}\end{equation}
\[
\mathrm{where}\quad r_{i}=\sum _{j=1}^{s}c_{ij}y_{j}-\phi y_{i}.\]
 For simplicity I retain the form of (\ref{ydot}) although a more
realistic chemical interpretation would require $\dot{y}_{i}$ to
be proportional to $y_{i}$ for inhibitory reactions. 

The relative population dynamics that follows from (\ref{ydot-})
using $x_{i}=y_{i}/\sum _{j=1}^{s}y_{j}$ is: \begin{equation}
\dot{x}_{i}=\left\{ \begin{array}{l}
 f_{i}\quad \mathrm{if}\; x_{i}>0\; \mathrm{or}\; f_{i}\ge 0\\
 0\quad \mathrm{if}\; x_{i}=0\; \mathrm{and}\; f_{i}<0\end{array}\right.\label{xdot-}\end{equation}
\[
\mathrm{where}\quad f_{i}=\sum _{j=1}^{s}c_{ij}x_{j}-x_{i}\sum _{k,j=1}^{s}c_{kj}x_{j}.\]

\begin{figure}
\begin{center}\includegraphics{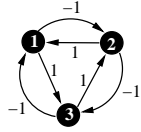}\end{center}

\caption{A graph that has a limit cycle attractor. The negative links, each
of strength -1, form a cycle. The positive links, each of strength
1, form another cycle running in the opposite direction. In the attractor
each $x_{i}$ oscillates around a mean value $1/3$ with a frequency
$2\pi \sqrt{3}$. The amplitude of oscillation depends on the initial
condition but is the same for each $x_{i}$.\label{cap:limitcycle}}\lyxline{\normalsize}

\end{figure}

The cutoff in the population dynamics makes it a nonlinear equation.
As a result now the attractor need not be only a fixed point. For
most graphs I have encountered in simulations the attractor of (\ref{xdot-})
has been a fixed point. In less than 0.2\% cases the attractor has
been a limit cycle where the relative populations of the nodes oscillate.
I have not observed any other types of attractors. The simplest type
of graph having a limit cycle attractor consists of 3 nodes and is
displayed in Figure \ref{cap:limitcycle}. For this graph each $x_{i}$
oscillates around a mean value of $1/3$ with a frequency $2\pi \sqrt{3}$.
The amplitude of oscillation depends on the initial condition but
is always equal for the 3 nodes. If the amplitudes are smaller than
$1/3$ then none of the $x_{i}$ hit zero and the cutoff in (\ref{xdot-})
plays no role in the oscillations. However if arbitrarily random small
changes are made to the interaction strengths of the graph the oscillations
of the $x_{i}$ start hitting zero and the cutoff comes into play.
In general the cutoff seems to play an important role in limit cycle
attractors. It is only in some special cases and for particular initial
conditions that the relative populations of nodes oscillate without
hitting zero. Numerical observations indicate that the property of
having a limit cycle attractor is structurally stable, i.e., a limit
cycle attractor remains if sufficiently small changes are made to
the interaction strengths of a given graph having a limit cycle attractor.

\begin{figure}
\textbf{a)}

\includegraphics[  width=12cm,
  keepaspectratio]{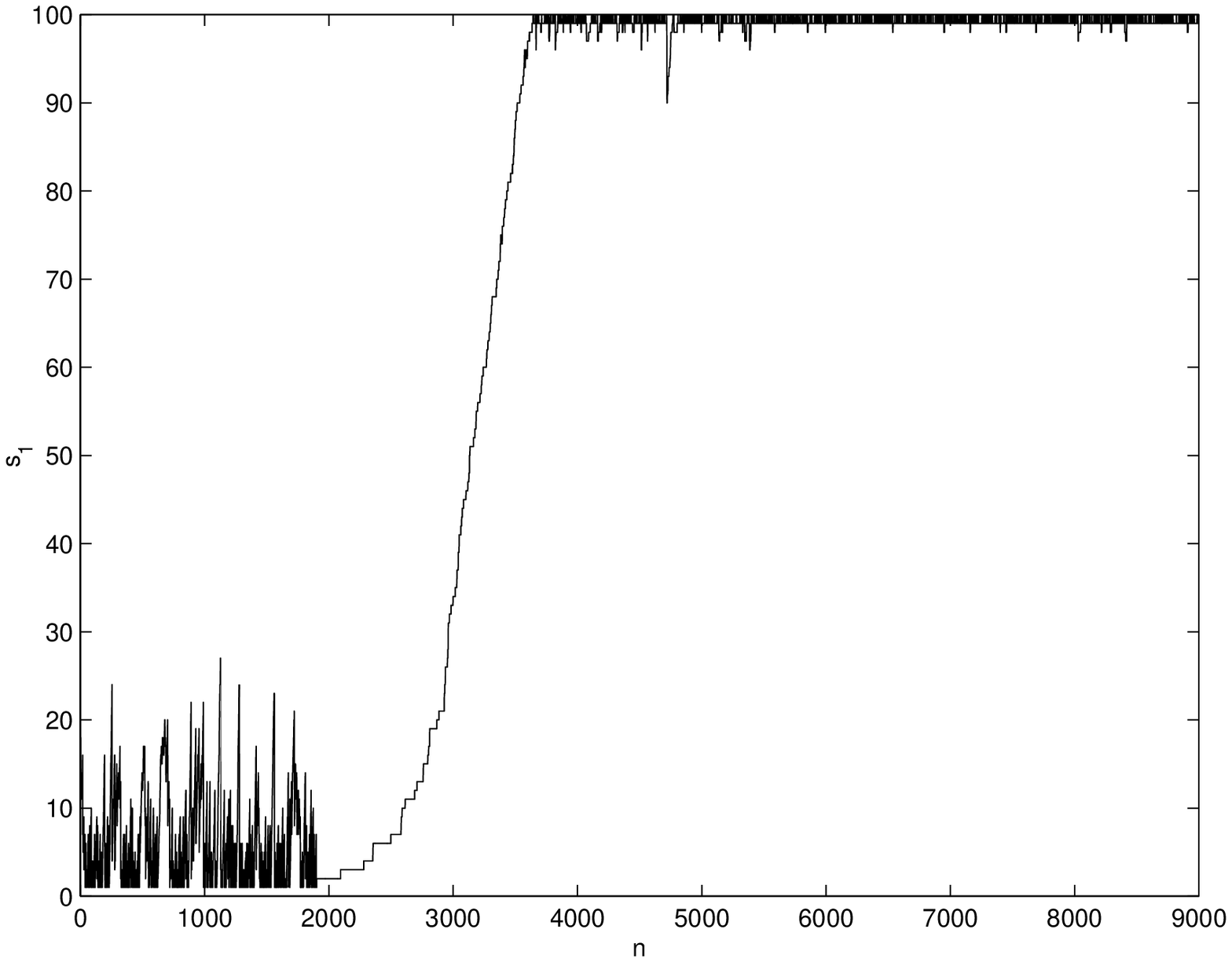}

\textbf{b)}

\includegraphics[  width=12cm,
  keepaspectratio]{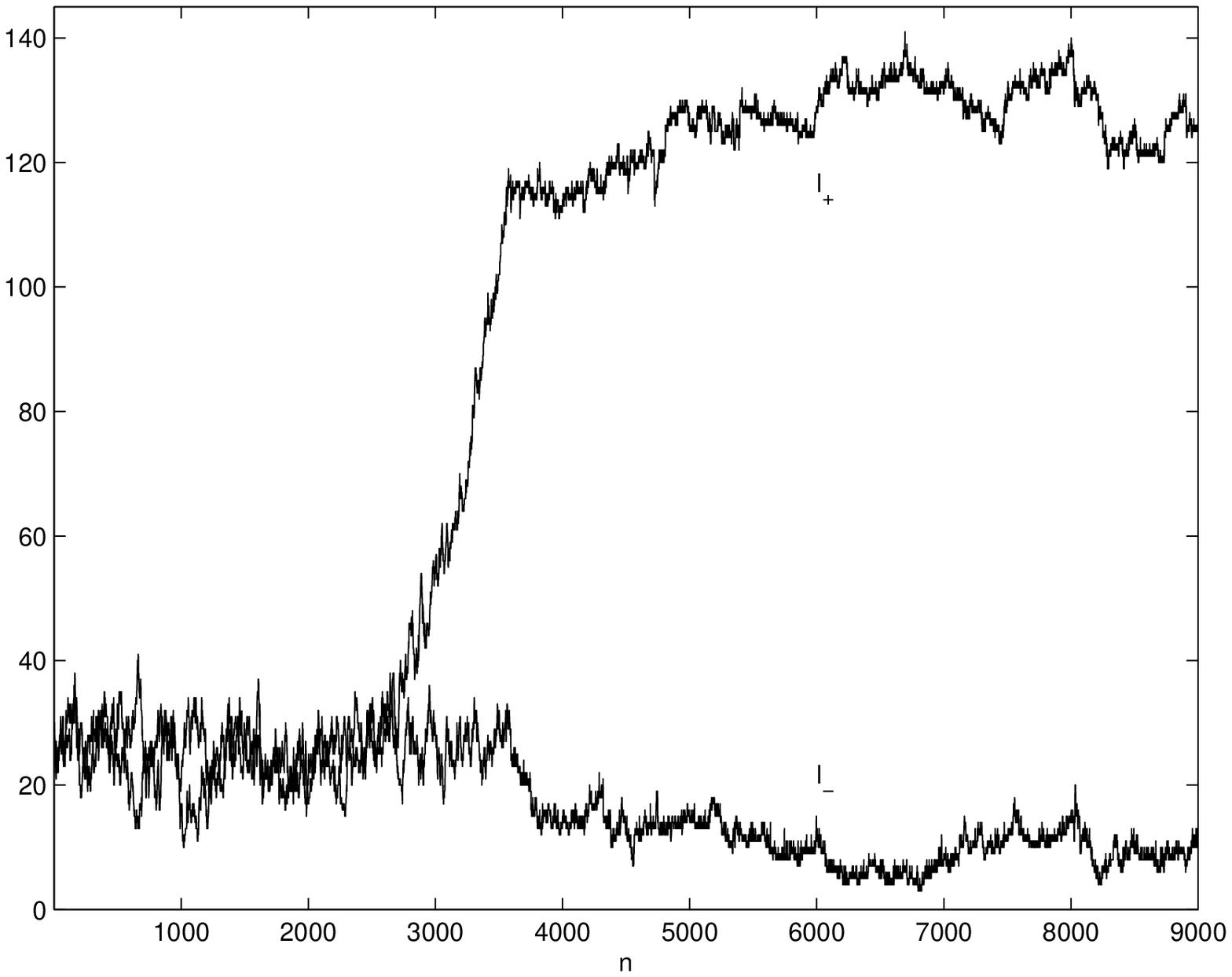}

\caption{A run with $s=100,p=0.005$ where negative links are allowed. \textbf{a.}
the number of nodes with $X_{i}>0$, $s_{1}$, versus time, $n$.
\textbf{b.} the number of positive, $l_{+}$, and negative links,
$l_{-}$, versus $n$.\label{cap:negfig}}
\end{figure}

The theorems used to analyze the attractors of the original model
depended on the matrix $C$ being non-negative and hence no longer
hold for the system with negative links. However numerical simulations
of the system show behaviour very similar to the original system \citep{JKcoop}
despite the slight bias toward negative links (because self-inhibitors
are allowed but not self-replicators). Figure \ref{cap:negfig} shows
the results of a run with $s=100$ and $p=0.005$. The number of nodes
with $X_{i}\ne 0$, $s_{1}$, after the $n^{th}$ addition of a new
node, the number of positive ($c_{ij}>0$), $l_{+}$, and negative
links ($c_{ij}<0$), $l_{-}$, are displayed. Similar to the original
model, the curves have three distinct regions. Initially $s_{1}$
is small; most of the nodes have zero $X_{i}$. $l_{+}$ and $l_{-}$
also do not vary much from their initial value ($\approx ps^{2}/2=25$)
and remain approximately equal. In the second region $s_{1}$ and
$l_{+}$ show a sharp increase while $l_{-}$ decreases. In the third
region the growth of $s_{1}$ and $l_{+}$ stops and almost all nodes
have non-zero $X_{i}$ in contrast to the initial period. The number
of positive links have increased to several times their initial value
while the number of negative links has declined to about $10$. I
have observed the same qualitative behaviour in several runs with
$p$ values ranging from $0.00002$ to $0.01$ and $s=100,150,200$.

Again the explanation for the above behaviour lies in the formation
and growth of autocatalytic sets. The definition of an ACS has to
be slightly modified: an ACS is a subgraph, each of whose nodes has
at least one incoming \textit{positive} link from a node belonging
to the same subgraph. Proposition 3(i) also has to be similarly modified:
An ACS must contain a cycle of \textit{positive} links.

Upto $n=n_{a}=1904$, there is no ACS; the graph has no cycles of
positive links. For such graphs, I find that $X_{i}=0$ for most nodes
except a few at which the longest paths of positive links terminate,
with intermediate nodes not having incoming negative links. This is
a numerical observation that is similar to proposition 4.2. Thus most
nodes have $X_{i}=0$. Nodes with $X_{i}\ne 0$ will also join the
set $\mathcal{L}$ of nodes with least $X_{i}$ if one of their supporting
nodes is picked for removal from the system. Hence, all nodes are
susceptible to be removed sooner or later; the graph remains as random
as the initial graph and so $s_{1}$, $l_{+}$ and $l_{-}$ hover
around their initial values. The dramatic increases in $s_{1}$ and
$l_{+}$ are triggered by the chance formation of an ACS at $n=n_{a}$.
In this system too an ACS will inevitably form by pure chance sooner
or later. The `time of appearance' of an ACS can again be approximated
by the time of formation of a 2-cycle of positive links. As one wants
the 2-cycle to contain positive links only, $p$ in the previous formula
for $\tau _{a}$ has to be replaced by $p/2$. Thus $\tau _{a}\approx 4/p^{2}s(=1600$
for $p=0.005$ and $s=100$) and $P(n_{a})=\frac{p^{2}s}{4}(1-\frac{p^{2}s}{4})^{n_{a}-1}$.

From $n=1904$ to $n=3643$ (when $s_{1}$ first reaches $s$), I
again find that the set of nodes with $X_{i}\ne 0$ is an ACS; the
`dominant ACS'. This is also a numerical observation. The proof of
proposition 4.3 no longer holds but the system nevertheless shows
identical behaviour in this respect to the system with only positive
links. Hence, as long as the dominant ACS does not span the entire
graph, i.e. $s_{1}<s$, there are nodes outside it with $X_{i}=0$
which form the set $\mathcal{L}$. Therefore, none of the nodes in
the dominant ACS can be removed in the next graph update.

\begin{figure}
\begin{center}\includegraphics{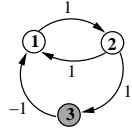}\end{center}

\caption{Example of a destructive parasite. A new node, 3, added to an existing
2-cycle, destroys the 2-cycle because of the negative link to node
1. The attractor is $\mathbf{X}=(0,0,1)^{T}$.\label{cap:dparasite}}\lyxline{\normalsize}

\end{figure}

All the processes that can happen in the original model when a new
node is brought in (numbered 1, 2 and 3 in section \ref{sec:The-growth-phase})
happen in this system with the same effects. The only new possibility
to which the presence of negative links leads is when the new node
joins the dominant ACS and some of the other nodes in the dominant
ACS receive negative links from it. Then those nodes may become extinct
causing $s_{1}$ to decrease. An example of this is shown in Figure
\ref{cap:dparasite}. A new node, 3, is added to a previously existing
2-cycle. It gets a positive incoming link from node 2 and a negative
outgoing link to node 1. This negative link destroys the ACS. In the
attractor only node 3 is non-zero while node 1 and 2 have $X_{i}=0$.
Node 3 is like a destructive parasite that feeds on the the host ACS
(nodes 1 and 2) to increase its own relative population, in the process
destroying the host. The negative link is essential for the destructive
property of such parasites; in the system with only positive links
only benign parasites are possible that do not harm the ACS upon which
they are feeding. This example also shows that propositions 4.3 and
6.1 are no longer true; the set of nodes with $X_{i}>0$ need not
be an ACS even if an ACS exists in the graph and $\lambda _{1}$ can
decrease while $s_{1}<s$. However, destructive parasites occur rarely
and do not reverse the trend of increasing $s_{1}$ -- in the displayed
run they formed $6$ times at $n=3388,3478,3576,3579,3592$ and $3613$,
and in each case only resulted in $s_{1}$ decreasing by $1$. Because
the dominant ACS grows by adding positive links to the existing dominant
ACS, the number of positive links increases as the ACS grows. On the
other hand nodes receiving negative links often have $X_{i}=0$, hence
negative links get eliminated when these nodes are removed.

\begin{figure}
\noindent \begin{center}\includegraphics[  width=7cm,
  keepaspectratio]{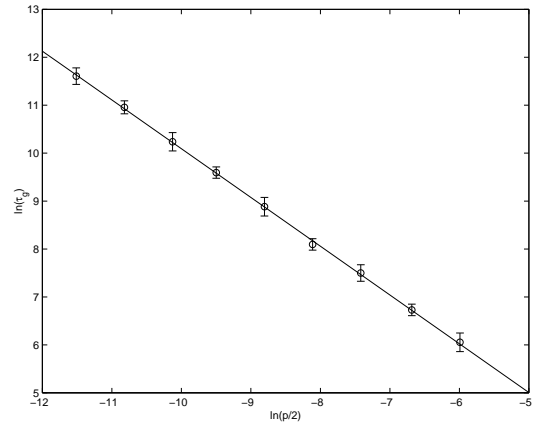} \end{center}

\caption{\noindent \textit{\emph{Dependence of $\tau _{g}$ on $p$ for the
model with negative links. Each data point shows the average of $\ln \tau _{g}$
over 10 different runs with $s=100$, $x_{0}=10^{-4}$ and the given
$p$ value. The error bars correspond to one standard deviation of
$\ln \tau _{g}$ values for each $p$. The best fit line has slope
$-1.02\pm 0.03$ and intercept $-0.08\pm 0.27$, consistent with the
expected slope $-1$ and intercept $0$.\label{cap:negtaug}}}}\lyxline{\normalsize}

\end{figure}

One can calculate the growth rate of the dominant ACS in exactly the
same way as for the system with positive links because destructive
parasites appear rarely. The only difference is that $p$ must be
replaced by $p/2$ because that is the probability for the added node
to get a \textit{positive} link to/from any other node. Thus once
an ACS forms, $s_{1}$ locally averaged in time grows exponentially
with a characteristic timescale $\tau _{g}=2/p$. This is confirmed
by numerical simulations (Figure \ref{cap:negtaug}).

This variant of the model, apart from being an important test of the
robustness of the ACS mechanism, also allows one to address questions
concerning the emergence of `cooperation' because negative links can
arise in the network. The negative links in the network can be considered
`destructive' links as they cause one species to decrease the population
growth rate of another. In contrast, the positive links in the network
can be considered `cooperative' links. The increase of the ratio of
positive to negative links, from its initial value of unity to value
more than an order of magnitude larger in Figure \ref{cap:negfig},
is evidence of the emergence of cooperation during the run.

\newpage
\section{Self-replicators}

\begin{figure}
\includegraphics[  width=14cm,
  keepaspectratio]{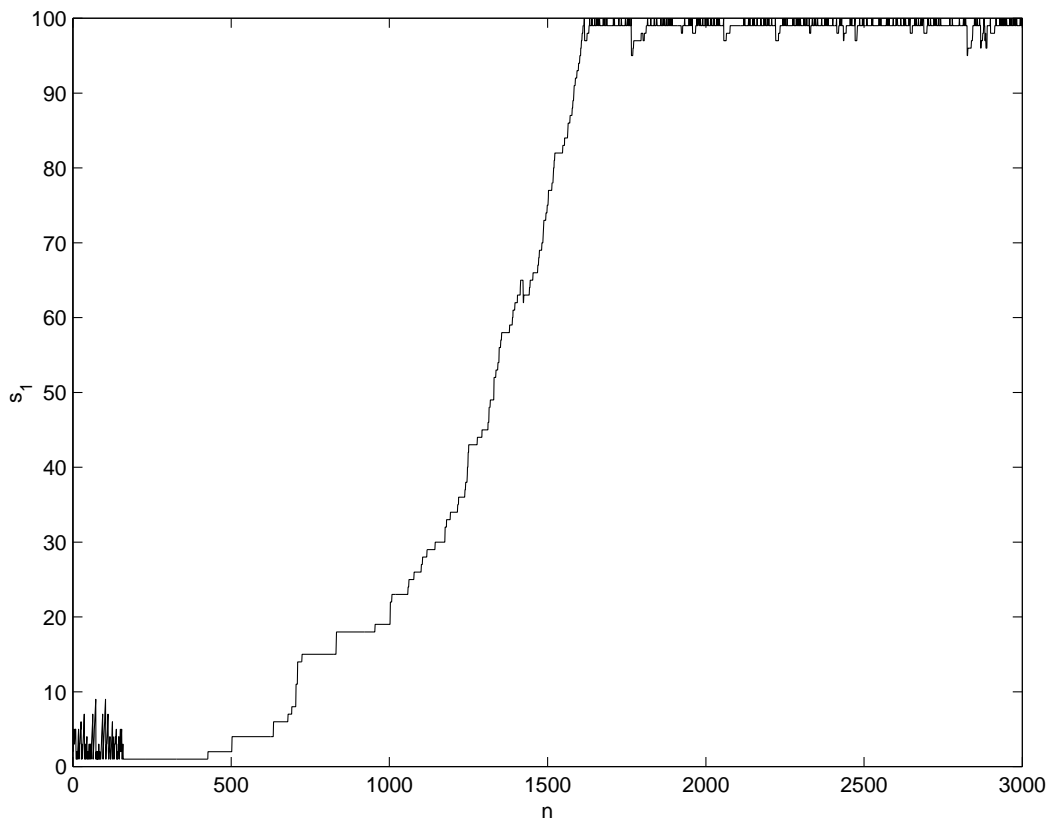}

\caption{A run where self-replicators are allowed. Diagonal elements $c_{ii}$,
when non-zero, are chosen randomly from the interval $[-1,1]$. The
run has $s=100,p=0.005$.\label{cap:selfrep}}\lyxline{\normalsize}

\end{figure}

In the original model, and all the variants discussed so far, self-replicators,
i.e., self-links from a node to itself, are disallowed. As discussed
in section \ref{sub:Absence-of-self-replicators}, this was done to
see whether structured chemical networks can arise even in the (forced)
absence of individual self-replicators. As shown in chapter \ref{cha:Formation-and-Growth},
structured networks do indeed form. The question that now arises is:
Will allowing self-links interfere with the formation and growth of
other ACSs? In fact, it does not, as evidenced by Figure \ref{cap:selfrep},
which shows a run with negative links in which self-replicators were
allowed, i.e., for the initial graph and for the new node $c_{ii}$
(when non-zero) was chosen randomly from the interval $[-1,1]$. The
probability for a self-replicator to form in a graph update event
is $p/2$. The probability for a 2-cycle to form is $p^{2}s/4$. These
are comparable when $p\approx 2/s$. Thus for small enough $p$, self-replicators
do form, and often trigger the growth phase, but their role in the
network dynamics is similar to that of any other ACS. Self-replicators
have a Perron-Frobenius eigenvalue $\lambda _{1}=1$ and are, therefore,
no more robust than cycles. ACSs with larger Perron-Frobenius eigenvalues
out-compete self-replicators.

\section{Non-extremal selection }

\begin{figure}
\textbf{a)}

\includegraphics[  width=12cm,
  keepaspectratio]{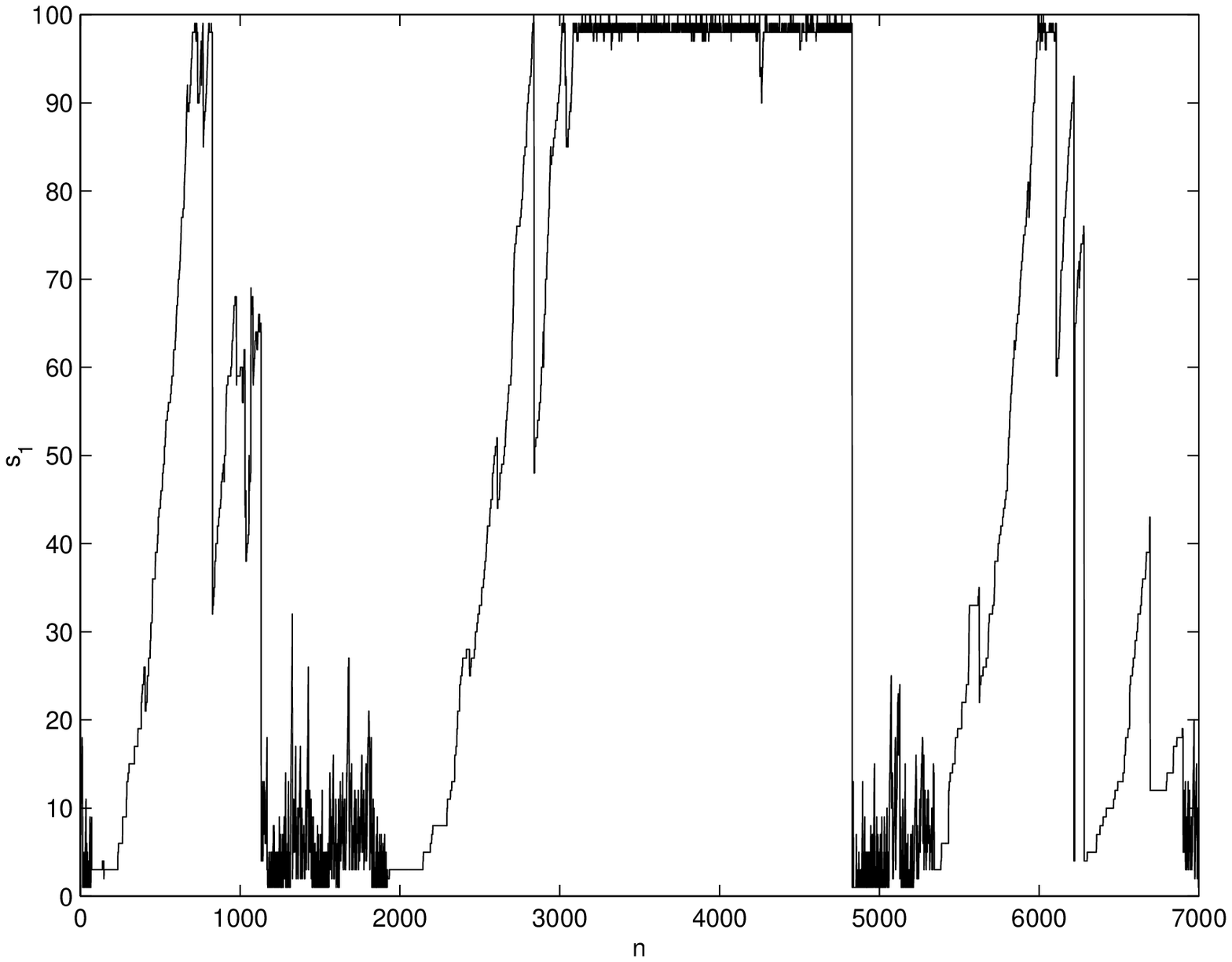}

\textbf{b)}

\includegraphics[  width=12cm,
  keepaspectratio]{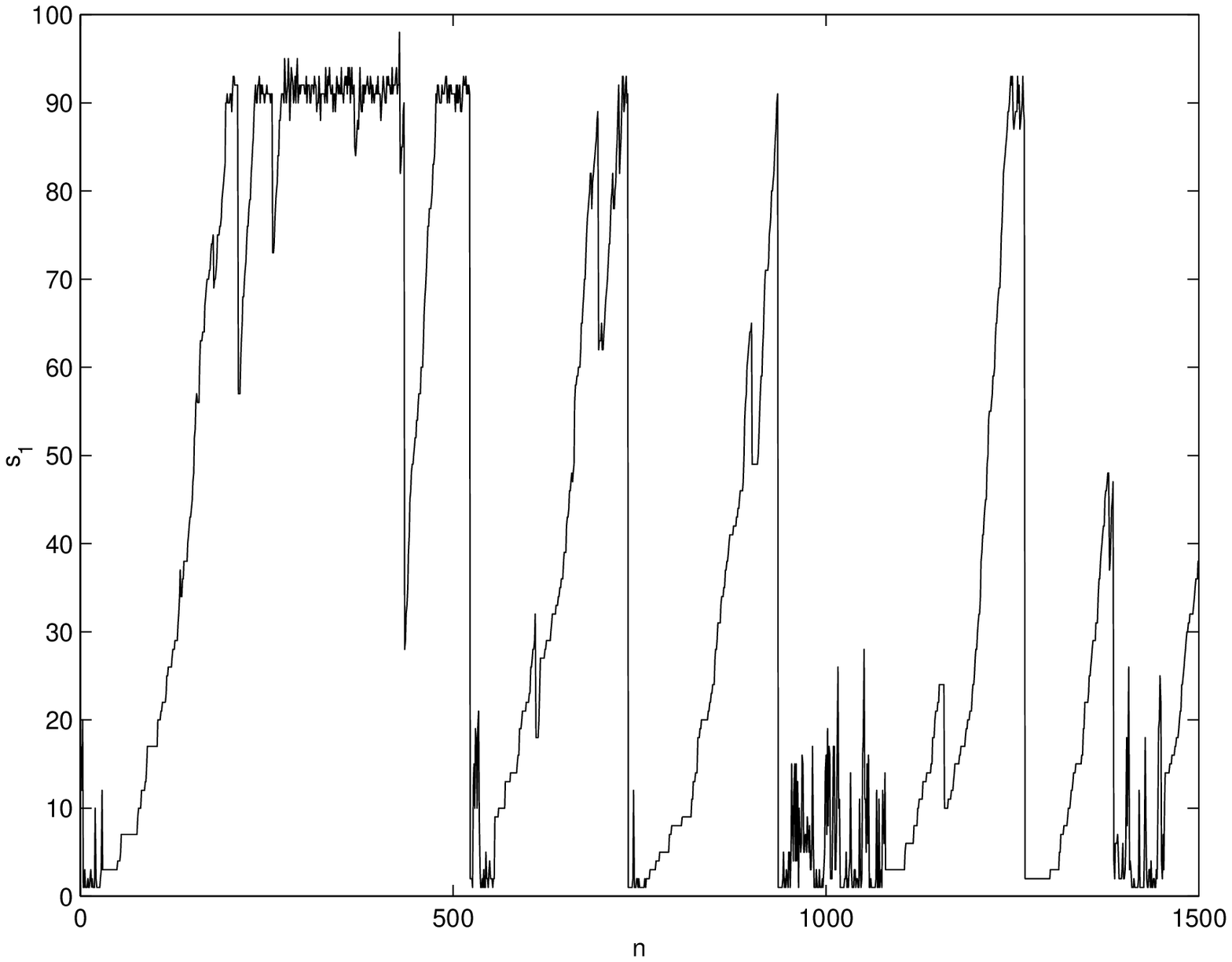}

\caption{Non-extremal runs of type 1 with $s=100,p=0.005$. \textbf{a.} at
each time step 2 nodes are replaced. \textbf{b.} at each time step
10 nodes are replaced.\label{cap:manyhit}}
\end{figure}

\begin{figure}
\includegraphics[  width=14cm,
  keepaspectratio]{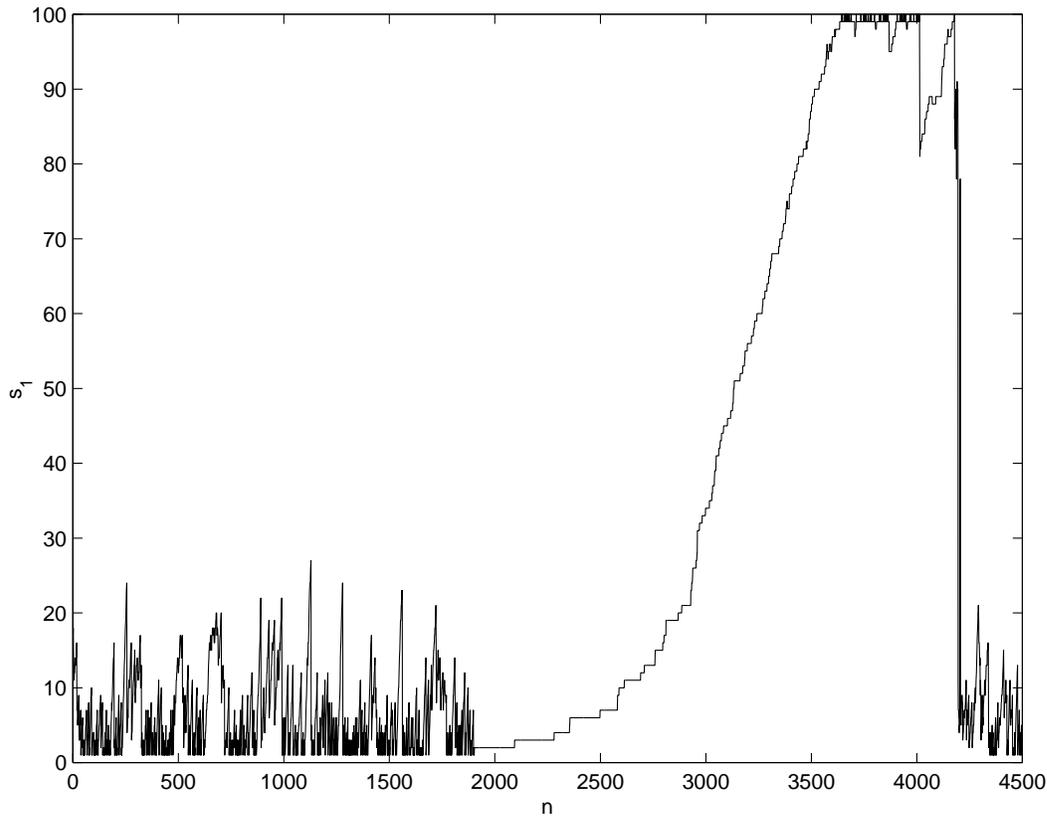}

\caption{Non-extremal run of type 2 with $s=100,p=0.005$. At each time step
a single node is replaced. The node is chosen with a probability proportional
to $1/X_{i}$. If some nodes have $X_{i}=0$ then one of them is randomly
chosen.\label{cap:onebyx}}\lyxline{\normalsize}

\end{figure}

The rule for removing a node can be made non-extremal in various ways.
Instead of removing a single node with the least $X_{i}$ one can
remove more than one node or remove nodes with higher $X_{i}$ also.
However, even when removing nodes which do not have the least $X_{i}$
it is reasonable to allow some bias in favour of nodes with a higher
$X_{i}$. I consider the following possibilities: 

\begin{enumerate}
\item A fixed fraction $f$ of the nodes with the least $X_{i}$ are replaced
at each graph update.
\item A single node is picked for removal, with a probability proportional
to $1/X_{i}$, at each graph update. When some nodes have $X_{i}=0$
then one of them is picked randomly.
\item The selection depends on a parameter, denoted $q$. At each graph
update, the node to be removed is selected from the set of nodes with
least $X_{i}$ with a probability $q$, or randomly from all nodes
with a probability $1-q$.
\end{enumerate}
Figure \ref{cap:manyhit}a shows a run with $s=100,p=0.005$ with
2 nodes removed at each graph update; case 1 with $f=0.02$. Figure
\ref{cap:manyhit}b shows a similar run with 10 nodes removed at each
graph update. Figure \ref{cap:onebyx} shows a run with $s=100,p=0.005$
for case 2. In these three runs the behaviour is similar to the earlier
models. An ACS forms by chance and grows until most of the graph is
autocatalytic. In these cases the nodes comprising the dominant ACS
are protected from removal until the size of the ACS becomes close
to $s$. Only after the ACS has become more than $s-2$ for Figure
\ref{cap:manyhit}a, more than $s-10$ for Figure \ref{cap:manyhit}b,
or has reached $s$ for Figure \ref{cap:onebyx} do nodes of the dominant
ACS start getting removed in the graph updates.

\begin{figure}
\textbf{a)}

\includegraphics[  width=8cm,
  keepaspectratio]{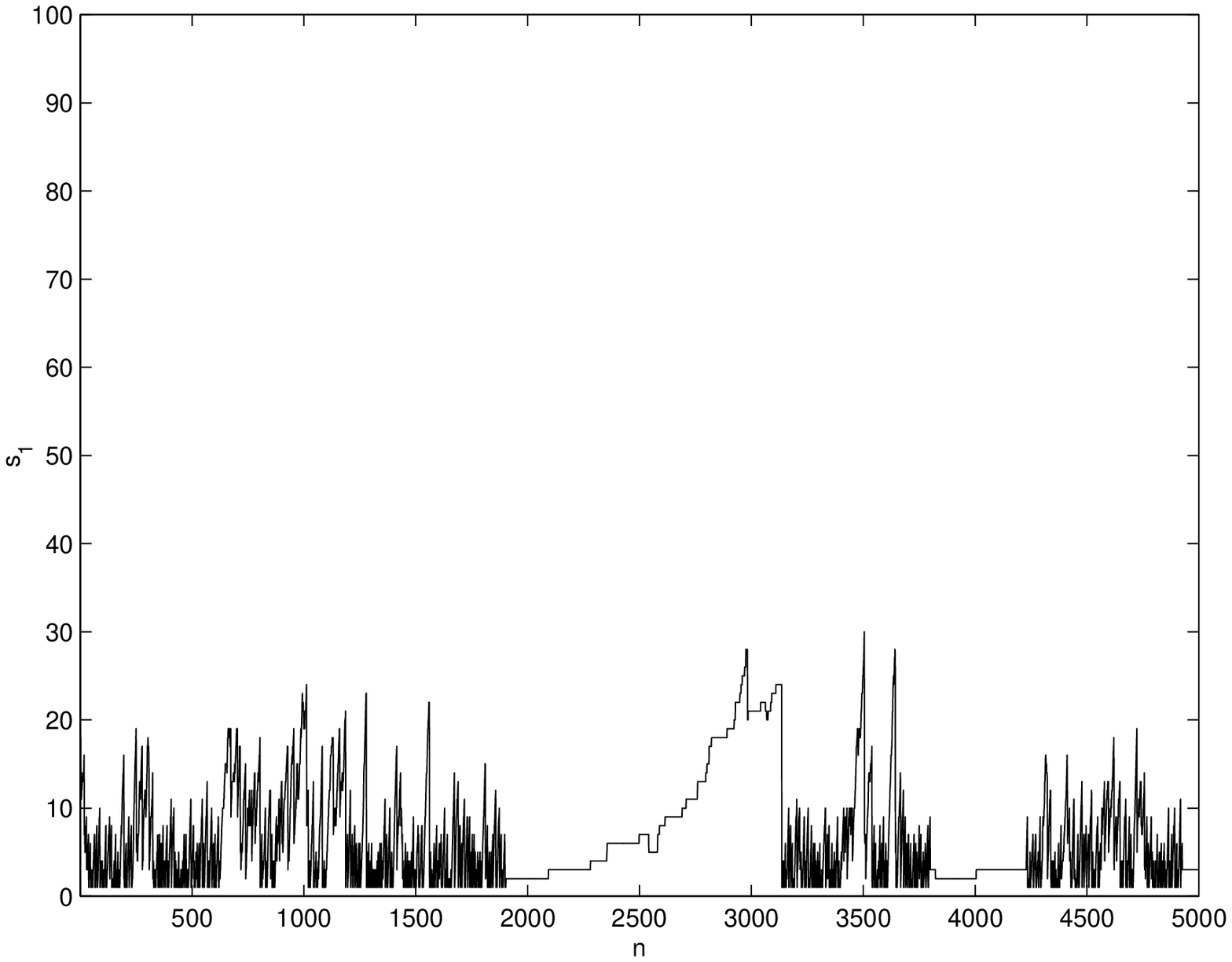}

\textbf{b)}

\includegraphics[  width=8cm,
  keepaspectratio]{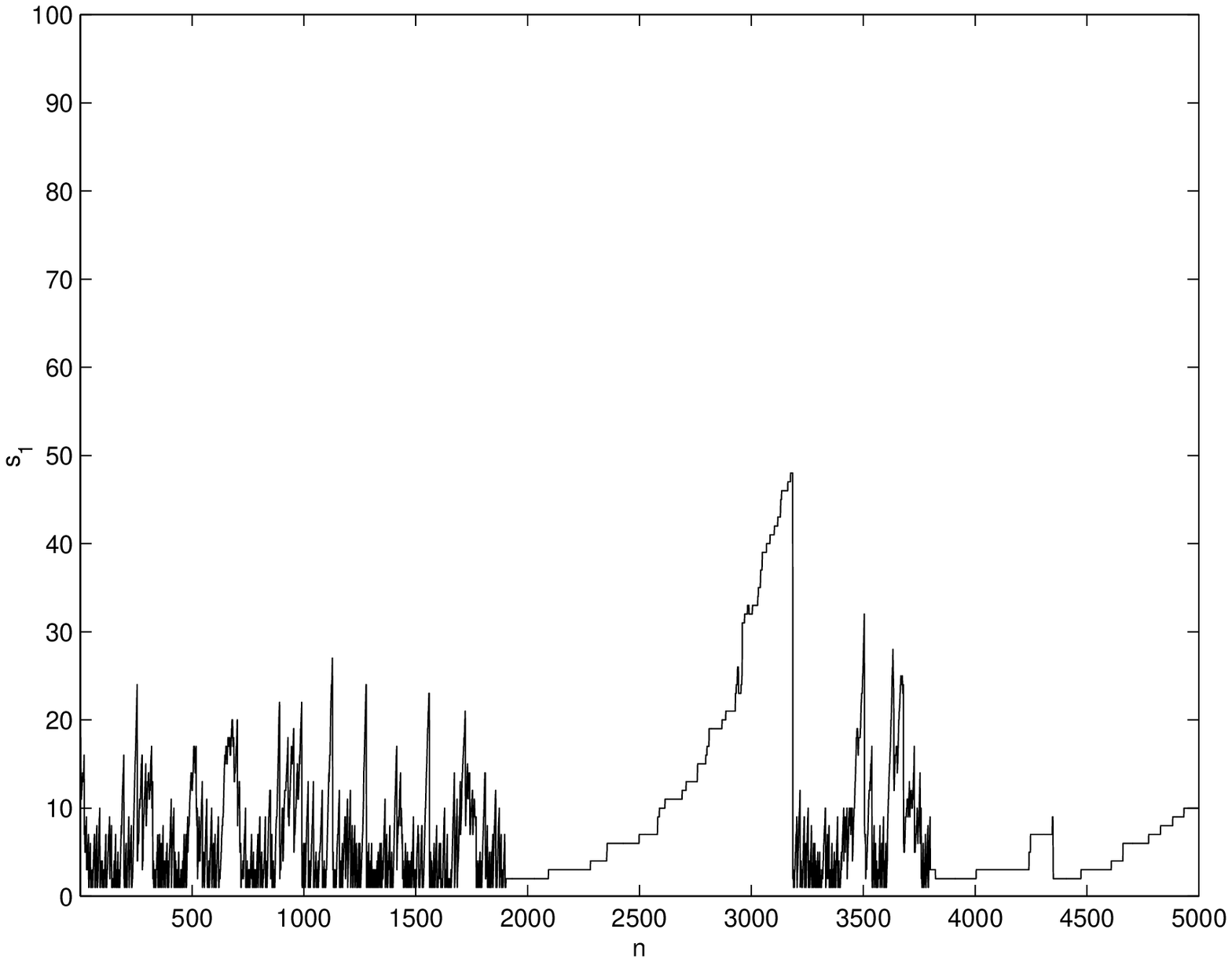}

\textbf{c)}

\includegraphics[  width=8cm,
  keepaspectratio]{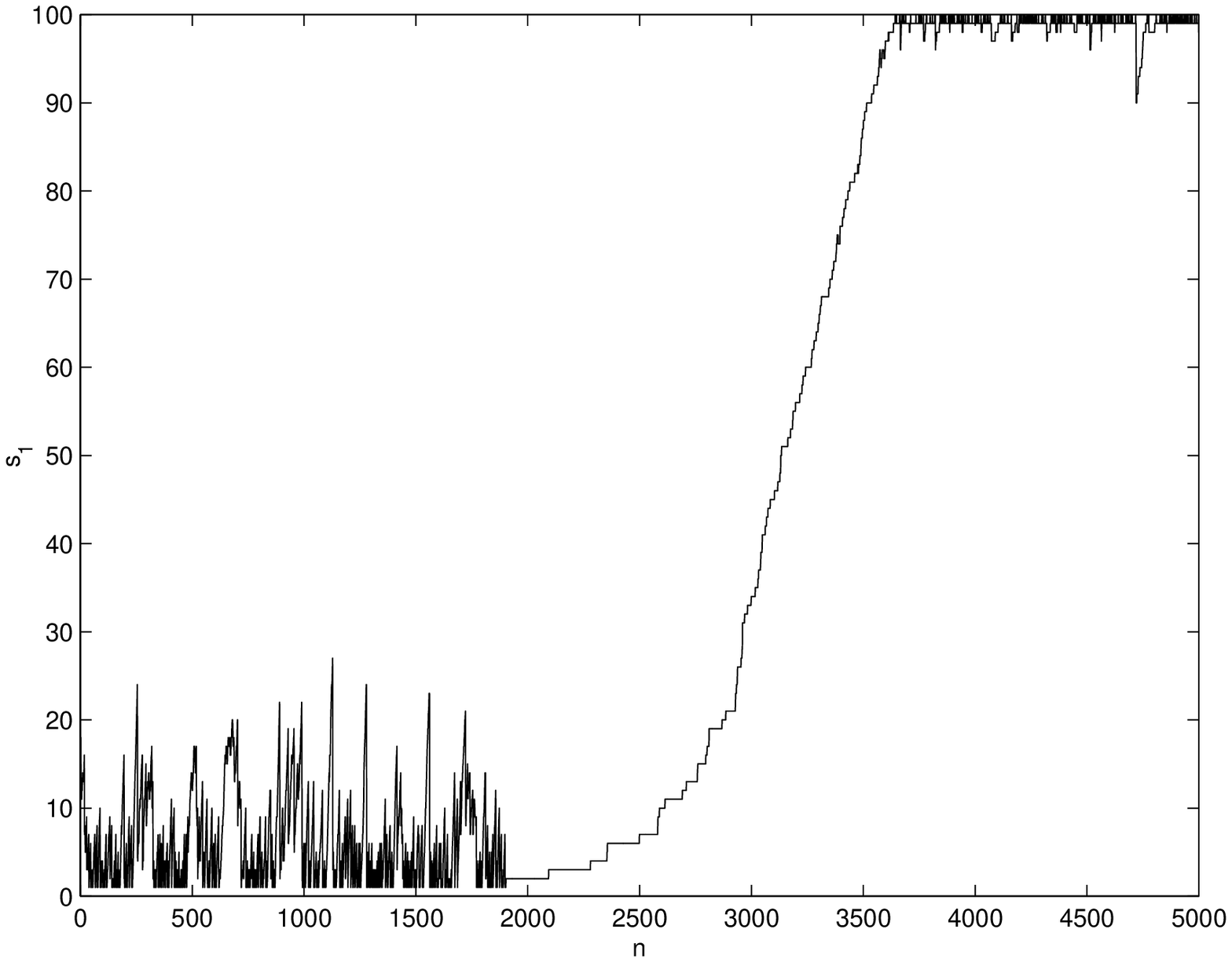}

\caption{Non-extremal run of type 3 with $s=100,p=0.0025$. \textbf{a.} $q=0.95$,
\textbf{b.} $q=0.99$, \textbf{c.} $q=0.99999$.\label{cap:qnonextr}}
\end{figure}

Case 3 is designed to interpolate between the extremal selection of
the original model ($q=1$) and the completely random run where there
is no selection ($q=0$, see black curve in Figure \ref{cap:links}).
Figure \ref{cap:qnonextr} displays what happens for different values
of $q$. The basic mechanism of an ACS forming and growing remains
but as $q$ is decreased the ACS becomes increasingly short lived
and susceptible to destruction. Suppose the ACS has $z$ indispensable
nodes, any of whose removal would destroy the ACS. Then the probability
that an indispensable node will be removed is $(1-q)z/s$. Hence,
there exists a timescale $\tau _{ne}=s/\{z(1-q)\}$ over which an
ACS will get destroyed due to non-extremality. On the other hand for
small $p$ ACSs get destroyed even with extremal dynamics over a timescale
$\tau _{s}$. $z$ does change with time but by choosing $q$ sufficiently
close to 1, one can make $\tau _{ne}\gg \tau _{s}$. In that case
the behaviour would be the same as in the case of extremal dynamics.
This is confirmed by the run in Figure \ref{cap:qnonextr}c, where
an ACS forms and grows until it spans the entire graph. After spanning
it does not get destroyed for a substantial period of time. 

\newpage
These non-extremal runs seem to indicate that the ACS mechanism is
stable to the introduction of non-extremality. In an ACS, organizationally
important nodes, such as the core nodes, tend to have high relative
populations and, therefore, these nodes are protected if there is
a selective bias that favours removing nodes with less relative population.
When the mechanism for choosing a node for removal is not based on
the relative population of nodes, as (partially) in case 3, this protection
is lost and the ACS is not as robust a structure.

\newpage
\section{Variable number of nodes}

Another idealization in the original model is that at each time step
exactly one node is removed and one node is introduced leaving the
total number of nodes constant. A variant in which the number of nodes
is not constant is the following: at each time step, all nodes with
$X_{i}$ below a fixed threshold, $x_{t}$, are removed (more than
one node can get deleted and the number of nodes is not constant).
The rest of the procedure remains the same; one new node is added,
the population dynamics is allowed to run its course with the new
graph, and this is iterated many times. Notice that it is also possible
for none of the nodes to be removed in a particular time step if they
all have relative populations above $x_{t}$. This is not possible
in the original model.

\begin{figure}
\textbf{a)}

\includegraphics[  width=12cm,
  keepaspectratio]{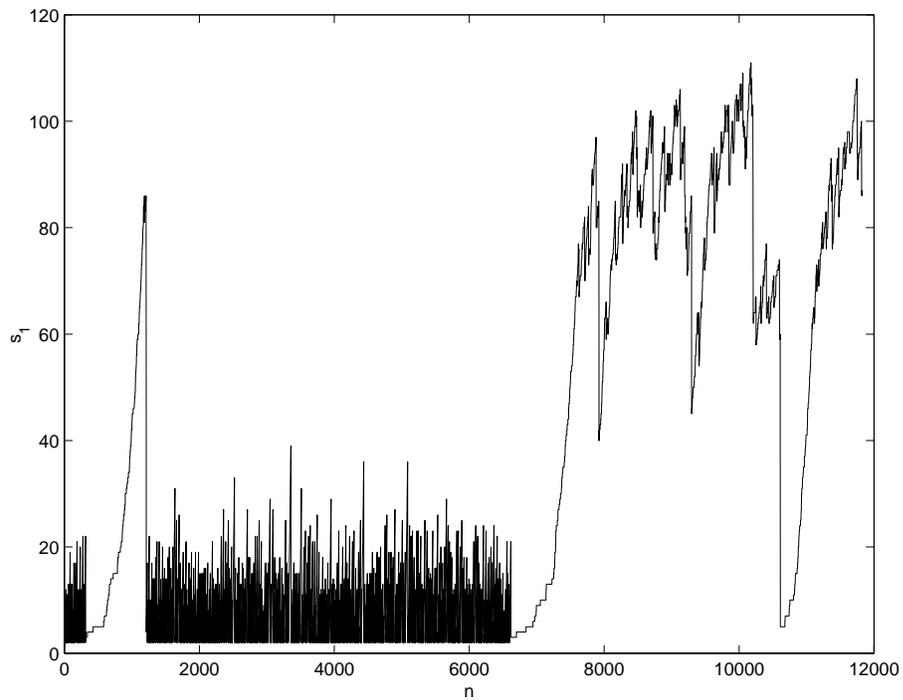}

\textbf{b)}

\includegraphics[  width=12cm,
  keepaspectratio]{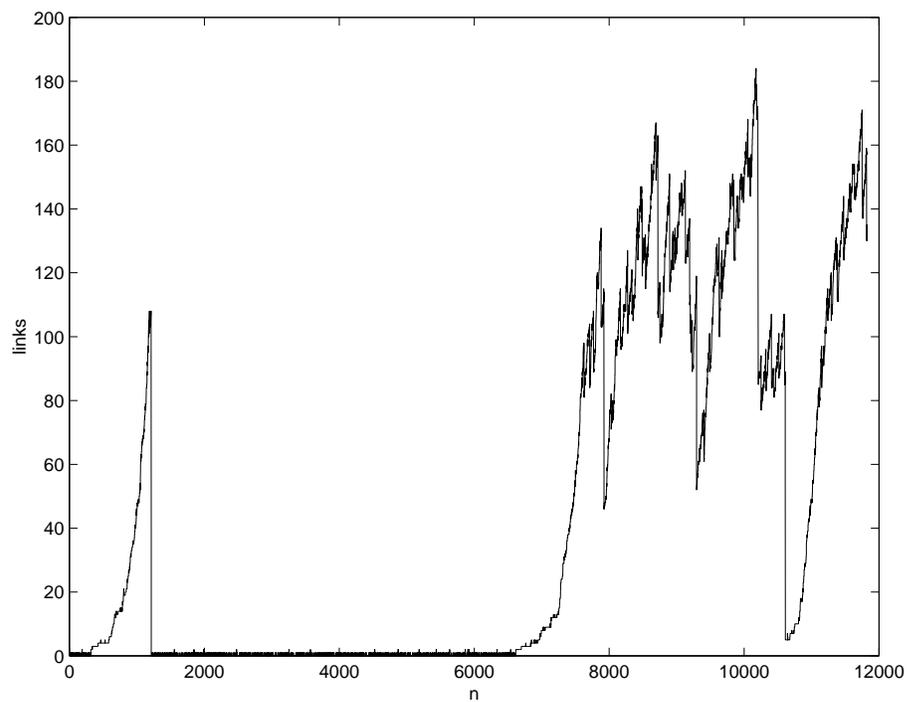}

\caption{\noindent \textit{\emph{A run with variable number of nodes. At each
time step all nodes with $X_{i}<0.005$ are removed. A single new
node is added and assigned links with existing nodes with a probability
$p=0.005$.}} \textbf{\textit{\emph{a.}}} \textit{\emph{the number
of nodes in the graph, $s_{1}$, as a function of time, $n$.}} \textbf{\textit{\emph{b.}}}
\textit{\emph{the total number of links in the graph as a function
of $n$.}}\textit{\label{cap:varsfig}}}
\end{figure}

This variant too shows the same qualitative behaviour (see Figure
\ref{cap:varsfig}). When there is no ACS in the graph, none of the
nodes remain above the threshold for long. The graph remains small,
with few nodes and links. The chance formation of an ACS changes things.
Again the number of links and the number of nodes are seen to rise
sharply as more and more nodes get attached to the dominant ACS. A
third region analogous to the organized phase can also be seen, once
the ACS grows large enough for some of its nodes to have $X_{i}$
values below the threshold, but the transition is not so clear because
there is no upper cutoff for the number of nodes in the graph.

Because only positive links are allowed all the analytical results,
including theorems 3.1 and 4.1, hold for this variant too. In the
initial graph, if there is no ACS, most of the nodes will have $X_{i}=0$.
All of these nodes will be removed from the graph leaving a graph
with very few nodes and only those links assigned to the new node.
Throughout the random phase the graph will consist of only very few
nodes. As soon as the first ACS forms the graph will consist only
of that ACS and the added node. The ACS grows whenever the new node
joins it. Here there is a slight difference from the original model.
The rare events where several nodes, including the new node, simultaneously
join the dominant ACS are not possible in this variant. It is also
not possible for an ACS that went extinct for some reason to be resurrected
(as happened in Figures \ref{cap:snapshots}j,u) -- in this variant
the extinct nodes would all be removed from the graph immediately.

Despite the absence of an explicit upper cutoff, the number of nodes
cannot grow indefinitely. Because $\sum _{i=1}^{s}x_{i}=1$ always,
the number of nodes cannot grow much larger than $1/x_{t}$. In the
run shown above the threshold is $x_{t}=0.005$ and the number of
nodes remains well below $200=1/0.005$.

\newpage
\section{Different population dynamics }

\textbf{Replicator equation.}

A variant where the qualitative behaviour does change drastically
is one where equation (\ref{xdot}) is replaced by the replicator
equation \citep{HSig}:\begin{equation}
\dot{x}_{i}=x_{i}\left[\sum _{j=1}^{s}c_{ij}x_{j}-\sum _{k,j=1}^{s}x_{k}c_{kj}x_{j}\right].\label{eq:replicator}\end{equation}

The name `replicator' comes from the fact that the equation models
self-replicating species, for example in model ecosystems \citep{HSig,DMc}.
Because $\dot{x_{i}}$ is proportional to $x_{i}$, the relative population
of species $i$ can only grow if it is non-zero in the first place;
once $x_{i}=0$ it will always stay zero, as expected for self-replicators.
The replicator equation can have a variety of attractors depending
on the coupling constants $c_{ij}$ and the initial conditions --
fixed points, limit cycles, heteroclinic cycles and chaotic attractors
have been observed \citep{HSig}. The replicator equation has also
been used in the hypercycle model of the origin of life \citep{ES}.

\begin{figure}
\includegraphics[  width=14cm,
  keepaspectratio]{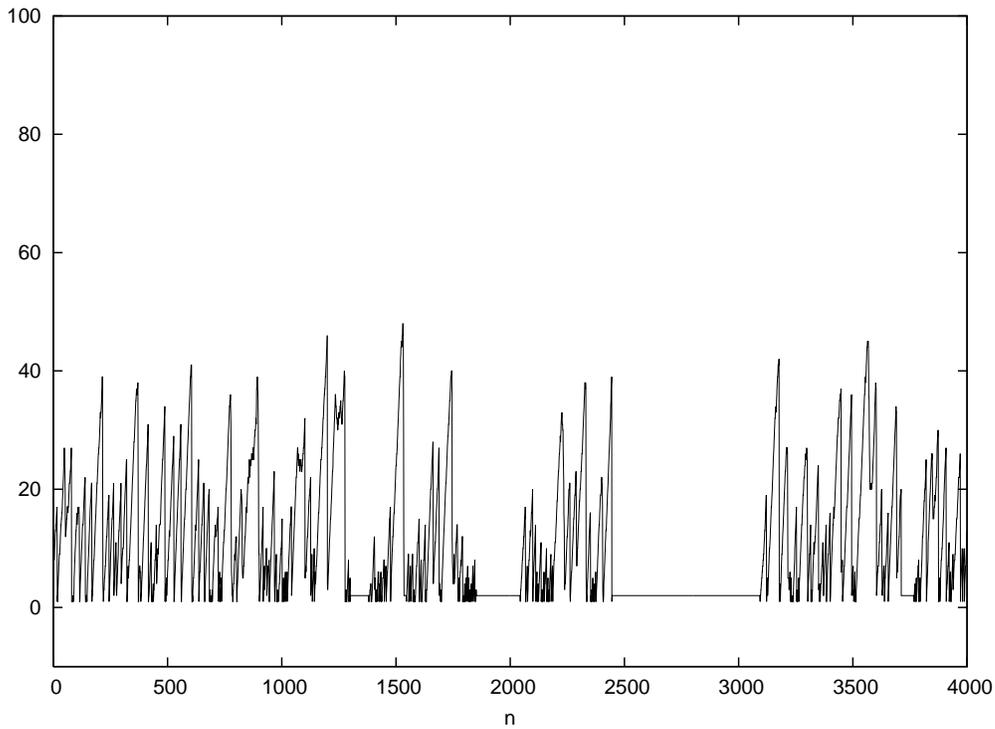}

\caption{Number of nodes with $X_{i}>0$, $s_{1}$, versus time, $n$, for
a run with the replicator equation used for the population dynamics.
For this run, $s=100,p=0.0025$.\label{cap:prd}}
\end{figure}

The other rules of the model are unchanged, making it an evolving
version of the hypercycle model. Figure \ref{cap:prd} shows a typical
run. When there is no ACS the number of nodes with non-zero $X_{i}$
fluctuates wildly, even reaching as high as 40 or more for short periods
of time. However, no graph structure is stable for very long and the
graph remains random. The times when there is an ACS are the periods
in Figure \ref{cap:prd} when $s_{1}$ remains at some small constant
value for a large number of graph updates. Yet, nothing resembling
a growth phase is seen and the system eventually relapses into the
random phase. What is happening is that small (hyper)cycles do form
but quickly get destroyed when a graph update creates a parasite.
This observed behaviour is consistent with existing work on well-stirred
hypercycles and replicator networks \citep{NHB,HS,TY2}.

\begin{figure}
\begin{center}\includegraphics[  width=6cm,
  keepaspectratio]{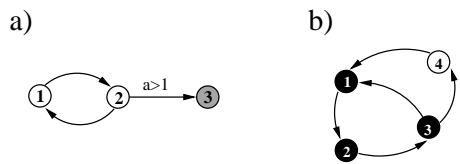}\end{center}

\caption{\textbf{a.} Example of the parasite instability of hypercycles. The
links between nodes 1 and 2 are of strength unity. When $a>1$ the
attractor of the replicator equation for this graph is $\mathbf{X}=(0,0,1)^{T}$
for generic initial conditions. \textbf{b.} Example of the short-circuit
instability of hypercycles. All link strengths are unity. The attractor
of the replicator equation for this graph is $\mathbf{X}=(1,1,1,0)^{T}/3$
for generic initial conditions.\label{cap:parasite}}
\end{figure}

\cite{NHB} have identified three instabilities to which hypercycles
are susceptible. Two are of relevance here: Figure \ref{cap:parasite}a
illustrates the parasite instability mentioned above. For the graph
shown here, if $a>1$, the attractor of equation (\ref{eq:replicator})
is $\mathbf{X}=(0,0,1)^{T}$ for generic initial conditions. Thus,
the nodes of the core cycle have the least $X_{i}$. Figure \ref{cap:parasite}b
illustrates the short-circuit instability. For generic initial conditions
the system always reaches an attractor where node 4, part of the longer
cycle, is zero -- in general, when long hypercycles are short-circuited
by a new link, only the smaller hypercycle survives. These two instabilities
are responsible for preventing a large, structured network from forming
when the replicator equation (\ref{eq:replicator}) is used in place
of equation (\ref{xdot}) for the population dynamics.

\newpage
\noindent \textbf{A generalization of the replicator equation.}

The replicator equation is a special case of a more general equation
in which the first term is quadratic in the relative population variables:

\begin{equation}
\dot{x}_{i}=\sum _{j,k=1}^{s}A_{ijk}x_{j}x_{k}-x_{i}\sum _{l,j,k=1}^{s}A_{ljk}x_{j}x_{k}.\label{eq:quadratic}\end{equation}

This equation preserves the normalization of $x_{i}$ because $\sum _{i=1}^{s}\dot{x}_{i}=0$.
The equation models a chemistry where two catalysts are (simultaneously)
required for the production of each molecular species. It reduces
to equation (\ref{eq:replicator}) in the special case where one of
the catalysts is the product itself: $A_{ijk}=\delta _{ij}c_{jk}$,
where $\delta _{ij}=1$ if $i=j$ and zero if $i\ne j$. 

In this case, the simplest ACS would consist of three species: species
1 and 2 catalyze species 3, species 2 and 3 catalyze species 1, and
species 3 and 1 catalyze species 2 . Consider the case where one periphery
node is added to this ACS -- let species 1 and 2 also catalyze the
production of species 4, and let species 4 not be a catalyst for any
of the other species. Numerical simulations show that, unlike the
replicator case, here both the core and the periphery nodes have non-zero
relative populations in the attractor for generic initial conditions.
The ratio of the relative populations of the periphery to the core
nodes depends on the link strengths but the core nodes 1, 2 and 3
remain with non-zero relative populations even if the link to the
periphery node is stronger than the links between core nodes. Therefore,
unlike the special case of the replicator equation, when equation
(\ref{eq:quadratic}) is used for the population dynamics it is possible
for small ACSs to grow by adding periphery nodes.

\chapter{\label{cha:Concluding-Remarks}Concluding Remarks}

In this thesis I have studied a model of an evolving catalytic network.
I have discussed some of the issues and questions of interest related
to the evolution of networks in section \ref{sec:Dynamics-of-networks}.
Here, I summarize the interesting phenomena observed in the model,
their implications for some of those questions, as well as the limitations
and possible extensions of the model.

\section{\label{sec:Interesting-features-of}Interesting features of the model}

\begin{enumerate}
\item \textbf{Rich dynamical behaviour: three phases and multiple timescales.}
For small $p$ the system always appears to be moving from one unstable,
or quasi-stable, state to another. It never appears to reach a stable
state. Three `phases' of behaviour are observed (see section \ref{sec:Results-of-graph}).
A `random phase' in which the graph remains sparse and consists only
of trees, chains and isolated nodes. The `growth phase' is triggered
by the chance formation of a small ACS which then grows by accreting
nodes to itself. Finally, the `organized phase' begins when the ACS
spans the entire graph. There are transitions from each phase to other
phases (except that there is never a transition from the growth phase
to the random phase). The appearance of different structures dynamically
generates multiple timescales: $\tau _{a}$ in the random phase, $\tau _{g}$
in the growth phase and $\tau _{s}$ in the organized phase (see sections
\ref{sec:The-growth-phase} and \ref{sec:Timescale-of-crashes}).
\item \textbf{Growth of structure, interdependence and cooperation.} Starting
from an initial random graph, the system spontaneously self-organizes
into a highly structured graph. This process is triggered by the chance,
but inevitable, formation of an ACS, which then grows by accreting
other nodes to it, until eventually the entire graph becomes an ACS
(section \ref{sec:The-growth-phase}). The resultant fully autocatalytic
graph is highly non-random, occupying an exponentially small (as a
function of $s$) volume of graph space (section \ref{sub:Probability-of-a}).
Nevertheless starting from a sparse graph the dynamics inevitably
leads the system into this exponentially small volume in a time that
grows only as $\ln s$. The growth of exponentially unlikely chemical
organizations on the prebiotic Earth in a short time is a puzzle that
has been discussed in section \ref{sec:The-origin-of}. The growth
of autocatalytic sets in this model might shed some light on this
puzzle -- the implications for the origin of life are discussed further
in section \ref{sec:Implications-for-the}. Along with the growth
of non-random structure, the interdependence of the species also grows
with the ACS. The fully autocatalytic sets that form have an interdependency
almost two orders of magnitude larger than that of random phase graphs
(sections \ref{sec:Results-of-graph} and \ref{sub:Degree-distribution}).
A variant of the model, where negative links are also allowed, exhibits
the emergence of cooperation. The formation and growth of ACSs in
this variant is accompanied by an increase in the number of cooperative,
positive links between species and a decrease in the number of destructive,
negative links. The ratio of cooperative to destructive links rises
from unity, in the random phase, to an order of magnitude larger in
the organized phase (section \ref{sec:Negative-links}). The formation
and growth of ACS might be one of the mechanisms responsible for the
spontaneous emergence of cooperation in a variety of contexts.
\item \textbf{Structure of fully autocatalytic graphs.} The graphs produced
at the end of the growth phase have several interesting structural
features. Firstly, they are fully autocatalytic. However, they are
not random fully autocatalytic graphs, as evidenced by their broad,
though not scale-free, out-degree distribution which is displayed
in Figures \ref{cap:facsdeg} and \ref{cap:facsdegout} (however,
whether runs from other parameter ranges produce scale-free graphs
or not remains an open question). An interesting feature is that the
in-degree distribution spans a much smaller range than the out-degree
distribution. While the fully autocatalytic graphs have a small clustering
coefficient (section \ref{sub:Clustering-coefficient}), they do consist
of an irreducible core that is more clustered, from which grow the
chains and trees that form the periphery. These observations suggest
that, in addition to the commonly studied properties of small-worldness
and scale-freeness, it might be interesting to examine real networks
for properties such as autocatalysis, a core-and-periphery structure,
a difference between the in- and out-degree distributions, etc.
\item \textbf{Large extinctions: destruction of ACSs.} ACSs not only form
and grow, but over long times also get destroyed. The destruction
of an ACS is usually sudden and accompanied by the extinction of a
large number of species, a `crash'. The asymmetry between rises and
drops of $s_{1}$ as well as fat tails in the distribution of crash
sizes (Figure \ref{cap:dropdist}) are seen in the fossil record \citep{NP}
and in financial markets \citep{Bouchad,JS}. An interesting feature
of the distribution of drops in $s_{1}$ is it's bimodal structure
for low $p$ -- the probability of large drops in $s_{1}$ is an order
of magnitude greater than intermediate size drops. A similar bimodality
is observed in distribution of earthquake sizes in an earthquake model
described by \citet{CL}, but the mechanism causing it is different.
\item \textbf{Causes of large extinctions: core-shifts, innovations and
keystone species.} A variety of mechanisms have been suggested as
causes of large extinction events in macroevolution \citep{MSmith,Glen}
and finance \citep{Bouchad,JS}. The largest of the extinction events
in this model are usually core-shifts -- a discontinuous change of
the core of the dominant ACS in a single graph update. It is interesting
that such core-shifts can be caused both by the removal of an important
`keystone' species as well as by graph structures created by the new
species, `innovations' (section \ref{sec:Classification-of-core-shifts}).
One can distinguish two processes involving innovations, both having
analogues in the real world. One is exemplified by the appearance
of the automobile that made the horse drawn carriage and its ancillary
industries obsolete. This is like the example of the core-transforming
innovation shown in Figure \ref{cap:6061,6062} where the graph update
produced an irreducible structure that was `stronger' than the existing
core. This structure became the new core, driving the old core and
nodes dependent on it to extinction. The subsequent development of
other industries dependent on the automobile mirrors the growth of
the ACS around the new core. The second process is exemplified by
the emergence of the body plans of several phyla that are dominant
today. It is believed that while these body plans originated in the
Cambrian era more than 520 million years ago \citep{VJE}, the organisms
with these body plans played no major role till about 250 million
years ago. They started flourishing only when the Permian extinction
depleted the other species that were dominant till that time \citep{Erwin}.
This is similar to the events shown in Figure \ref{cap:5041,5042}
where an earlier innovation had lain dormant for a while without disturbing
the existing core, but when the latter became sufficiently weak, took
over as the new core.
\item \textbf{Predicting crashes.} Certain graph theoretic markers could
possibly be used to predict whether a crash is imminent (section \ref{sec:Structure-of-the}).
The Perron-Frobenius eigenvalue, $\lambda _{1}$, is one such marker.
Most of the large extinctions are core-shifts, caused by the removal
of keystone species, takeover by core-transforming innovations, or
takeover by dormant innovations. The graph is most susceptible to
these core-shifts when its $\lambda _{1}=1$ or close to it. The dependency
distribution is another marker: it is bimodal for graphs preceding
a crash. The in and out-degree distributions also appear to fall off
faster, as a function of degree, for graphs just before a crash, than
for the fully autocatalytic graphs produced in the same runs. In combination,
these different markers might indicate whether a given graph has an
enhanced probability of suffering a crash. 
\item \textbf{Analytical tractability of the model.} The analytical tractability
of the model is a result of the linearity of the $y_{i}$ dynamics.
Equation (\ref{xdot}) is nonlinear but, as it originates via a nonlinear
change of variables from a linear equation, equation (\ref{ydot}),
its attractors can be easily analyzed in terms of the underlying linear
system. The attractors are always fixed points and are just the Perron-Frobenius
eigenvectors of the adjacency matrix of the graph. Theorems 3.1 (the
eigenvector profile theorem) and 4.1 (the attractor profile theorem)
form the backbone of the analytic results about the population dynamics.
Note that while the $y_{i}$ dynamics in the present model is essentially
linear as long as the graph is fixed, it is highly nonlinear over
long timescales. Because of the coupling of the population and graph
dynamics, over long timescales the `coupling constants' $c_{ij}$
in equation (\ref{ydot}) are not constant but implicitly depend upon
the $y_{i}$, thus making the evolution highly nonlinear. The simplifying
device of widely separated timescales for the graph dynamics and the
population dynamics (the population variables reach their attractor
before the graph is modified) results in a piecewise linear $y_{i}$
dynamics that is easier to analyze. The analytic results about the
population dynamics have been used in the analysis of the graph dynamics:
results such as proposition 6.1, as well as average timescales of
appearance and growth of an ACS use insight about the attractors of
the population dynamics to predict aspects of the graph dynamics.
\item \textbf{Predictability and historicity in the network evolution.}
There is an interesting mixture of predictability and historicity
in the model dynamics. The detailed structure of the dominant ACS
is dependent on the chance events that happen throughout the run.
Many complicated processes are possible, such as formation of new
disconnected ACSs or the reinforcement of an old overshadowed ACS
before it was completely broken up (Figures \ref{cap:snapshots}j,u).
Thus, the actual growth of the ACS is usually very complicated with
the dominant ACS often undergoing drastic changes in structure caused
purely by chance events. However, despite the historical particularities
of a given run several aspects are predictable. The non-decreasing
nature of $\lambda _{1}$ in the growth phase is a rule that holds
for every run with any values of the parameters $s$ and $p$ (see
proposition 6.1, section \ref{sec:The-growth-phase}). Other results,
such as the timescales of appearance and growth of the ACS, are predictable
for ensembles of runs with the same value of $s$ and $p$ (section
\ref{sec:The-growth-phase}). Also, as discussed above, the transitions
from the organized to the growth or random phases that involve the
sudden extinction of a large number of species, might be predictable. 
\item \textbf{Correlation between graph structure of perturbations and their
impact on the evolution of the network.} The network evolves by periodic
perturbations, that involve the removal and addition of nodes, and
there is a correlation between the graph theoretic nature of the perturbation
and its short and long term impact. The perturbations can be broadly
placed in two classes based on their effect on $s_{1}$:\\
 (i) `Constructive perturbations': these include the birth of a new
organization (an innovation of type 6), the attachment of a new node
to the core (an innovation of type 4) and an attachment of a new node
to the periphery of the dominant ACS (an innovation of type 2).\\
 (ii) `Destructive perturbations': these include complete crashes
and takeovers by dormant innovations (both caused by the loss of a
keystone node), and takeovers by core-transforming innovations (innovations
of type 5). The word `destructive' is used only in the sense that
several species become extinct on a short timescale (a single graph
update) after such a perturbation. In fact, over a longer timescale
(ranging from a few to several hundred graph updates), the `destructive'
takeovers by innovations usually trigger a new round of `constructive'
events like incremental innovations (type 2) and core enhancing innovations
(type 4).\\
\\
 The maximum upheaval is caused by those perturbations that introduce
new irreducible structures in the graph (innovations of type 4, 5
and 6) or those that destroy the existing irreducible structure. For
example, the creation of the first ACS at $n=2854$ triggered the
growth phase in the run of Figure \ref{cap:s1lambda}b. Other examples
of large upheavals are core-shifts due to takeover by a core-transforming
innovation at $n=6061$, takeover by a dormant innovation at $n=5041$,
and a complete crash at $n=8233$. 
\item \textbf{Context dependence of the effective dynamics of the network
evolution.} It is characteristic of evolutionary systems, such as
this model, that as different (graph) structures spontaneously appear,
the nature of selective pressure on existing structures and hence
the effective dynamics changes. In the present model the effective
dynamics appears to be different in each of the phases even though
the microscopic rules remain the same. Thus, in the random phase,
because there is no ACS, the graph updates are almost random. In the
growth phase there is competition between ACS and non-ACS structures.
The dynamics essentially involves the growth of the dominant ACS by
the accretion of nodes outside it. Once the entire graph becomes an
ACS and the organized phase begins, the effective dynamics once again
changes. Now there are no nodes outside the dominant ACS, yet some
node has to be removed at each graph update. The competition now shifts
to within the ACS -- between core and periphery nodes. Typically the
core nodes are protected by a higher relative population, therefore
most of the graph updates involve periphery nodes. As the periphery
realigns itself around stronger core nodes it can happen that a weak
core node may be removed at some graph update. Over a long period
of time such events may cause the core to become fragile and susceptible
to core-shifts. Thus ACSs, which were robust in one phase, become
fragile in another. Robust and fragile structures are also found in
highly designed systems \citep{CD}. A different system, in which
the selective pressures and effective dynamics change as different
structures arise, is discussed by \cite{CRA}.
\item \textbf{Robustness of the ACS growth mechanism to changes in model
rules.} The basic model has a number of simplifying features that
depart from realism but enhance analytical tractability. These are
listed in chapter \ref{cha:Variants-of-model} along with variants
of the model that relax those simplifications. As shown in that chapter,
in most cases relaxing the various idealizations does not change the
qualitative behaviour. A variant where the qualitative behaviour does
change drastically is one where equation (\ref{xdot}) is replaced
by the replicator equation, making it an evolving version of the hypercycle
model \citep{ES}. In this case small (hyper)cycles do form but quickly
get destroyed when a graph update creates a parasite. Consequently,
large complex networks cannot form. However, the replicator equation
is a special case of a more general equation. In the more general
case, unlike the hypercycle model, the ACSs that arise do not get
destroyed by parasites (nodes belonging to the periphery). This robustness,
one expects, would allow them to evolve into complex networks, as
in the original model. 
\end{enumerate}

\section{\label{sec:Implications-for-the}Implications for the origin of life}

As discussed in chapter \ref{cha:Introduction-and-Summary}, one of
the possible scenarios for which the model might be relevant is for
the evolution of a chemical network in a pool on the prebiotic Earth.
From the point of view of the origin of life problem the main conclusions
are: 

\begin{enumerate}
\item The model shows the emergence of a chemical organization where none
exists. A small ACS emerges spontaneously by random processes and
then grows until the entire system is autocatalytic. 
\item The graph structures that emerge (ACSs) are collective self-replicators
even though none of their components (individual nodes) are self-replicators.
\end{enumerate}
As mentioned in section \ref{sec:The-origin-of}, one of the puzzles
of the origin of life is the emergence of structured non-random chemical
organizations in a relatively short time. A highly structured organization,
whose timescale of forming by pure chance is exponentially large (as
a function of the size of the system), forms in this model on a very
short timescale that grows only logarithmically with the size of the
system. Under the assumption that the present model captures what
happens in a prebiotic pool, the timescale for a fully autocatalytic
chemical organization to grow in the pool is $\tau _{g}=1/p$, in
units of the graph update time step. This time unit corresponds to
the periodicity of the influx of new molecular species, hence it ranges
from a day (for tides) to a year (for floods). Further, in the present
model the `catalytic probability' $p$ is the probability that a random
small peptide or RNA molecule will catalyze the production of another,
and this has been estimated by \cite{Kauffman3} as being in the range
$10^{-5}$ to $10^{-10}$. With this range of values for $p$, the
timescale, $\tau _{g}$, for a fully autocatalytic chemical organization
to grow in the pool lies in the range $10^{3}$ to $10^{10}$ years.
The time taken for the first cells to arise on Earth, a few hundred
million years after the oceans condensed \citep{Joyce,Schopf}, lies
within this range. However, $10^{3}$ to $10^{10}$ years is a very
wide interval. A tighter prediction, and therefore a better test of
the model, would require a more accurate value for the catalytic probability
$p$ for peptides and catalytic RNA, and a chemically more realistic
model.

These speculations might be complementary to some other approaches
to the origin of life. Autocatalytic organizations of polypeptides
could enter into symbiosis with the autocatalytic citric acid cycle
proposed by \citet{Wacht} and \citet{MKYC}. The latter would help
produce, among other things, amino acid monomers needed by the former;
the former would provide catalysts for the latter. It is conceivable
that lipid membranes (that have been argued by \citealp{SBL,SBDL},
to have their own catalytic dynamics) could form and surround autocatalytic
sets, of the kind discussed here, in an enclosure. These `cells' may
contain different ACSs, or different parts of the same ACS, thereby
endowing them with different fitness levels; such a population could
evolve by natural selection. It is also conceivable that such autocatalytic
sets of polypeptides formed an enabling environment for the formation
and maintenance of self-replicating molecules such as those needed
for an RNA world.

\section{\label{sec:Limitations-of-the}Limitations of the model as a description
of a chemical network}

I have interpreted the model as a chemical system, with nodes representing
molecular species and the links representing catalytic interactions.
This interpretation raises some questions. For example, what does
it mean to remove a node from the graph? This effectively means that
that molecular species cannot be produced again for a finite time
until, at some later graph update, a new node comes in with exactly
the same links as the one removed. If the species to be removed has
a catalyst, then this is inconsistent with the assumption that there
is a constant supply of all required reactants. In the random and
growth phases this does not happen -- in all cases the removed node
either has no incoming links, or the $X_{i}$ values of all its catalysts
are zero. In the organized phase too, for runs with small $p$ ($ps<1$),
a large fraction of graphs have $s_{1}<s$ in which case the node
that is removed (one of the nodes outside the dominant ACS) either
has no incoming links or, if it does, all its catalysts have $X_{i}=0$
(otherwise the node would be part of the dominant ACS). However, when
$s_{1}=s$, i.e., all nodes have non-zero $X_{i}$ values, then the
removed node always has a catalyst with a non-zero $X_{i}$ value.
Most of the time such removals do not significantly disturb the graph
structure. However, in some cases they have drastic effects -- for
instance, complete crashes (section \ref{sub:Complete-crashes}).

Another question concerns the structure of the molecules in the system.
In this model, the identity of the molecular species is defined solely
by their interactions and their structure is not specified at all.
In reality, it is the structure of molecules that determines their
chemical properties. The absence of structure in the model precludes
its use in addressing a different problem of the origin of life: How
did long proteins with strong and highly specific catalytic properties
form from the short, weakly catalytic polypeptides existing initially? 

Other weaknesses of the interpretation of the model as a chemical
system lie in the assumptions behind the functional form of $f_{i}$,
in particular, the dependence of $\dot{x_{i}}$ only on the relative
populations of catalysts of $i$ and not on the reactants. Such an
assumption implies that the reactants are all present and have unchanging
concentrations, i.e., they are buffered and there is effectively a
constant supply of reactants. This can lead to the uncontrolled growth
in the population of some, or all, molecular species. In reality,
the reactants will eventually get consumed thereby truncating the
growth of population. Further, the dependence of the growth rate on
reactants would make the dynamics nonlinear and hence oscillations
and chaotic dynamics might occur. The form of $f_{i}$ precludes the
possibility of this kind of dynamics. The choice of $f_{i}$ also
does not allow for the possibility of other types of chemical reactions,
such as conversion of one substrate to another, or cleavage and ligation
of molecules, which might be important for the prebiotic pool scenario. 

\newpage
Another limitation is the absence of spatial degrees of freedom. Thus,
the model applies only to a well-stirred chemical reactor and cannot
address the emergence of spatial structure and its effect on the dynamics
of the system. As prebiotic pools need not be well-stirred it would
be more appropriate to use reaction-diffusion equations to model the
$x_{i}$ dynamics \citep{Bhalla2}. 

These are the most serious limitations. Apart from them, the model
has several simplifications that aid in the analysis of its dynamical
behaviour. These are listed below:

\begin{enumerate}
\item All catalysts have the same catalytic efficiency.
\item Inhibitors -- negative links -- are not present in the model.
\item Self-replicators are not allowed.
\item The selection is extremal; the node with the \emph{least} $X_{i}$
is removed at each graph update.
\item The total number of nodes is fixed; the rate of node removals and
node additions is exactly the same.
\end{enumerate}
Variants of the model, in which the rules are modified to relax these
simplifications, were described in chapter \ref{cha:Variants-of-model}
and show the same qualitative behaviour as the original model. Therefore,
these are not serious limitations.

\section{Possible extensions of the model}

The above limitations of the model suggest possible directions in
which it would be worthwhile to extend the model. Firstly, the structure
of the molecular species could be explicitly introduced into the model.
This would require specifying some rules to determine the catalytic
properties of a molecule from its structure. The structure of the
molecular species should determine the reactions in which the molecule
will participate. Then there would be no need to remove a node from
the graph. In such a model one could explicitly study the evolution
of certain structural features of molecules in the system -- for instance
their length.

Another direction in which to extend the model as a description of
a chemical system would be to introduce the reactants explicitly,
and also include other kinds of reactions, e.g., reactions in which
one molecular species is converted to another type of molecular species.
In the reductive citric acid cycle discussed above, a crucial reaction
involves the splitting of citrate into two smaller molecules, which
then undergo various reactions until each is re-converted into citrate.
Therefore, including these kind of reactions is necessary to study
the emergence of this kind of autocatalytic set in a chemical network.
These changes fit within the general framework described in section
\ref{sec:Framework-of-a}.

\newpage
Going beyond that framework, spatial degrees of freedom could be added
to the model by using reaction-diffusion equations for the $x_{i}$
dynamics. However, it is computationally difficult to solve such equations
for complicated geometries \citep{Numerical}. A compromise might
be achieved by simulating a finite number of compartments, within
each of which the system is well mixed, and allowing some exchange
or diffusion across compartment boundaries \citep{Bhalla2}. Then
different ACSs could exist in different compartments and interact
with each other. This is likely to be computationally easier to simulate
and at the same time allows some spatial differentiation.

\section{Modeling other systems within the same framework}

The general framework described in section \ref{sec:Framework-of-a}
can be used to model systems other than chemical networks. For example,
an ecosystem could be modeled. In that case the nodes would represent
biological species while the links would represent interactions such
as prey-predator, competitive or symbiotic interactions between species.
Here removing a node that has become extinct is not a problem because
the species are self-replicators (once a species becomes extinct its
chance of being recreated can be neglected). However, the self-replicating
character of the species in this interpretation must also be reflected
in the population dynamics. The Lotka-Volterra equation, replicator
equation or other more complicated dynamical equations may be tried
\citep{DMc}. In modeling an ecosystem one might also wish to modify
the way the new node is assigned links. While a random assignment
might be interpreted as the invasion of the ecosystem by an entirely
alien species, one might also want to include the possibility of an
existing species mutating.

Another system that might be modeled in this manner is a genetic regulatory
network. The nodes would represent the genes. The $x_{i}$ would represent
the level of expression of a gene, and the links would represent regulatory
interactions between the protein product of one gene and another gene.
Here again one might wish to allow a new node (a new gene) to be a
mutated copy of an existing gene. One might also wish to include other
processes that change the graph such as gene duplications.

A third way to modify the graph dynamics would be to allow the rewiring
of existing links. This might be most relevant when using the model
to describe an economic system, with the nodes representing economic
`agents' such as buyers and sellers, or companies, or nations. 

\appendix

\chapter{Proofs of Propositions}

\begin{description}
\item [Proposition~2.1:]\textit{\emph{If a graph, $C$,}}\emph{}\\
 \emph{}(i) \emph{}\textit{\emph{has no closed walk then $\lambda _{1}(C)=0$}}\emph{,}\\
 \emph{}(ii) \emph{}\textit{\emph{has a closed walk then $\lambda _{1}(C)\geq 1$.}}
\end{description}
(i) If a graph has no closed walk then all walks are of finite length.
Let the length of the longest walk of the graph be denoted $r$. If
$C$ is the adjacency matrix of a graph then $(C^{k})_{ij}$ equals
the number of distinct walks of length $k$ from node $j$ to node
$i$. Clearly $C^{m}=0$ for $m>r$. Therefore all eigenvalues of
$C^{m}$ are zero. If $\lambda _{i}$ are the eigenvalues of $C$
then $\lambda _{i}^{k}$ are the eigenvalues of $C^{k}$. Hence, all
eigenvalues of $C$ are zero, which implies $\lambda _{1}=0$. This
proof was supplied by V. S. Borkar.\\
(ii) If a graph has a closed walk then there is some node $i$ that
has at least one closed walk to itself, i.e. $(C^{k})_{ii}\geq 1$,
for infinitely many values of $k$. Because the trace of a matrix
equals the sum of the eigenvalues of the matrix, $\sum _{i=1}^{s}(C^{k})_{ii}=\sum _{i=1}^{s}\lambda _{i}^{k}$,
where $\lambda _{i}$ are the eigenvalues of $C$. Thus, $\sum _{i=1}^{s}\lambda _{i}^{k}\ge 1$,
for infinitely many values of $k$. This is only possible if one of
the eigenvalues $\lambda _{i}$ has a modulus $\geq 1$. By the Perron-Frobenius
theorem, $\lambda _{1}$ is the eigenvalue with the largest modulus,
hence $\lambda _{1}\geq 1$. This proof was supplied by R. Hariharan.
$\hfill \square $\\

\begin{description}
\item [Proposition~2.2:]If all the basic subgraphs of a graph $C$ are
cycles then $\lambda _{1}(C)=1$, and vice versa.
\end{description}
For any cycle of $k$ nodes whose adjacency matrix is $A$, it follows
that $A^{k}$ is the identity matrix $(A^{k})_{ij}=\delta _{ij}$
(for each $i$ there is a closed walk of length $k$ from node $i$
to itself, and there are no other walks of length $k$). Therefore,
each eigenvalue, $\lambda $, of $A$ satisfies $\lambda ^{k}=1$,
i.e., $|\lambda |=1$. The Perron-Frobenius eigenvalue of $A$ must
be real, therefore, $\lambda _{1}(A)=1$. Thus, if all basic subgraphs
of a graph $C$ are cycles then $\lambda _{1}(C)=1$.

If $\lambda _{1}(C)=1$ then the basic subgraphs must be irreducible,
they cannot be single nodes. Assume that one of the basic subgraphs
is not a cycle. Let its adjacency matrix be $A$. Because $A$ is
irreducible, every row has at least one non-zero entry. Construct
$A'$ by removing, from each row of $A$, all non-zero entries except
one that can be chosen arbitrarily. Thus $A'$ has exactly one non-zero
entry in each row. Clearly the column vector $\mathbf{x}=(1,1,\ldots ,1)^{T}$
is a right eigenvector of $A'$ with eigenvalue 1 and hence $\lambda _{1}(A')\ge 1$.
Also $A>A'$, therefore, by the Perron-Frobenius theorem for irreducible
matrices, $\lambda _{1}(A)>\lambda _{1}(A')\Rightarrow \lambda _{1}(A)>1$.
But this contradicts the earlier assumption that $\lambda _{1}(A)=1$.
Hence, if $\lambda _{1}(C)=1$, all the basic subgraphs must be cycles.$\hfill \square $\\

\begin{figure}
\begin{center}\includegraphics{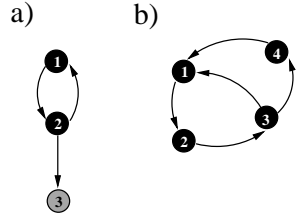}\end{center}

\caption{\textbf{a.} A graph that is an ACS, but not irreducible (e.g. there
is no path from node 3 to node 1). \textbf{b.} A graph that is irreducible
but is not a cycle. Removing the link from node 3 to node 1 would
convert it into a cycle.\label{cap:counteregs}}\lyxline{\normalsize}

\end{figure}

\begin{description}
\item [Proposition~3.1:](i) All cycles are irreducible graphs and all
irreducible graphs are ACSs.\\
(ii) Not all ACSs are irreducible graphs and not all irreducible graphs
are cycles.
\end{description}
(i) Clearly in a cycle, there is a path from every node $i$ to any
other node $j$ hence all cycles are irreducible. In every irreducible
graph each node must have an incoming link from one of the other nodes.
Hence all irreducible graphs are ACSs.\\
(ii) This is proved by providing two counterexamples. Figure \ref{cap:counteregs}a
shows an ACS that is not an irreducible graph, and Figure \ref{cap:counteregs}b
shows an irreducible graph that is not a cycle. $\hfill \square $\\

\begin{description}
\item [Proposition~3.2:](i) \textit{\emph{An ACS must contain a}} closed
walk\emph{.} Consequently,\\
 (ii) \textit{\emph{If a graph $C$ has no ACS then}} \textit{$\lambda _{1}(C)=0$}.\\
 (iii) \textit{\emph{If a graph $C$ has an ACS then}} \textit{$\lambda _{1}(C)\geq 1$}.
\end{description}
(i) Let $A$ be the adjacency matrix of a graph that is an ACS. Then
by definition, every row of $A$ has at least one non-zero entry.
Construct $A'$ by removing, from each row of $A$, all non-zero entries
except one that can be chosen arbitrarily. Thus $A'$ has exactly
one non-zero entry in each row. Clearly the column vector ${\textbf {x}}=(1,1,\ldots ,1)^{T}$
is an eigenvector of $A'$ with eigenvalue 1 and hence $\lambda _{1}(A')\ge 1$.
Proposition 2.1 therefore implies that $A'$ contains a closed walk.
Because the construction of $A'$ from $A$ involved only removal
of some links, it follows that $A$ must also contain a closed walk.
(ii) and (iii) follow from part (i) and propositions 2.1(i) and (ii),
respectively. $\hfill \square $

\begin{description}
\newpage
\item [Proposition~3.3:]\textit{\emph{If $\lambda _{1}(C)\geq 1$, then
the subgraph of any PFE of $C$ is an ACS}}\emph{. }
\end{description}
Let ${\textbf {x}}$ be a PFE of a graph. Renumber the nodes of the
graph so that $x_{i}>0$ only for $i=1,\ldots ,k$. Let $C$ be the
adjacency matrix of this graph. As ${\textbf {x}}$ is an eigenvector
of the matrix $C$ \[
\sum _{j=1}^{s}c_{ij}x_{j}=\lambda _{1}x_{i},\]
\[
\Rightarrow \sum _{j=1}^{k}c_{ij}x_{j}=\lambda _{1}x_{i}.\]
 Because $x_{i}>0$ for $i=1,\ldots ,k$ and $\lambda _{1}\ne 0$,
it follows that $\sum _{j=1}^{k}c_{ij}x_{j}>0$ for each $i\in \{1,\ldots ,k\}$.
Therefore, for each $i\in \{1,\ldots ,k\}$ there exists a $j$ such
that $c_{ij}>0$. Hence the $k\times k$ submatrix $C'\equiv (c_{ij})$,
$i,j=1,\ldots ,k$, has at least one non-zero entry in each row. Thus
each node of the subgraph corresponding to this submatrix has an incoming
link from one of the other nodes in the subgraph. Hence the subgraph
is an ACS. $\hfill \square $\\

\begin{description}
\item [Lemma~A:]\textit{\emph{Let $A$ be a submatrix of a non-negative
matrix $B$. Denote the Perron-Frobenius eigenvalue of $A$ by $\lambda _{1}(A)$
and that of $B$ by $\lambda _{1}(B)$. Then $\lambda _{1}(A)\leq \lambda _{1}(B)$}}\emph{.} 
\end{description}
Renumber the nodes so that $B$ can be written as: \[
B=\left(\begin{array}{ccc|ccc}
  &  &  &  &  & \\
  & A &  &  & C_{1} & \\
  &  &  &  &  & \\
\hline  &  &  &  &  & \\
  & C_{2} &  &  & C_{3} & \end{array}\right).\]

Consider the matrix \[
B'=\left(\begin{array}{ccc|ccc}
  &  &  &  &  & \\
  & A &  &  & 0 & \\
  &  &  &  &  & \\
\hline  &  &  &  &  & \\
  & 0 &  &  & 0 & \end{array}\right),\]
 and denote its Perron-Frobenius eigenvalue by $\lambda _{1}(B')$.

Because $B'\leq B$ we have $\lambda _{1}(B')\leq \lambda _{1}(B)$
from the Perron-Frobenius theorem.\\
 Also $\lambda _{1}(B')=\mathrm{max}\{0,\lambda _{1}(A)\}$.\\
 Hence, $\lambda _{1}(B')=\lambda _{1}(A)$.\\
 Therefore $\lambda _{1}(A)\leq \lambda _{1}(B)$. $\hfill \square $\\

\begin{description}
\item [Proposition~3.4:]Let ${\textbf {x}}$ be a PFE of a graph $C$,
and let $C'$ denote the adjacency matrix of the subgraph of ${\textbf {x}}$.
Let $\lambda _{1}(C')$ denote the Perron-Frobenius eigenvalue of
$C'$. Then $\lambda _{1}(C')=\lambda _{1}(C)$ and $C'$ must contain
at least one of the basic subgraphs of $C$.
\end{description}
Let the nodes be numbered such that $x_{i}>0$ only for $i=1,\ldots ,k$.
Because ${\textbf {x}}$ is an eigenvector of the matrix $C$ we have
$\sum _{j=1}^{s}c_{ij}x_{j}=\lambda _{1}x_{i};\; i=1,\ldots ,s$ ($\lambda _{1}(C)$
has been abbreviated to $\lambda _{1}$).\\
 $\Rightarrow \sum _{j=1}^{k}c_{ij}x_{j}=\lambda _{1}x_{i};\; i=1,\ldots ,k$.\\
 Therefore $(x_{1},x_{2},\ldots ,x_{k})^{T}$ is an eigenvector of
the submatrix $C'=(c_{ij}),\; i,j=1,\ldots ,k$, with eigenvalue $\lambda _{1}$.
Also by lemma A, $\lambda _{1}(C')\le \lambda _{1}$. Therefore $\lambda _{1}(C')=\lambda _{1}$.
It follows that the subgraph $C'$ must contain a strong component
with Perron-Frobenius eigenvalue $\lambda _{1}$. That strong component
would be present in $C$ also and would be one of it's basic subgraphs.
Hence there exists at least one basic subgraph of $C$ that is non-zero
in any PFE of $C$. $\hfill \square $\\

\begin{description}
\item [Lemma~B:]\textit{\emph{Let $C$ be an $s\times s$ non-negative
irreducible matrix with Perron-Frobenius eigenvalue $\lambda _{1}$
and let $E=Adj(C-\lambda _{1}I)$. Then either $E>0$ or $E<0$.}} \emph{}
\end{description}
As $E$ is the adjoint of $C-\lambda _{1}I$ and $|C-\lambda _{1}I|=0$,
it follows that $E(C-\lambda _{1}I)=0$ and $(C-\lambda _{1}I)E=0$.
Hence any column of $E$ is either a PFE of $C$ or is a column of
zeroes. If it is a PFE then by the Perron-Frobenius theorem it is
either strictly positive or strictly negative. A similar result holds
for the rows of $E$. Hence either $E=0$ or $E$ has all elements
non-zero. If $E$ has all elements non-zero then it follows that either
$E>0$ or $E<0$.

I now show that one element of $E$ is non-zero. The $(s,s)$ element
of $E$ is $|D-\lambda _{1}I|$ where $D$ is the matrix $C$ with
the last row and column deleted. Consider the matrix \[
C'=\left(\begin{array}{ccc|c}
  &  &  & \\
  & D &  & 0\\
  &  &  & \\
\hline  & 0 &  & 0\end{array}\right).\]
 Clearly $C'\leq C$. Because $C$ is irreducible it cannot have a
zero column or row; if it did there would be no path to or from that
node to other nodes. Hence $C'\ne C$, which implies, as a consequence
of the Perron-Frobenius theorem for irreducible matrices, that $\lambda _{1}$
is not an eigenvalue of $C'$ and, therefore, is not an eigenvalue
of $D$ either. Hence $E_{ss}=|D-\lambda _{1}I|\ne 0$.

Therefore, either $E>0$ or $E<0$. $\hfill \square $\\

\begin{description}
\newpage
\item [Proposition~3.5:]\textit{\emph{If a node is non-zero in a PFE then
all nodes it has access to are non-zero in that PFE.}}
\end{description}
Let $i$ be a node that is non-zero in the PFE $\mathbf{x}$. Let
$j$ be any node downstream from it. Therefore there exists a positive
integer $k$ such that $(C^{k})_{ji}>0$.\\
 Assume that $x_{j}=0$. As $\sum _{l=1}^{s}(C^{k})_{jl}x_{l}=\lambda _{1}^{k}x_{j}$,\\
 $\Rightarrow \sum _{l=1}^{s}(C^{k})_{jl}x_{l}=0$,\\
 $\Rightarrow (C^{k})_{jl}x_{l}=0$, for each $l=1,\ldots ,s$. This
is a contradiction because $x_{i}>0$ and $(C^{k})_{ji}>0$. Hence
$x_{j}>0$. $\hfill \square $\\

\begin{description}
\item [Proposition~3.6:]\textit{\emph{If a node is upstream from a basic
subgraph then that node is zero in any PFE.}} \textit{}
\end{description}
Let the nodes be renumbered so that the last few -- nodes $r,r+1,\ldots ,s$
-- form a basic subgraph of the graph $C$. Let node $p$ be some
other node that is upstream from this basic subgraph, i.e., there
exists a positive integer $k$ such that $(C^{k})_{rp}>0$. Let $\mathbf{x}$
be a PFE of $C$. Then: \[
\sum _{j=1}^{s}(C^{k})_{ij}x_{j}=\lambda _{1}^{k}x_{i};\quad i=1,\ldots ,s.\]
 The equations for $i=r,r+1,\ldots ,s$ can be written in matrix form
as follows: \[
A\left(\begin{array}{c}
 x_{1}\\
 \vdots \\
 x_{r-1}\end{array}\right)+(D^{k}-\lambda _{1}^{k}I)\left(\begin{array}{c}
 x_{r}\\
 \vdots \\
 x_{s}\end{array}\right)=0,\]
 where $A_{ij}=(C^{k})_{i+r-1,j}\; (i=1,\ldots ,m;j=1,\ldots ,r-1;m\equiv s-r+1)$,
\\
and $D_{ij}=C_{i+r-1,j+r-1}\; (i,j=1,\ldots ,m)$ is the adjacency
matrix of the basic subgraph.

\begin{equation}
\Rightarrow (D^{k}-\lambda _{1}^{k}I)\left(\begin{array}{c}
 x_{r}\\
 \vdots \\
 x_{s}\end{array}\right)=\left(\begin{array}{c}
 b_{1}\\
 \vdots \\
 b_{m}\end{array}\right),\label{14}\end{equation}
 with ${\textbf {b}}\equiv (b_{1},b_{2},\ldots ,b_{m})^{T}\leq 0,\ne 0$.

Let us assume that node $p$ is non-zero in the PFE, i.e. $x_{p}>0$.
Also $(A)_{1p}\equiv (C^{k})_{rp}>0$, therefore $b_{1}<0$.

Let $R_{i}$ denote the $i^{th}$ row of $D^{k}-\lambda _{1}^{k}I$.\\
 Because $|D^{k}-\lambda _{1}^{k}I|=0$,\begin{equation}
\Rightarrow R_{1}=\sum _{j=2}^{m}\alpha _{j}R_{j},\label{15}\end{equation}
 with at least one $\alpha _{j}$ being non-zero.

(\ref{14}) gives \[
\sum _{i=1}^{m}(R_{j})_{i}\, x_{i+r-1}=b_{j};\quad j=1,\ldots ,m.\]

Multiply the above equation by $\alpha _{j}$ and sum from $j=2$
to $j=m$. \[
\Rightarrow \sum _{i=1}^{m}\sum _{j=2}^{m}\alpha _{j}(R_{j})_{i}\, x_{i+r-1}=\sum _{j=2}^{m}\alpha _{j}b_{j},\]
\[
\Rightarrow \sum _{i=1}^{m}(R_{1})_{i}\, x_{i+r-1}=\sum _{j=2}^{m}\alpha _{j}b_{j},\]
\begin{equation}
\Rightarrow b_{1}=\sum _{j=2}^{m}\alpha _{j}b_{j}.\label{18}\end{equation}

From (\ref{15}) we have \[
(R_{1})_{i}=\sum _{j=2}^{m}\alpha _{j}(R_{j})_{i},\]
\begin{equation}
\Rightarrow (D^{k}-\lambda _{1}^{k}I)_{1i}=\sum _{j=2}^{m}\alpha _{j}(D^{k}-\lambda _{1}^{k}I)_{ji}.\label{19}\end{equation}

Let $E=Adj(D^{k}-\lambda _{1}^{k}I)$. By definition \[
\sum _{j=1}^{m}E_{ij}(D^{k}-\lambda _{1}^{k}I)_{ji}=|D^{k}-\lambda _{1}^{k}I|=0;\quad i=1,\ldots ,m,\]
\[
\Rightarrow (D^{k}-\lambda _{1}^{k}I)_{1i}=-\sum _{j=2}^{m}\frac{E_{ij}}{E_{i1}}(D^{k}-\lambda _{1}^{k}I)_{ji};\quad i=1,\ldots ,m.\]

Comparing with (\ref{19}) we have \[
\alpha _{j}=-\frac{E_{ij}}{E_{i1}}<0,\]
 because from lemma B either $E>0$ or $E<0$ (as $D$ is irreducible
it follows that $D^{k}$ is irreducible hence lemma B is applicable).

\[
\Rightarrow \sum _{j=2}^{m}\alpha _{j}b_{j}\geq 0.\]

Therefore from (\ref{18}) \[
\Rightarrow b_{1}\geq 0.\]
 However, we are given that $b_{1}<0$. Hence the assumption that
$x_{p}>0$ has led to a contradiction. The only other possibility
is that $x_{p}=0$ in the PFE. $\hfill \square $

\begin{description}
\item [Proposition\textit{~}\textit{\emph{3.7:}}]\textit{\emph{If a node
does not have access from a basic subgraph then it is zero in any
PFE.}}
\end{description}
Let the $k\times k$ submatrix $A$ consist of all the strong components
that do not have access from any basic subgraph of $C$. The matrix
$C$ can be written as: \[
C=\left(\begin{array}{ccc|ccc}
  &  &  &  &  & \\
  & A &  &  & 0 & \\
  &  &  &  &  & \\
\hline  &  &  &  &  & \\
  & R &  &  & B & \\
  &  &  &  &  & \end{array}\right),\]
 where $B$ consists of all the basic subgraphs of $C$ and all the
nodes having access from them. Let $\mathbf{x}$ be a PFE of $C$.
The eigenvector equations give: \begin{equation}
(A-\lambda _{1}I)\left(\begin{array}{c}
 x_{1}\\
 \vdots \\
 x_{k}\end{array}\right)=0.\label{a}\end{equation}
 $A$ has no basic subgraphs of $C$, hence $\lambda _{1}$ is not
an eigenvalue of $A$. Therefore (\ref{a}) has only the trivial solution
$x_{i}=0;\; i=1,\ldots ,k$. $\hfill \square $\\

\begin{description}
\item [Theorem~3.1:]Eigenvector profile theorem\\
For any graph $C$, determine all the basic subgraphs of $C$. Denote
them by $D_{1},\ldots ,D_{K}$. Determine which of these does not
have any other $D_{i}$ downstream from it. Denote these by $E_{1},\ldots ,E_{N}$.
Then: \\
\textit{\emph{(i) For each $i=1,\ldots ,N$ there exists a unique
(upto constant multiples) PFE in which the nodes of $E_{i}$ and all
nodes having access from them are non-zero and all other nodes are
zero. It is evident that these PFEs are simple. Moreover these are
the only simple PFEs of the graph $C$. }}\\
\textit{\emph{(ii) Any PFE is a linear combination of these $N$ simple
PFEs.}}
\end{description}
The first statement follows from propositions 3.4--3.7. To prove the
second statement, I first write the matrix as follows: \[
C=\left(\begin{array}{ccc|ccccc|ccc}
  &  &  &  &  &  &  &  &  &  & \\
  & A &  &  &  & 0 &  &  &  & 0 & \\
  &  &  &  &  &  &  &  &  &  & \\
\hline  &  &  & E_{1} &  &  &  &  &  &  & \\
  &  &  &  & . &  &  &  &  &  & \\
  & P &  &  &  & . &  &  &  & 0 & \\
  &  &  &  &  &  & . &  &  &  & \\
  &  &  &  &  &  &  & E_{N} &  &  & \\
\hline  &  &  &  &  &  &  &  &  &  & \\
  & Q &  &  &  & R &  &  &  & D & \\
  &  &  &  &  &  &  &  &  &  & \end{array}\right).\]
 $A$ contains all the nodes $i=1,\ldots ,p-1$ that don't have access
from any $E_{i}$. Thus all nodes in $A$ will be zero in any PFE
$\mathbf{x}$. $D$ contains all nodes $i=p+m+1,\ldots ,s$ having
access from some $E_{i}$, i.e. $R\ne 0$. As $x_{1},\ldots ,x_{p-1}=0$
the eigenvector equations for each $E_{i}$ are: \[
(E_{i}-\lambda _{1}I)\left(\begin{array}{c}
 x_{l_{i}}\\
 \vdots \\
 x_{l_{i}+m_{i}}\end{array}\right)=0.\]
 Because each $E_{i}$ is irreducible the above equations have solutions
that are unique upto constant multiples. Denote these solutions by
$x_{j}^{(i)};\; j=l_{i},\ldots ,l_{i}+m_{i}$. The equations for $D$
are: \[
R\left(\begin{array}{c}
 x_{p}\\
 \vdots \\
 x_{p+m}\end{array}\right)+(D-\lambda _{1}I)\left(\begin{array}{c}
 x_{p+m+1}\\
 \vdots \\
 x_{s}\end{array}\right)=0,\]
\[
\Rightarrow (D-\lambda _{1}I)\left(\begin{array}{c}
 x_{p+m+1}\\
 \vdots \\
 x_{s}\end{array}\right)={\textbf {b}},\]
 where ${\textbf {b}}\leq 0,\ne 0$.

Consider the case where only one of the $E_{i}$ is non-zero. Denote
the ${\textbf {b}}$ for that case by ${\textbf {b}}^{(i)}$ and the
corresponding (unique) solution of the above equations by $x_{j}^{(i)};\; j=p+m+1,\ldots ,s$.
For the general case where more than one $E_{i}$ is non-zero, ${\textbf {b}}$
will be a linear combination of the ${\textbf {b}}^{(i)}$ because
$x_{k}$, for each node $k$ in each $E_{i}$, is unique upto constant
multiples. Hence the general solution for $x_{p+m+1},\ldots ,x_{s}$
will also be a linear combination of $x_{p+m+1}^{(i)},\ldots ,x_{s}^{(i)}$.
\cite{Rothblum} provides a different proof that uses the principle
of mathematical induction. $\hfill \square $

\begin{description}
\item [Proposition~3.8:]\textit{\emph{Every node in the core of a simple
PFE has access to every other node of the PFE subgraph. No periphery
node has access to any core node.}} 
\end{description}
From theorem 3.1 it follows that the subgraph of any simple PFE is
the subgraph induced by the nodes of one of the basic subgraphs $E_{i}$
(defined in the statement of theorem 3.1) and all nodes having access
from $E_{i}$. The core nodes are the nodes of $E_{i}.$ Because $E_{i}$
is irreducible, it is evident that every core node has access to every
other core node. Further, every periphery node has access from some,
and therefore all, nodes of $E_{i}$. Thus, each core node has access
to every other node of the PFE subgraph.

Now assume there is some periphery node $p$ that has access to some
core node $r$. Also $r$ has access to $p$, therefore $p$ is strongly
connected to $r$. In that case $p$ would also be part of the basic
subgraph $E_{i}$, i.e., $p$ would be a core node, not a periphery
node. Thus, no periphery node can have access to any core node. $\hfill \square $\\

\begin{description}
\item [Proposition~4.1:]\textit{\emph{For any graph $C$,}} \emph{}\\
 \emph{}(i) \emph{}\textit{\emph{Every eigenvector of $C$ that belongs
to the simplex $J$ is a fixed point of equation (\ref{xdot}), and
vice versa.}}\emph{}\\
 (ii) \emph{}\textit{\emph{Starting from any initial condition in
the simplex $J$, the trajectory converges to some fixed point (generically
denoted ${\textbf {X}}$) in $J$.}}\emph{}\\
 (iii) \emph{}\textit{\emph{For generic initial conditions in $J$,
${\textbf {X}}$ is a Perron-Frobenius eigenvector (PFE) of $C$.}}\emph{}\\
(iv) \emph{}\textit{\emph{If $C$ has a unique (upto constant multiples)
PFE, it is the unique stable attractor of equation (\ref{xdot}).}}
\emph{}\\
 (v) \emph{}\textit{\emph{If $C$ has more than one linearly independent
PFE, then ${\textbf {X}}$ can depend upon the initial conditions.
The set of allowed ${\textbf {X}}$ is a linear combination of a subset
of the PFEs.}} The interior of this set in $J$ may then be said to
be the `attractor' of (\ref{xdot}), in the sense that for generic
initial conditions all trajectories converge to a point in this set.
 \emph{}
\end{description}
The proof uses the Jordan canonical form of the matrix $C$ \citep{HJ}:
\[
J=\left(\begin{array}{cccccc}
 J_{s_{1}}(\lambda _{1}) &  &  &  &  & 0\\
  & J_{s_{2}}(\lambda _{2}) &  &  &  & \\
  &  & . &  &  & \\
  &  &  & . &  & \\
  &  &  &  & . & \\
 0 &  &  &  &  & J_{s_{k}}(\lambda _{k})\end{array}\right),\]

with $s_{1}+s_{2}+\ldots +s_{k}=s$. 

Each $J_{s_{i}}(\lambda _{i})$ is called a Jordan block and is an
$s_{i}\times s_{i}$ matrix of the form: \[
J_{s_{i}}(\lambda _{i})=\left(\begin{array}{ccccccc}
 \lambda _{i} & 1 &  &  &  &  & 0\\
  & \lambda _{i} & 1 &  &  &  & \\
  &  & \lambda _{i} & . &  &  & \\
  &  &  & . & . &  & \\
  &  &  &  & . & . & \\
  &  &  &  &  & . & 1\\
 0 &  &  &  &  &  & \lambda _{i}\end{array}\right).\]

The $\lambda _{i}$ are the eigenvalues of $C$. The $\lambda _{i}$
need not be distinct; there may be more than one Jordan block with
the same diagonal. It is known that any complex matrix can be put
in the above form by a similarity transformation, i.e. there exists
a non-singular matrix $P$ such that $C=PJP^{-1}$, and moreover,
the form of $J$ is unique barring rearrangements of the Jordan blocks
\citep{HJ}. The columns ${\textbf {p}}_{i}$ of $P$ form $k$ `Jordan
chains' $\{{\textbf {p}}_{1},\ldots ,{\textbf {p}}_{s_{1}}\},\{{\textbf {p}}_{s_{1}+1},\ldots ,{\textbf {p}}_{s_{1}+s_{2}}\},\ldots ,$
$\{{\textbf {p}}_{s-s_{k}+1},\ldots ,{\textbf {p}}_{s}\}$ that satisfy
the following equations:\[
\begin{array}{l}
 C{\textbf {p}}_{1}=\lambda _{1}{\textbf {p}}_{1,}\\
 C{\textbf {p}}_{i}=\lambda _{1}{\textbf {p}}_{i}+{\textbf {p}}_{i-1};\quad i=2,\ldots ,s_{1},\\
 C{\textbf {p}}_{s_{1}+1}=\lambda _{2}{\textbf {p}}_{s_{1}+1},\\
 C{\textbf {p}}_{s_{1}+i}=\lambda _{2}{\textbf {p}}_{s_{1}+i}+{\textbf {p}}_{s_{1}+i-1};\quad i=2,\ldots ,s_{2},\\
 \vdots \\
 C{\textbf {p}}_{s-s_{k}+1}=\lambda _{k}{\textbf {p}}_{s-s_{k}+1},\\
 C{\textbf {p}}_{s-s_{k}+i}=\lambda _{k}{\textbf {p}}_{s-s_{k}+i}+{\textbf {p}}_{s-s_{k}+i-1};\quad i=2,\ldots ,s_{k}.\end{array}\]
 Thus ${\textbf {p}}_{1},{\textbf {p}}_{s_{1}+1},\ldots ,{\textbf {p}}_{s_{k-1}+1}$
are eigenvectors of $C$ with eigenvalues $\lambda _{1},\lambda _{2},\ldots ,\lambda _{k}$
respectively. It can be shown that there is one and only one linearly
independent eigenvector of $C$ associated with each Jordan block
and vice versa \citep{HJ}. The other columns of $P$ are called \textit{\emph{`generalized
eigenvectors'}} of $C$.

Now consider the underlying dynamics (\ref{ydot}) from which (\ref{xdot})
is derived: Because (\ref{xdot}) is independent of $\phi $, we can
set $\phi =0$ in (\ref{ydot}) without any loss of generality. With
$\phi =0$ the general solution of (\ref{ydot}), which is a linear
system, can be schematically written as: \[
{\textbf {y}}(t)=e^{Ct}{\textbf {y}}(0),\]
 where ${\textbf {y}}(0)$ and ${\textbf {y}}(t)$ are column vectors. 

\newpage
Using the Jordan canonical form of $C$ in the general solution of
(\ref{ydot}) we find: \[
{\textbf {y}}(t)=Pe^{Jt}P^{-1}{\textbf {y}}(0).\]

Now \[
e^{Jt}=\left(\begin{array}{cccccc}
 e^{J_{s_{1}}(\lambda _{1})t} &  &  &  &  & 0\\
  & e^{J_{s_{2}}(\lambda _{2})t} &  &  &  & \\
  &  & . &  &  & \\
  &  &  & . &  & \\
  &  &  &  & . & \\
 0 &  &  &  &  & e^{J_{s_{k}}(\lambda _{k})t}\end{array}\right),\]

where \[
e^{J_{s_{i}}(\lambda _{i})t}=e^{\lambda _{i}t}\left(\begin{array}{cccccc}
 1 & t & \frac{t^{2}}{2} & . & . & \frac{t^{s_{i}-1}}{(s_{i}-1)!}\\
  & 1 & t & . & . & \frac{t^{s_{i}-2}}{(s_{i}-2)!}\\
  &  & . &  &  & .\\
  &  &  & . &  & .\\
  &  &  &  & . & t\\
 0 &  &  &  &  & 1\end{array}\right).\]
\\

Let \[
\mathbf{Y}=P\lim _{t\rightarrow \infty }e^{Jt}P^{-1}{\textbf {y}}(0).\]
 \\
Then the attractor of (\ref{xdot}) will be $\mathbf{X}=\mathbf{Y}/\sum _{i=1}^{s}Y_{i}$
and only the fastest growing components in $\mathbf{Y}$ will contribute
to $\mathbf{X}$. Therefore, in calculating $\mathbf{Y}$, I will
keep only the fastest growing terms and set all others to zero for
convenience. 

First consider the case where $\mathbf{y}(0)$ is unrestricted --
any initial condition is allowed. Then some components of $P^{-1}\mathbf{y}(0)$
could be zero. If all the components corresponding to a particular
Jordan block are zero in $P^{-1}\mathbf{y}(0)$ then that Jordan block
will not contribute to $e^{Jt}P^{-1}\mathbf{y}(0)$. Further, among
the Jordan blocks that do have some non-zero components in $P^{-1}\mathbf{y}(0)$,
only the ones with the largest $\lambda _{i}$ will survive in the
$t\rightarrow \infty $ limit because of the $e^{\lambda _{i}t}$
term. Let the Jordan blocks that survive be the first $K$ ones (the
nodes can always be renumbered to achieve this). From the structure
of $e^{J_{s_{i}}(\lambda _{i})t}$ it then follows that the only non-zero
components of $e^{Jt}P^{-1}\mathbf{y}(0)$ will be the components
$1,1+s_{1},1+s_{1}+s_{2},\ldots ,1+\sum _{j=1}^{K-1}s_{j}$. Therefore
$\mathbf{Y}=t^{r}e^{\lambda t}\sum _{i}^{'}c_{i}\mathbf{p}_{i}$,
with $i$ being summed over the set ${\{1,1+s_{1},1+s_{1}+s_{2},\ldots ,1+\sum _{j=1}^{K-1}s_{j}\}}$.
$c_{i},r,\lambda $ are constants and $\mathbf{p}_{i}$ is the $i$th
column of $P$. Then, the attractor is $\mathbf{X}=c_{0}(c_{1}\mathbf{p}_{1}+c_{1+s_{1}}\mathbf{p}_{1+s_{1}}+\ldots +c_{k}\mathbf{p}_{k})$,
where $c_{0}$ is a constant chosen to make $\sum _{i=1}^{s}X_{i}=1$,
and $k=1+\sum _{j=1}^{K-1}s_{j}$. Notice that $\mathbf{p}_{1},\mathbf{p}_{1+s_{1}},\ldots ,\mathbf{p}_{k}$
are all eigenvectors of $C$ with eigenvalue $\lambda .$ Therefore
$\mathbf{X}$ is a fixed point, and is an eigenvector of $C$. This
proves (i) and (ii).

Now consider the case of generic initial conditions. In this case
$P^{-1}\mathbf{y}(0)>0$. The only terms to survive will be in the
Jordan blocks corresponding to the Perron-Frobenius eigenvalue, $\lambda _{1}$,
of $C$. Moreover, of these blocks only the largest sized blocks will
survive. Suppose the largest Jordan blocks corresponding to eigenvalue
$\lambda _{1}$ have size $r$. Let the Jordan blocks be arranged
such that these blocks are the first few. Then, in the limit $t\rightarrow \infty $,
$e^{Jt}$ will have the following form: \[
e^{Jt}\sim \frac{t^{r}}{(r-1)!}e^{\lambda _{1}t}\left(\begin{array}{ccccccccccccccc}
 0 & \ldots  & 1 &  &  &  &  &  &  &  &  &  &  &  & 0\\
 \vdots  &  & \vdots  &  &  &  &  &  &  &  &  &  &  &  & \\
 0 & \ldots  & 0 &  &  &  &  &  &  &  &  &  &  &  & \\
  &  &  & . &  &  &  &  &  &  &  &  &  &  & \\
  &  &  &  & . &  &  &  &  &  &  &  &  &  & \\
  &  &  &  &  & . &  &  &  &  &  &  &  &  & \\
  &  &  &  &  &  & 0 & \ldots  & 1 &  &  &  &  &  & \\
  &  &  &  &  &  & \vdots  &  & \vdots  &  &  &  &  &  & \\
  &  &  &  &  &  & 0 & \ldots  & 0 &  &  &  &  &  & \\
  &  &  &  &  &  &  &  &  & 0 & \ldots  & 0 &  &  & \\
  &  &  &  &  &  &  &  &  & \vdots  &  & \vdots  &  &  & \\
  &  &  &  &  &  &  &  &  & 0 & \ldots  & 0 &  &  & \\
  &  &  &  &  &  &  &  &  &  &  &  & . &  & \\
  &  &  &  &  &  &  &  &  &  &  &  &  & . & \\
 0 &  &  &  &  &  &  &  &  &  &  &  &  &  & .\end{array}\right).\]

Thus if there are $l$ Jordan blocks of size $r$ corresponding to
eigenvalue $\lambda _{1}$ then, for the purpose of calculating $\mathbf{X}$,
all elements of $e^{Jt}$ can be set to zero except the elements $(1,r),(r+1,2r),\ldots ,$
$((l-1)r+1,lr)$ which will be equal to $\frac{t^{r}}{(r-1)!}e^{\lambda _{1}t}$.
Therefore the only non-zero elements of $e^{Jt}P^{-1}{\textbf {y}}(0)$,
in the $t\rightarrow \infty $ limit, will have values $c_{i}t^{r}e^{\lambda t}/(r-1)!$
for $i=1,r+1,\ldots ,(l-1)r+1$, where $c_{i}$ are real constants
that depend on $P^{-1}{\textbf {y}}(0)$. Thus, \[
\mathbf{Y}=P\lim _{t\rightarrow \infty }e^{Jt}P^{-1}\mathbf{y}(0)\sim \frac{t^{r}e^{\lambda _{t}}}{(r-1)!}(c_{1}\mathbf{p}_{1}+c_{r+1}\mathbf{p}_{r+1}+\ldots +c_{(l-1)r+1}\mathbf{p}_{(l-1)r+1}).\]
 Therefore,\[
\mathbf{X}=c_{0}(c_{1}\mathbf{p}_{1}+c_{r+1}\mathbf{p}_{r+1}+\ldots +c_{(l-1)r+1}\mathbf{p}_{(l-1)r+1}),\]
where $c_{0}$ is a constant chosen to make $\sum _{i=1}^{s}X_{i}=1$.
Thus $\mathbf{X}$ is a linear combination of those PFEs of $C$ that
correspond to the largest Jordan blocks with eigenvalue $\lambda _{1}$.
This proves (iii) and (iv). $\hfill \square $

\begin{description}
\newpage
\item [Theorem~4.1:]Attractor profile theorem\\
 1) Determine all the strong components of the given graph $C$. Denote
these by $C_{1},\ldots ,C_{M}$. \\
 2) Determine which of these are basic subgraphs. Denote them by $D_{1},\ldots ,D_{K}$.\\
3) Construct a graph, denoted $D^{*}$, with $K$ nodes, representing
the $D_{i}$, and a link from node $j$ to node $i$ if there is a
path from any node of $D_{j}$ to any node of $D_{i}$ that does not
contain a node of any other basic subgraph.\\
 4) Determine which of the $D_{i}$ are at the ends of the longest
paths in the above graph $D^{*}$. Denote these by $F_{i},i=1,\ldots ,N$.\textit{}\\
\textit{\emph{For each $i=1,\ldots ,N$ there exists a unique (upto
constant multiples) PFE that has only nodes of $F_{i}$ and all nodes
having access from them non-zero, and all other nodes zero. The attractor
set consists of all linear combinations of these PFEs that lie on
the simplex $J$.}}
\end{description}
It has already been shown (proposition 4.1) that the attractor $\mathbf{X}$
is a linear combination of those PFEs of $C$ that correspond to the
largest Jordan blocks with eigenvalue $\lambda _{1}$. Therefore to
prove this theorem it suffices to show that the PFEs corresponding
to the largest Jordan blocks with eigenvalue $\lambda _{1}$ are precisely
those PFEs corresponding to the subgraphs $F_{i}$ that are at the
ends of the longest paths in $D^{*}$.

It can be shown that the size of the largest Jordan block with eigenvalue
$\lambda _{1}$ is equal to the length of the longest path in the
graph $D^{*}$ (see \citealp{Rothblum}, where it is called the index
of $C$). It is also shown by Rothblum that for each basic subgraph
$D_{i}$ there is a sequence of generalized eigenvectors ${\textbf {y}}^{1},\ldots ,{\textbf {y}}^{h}$,
where $h$ is the length of the longest path in $D^{*}$ starting
at $D_{i}$, such that $(C-\lambda _{1}I){\textbf {y}}^{1}=0$, $(C-\lambda _{1}I){\textbf {y}}^{k}={\textbf {y}}^{k-1}$
(where $k=2,\ldots ,h$) and $y_{j}^{k}>0$ if and only if the longest
path from a node of $D_{i}$ to node $j$ when projected onto the
graph $D^{*}$ has a length $k$.

Let the length of the longest path in $D^{*}$ be $r$ and consider
one such path starting at, say, $D_{j}$ and ending at some $F_{i}$.
Then, corresponding to this path, there is a sequence of $r$ generalized
eigenvectors that includes the PFE in which only $F_{i}$ and nodes
having access from it are non-zero. Thus there is a Jordan block of
size $r$ corresponding to the PFE in which only $F_{i}$ and all
nodes having access from it are non-zero. This is also the largest
Jordan block. Therefore the PFEs corresponding to the largest Jordan
blocks with eigenvalue $\lambda _{1}$ are precisely those PFEs corresponding
to the subgraphs $F_{i}$ that are at the ends of the longest paths
in $D^{*}$. The attractor set will consist of linear combinations
of these PFEs.$\hfill \square $

\begin{description}
\newpage
\item [Proposition~4.2:]For any graph with $\lambda _{1}(C)=0$, in the
attractor only the nodes at the ends of the longest paths are non-zero.
All other nodes are zero.
\end{description}
Consider a graph consisting only of a linear chain of $r+1$ nodes,
with $r$ links, pointing from node 1 to node 2, node 2 to 3, etc.
The node 1 (to which there is no incoming link) has a constant population
$y_{1}$ because the r.h.s of (\ref{ydot}) vanishes for $i=1$ (taking
$\phi =0$). For node 2, we get $\dot{y_{2}}=y_{1}$, hence $y_{2}(t)=y_{2}(0)+y_{1}t\sim t$
for large $t$. Similarly, it can be seen that $y_{k}$ grows as $t^{k-1}$.
In general, it is clear that for a graph with no cycles, $y_{i}\sim t^{r}$
for large $t$ (when $\phi =0$), where $r$ is the length of the
longest path terminating at node $i$. Thus, nodes with the largest
$r$ dominate for sufficiently large $t$. Because the dynamics (\ref{xdot})
does not depend upon the choice of $\phi $, $X_{i}=0$ for all $i$
except the nodes at which the longest paths in the graph terminate.
$\hfill \square $\\

\begin{description}
\item [Proposition~4.3:]\textit{\emph{For any graph $C$,}} \emph{}\\
 \emph{}(i) \emph{}\textit{\emph{For every ${\textbf {X}}$ belonging
to the attractor set, the set of nodes $i$ for which $X_{i}>0$ is
the same and is uniquely determined by $C$. The subgraph formed by
this set of nodes will be called the `subgraph of the attractor' of
(\ref{xdot}) for the graph $C$.}} \\
 (ii) \textit{\emph{If $\lambda _{1}(C)\ge 1$, the subgraph of the
attractor is an ACS.}} \emph{}
\end{description}

This is a corollary to theorem 4.1. The set of nodes $j$ for which
$X_{j}>0$ consists of nodes in all the subgraphs $F_{i}$ described
in the statement of theorem 4.1, and all nodes having access from
them. This set evidently depends only on $C$. Further $\mathbf{X}$
is a PFE of $C$. Therefore, from proposition 3.3 the subgraph of
the attractor $\mathbf{X}$ is an ACS if $\lambda _{1}(C)\ge 1$.
$\hfill \square $\\

\begin{description}
\item [Proposition~6.1:]\textit{\emph{$\lambda _{1}$ is a non-decreasing
function of $n$ as long as $s_{1}<s$}}\emph{. }
\end{description}
Let the adjacency matrix of the graph at time step $n$ be denoted
$C_{n}$ and its Perron-Frobenius eigenvalue be denoted $\lambda _{1}(C_{n})$.
Removing a node with the least $X_{i}$ from this matrix gives an
$(s-1)\times (s-1)$ matrix that I denote $T$. 

From lemma B, $\lambda _{1}(T)\leq \lambda _{1}(C_{n})$.\\
Because $s_{1}<s$ the removal of the node with the least $X_{i}$
leaves the dominant ACS (denoted $D_{n}$) unaffected, i.e., $D_{n}$
is a subgraph of $T$.\\
 $\Rightarrow \lambda _{1}(D_{n})\leq \lambda _{1}(T)\leq \lambda _{1}(C_{n}).$

However $\lambda _{1}(C_{n})$ is an eigenvalue of $D_{n}$. Hence,
$\lambda _{1}(D_{n})=\lambda _{1}(C_{n})$,\\
 $\Rightarrow \lambda _{1}(T)=\lambda _{1}(C_{n})$.

Now a new node is added to $T$ to get $C_{n+1}$. Because $T$ is
a subgraph of $C_{n+1}$, it follows that $\lambda _{1}(C_{n+1})\geq \lambda _{1}(T)$,\\
 $\Rightarrow \lambda _{1}(C_{n+1})\geq \lambda _{1}(C_{n})$. $\hfill \square $
\\

\begin{description}
\item [Proposition~7.1:]Let $N_{n}$ denote the maximal new irreducible
subgraph that includes the new node at time step $n$. $N_{n}$ will
become the new core of the graph, replacing the old core $Q_{n-1}$,
whenever either of the following conditions are true: \\
 (a) $\lambda _{1}(N_{n})>\lambda _{1}(Q'_{n})$ or, \\
 (b) $\lambda _{1}(N_{n})=\lambda _{1}(Q'_{n})$ and $N_{n}$ is downstream
of $Q'_{n}$.
\end{description}
After a node is removed from the graph $C_{n-1}$, the resultant graph
is $C'_{n}$ which has the core $Q'_{n}$. Now a new node is added
to this graph to get $C_{n}$. This graph contains the irreducible
subgraph $N_{n}$. Clearly if $\lambda _{1}(N_{n})>\lambda _{1}(Q'_{n})$
then $N_{n}$ is the only basic subgraph of $C_{n}$ and will therefore
be its core. But if $\lambda _{1}(N_{n})=\lambda _{1}(Q'_{n})$ then
it is only one of the basic subgraphs of $C_{n}$. By definition $Q'_{n}$,
because it is the core of $C'_{n-1}$, contains those basic subgraphs
that are at the ends of the longest chains of basic subgraphs in $C'_{n-1}$.
Therefore, $N_{n}$ will be a basic subgraph at the end of the longest
chain of basic subgraphs in $C_{n}$ only if it is downstream from
some node of $Q'_{n}$. If this is so, then it will be the single
basic subgraph at the end of the longest chain, and therefore will
be the core of $C_{n}$. $\hfill \square $

\chapter{Programs to simulate the model}

This appendix describes two programs, included in the attached CD,
that can be used to simulate the model and its variants. The programs
differ in the way the attractor is found at each time step. The first
program uses theorem 4.1. The second numerically integrates equation
(\ref{xdot}) to find the attractor. These different methods will
be described below, in sections \ref{sec:Program-which-uses} and
\ref{sec:Program-which-finds}, but first I will describe what is
common to the two programs. Both programs are written in C++, however
they can be easily translated to C because I have used only certain
small convenient features of C++ and no object oriented code.

\section{Variables and data types}

The two parameters of the model are $s$, the total number of nodes
of the graph, which is specified by the variable named \texttt{n}
of type \texttt{int} (integer), and $p$, the catalytic probability,
which is specified by the variable named \texttt{lprob} of type \texttt{double}
(double precision real number).

The dynamical variables are the graph $C$, whose adjacency matrix
is stored as a two dimensional $s\times s$ array of type \texttt{double},
and the relative populations $x_{i}$ which are stored as a one dimensional
array of type \texttt{double}.

\section{Creating the initial random graph}

The initial random graph is created in the function \texttt{makec4.}
For each $i,j\in \{1,\ldots ,s\}$, I generate a (pseudo)random number
of type \texttt{double} uniformly distributed in the interval $[0,1]$.
If the random number is less than $p$ then $c_{ij}$ is assigned
unity, otherwise it is assigned zero. For generating (pseudo)random
numbers uniformly distributed in the interval $[0,1]$ I have used
the routine \texttt{ran1} from Numerical Recipes \citep{Numerical}.
However, as Press et al. do not allow Numerical Recipes routines to
be distributed publicly, I have replaced calls to \texttt{ran1} in
the programs included in the attached CD by calls to the internal
random number generator \texttt{drand48}. The function call is \texttt{drand48()}
and returns a number of type \texttt{double} uniformly distributed
in the interval $[0,1]$. The variable \texttt{idum} of type \texttt{long}
(long integer) is used to set the initial seed of the random number
generator using the function \texttt{srand48}.

\section{The graph update}

Once the attractor is known, the set $\mathcal{L}$ of nodes with
the least $X_{i}$ has to be determined. Because the $X_{i}$ are
stored as data type \texttt{double} it makes sense to choose a threshold
such that two numbers differing by less than the threshold will be
considered equal. In the programs, the threshold has been chosen to
be $10^{-5}$. Note that in the case where theorem 4.1 is used to
determine the attractor, the set $\mathcal{L}$ is known exactly in
most cases; there are few situations where the threshold is actually
required.

After the set $\mathcal{L}$ is determined, one picks an arbitrary
node out of this set for removing from the graph. This node is picked
by choosing a random number from the set $\{0,1,\ldots ,l-1\}$, where
$l$ is the size of the set $\mathcal{L}$ (if $r$ is a random number
in the interval $[0,1]$ generated using \texttt{ran1} or \texttt{drand48},
then \texttt{int(l{*}r)} is a random integer between 0 and $l$, inclusive).

Once the node to be removed is decided, say node $k$, the graph update
is done by the function call \texttt{update4(k}). All the elements
of the matrix in that row and column are reassigned to 0 or 1 by the
same procedure used to create the initial matrix, i.e., a random number
is generated and for each $i\ne k$, $c_{ik}$ is assigned unity if
the random number is less than $p$, and is assigned zero otherwise,
and similarly for $c_{ki}$. The function \texttt{update4} also sets
$x_{k}$ to $x_{0}$ (the variable \texttt{xin} of type \texttt{double})
and rescales all other $x_{i}$ so that $\sum _{i=1}^{s}x_{i}=1$.

\addtocontents{toc}{\newpage}

\section{\label{sec:Program-which-uses}Program that uses the attractor profile
theorem }

\subsection{Determining which node to remove}

The crucial step involves the determination of the set, $\mathcal{L}$,
of nodes with the least $X_{i}$ at each time step. The results from
chapter \ref{cha:Population-Dynamics}, in particular theorem 4.1,
are used to do this.

\begin{enumerate}
\item Determine all the strong components of the given graph. Denote these
by $C_{1},C_{2},\ldots ,C_{M}$. 
\item Determine which of these are basic subgraphs. Denote them by $D_{1},\ldots ,D_{K}$. 
\item Construct a graph, $D^{*}$, of $K$ nodes representing the $D_{i}$
with a link from node $j$ to node $i$ if there is a path from any
node of $D_{j}$ to any node of $D_{i}$ that does not contain a node
of any other basic subgraph.
\item Determine which of the $D_{i}$ are at the ends of the longest paths
in the above graph $D^{*}$. Denote these by $F_{i}$.
\item For each $F_{i}$ construct a vector with (arbitrary) non-zero values
assigned to those components corresponding to nodes in $F_{i}$ and
all nodes having access from them. Set all other components to zero.
\item (a) If there are any nodes that are zero in all these vectors then
they form the unique set $\mathcal{L}$ for generic initial conditions.\\
(b) If no nodes are zero, then first the PFE corresponding to each
$F_{i}$ is determined using Matlab. This is done by constructing
the adjacency matrix of the subgraph induced by all nodes in $F_{i}$
and all nodes having access from them and using Matlab's \texttt{eig}
function to determine the (unique) PFE of this matrix. Then,\\
If there is only one $F_{i}$, its PFE is used to determine $\mathcal{L}$.
Because its PFE is unique upto constant multiples, the set $\mathcal{L}$
is also unique and independent of the (generic) initial conditions.\\
If there is more than one $F_{i}$, then $\mathcal{L}$ depends on
the initial conditions. In this case $X_{i}$ is taken to be a linear
combination of the PFEs corresponding to each $F_{i}$, with the coefficients
of the linear combination being random numbers generated by \texttt{ran1}
or \texttt{drand48}. The set $\mathcal{L}$ is then determined from
these values of $X_{i}$.
\end{enumerate}
This last case only occurs for graphs in which several ACSs coexist
and the whole graph is spanned by these ACSs. Such a situation occurs
very rarely. Thus when $s_{1}<s$ the set $\mathcal{L}$ is unique
and independent of initial conditions. When $s_{1}=s$ most of the
time there is only one $F_{i}$ in which case again $\mathcal{L}$
is unique and independent of initial conditions but has to be determined
numerically (using Matlab). Finally, very rarely, when $s_{1}=s$
and there is more than one $F_{i}$ in the graph then the set $\mathcal{L}$
depends on the initial conditions and an arbitrary linear combination
of PFEs is taken as the attractor. The function that implements this
algorithm in the program is called \texttt{thmeig}. 

Note that the code was written to work with Matlab 5.2 and therefore
may require some modifications to work with other versions of Matlab.

\subsection{Output files}

Having specified the values of the variables \texttt{n} (the size
of the graph), \texttt{lprob} (the catalytic probability), \texttt{xin}
(the relative population assigned to each new node) and \texttt{idum}
(the initial random seed), the program produces a run with the specified
number of graph updates. Several output files are created, which consist
of one or more columns of data. Each row corresponds to an iteration
of the graph dynamics and different filenames are used for storing
different quantities. The list below shows the naming scheme I have
used for the output files:

\begin{enumerate}
\item \texttt{{*}.l} : the total number of links.
\item \texttt{{*}.px} : column 1 has the number of nodes with $X_{i}>10^{-5}$,
column 2 has the number of nodes with $X_{i}>0$ calculated (using
theorem 4.1) in \texttt{thmeig} if there is an ACS, or -1 if there
is no ACS in the graph.
\item \texttt{{*}.lm} : the Perron-Frobenius eigenvalue of the graph.
\item \texttt{{*}.min} : information about each graph update, which node
was replaced, which new links were assigned. This can be used to reconstruct
all the graphs later using the program \texttt{mtx.cpp} included in
the attached CD (see appendix C).
\item \texttt{{*}.clno} : the number of irreducible subgraphs of the graph.
\item \texttt{{*}.clev} : the Perron-Frobenius eigenvalue of each irreducible
subgraph.
\item \texttt{{*}.cn} : number of core nodes.
\item \texttt{{*}.cs} : list of core nodes.
\item \texttt{{*}.0num} : number of nodes with $X_{i}=0$.
\item \texttt{{*}.0nodes} : list of nodes with $X_{i}=0$.
\end{enumerate}
Any filename can be substituted for \texttt{{*}}. I typically use
a name that contains information about the values given to \texttt{n},
\texttt{lprob} and \texttt{idum}, thereby allowing me to distinguish
one type of run from another.

\section{\label{sec:Program-which-finds}Program that finds the attractor
numerically}

\subsection{Numerical method for determining the attractor}

As an alternative to the above program that uses theorem 4.1, I have
included, in the attached CD, a program that numerically integrates
equation (\ref{xdot}) to find its attractor. The program uses a 5th
order adaptive Runge-Kutta scheme based on the algorithm given in
Numerical Recipes \citep{Numerical}. The function that implements
this in the program is called \texttt{dyneig}. A call to \texttt{dyneig}
replaces the call to \texttt{thmeig} in the previous program.

I have included this alternate program for two reasons. One, for several
variants of the model, such as the variant with negative links (section
\ref{sec:Negative-links}), theorems 3.1 and 4.1 do not hold. Then
the only option is to use a numerical method to find the attractor
-- many of the variants discussed in chapter \ref{cha:Variants-of-model}
have been simulated using this method.

The second reason is historical. When I initially began working on
this model, theorems 3.1 and 4.1 did not exist (in my mind). All the
initial runs were done using the numerical method. Some patterns were
consistently observed in these runs. For example, whenever there was
a 2-cycle downstream from another, and these were the only basic subgraphs,
then the upstream 2-cycle was always zero. These observations were
the stepping stones that led to the theorems.

\newpage
\subsection{Output files}

As in the previous program, the following variables must be set before
running the program: \texttt{n} (the size of the graph), \texttt{lprob}
(the catalytic probability), \texttt{xin} (the relative population
assigned to each new node) and \texttt{idum} (the initial random seed).
For a given run, the program produces four output files, each of which
consist of one or more columns of data, and are named as follows:

\begin{enumerate}
\item \texttt{{*}.l} : the total number of links.
\item \texttt{{*}.px} : column 1 has the number of nodes with $X_{i}>10^{-5}$,
column 2 has a flag that is 1 if the set of nodes with $X_{i}>10^{-5}$
is an ACS and 0 otherwise.
\item \noindent \texttt{{*}.S1} : list of nodes having $X_{i}>10^{-5}$.
\item \texttt{{*}.min} : information about each graph update, which node
was replaced, which new links were assigned. This can be used to reconstruct
all the graphs later using the program \texttt{mtx.cpp} included in
the attached CD (see appendix C).
\end{enumerate}

\chapter{Contents of the attached CD}

The attached CD contains the following files:

\begin{itemize}
\item \texttt{contents}: The contents of the CD; this list.
\item \texttt{thesis.pdf}: This thesis in PDF format.
\item \texttt{thesis.ps}: This thesis in Postscript format.
\item \texttt{Programs/thmruns.cpp}: A program to simulate the model, that
uses theorem 4.1 (see section \ref{sec:Program-which-uses}).
\item \texttt{Programs/rk5runs.cpp}: A program to simulate the model, that
uses a 5th order Runge-Kutta method to integrate equation (\ref{xdot})
(see section \ref{sec:Program-which-finds}).
\item \texttt{Programs/mtx.cpp}: A program to extract the adjacency matrix
of the graph at any given iteration, from the \texttt{{*}.min} file
produced by \texttt{thmruns.cpp} or \texttt{rk5runs.cpp}. The program
requires the seed used to initialize the random number generator for
that run.
\item \texttt{Programs/mtx2gw.cpp}: A program to convert the adjacency matrix
files produced by \texttt{mtx.cpp} to the LEDA format.
\item \texttt{Programs/calcdeg.cpp}: A program to calculate the in, out
and total degree distributions for some or all graphs of a run from
the \texttt{{*}.min} file produced by \texttt{thmruns.cpp} or \texttt{rk5runs.cpp}.
The program requires the seed used to initialize the random number
generator for that run.
\item \texttt{Programs/calcdep.cpp}: A program to calculate the interdependency
and dependency distribution for some or all graphs of a run from the
\texttt{{*}.min} file produced by \texttt{thmruns.cpp} or \texttt{rk5runs.cpp}.
The program requires the seed used to initialize the random number
generator for that run.
\item \texttt{Programs/rndgraph.cpp}: A program to produce a set of graphs
drawn from the ensemble $G_{s}^{p}$, and to calculate the degree
and dependency distribution.
\item \texttt{Data/t01s1n100.{*}}: The data files produced by \texttt{thmruns.cpp}
for a run with $s=100,p=0.001$. This is the run displayed in chapter
\ref{cha:Graph-Dynamics} (see Figures \ref{cap:links} and \ref{cap:s1lambda}a). 
\item \texttt{Data/t025s{*}n100.{*}}: The data files produced by thmruns.cpp
for runs with $s=100,p=0.0025$ and different seeds. These are the
set of runs, totaling 1.55 million iterations, mentioned in sections
\ref{sub:Clustering-coefficient}, \ref{sub:Degree-distribution}
and \ref{sec:Crashes-and-core-shifts}. \texttt{t025s1n100.{*}} is
the run displayed in chapter \ref{cha:Graph-Dynamics} (see Figures
\ref{cap:links} and \ref{cap:s1lambda}b).
\item \texttt{Data/t05s3n100.{*}}: The data files produced by \texttt{thmruns.cpp}
for a run with $s=100,p=0.005$. This is the run displayed in chapter
\ref{cha:Graph-Dynamics} (see Figures \ref{cap:links} and \ref{cap:s1lambda}c).
Note that these data files, and the ones above, were produced by a
version of \texttt{thmruns.cpp} that uses \texttt{ran1} as the random
number generator and, therefore, are likely to differ from runs with
the same random seed that are produced by the program provided in
the CD, which uses \texttt{drand48}.
\item \texttt{Data/crashes.txt}: A list of iterations, for each of the runs
\texttt{t025s{*}n100.{*}} with $s=100,p=0.0025$, at which the graph
update resulted in a crash. The core overlap for each crash is also
given.
\item \texttt{Graphs/n{*}.gw}: The graphs of Figure \ref{cap:snapshots}
in LEDA format.
\item \texttt{Graphs/n{*}.mtx}: The adjacency matrices of the graphs shown
in Figure \ref{cap:snapshots}.\newpage

\end{itemize}
\addcontentsline{toc}{chapter}{\numberline{}Bibliography}

\newpage\vspace*{2.5cm}
\textsf{\textbf{\huge Acknowledgments}}\thispagestyle{plain}

\markboth{Acknowledgments}{Acknowledgments}
\vspace*{1.5cm}

\addcontentsline{toc}{chapter}{\numberline{}Acknowledgements}

I would like to thank Sanjay Jain and Chandan Dasgupta for being my
PhD advisors, and Diptiman Sen for being my acting advisor for a year.
I thank Sanjay Jain, in collaboration with whom this work was done,
for suggesting the models discussed in this thesis. Chandan Dasgupta
has always been ready to provide advice when I wanted it and has not
seemed to mind if I didn't follow the advice. Sriram Shastry's constant
interest in the details of my work and the opportunities he has provided
me outweigh a rather heavy suitcase. I thank V. Vinay, V. Nanjundiah
and H. R. Krishnamurty for serving on my comprehensive examination
committee.

Ramesh Hariharan and Vivek Borkar provided proofs for two propositions
used in the analysis of the model. Thanks are due to William Cheswick,
Dennis Bray and Hawoong Jeong who kindly gave me permission to use
their images in this thesis, and to the COSIN project that has made
data about a number of real networks publicly available. Many thanks
to the Lyx team for an excellent free software that made the process
of writing this thesis easier and faster.

I thank the Harishchandra Research Institute, NBI/NORDITA, Santa Fe
Institute, Bell Laboratories, Rutgers Univ., Rockefeller Univ. and
Princeton Univ. for their hospitality. 

G. Rangarajan, Kim Sneppen, Maya Paczuski, Andrei Ruckenstein, Boris
Shraiman, Anirvan Sengupta, members of the Bell Labs Biology Bandwagon,
Vikram Soni, Ravi Mehrotra, Eric Siggia and William Bialek provided
invaluable comments and complaints about this work, that have helped
shape my understanding of it.

I am obliged to the Council of Scientific and Industrial Research,
Govt. of India, for a research fellowship. Fortunately, circumstances
never resulted in my testing their condition that, being unmarried,
I would never marry again while my first wife was still alive -- the
bizarreness of which condition is only exceeded by their classification
of fellows into the three categories: unmarried, married and female.

Thanks to the students, staff and faculty (in particular Raghu, Ritesh,
Poulose and Aavishkar the blue-nosed porcupine) for maintaining a
breathable atmosphere at CTS. Special thanks are due to Janardhana
who made life in CTS vastly easier by efficiently and invisibly taking
care of numerous small tasks.

\newpage
Working with Dhruba, Randhir, Joby, Pinaki and others in the SCouncil
has been very interesting --- not least in illuminating the essentially
feudal structure of IISc, designed to maintain, as the bottom rung,
an inexpensive and \emph{grateful} labour force. In addition, during
my stint in the Scouncil, I benefited from four crash courses: Beautifying
IISc in 30 Days, Eradicating Child Labour from IISc in Six Easy Steps,
Practical Lessons for the Modern Witch-hunter, and How to Talk at
Important Meetings for Dummies.

If I survived coursework, it is both because of and despite my Integrated
PhD classmates. Their names, in descending order of weirdness:

Shubra > Prabudhha > Ayash > Dhruba > Pinaki > Nandan > Rangeet.

This list misses out Sundar, who is no less weird, but his weirdness
is orthogonal to the others. Thanks for (among many many other things
that I can't remember right now) teaching me how to cycle.

In these six years many friends have left for other places. I remember
them with fondness and miss their company: Aditi, Niruj, Arti, Rjoy,
Dandy, Sumana, Devyani, Sundar, Raju, Baliga, Poulose, Prithu, Raghavan,
Subroto and Rsidd. Fortunately, new friendships have also formed:
Osho, Ayesha, Moushumi and Natasha will be remembered for helping
me and others during CTS parties, impromptu chocolate pastries, agreeing
to be a murder victim and dropping a snake onto (well, close to) my
foot, respectively.

My memories of IISc will also include: heavy metal education at Styx,
courtesy Niruj; Dandy taking us to the edge of urbanity; Vatsa's views
on mixing quantum field theory with alchohol; Raju and Negi frightening
me out of my wits by mistaking a cow for an elephant; and the pool
incident, which deserved a quieter night.

Much of my time recently has been spent in the welcome company of
Bhavtosh, Moyna, Raghu, Joby, Toby and Vivek. Thanks to Vivek for
catalyzing puns and other wordsmithings. Perhaps one day we'll finish
the movie, the play and some junk we started. Toby and I share a common
love for Calvin and Hobbes. From Chembra to World Cups to Rallyk and
`Walk in the Woods', he's always been great company and great competition.
Joby has made a positive difference to the lives of many people and
he can count me among them. The small amount of time I have spent
at the tmsc classes and the construction colony with him have been
more educative than all the coursework I did in IISc. I will miss
Joby every time I see a bird that I can't identify, or hear a bird
that I can't see.

It's difficult to adequately describe the kind of crucial life support
those closest to me have provided. I hope they will extrapolate from
the few words I have managed to put down. My parents have, successfully,
tread the narrow path between pressurizing me with their opinions
and abandoning me to my own decisions. Bhavtosh, Moyna and Raghu:
If at all we drift apart, I hope it will only be geographically.

\end{document}